\documentclass[twoside,12pt]{article} \usepackage{epsfig}

\def \with respect to { with respect to }

\def\L {\Lambda }
\def\l {\lambda } 
 
\def \t {\theta }

\def\a {\alpha }
\def\dh {\partial }
\def \d {\delta }
\def \D {\Delta }

\def \g {\gamma }
\def \G {\Gamma }
\def \O {\Omega }
\def \o {\omega }
\def \b {\beta }
\def \S {\Sigma }
\def \s {\sigma }
\def \e {\epsilon }

\def \ud { {1 \over 2} }

\def \cddd { {\cal D } }
\def \cala { {\cal A } }

\def \calp { {\cal P } }

\def \caln { {\cal N } }

\def \calu { {\cal U } }
\def \calv { {\cal V } }
\def \tchi  {{\tilde \chi }}
\def \Eslash {E \kern-.5em\slash}
\def \pslash {p \kern-.5em\slash}
\def \kslash {k \kern-.5em\slash}
\def \Dslash {D \kern-.5em\slash}
\def \hslash {h \kern-.5em\slash}
\def \dslash {\partial \kern-.5em\slash}
\def \vslash {v \kern-.5em\slash}

\def \text {\mbox}

\newcommand{\be}{\begin{equation}} 
\newcommand{\ee}{\end{equation}} 
\newcommand{\ba}{\begin{array}}
\newcommand{\ea}{\end{array}}
\newcommand{\bea}{\begin{eqnarray}} 
\newcommand{\eea}{\end{eqnarray}} 
\newcommand{\bsea}{\begin{subeqnarray}} 
\newcommand{\esea}{\end{subeqnarray}}

\def \np { Nucl. Phys. }

\begin{document}

\title{PHENOMENOLOGICAL CONSTRAINTS  ON BROKEN R PARITY SYMMETRY IN 
SUPERSYMMETRY MODELS
\thanks {\it Supported by the 
Laboratoire de la Direction des Sciences de la Mati\`ere du
Commissariat \`a l'Energie Atomique } }
\author{Marc  Chemtob \\
CEA/Saclay, Service de Physique Th\'eorique 
\\ F-91191 Gif-sur-Yvette  Cedex  FRANCE }
\date{\today}
\maketitle

\begin{abstract}
The R parity odd renormalizable Yukawa interactions of quarks and
leptons with the scalar superpartners have the ability to violate the
baryon and lepton numbers, change the hadron and lepton flavors and
make the lightest supersymmetric particle unstable.  The existence of
an approximate R parity symmetry would thus affect in a deep way the
conventional framework of the Minimal Supersymmetric Standard Model
where an exact R parity symmetry is built-in by assumption.  The
purpose of the present review is to survey in a systematic way the
direct experimental constraints set on the R parity violating
couplings by the low and intermediate energy physics processes.  We
consider first the option of bilinear R parity violation and
spontaneously broken R parity symmetry and proceed next to the
trilinear R parity violating interactions. The discussion aims at
surveying the indirect coupling constant bounds derived from
fundamental tests of the Standard Model and the variety of scattering
and rare decay processes.  We also discuss the constraints imposed by
the renormalization group scale evolution and the cosmological and
astrophysical phenomenology.
\end{abstract}


\section{ Introduction}
\label{sec:intro}
The multiplicative $Z_2$ symmetry, known as R parity, fulfills a
central function in supersymmetry physics. Without R parity symmetry,
the Minimal Supersymmetric Standard Model (MSSM) would include
bilinear and trilinear renormalizable superpotential terms, coupling
the quarks and leptons to their scalar superpartners, which have the
ability to initiate fast nucleon decay, large neutrinos masses and the
LSP (lightest supersymmetric particle) disintegration into ordinary
particles.  Thus apart from threatening the nucleon and neutrino
stability and contributing to the hadron and lepton flavor changing
processes and to neutrino masses and flavor mixing, the R parity odd
Yukawa interactions would also disallow any of the supersymmetric
cosmic relic particles to contribute to the Universe dark matter
component.

The R parity symmetry was first introduced in a 1978 work by Farrar
and Fayet~\cite{farrarfayet}, as part of attempts towards building a
realistic particle physics phenomenology of
supersymmetry~\cite{fayetsusy,fayet78,fayet79}.  This step followed
closely in time the major developments in years 1974-1975 which
culminated in the construction of supersymmetric field
theories~\cite{wesszum74,ferrara74,salam75}, the implementation of
spontaneous supersymmetry breaking schemes~\cite{fi74,raif75} and the
discovery of R symmetries~\cite{salam}.  The collection of reprints by
Ferrara~\cite{ferrara} offers a valuable grasp on the progress of
ideas from the early period until the middle 1980's.  The subsequent
developments are discussed in the review
articles~\cite{niuwen81,sohn85} and the collection of preprints by
Salam and Sezgin~\cite{sezgin}.  As an historical aside on the origin
of R symmetries, we note that Wess and Zumino~\cite{wesszum74}
introduced an R quantum number as a weighting index labeling distinct
representations of the supersymmetry algebra. While an early
consideration of R symmetries figures in works by Fayet and
O'Raifeartaigh~\cite{raif75}, the first explicit discussion,
identifying R symmetry with a generalized version of fermion number
conservation, appears in the work by Salam and Strathdee~\cite{salam}.

The conventional definition of the R parity quantum number, $R=
(-1)^{3B+L+2s} = (-1)^{3(B-L)+2s} $, combines the baryon, lepton and
spin $(B,\ L ,\ s)$ of the MSSM particles in such a way that quarks
and leptons have opposite parity to Higgs bosons, $R(Q,
U^c,D^c,L,E^c)=-1,\ R(H_u,H_d)=1$, with fermionic and bosonic
superpartners having opposite parities.  The spin independent $ Z_2$
symmetry, $(-1)^{3B+L} $, known as matter (or quark-lepton) parity, is
operationally equivalent to R parity, since the spin dependent phase
factor, $ (-1)^{2s} $, corresponding to a $2\pi $ spatial rotation,
always equals unity upon acting on the Lagrangian terms involving
fermion fields in pairs.  Less restrictive symmetries have also been
considered, corresponding to the lepton or baryon R parities,
$(-1)^{L+2s} $ or $ (-1)^{3B+ 2s} $.

Without R parity symmetry, the superpotential would include
renormalizable bilinear and trilinear Lagrangian terms inducing
higgsino-lepton field mixings and Yukawa couplings between the
ordinary quarks and leptons matter particles and the squarks and
sleptons superpartner particles. The R parity odd interactions can
violate the baryon and lepton numbers as well as couple the different
generations or flavors of quarks and leptons.  For comprehensiveness,
we recall at this point the general formulas for the R parity odd
renormalizable bilinear and trilinear superpotential and for the
corresponding Yukawa couplings of fermions with sfermions, \bea &&
W_{RPV}=\sum_i \mu_i L_i H_u + \sum_{i,j,k} \bigg (\ud \l _{ijk}
L_iL_j E^c_k+ \l ' _{ijk} L_i Q_j D^c_k+ \ud \l '' _{ijk}
U_i^cD_j^cD_k^c \bigg ) ,\cr && L_{RPV} = \sum _i \mu _i (\bar \nu
_{iR} \tilde H _{uL} ^{0c} - \bar e_{iR} \tilde H _{uL} ^{+c} ) +
\sum_{i,j,k} \bigg [ \ud \l_{ijk}[\tilde \nu_{iL}\bar e_{kR}e_{jL} +
\tilde e_{jL}\bar e_{kR}\nu_{iL} + \tilde e^\star _{kR}\bar \nu^c_{iR}
e_{jL} -(i\to j) ] \cr &+&\l '_{ijk}[\tilde \nu_{iL}\bar d_{kR}d_{jL}
+ \tilde d_{jL}\bar d_{kR}\nu_{iL} + \tilde d^\star _{kR}\bar
\nu^c_{iR} d_{jL} -\tilde e_{iL}\bar d_{kR}u_{jL} - \tilde u_{jL}\bar
d_{kR}e_{iL} - \tilde d^\star _{kR}\bar e^c_{iR} u_{jL} ] \cr &+& \ud
\l ''_{ijk}\e_{\a \b \g }[\tilde u^\star _{i\a R}\bar d_{j\b
R}d^c_{k\g L} + \tilde d^\star _{j\b R}\bar u_{i\a R}d^c_{k\g L} +
\tilde d^\star _{k\g R}\bar u_{i\a R}d^c_{j\b L} ] \bigg ]+ \ H. \
c. , \label{eqi1} \eea with the indices $i,j,k = (1,2,3) $ labeling
the quarks and leptons generations of the quarks and leptons chiral
superfields, $Q_i, \ U^c_i,\ D^c_i,\ L_i,\ E^c_i$.  The R parity
violating (RPV) superpotential introduces a total of $48$ coupling
constants consisting of $3$ dimensional coupling constants $\mu_i $,
describing lepton number violation mass mixing terms between the up
Higgs boson and leptons superfields, two sets of $9 $ and $27 $
dimensionless coupling constants, $\l_{ijk} = -\l_{jik}$ and $ \l
'_{ijk} $, responsible for lepton number violation and a set of $9 $
coupling constants, $ \l ''_{ijk} =- \l ''_{ikj}$, responsible for
baryon number violation.  (We shall use occasionally the notation
$\hat \l_{ijk} $ to designate a generic trilinear coupling constant.)
Once R parity symmetry is broken, one naturally expects that the
sector responsible for supersymmetry breaking would induce R parity
odd soft supersymmetry breaking interactions in addition to the
regular interactions represented by the gauginos Majorana mass terms,
the sfermions hermitian mass terms and the bilinear and trilinear
sfermions self couplings.  The general bilinear and trilinear RPV soft
supersymmetry breaking operators of dimension $\cddd \le 4$ are
expressed by the effective scalar potential,
\begin{eqnarray} &&  
V _{RPV} = \sum _i [\mu_{ui}^2 \tilde L_i H_u + \tilde m ^2_{di} H_d
 ^\dagger \tilde L_i ] \cr && + \sum _{i,j,k} m_{\tilde G } [\ud A^\l
 _{ijk} \l _{ijk} \tilde L_i \tilde L_j \tilde E^c_k + A^{ \l '}
 _{ijk} \l ' _{ijk} \tilde L_i \tilde Q_j \tilde D^c_k +\ud A^{ \l ''}
 _{ijk} \l '' _{ijk} \tilde U^c_i \tilde D^c_j \tilde D^c_k ] +H. \
 c. \label{eqx16p} \end{eqnarray} The parameters associated with the
 holomorphic type operators coupling the $H_u$ Higgs boson with
 sleptons are sometimes denoted as, $ \mu_{ui}^2 = B_i m_{\tilde G }
 \mu_i $, although the main contributions to these operators may arise
 from the heavy mass threshold effects rather than from the familiar
 supergravity mediated tree level effects.  Borrowing from the
 conventional notations used in the supergravity approach to
 supersymmetry breaking, we have factored out the gravitino mass
 parameter, $m_{\tilde G }\equiv m_{3/2} $, and introduced the free
 dimensionless parameters, $ B_i, \ A^{\l ,\l ' ,\l''} _{ijk}$, which
 are of natural order unity. However, the above parameterization is
 clearly not restricted to just the supergravity approach.  In the
 case of a generation universal supersymmetry breaking, the
 parameters, $ B_i, \ A^{\l ,\l ' ,\l''} _{ijk}$, are taken to be
 independent of the generation indices.  For definiteness, we
 summarize our notational conventions by quoting the formulas for the
 regular R parity conserving (RPC) superpotential and the soft
 supersymmetry breaking Lagrangian terms in the MSSM extended by the
 addition of a right chirality neutrino superfield, $ W_{RPC} = \mu
 H_d H_u + \sum _{i,j}\l ^u _{ij} Q_i H_u U_j^c +\l ^d _{ij} Q_i H_d
 D_j^c + \l ^e _{ij} L_i H_d E_j^c + \l ^\nu _{ij} L_i H_u \nu_j^c ,
 \quad V^{RPC} _{soft} = \ud M_a \l _a \l _a + \mu _{ud} ^2 H_d H_u +
 m_{3/2} [ A^u _{ij} \l ^u _{ij} \tilde Q_i H_u \tilde D^c_j+ A_{ij}
 ^d \l ^d_{ij} \tilde Q_i H_d \tilde D^c_j+ A ^e_{ij} \l ^e_{ij}
 \tilde L_iH_d \tilde E^c_j + A ^\nu _{ij} \l ^\nu _{ij} \tilde L_i
 H_u \tilde \nu ^c_j ] + \sum _{\tilde f } m^{2}_ { \tilde f_i \tilde
 f_j} \tilde f _i ^\dagger \tilde f _j + H.\ c. , $ with $ \tilde f =
 [\tilde Q, \tilde Q^c, \tilde L, \tilde E^c, H_u, H_d] $, and noting
 that the Standard Model (SM) gauge interactions coupling constants
 are denoted by $ g_1, g_2, g_3$, and the neutral Higgs bosons VEVs
 ratio by $\tan \b = <H_u>/<H_d> = v_u /v_d$.

\subsection{Status of R parity symmetry} 

In a single strike, R parity symmetry protects the MSSM against
renormalizable $ B $ and $ L $ number violating interactions while
forbidding the LSP to decay.  The remarkable effectiveness of R parity
symmetry contrasts with its indefinite theoretical status. Is R parity
a low energy remnant of an extended symmetry which has undergone
spontaneous breaking at some higher mass scale?  Is it realized in the
conventional way or rather as a higher order cyclic symmetry or
possibly as a non-abelian discrete symmetry?  Is the underlying
symmetry of global or local kind, or of ordinary or R type? General
considerations from quantum gravity suggest that only gauge symmetries
deserve a fundamental status while global symmetries are only
acceptable as accidental symmetries of the non-renormalizable part of
the effective Lagrangian. This observation makes it clear that the
physical implications of an ordinary or extended R parity symmetry
would sensitively depend on whether it arises via gauge, flavor or
compactified string dynamics.

The recent interest in exploring the consequences of a broken R parity
symmetry is motivated to a large extent by the increased favor
acquired by supersymmetry over compositeness as a viable option for
the new physics beyond the SM.  The farthest reaching consequence
clearly stems from the presence of the renormalizable level $B$ and $
L$ number violation.  The renormalizable and non-renormalizable
operators of dimensions $\cddd \le 4$ and $\cddd \ge 5$, have quite
different status in the effective Lagrangian, since the latter are
explicitly suppressed by powers of the underlying theory mass scale,
$\L $, while the former may in principle arise with $O(1)$ coupling
constants.  The classification of gauge invariant dangerous operators
in the Standard Model~\cite{weinbergs,wilczee,weinberg80,langacker}
reveals that lepton number can be violated starting from dimension $5
$, baryon number from dimension $6$, CP symmetry from dimension $ 6$,
and quarks and leptons flavor from dimension $6$.  According to the 't
Hooft naturalness criterion~\cite{thooft}, a local operator of
dimension $\cddd $ in the effective action can arise with a reduced
coupling constant of natural magnitude well below, $ O(1)/ \L ^{\cddd
-4} $, only to the extent that the theory acquires an enhanced
symmetry when this is set to zero.

The phenomenology of baryon and lepton number violation is much of a
puzzle.  Despite the lack of evidence of nucleon decay from laboratory
searches, the observation of matter-antimatter asymmetry in the
Universe leads inescapably to the conclusion that baryon and/or lepton
numbers cannot be absolutely conserved.  While the cosmological baryon
asymmetry could possibly be generated by non-perturbative anomalous
thermal processes at the electroweak symmetry phase transition, the
condition that the transition be strongly first order is not naturally
satisfied in the SM and is marginally satisfied in the MSSM.
Regarding the issue of lepton number violation, here too the
observation of the neutrinoless double beta decay reaction is still
lacking and the experimental information on neutrino flavor
oscillations have not conclusively established the presence of a
lepton number violating Majorana mass component.  However, the
possibility that a primordial leptogenesis, requiring $L$ number
non-conserving interactions, is responsible for the cosmological
baryon asymmetry stands as a viable attractive option.  The transfer
of the lepton asymmetry into baryon asymmetry would then be achieved
by the well controlled effects associated with the $B+L$ number
violating fast thermal reactions in the cosmic bath induced by the
anomaly related electroweak sphaleron solution.

The broken R parity symmetry option might offer a promising
alternative to the currently favored mechanism of baryogenesis or
leptogenesis involving the out-of-equilibrium decay of ultra-massive
relic particles, which requires new physics at intermediate or
unification mass scales.  The relation of RPV interactions with
baryogenesis is two-sided.  There is a passive side associated with
the non-dilution constraints of the $\D B $ and $ \D L$ cosmic
asymmetries imposed on the RPV interactions upon requiring that the
induced thermal reactions remain out-of-equilibrium.  There is also an
active side stemming from the availability of attractive production
mechanisms of the $\D B $ or $ \D L$ cosmic asymmetries at the
supersymmetry breaking mass scale.

\subsection{Bilinear and trilinear options of R parity violation}

The MSSM without R parity symmetry specifically designates the option
where the R parity violating interactions are considered on their own,
exclusive of additional degrees of freedom from new physics.  It is
convenient to group together the cases of spontaneous R parity
symmetry breaking and explicit symmetry breaking by bilinear
interactions only.  The distinction with the case involving the
trilinear or the combined trilinear and bilinear RPV interactions
defines a natural dividing line in the discussion of phenomenological
applications.

The separate study of bilinear R parity violation is motivated by the
fact that this constitutes a predictive and strongly constrained
option on its own.  Only lepton number violation is then of concern.
An independent discussion of the bilinear interactions is also useful
on methodological grounds for exposing the interplay between the
spontaneous and explicit breaking cases.  The spontaneous breaking is
characterized by an unbroken symmetry at the level of the effective
action, accompanied by non-vanishing VEVs for R parity odd fields,
such as the sneutrinos, for example, $<\tilde \nu _i > \ne 0 $, which
entail the spontaneous breakdown of the lepton number symmetry, $U(1)
_L$.  By contrast, the explicit breaking case is signaled by the
presence of R parity odd interaction terms in the action, accompanied
or not by finite sneutrino fields VEVs.  The bilinear interaction
superpotential between the leptons and Higgs bosons, $ W= ( \mu _iL_i
+ \mu H_d) H_u $, has the ability to initiate at the tree level
additional $Z$-boson decay modes along with contributions to the
neutrinos Majorana mass matrix.  The experimental constraints are
expressed as bounds on the dimensional bilinear coupling constants
which may also depend on the sneutrinos and down type Higgs boson
VEVs.  The spontaneous R parity violation case is distinguished by the
presence of the massless Nambu-Goldstone majoron boson, which is often
accompanied by a light scalar particle.  Strong experimental
constraints are set on these scalar particles coupling parameters by
the $Z$-boson invisible width and the Compton like scattering process
controlling the stellar cooling rates.  These can be evaded, however,
by allowing for a very small explicit symmetry breaking.
The proposals range from the MSSM subject to a constrained parameter
space,
to extended models involving a single electroweak singlet superfield,
a right chirality neutrino superfield or a combination of both
superfield types.

The trilinear interactions can initiate a variety of processes
involving single, pair, or more factors of the RPV coupling
constants. Including the bilinear couplings on top of the trilinear
ones appears more natural from the point of view of the
renormalization group than the reverse.  It must be realized, however,
that considering the bilinear and trilinear interactions
simultaneously may run into a redundancy, since one can absorb the
former inside the latter by means of a suitable field transformation.
The removal of bilinear interactions depends, however, on the
renormalization scale at which the fields redefinition is performed,
since the bilinear interactions are radiatively induced from the
trilinear ones through the renormalization.  Another obstruction to
the cancellation of bilinear RPV interactions arises in the presence
of generation non-universal soft supersymmetry breaking.

\subsection{Indirect bounds on RPV coupling constants}

The indirect bounds on the RPV coupling constants are inferred from
the experimental constraints associated with the low and intermediate
energy particle, astroparticle, nuclear and atomic physics. These
bounds lie at a crucial interface between theory and phenomenology.
Building up on the pioneering
works~\cite{farrarfayet,fayetsusy,weinberg83,weinbergfarrar,aulak,zwirner,hallsuz,lee,ellisvalle,rossvalle,dawson,moha,masiero,halldimo,dimohall,bargerg,dreiner1,reviewsm},
an important activity has developed in recent
years~\cite{reviews,reviews2,review99,royrev} on the issue of a broken
R parity symmetry.  Indirect manifestations of the R parity odd
interactions can take place at energies below the threshold for
production of single supersymmetric particles.  The processes involve
the quarks and/or leptons, from the lowest energies up to the highest
reachable ones.  The mechanisms of interest arise from tree or loop
level Feynman diagrams in which the superpartners of ordinary
particles propagate as internal off-shell particle lines.  The
perturbation theory contributions to the transition probability
amplitudes of loop order $ l$ are controlled by combinations of
coupling constants and superpartners masses of form, $ {\hat \l ^2
\over \tilde m^q } ({\hat e^2 \over (4\pi )^2 } )^l $, where the
symbols $\hat \l , \ \hat e , \ \tilde m $ denote generic coupling
constants for the RPV and gauge interactions and the superpartners
mass parameters. Note that the power index, $q $, is necessarily
positive, as suits the decoupling of supersymmetric degrees of freedom
in the ultraviolet.

The major fraction of indirect bounds is obtained by making use of the
so-called single coupling dominance hypothesis, where a single
coupling constant is assumed to dominate over all the others.  One can
rephrase this useful working hypothesis~\cite{dimohall,bargerg} by
saying that each of the RPV coupling constants is then assumed to
contribute one at a time.  Apart from a few isolated cases, the bounds
derived under the single coupling constant dominance hypothesis are of
typical orders of magnitude, $[ \l ,\ \l' ,\ \l '' ] < [O(10^{-1})\ -
\ O(10^{-2}) ] \times ( { \tilde m\over 100 \text{GeV} }) $, with the
linear dependence on the superpartner mass strictly holding in tree
level type mechanisms. The quoted correlation between the RPV coupling
constants and superpartners masses shows clearly that these
constraints would significantly relax if the supersymmetry breaking
mass scale were pushed out to the TeV range.  The observables
associated with the charged and neutral currents, neutrinoless double
beta decay and $ n-\bar n $ oscillation all yield individual coupling
constant bounds.  The largest number of constraints on the RPV
interactions are derived, however, by making use of the extended
hypotheses invoking the dominance of two or more coupling constants,
yielding coupling constant bounds of quadratic or quartic orders. The
quadratic coupling constant bounds are generally more stringent than
those derived by combining the corresponding pair of coupling constant
bounds obtained under the single coupling constant dominance
hypothesis.  The processes controlled by distinct pairs of RPV
coupling constants fall in the following four classes:

$\bullet \quad $ Hadron flavor changing effects, as observed through
the mixing parameters of the neutral light and heavy flavored mesons
and in the leptonic or semileptonic decays of $ K $ or $ B$ mesons, $
K ^0\to e_i + \bar e_j , \ K^+ \to \pi ^+ +\nu +\bar \nu $.

$\bullet \quad $ Lepton flavor changing effects, as observed in
muonium-antimuonium oscillation, $ Mu \to \overline{Mu} $, lepton
conversion nuclear processes, $ \mu ^- +N \to e^- +N $, or radiative
decays of charged leptons.

$\bullet \quad $ Lepton number violating effects, as observed in
neutrinoless double beta decay, neutrino Majorana masses and mixing,
or leptons three-body decays, $e_l^\pm \to e _m ^\pm+e _n^-+e ^+_p $.

$\bullet \quad $ Baryon number violating effects, as observed in
proton partial decay lifetimes, rare decays of heavy mesons or two
nucleon nuclei disintegration.

\subsection{Limitations on phenomenological studies} 
Any realistic discussion of the experimental constraints must be
guided by the use of some motivated simplifications.  An important
working assumption lies in the choice of basis for matter fields.
Although physical observables are not affected by field redefinitions,
it is still necessary to express the coupling constant bounds in some
definite field basis.  The problem, however, is that the single or
quadratic coupling constant dominance hypotheses are well defined only
once one specifies the matter fields bases.  Without R parity symmetry
another ambiguity also arises in assigning the lepton quantum number,
due to the impossibility to distinguish between the down type Higgs
boson and the leptons.

The current (gauge or flavor) and mass eigenstates fields for quarks
and leptons constitute two naturally distinguished field bases.  The
mass basis is naturally singled out for a comparison with experimental
results, while the gauge basis reflects in a more transparent way the
physics of the underlying theory.  Most applications make tacit use of
the RPV superpotential in the mass eigenstate basis.  On the other
hand, applying the coupling constant dominance hypotheses to the gauge
basis representation is more natural, since the presumed hierarchy in
coupling constants is likely to originate from high mass scales
physics.  An added interest of the gauge basis is in giving access to
derived bounds for the mass basis coupling constants related to the
dominant current basis coupling constant by the flavor transformation
matrices~\cite{ellis98}.  Upon introducing the dependence on the
Cabibbo-Kobayashi-Maskawa (CKM) flavor mixing matrix, one can derive
useful representations of the RPV interactions in which the hadron
flavor changing contributions arise in either the up-quark or the
down-quark sectors, even assuming the single dominant coupling
constant hypothesis.  The sfermion fields basis dependence also must
be taken into account, since the superfield bases for the squarks and
sleptons, obtained from the gauge basis fields by the same unitary
matrices, $ V_{L,R}^{u}, \ V_{L,R}^{d},\ V_{L,R}^{e} ,\ V_{L}^{\nu }
$, transforming the quarks and leptons current fields into mass
eigenstate fields, need not coincide with the mass bases of squarks
and sleptons.  The field basis independent description of RPV effects
is to be preferred in applications dealing with finite temperatures
processes where the renormalization corrections introduce temperature
dependent mass parameters for the particles.  The need of a consistent
procedure in dealing with thermal processes has motivated the
construction by Davidson and Ellis~\cite{davidson0,davidson,davidson1}
of basis invariant algebraic products of the RPV coupling constants.
The basis independent formalism has also been used in studies of the
bilinear interactions contributions to the neutrino
masses~\cite{banksnir} and the neutral scalar particles sector of
Higgs bosons and sneutrinos~\cite{davus00,groer01,grosshab03} and in
studies aimed at establishing the basis dependence of constraints
involving the sneutrino fields VEVs~\cite{ferrandis}.

The significance of constraints on new physics bears heavily on two
related issues: The sensitivity of experimental data and the
uncertainties on the SM predictions. To make a good use of the
experimental information on hadronic processes, it is necessary to
disentangle the long and short distance physics contributions in
transition amplitudes.  The latter include the perturbative physics
information of interest to particle physics, while the former include
still poorly understood non-perturbative hadronic physics associated
with coupled hadronic channels, final state rescattering effects and
absorptive parts reflecting the S-matrix unitarity. These are not,
however, the only source of limitations.  One must also watch out that
the inaccuracies on the SM parameters do not cause large theoretical
uncertainties on predictions which would invalidate a comparison
between theory and experiment.  The best of all possible cases is
clearly that in which the RPV contributions lie well above both the
theoretical uncertainties and the experimental errors.  For a given
process, however, the situation can evolve with time. This point is
well illustrated by the example of the $K$ meson decay reaction, $ K
^+ \to \pi ^+ +\nu +\bar \nu $, where the current experimental
sensitivity and theoretical uncertainties
are both standing at the same $O(10^{-10})$ level, thus momentarily
discouraging attempts to use this reaction in testing for new physics
effects.

\subsection{Scope  and general plan of review}

A wide spectrum of applications is opened up by the physics R parity
symmetry violation.  Our purpose in the present work is to present a
comprehensive review of the indirect bounds on the RPV coupling
constants which hopefully complements similar recently published
works~\cite{reviews,reviews2,review99,royrev}.  The main objective is
in identifying the physically interesting cases where progress is
possible and needed.  The contents are organized into six sections.
We begin the discussion with the bilinear R parity violation option,
including the case of spontaneous R parity symmetry breakdown.  We
proceed next with a discussion of the constraints on the trilinear
interactions coupling constants, derived from the variety of processes
exhibiting baryon and lepton number violation and hadron and lepton
flavor changes. We finally discuss the constraints inferred from the
renormalization group and from the astroparticle physics
phenomenology.  Due to lack of space, we have left aside the
discussion of high energy colliders tests.  The above thematic
discussion is complemented by a catalog of the main existing bounds
and the summary of our main conclusions.

We do not address in this review the theoretical implications of R
parity symmetry. This subject constitutes an essential part of the
naturalness problem in supersymmetry model building.  As is known,
worries about $B, \ L$ non-conservation in supersymmetric models do
not end with the renormalizable R parity odd operators.  One must
still care~\cite{weinberg82,sakai,dimorab82,ellisrud82} about the
gauge invariant R parity even F term operators of dimension $\cddd= 5
$.  Concerning the threat from higher dimension dangerous operators,
$\cddd \ge 6$, this may be averted or exploited, depending on one's
point of view, by reasoning along similar lines as in the
non-supersymmetric case~\cite{weinbergs,wilczee,langacker}.  The two
familiar extreme cures involve postulating generalized parity
symmetries~\cite{ibross,hinchliffe}, preferably associated with local
discrete symmetry groups~\cite{wilczek}, or a grand desert between the
SM and the gauge or string theory unification physics.  At this point
we should emphasize that the distinction between an approximate R
parity symmetry and new physics is an artificial one. In fact, the
renormalizable character of the R parity odd interactions may only be
apparent, reflecting non-renormalizable baryon and lepton number
non-conserving interactions introduced through some high energy
extension of the MSSM.

The contents of this review may be summarized as follows.  In
Section~\ref{secxxx1}, we focus on the options of spontaneous and
explicit breakdown of R parity symmetry by the bilinear interactions.
In Sections~\ref{secxxx2} and \ref{secxxx3}, we proceed to the
trilinear R parity interactions.  In Section~\ref{secxxx2}, we examine
the constraints from fundamental tests of the SM associated with high
precision observables, CP violation, neutrino masses and neutrino
instability. In Section~\ref{secxxx3}, we examine the constraints
inferred from the variety of low and intermediate energy scattering
and decay processes. We organize the discussion of the current
literature according to four distinct themes associated with the
violation of hadron and lepton flavors, respectively and the violation
of baryonic and leptonic numbers, respectively. In
Section~\ref{secin2}, we discuss the constraints resulting from the
renormalization group flow equations for the RPV Yukawa and soft
supersymmetry breaking coupling parameters. In Section~\ref{secxxx6},
we discuss the implications from cosmology and astrophysics on an
unstable lightest supersymmetric particle (LSP) and the baryon
asymmetry in the Universe.
Having in hand a sufficiently thorough compilation of the bounds, we
turn in Section~\ref{secin4} to a general discussion aimed at
selecting some of the strongest, most robust bounds for each of the
trilinear interactions coupling constants, $\l , \ \l ' $ and $ \l
''$.


In the presentation of bounds for the RPV coupling constants, we need
at times to distinguish between the first two quarks and leptons
families and the third.  When quoting numerical coupling constant
bounds, we assume the following conventions for the alphabetical
indices, $ l, m, n \in [1,2] $ and $ i, j , k \in [1,2,3]$.  Unless
otherwise stated, the mass of superpartners uniformly are set at the
reference value, $\tilde m = 100 \ \text{GeV} $.  To ease the writing
of numerical relations or bounds, we use abbreviated formulas of the
form, $ \l '_{ijk} < n\times \tilde d_{kR}^p $, to mean, $ \l '_{ijk}
< n\times ({ m_{\tilde d_{kR} }\over 100 \text{GeV} })^p $, with
similar conventions applying for other particle species.

We rely on the year 1998 Review of Particle Physics of the Particle
Data Group~\cite{pdg} as our source of information on experimental
data as well as for short reviews on the main particle physics
subjects. (The compilations of year 2000~\cite{pdg00} and beyond have
been used very scarcely.)  We adhere to the review of Haber and
Kane~\cite{haber} for the notations and conventions. We recommend some
familiarity with the textbook by Ross~\cite{rosstext} and the review
by Nilles~\cite{nilles84} on general particle physics theory and the
textbooks by Mohapatra and Pal~\cite{palmoha} and Barger and
Phillips~\cite{bargerphil} on particle physics phenomenology.
Throughout the text, we use the abbreviation APV for `atomic physics
parity violation', MSSM for `Minimal Supersymmetric Standard Model',
PQ for `Peccei-Quinn', RPV for `R parity violating', SM for `Standard
Model', and VEV for `vacuum expectation value'.


\section {Spontaneous breaking of R parity symmetry  and bilinear R parity
symmetry violation}
\label{secxxx1}

In the present section we focus on the bilinear and spontaneously
broken options of R parity symmetry.  The spontaneous breaking
scenario is characterized by the presence of a massless
Nambu-Goldstone particle, the majoron, which acquires a finite mass
only once includes explicit symmetry breaking terms in the Lagrangian.
The discussion of non-supersymmetric majoron models is given in
Refs.~\cite{chikashige,georgi,gelmini} and that of the supersymmetric
models in Refs.~\cite{aulak,rossvalle,ellisvalle,masierovalle}.  In
the first subsection below, we begin with a review of the situation in
the MSSM case and proceed next to the extended models implementing the
spontaneously breaking of R parity symmetry. In the second subsection,
we review the explicit symmetry breaking option including the RPV
bilinear interactions.  Several other phenomenological implications
are surveyed in the third subsection.

Before proceeding to the main discussion, we emphasize that the
bilinear and trilinear interactions are not mutually independent,
since these can be reshuffled by means of unitary transformations of
the $ H_d , \ L_i$ fields. Specifically, the infinitesimal superfield
transformations, $ H_d \to H_d -{\mu _i \over \mu } L_i ,\ L_i \to L_i
+ {\mu _i \over \mu } H_d $, remove away the RPV bilinear
superpotential in favor of the trilinear, at the price of inducing the
following $O ( \mu _i /\mu )$ shifts on the various parameters: $ \mu
_i \to 0, \ \l_{ijk} \to \l_{ijk} -2\l^e_{ik} {\mu_j\over \mu } , \ \l
'_{ijk}= \l '_{ijk} -\l^d_{ik} {\mu_j\over \mu } ;\ \mu_{ud } ^2 \to
\mu_{ud } ^2 +\mu_{ui} ^2 {\mu_i\over \mu }, \ \mu_{ui} ^2 \to
\mu_{ui} ^2 -\mu_{ud } ^2 {\mu_i\over \mu } ,\ \quad \tilde m ^2
_{H_dL_i} \to \tilde m ^2_{H_dL_i} + \tilde m^{2}_{L_jL_i} {\mu_j\over
\mu } - \tilde m _{H_d} ^2 {\mu_i\over \mu } , \ \tilde m^{2}_{L_iL_j}
\to \tilde m^{2} _{L_iL_j} - \tilde m ^2 _{H_dL_j} {\mu_i\over \mu } ,
\ \tilde m^2_{H_d} \to \tilde m^2_{H_d} + \tilde m ^2_{H_dL_i}
{\mu_i\over \mu } .  $ These relations can be used, in particular, to
transcribe any of the bounds on $\mu _i$ into bounds on the trilinear
interactions coupling constants, and conversely, by using the known
information on the regular Yukawa coupling constants.
The processes controlled by the bilinear and trilinear interactions
acting in concert, involving the parameters products, $ \hat \l_{ijk}
v_l $ or $\hat \l_{ijk} \mu_l$, are discussed in the subsequent
sections.

\subsection{Spontaneous  breaking of R parity}
\label{secxxx1b}

\subsubsection{\bf Constraints on MSSM from spontaneous R parity 
symmetry breakdown}

The spontaneous violation of lepton number initiated by sneutrinos
VEVs, $ v_i = \sqrt 2 <\tilde \nu _i> \ne 0$, gives rise to a single
massive neutrino Majorana field, $ \nu (x)= (\sum_j v_j^2)^{-\ud }
\sum_{i=1}^3 v_i \nu_i (x)$.  The mass generation mechanism is
represented by the Feynman diagram A.1 in Figure~\ref{figun}.  In
order to retain a small neutrino mass of size compatible with the
cosmological constraint, $ m_ \nu \leq O( 100) $ \text{eV}, the
sneutrino VEV must exhibit a wide hierarchy with the Higgs bosons
VEVs, $ v_i = v_L = O(1) \ \text{MeV} << v = 245 $ GeV.  In the
simplified case of a single non vanishing sneutrino VEV, identifed for
more safety with that of the tau-sneutrino, one then infers the
bound~\cite{barbieri}, $ v_{ \tau } < 5 \ \text{GeV} \ ({ m_{\nu_\tau
} \over 35 \ \text{ MeV} })^\ud .$ The case involving three
generations of sneutrinos leads to an analogous constraint involving
the quadratic form~\cite{hallsuz,lee,dawson,barbieri}: $(\sum_i v_i^2
)^\ud < (12\ - \ 24 ) \ \text{GeV}. $

The MSSM effective scalar potential depends in an essential way on the
supersymmetry breaking and the top quark Yukawa interactions.  Within
the minimal supergravity framework~\cite{gato}, the requirement of
ensuring vanishing sneutrino VEVs imposes additional constraints to
those of vacuum stability and radiative breaking of electroweak
symmetry.  In particular, one must require that the renormalization
momentum scale, $Q$, at which the finite sneutrino VEV develops must
be larger than that at which the symmetric vacuum becomes unstable.
This constraint becomes significant for top quark masses in the
physically relevant range, $ m_t > 55 $ GeV.  The domain of
supergravity parameter space compatible with no spontaneous breaking
of R parity symmetry has been studied by Comelli et
al.,~\cite{comelli}.  Using the effective scalar potential up to
one-loop order, one can avoid finite sneutrino VEVs only by excluding
the sfermion and gaugino mass parameters, $\tilde m_0 $ and $ M_\ud $,
from the roughly triangle shaped domain bounded by the maximal
values~\cite{comelli}, $ (\tilde m_0)_{max} = 50 \ \text{GeV}, \
(M_\ud )_{max} = 220 \ \text{GeV} $.  However, only a tiny amount of
explicit R parity symmetry violation, such as included, say, through
the soft scalar potential term, $ V_{soft} = \mu_{u \tau } ^2 \tilde
L_\tau H_u , $ suffices to contribute to the majoron, $J $, and its
light scalar companion, $\s $, large enough masses, $ m^2 _{J }\simeq
m^2 _{\s } \simeq \mu_{u \tau } ^2 v_u /v_L \sim O(m_Z) $, enough to
evade the experimental constraint on the $Z$-boson invisible width. We
recall that, $\G (Z \to J +\s ) \simeq \G ^Z _{inv}/ (2 N_\nu ) \simeq
{498.3\over 6} \text{ MeV} \simeq 83 $ MeV, in the physically relevant
case of three light neutrino species, $ N_\nu =3$.

The most comfortable and easiest way to achieve a spontaneous breaking
of R parity symmetry is through a finite tau-sneutrino VEV, as noted
initially by Ross and Valle~\cite{rossvalle}.  The finite VEV, $v_\tau
\ne 0$, violates $L_\tau $ lepton number only but leaves the more
tightly constrained $L_e,\ L_\mu $ numbers conserved.  An attractive
model has been developed along these lines by Ellis et
al.,~\cite{ellisvalle} by assuming a very tiny $\mu $ parameter, as
needed to ensure a light enough neutrino and to suppress the stellar
majoron emission process.  Assuming $ v_\tau /v = O(1)$, the
constraint on the $\nu _\tau $ mass is expressed by the bound, $ \mu <
100 \ \text{ eV} $.  Because of their feeble coupling with matter
particles, the majorons produced in the interior of stars through the
Compton scattering like reaction, $ \g + e \to J +e$, easily escape to
the stars surface causing their excessive cooling.  Whether the
stellar majoron emission is energetically allowed or not depends on
the star core temperature, $ T $, which sets the average value of the
initial electron energies. Typically, $ T \simeq 1.5\ \text{ keV} $
for the Sun and $T \simeq 1\ \text{ MeV} $ for the red giant stars.
The experimental constraint from the stellar cooling rates by majoron
emission sets the VEV bound, $ v _{\tau } < O(100 ) \ \text{ keV} $.
A light axion is also potentially present in the scenario assumed by
Ellis et al.,~\cite{ellisvalle} due to the approximate accidental
Peccei-Quinn symmetry, $U(1)_{PQ}$, caused by the small $\mu $
parameter.  The resulting constraints can be relaxed only if the
majoron and axion acquire some finite mass, which requires an explicit
breaking of R parity symmetry.  Minimally extended models including a
heavy right handed neutrino superfield or a heavy pure singlet
superfield whose scalar components acquire finite VEVs are discussed
by Ellis et al.,~\cite{ellisvalle}.  Detailed discussions of the
implications on neutrino masses and $Z$ gauge boson rare decay modes
are presented in works by Santamaria and Valle~\cite{santa87}.

\subsubsection{\bf Models of spontaneous R parity  symmetry breakdown}  

We discuss now the realization of the R-parity symmetry spontaneous
breaking in the framework of extended supersymmetry models.  One of
the earliest proposal, due to Aulakh and Mohapatra~\cite{aulak},
introduces an electroweak gauge singlet chiral superfield, $\Phi $,
interacting with the Higgs bosons through the superpotential, $ W_\Phi
= \l \Phi (H_d H_u -M^2) $. The majoron in this model is the scalar
superpartner of the neutrino massless eigenstate.  This is also
accompanied by a light real scalar companion particle, $ \s (x) $.
The constraint from the cooling rates of red giant stars by majoron
production translates into the sneutrino VEV bound, $ v_L < 800 \
\text{ keV}$.  The finite sneutrino VEV which mixes the higgsinos with
leptons, $\tilde H_d ^- -e $, is determined by the dimensionless
parameter, $\e _\nu \equiv {(\sqrt 2 G_F ) ^ \ud v_L }$. The
deviations from $e-\mu $ charged current universality, based on
experimental data for nuclear beta decay and muon decay rates, lead to
the VEV bound: $ \e _\nu < 10^{-2} \ \Longrightarrow \ \sqrt 2 <\tilde
\nu _L> = v_L < 2.35 \ \text{GeV}$.  The induced RPV Yukawa coupling,
$ \l ^d \tilde u_L (d^c e)$, can initiate the neutrinoless double beta
decay process, with a transition amplitude described by a Feynman
diagram involving the $t$-channel exchange of neutralino and $\tilde
u_L$ squark, similar to diagram I.2 in Figure~\ref{fig3}. The
resulting contribution turns out, however, to be ineffective.

Ensuring a scale hierarchy between the VEVs of sneutrinos or
additional singlet scalar fields and those of the electroweak Higgs
bosons can be achieved only at the cost of a fine tuning between
different parameters in the scalar potential.  This naturalness
problem may, however, be alleviated by introducing extra right
chirality and pure singlet neutrino superfields.  The extended model
of Masiero and Valle~\cite{masierovalle} realizes satisfactorily this
goal by including three generations of right neutrinos and pure
singlet chiral superfields, $\nu^c_i $ and $S_i $, along with a pure
singlet chiral superfield, $\Phi $, coupled through the interaction
superpotential, $ W= \l ( H_d H_u-M^2) \Phi + \l^\nu _{ij} \nu^c_i L_j
H_u + h_{ij} \Phi \nu_i ^c S_j $.  The experimental constraint from
the stellar energy loss due to the Compton scattering majoron emission
reaction, $e+\g \to e+J$, sets the strong bound on the ratio of left
and right chirality sneutrino VEVs, $ v_L^2/(v_R m_W) < 10^{-7} , \
[v_L = \sqrt 2 <\tilde \nu > , \ v_R = \sqrt 2 <\tilde \nu ^c> ]
$. Assuming, for instance, $v_R = O ( 10 ^3 ) \ \ \text{GeV} $,
translates the above quoted condition into the sneutrino VEV bound, $
v_L < O ( 100) $ \text{MeV}.  The Yukawa coupling constant matrix, $
\l ^\nu _{ij} $, is instrumental in realizing in a natural way the
requisite hierarchy between the VEVs, $ v_L, \ v_R$, which can be
achieved by setting the value of a generic element of the
neutrino-lepton Yukawa coupling constant matrix in the range, $ \l
^\nu _{ij} = O(10^{-1}) \ - \ O(10^{-10}) $.

The model of Masiero and Valle possesses several other attractive
features~\cite{roma097}. The predictions include a mass matrix for
neutrinos combining Dirac and Majorana mass components;
unstable neutrinos decaying through the two main two-body channels, $
\nu_\tau \to 3\nu $ and $ \nu_\tau \to \nu + J$;
charginos and leptons two-body decay modes with majoron emission, $
\tchi ^\pm \to \tau ^\pm + J, \ \tau ^\pm \to \mu ^\pm + J ,$ having
branching fractions of order, $ 10^{-3}\ - \ 10^{-4} $; and invisible
Higgs boson decay mode into majoron pairs, $ h\to J+J$.  Although the
majoron coupling to the $Z$ boson is tiny, its coupling to the Higgs
bosons, $ h, \ H$, is sizeable and may lead to a clear signal through
the cascade decays, $ q \to Z ^0 +h \to Z ^0 +J+J .$ The complete
version of the model, incorporating the explicit breaking of R parity
symmetry by bilinear interactions, leads to satisfactory predictions
which provide it with a more natural outlook~\cite{campos95,diaz98}.
The extensive applications of the extended model are presented in the
original references~\cite{masierovalle,joshinowak95,valle1} and in the
lecture notes by Rom\~ao~\cite{romao}.

The construction of  extended models realizing the spontaneous breaking 
of R parity symmetry has been actively pursued over the years. For the
left-right symmetric gauge theories, in general, it is found that the
scalar fields VEVs driving the spontaneous breaking of  space parity 
necessarily  cause the breaking of lepton number
symmetry~\cite{kuki93}, meaning that  R parity symmetry violation
is necessary for  space parity symmetry violation. 
A left-right  symmetric unified model
embodying an exact matter parity can still  be constructed~\cite{moha96} by
realizing the  unification group $SU(5)$ as the diagonal subgroup of
$SU(5)\times SU(5) $.    Puolamaki~\cite{puolo00} discusses four
variants of the minimal left-right symmetric gauge theories associated
with the abelian and non-abelian gauge symmetry groups, $ SU(3)_c
\times SU(2)_L \times U(1)_R \times U(1)_{B-L}$ and $SU(3)_c \times
SU(2)_L \times S(2)_R \times U(1)_{B-L}$.  The models with large scale
in comparison to the supersymmetry breaking mass scale, $ v_R >> M_S$,
have an automatic suppression of the RPV coupling constants, while
those with a comparable scale, $v_R \sim M_S$, are made safe by the
restricted subset of allowed lepton number violating couplings.
Kitano and Oda~\cite{kitano99} discuss an extended model with a low
left-right gauge symmetry breaking VEV, $v_R \sim M_S$, including $
\nu ^c$ and Higgs boson singlet chiral superfields.  The combined
presence of $\nu - \nu ^c$ Majorana masses and effective bilinear RPV
superpotential predict  mass matrices for the light neutrinos featuring either
hierarchical or non-hierarchical patterns.

\subsection{Bilinear R parity symmetry  violation}
\label{secxxx1c}

The option of bilinear R parity symmetry violation is attractive
mainly owing to its simplicity and predictive power. In spite of the
limited number of free parameters, the characteristic tree level field
mixing of Higgs bosons with sleptons and gauginos or higgsinos with
leptons leads to an especially rich phenomenology.

The VEV shifts of the left and right chirality sneutrino fields,
$\tilde \nu _{iL} $ and $\tilde \nu ^c _{iL} $, accompanying the
spontaneous breaking of R parity in the MSSM and its minimal
extensions, lead to coupled systems of mixed neutralinos-neutrinos and
charginos-leptons, $\tchi ^{0T} = (-i \tilde \g ,\ -i \tilde Z,\
\tilde H_u,\ \tilde H_d ,\ (\nu _i) _{1\times 3} )^T , \quad \tchi ^{-
T} = (-i\tilde W^-,\ \tilde H_d^-,\ (e_i) _{1\times 3})^T , \ \ \tchi
^{+T} = (-i\tilde W^+ , \ \tilde H_u^+,\ (e_i^c)_{1\times 3} ) ^T $.
The corresponding mass matrices, $ M_n ,\ M_c$, of dimension $7 $ and
$4$, have a typical see-saw structure with a zero diagonal lower
corner block and small off-diagonal blocks in comparison to the
diagonal upper corner block. The procedure used to diagonalize the
approximate see-saw type neutrino-neutralino and lepton-chargino mass
matrix has been discussed in several early works~\cite{schech82}.  The
unitary transformation matrices to the mass bases, $ (\tchi ^0
_l)_{mass} = \caln _{lm} \tchi ^0 _m,\ (\tchi ^ {\pm } _l)_{mass} =
[\calu _{lm} , \calv _{lm} ] \tchi ^{\pm } _m $, in the approximation
of small RPV couplings can be expressed in an closed form
representation~\cite{joshinowak95,nowak96,faesskov98,hirval99} in
terms of the usual transformation matrices for gauginos and leptons
and the characteristic alignment parameters, $\L _i = [ \mu <\tilde
\nu _{iL} > - \l ^\nu _{ij} <H_d^0> <\tilde \nu ^c_{jL} > - \mu _i
<H_d ^0> ] $.

\subsubsection{\bf Supersymmetry breaking  RPV effects}
\label{secxxx1c1}

The model proposed by Hall and Suzuki~\cite{hallsuz} uses the
supergravity framework with general generation universal soft
supersymmetry breaking and a bilinear superpotential. The latter is
transferred by the $(H_d,\ L_i) $ superfields transformation into a
trilinear lepton number violating superpotential of small size
controlled by the regular Yukawa coupling constants, $\l ^d _{ij} , \
\l ^e _{ij}$. The soft supersymmetry breaking mass mixing term, $
V_{soft} = \mu^2_{ui} \tilde L_i H_u + H.\ c.$, is instrumental in
raising finite sneutrino VEVs, $ v_i$, which initiate a single massive
Majorana neutrino of mass, $ m_\nu \simeq \mu (v_i /v_d) ^2 $.
Depending on whether this mass eigenstate dominates $\nu _e,\ \nu _\mu
$ or $ \nu _\tau $, the resulting correlation between $ v_i= [v_{e}
,v_{\mu } ,v_{\tau }] $ and the associated neutrino masses can be
expressed as~\cite{hallsuz}, $[ { v_ e \over 20 \ \text{MeV} } , \ {
v_ \mu \over 2 \ \text{GeV} } , \ { v_ \tau \over 40 \ \text{GeV} } ]=
({25 \ \text{GeV} \over \mu } ) ^{1/2} \times [ ({ m_ {\nu _e} \over
46 \ \text{eV} } )^{1/2}, \ ({ m_ {\nu _\mu } \over 520 \ \text{keV} }
)^{1/2}, \ ({m_{\nu_\tau } \over 250 \ \text{ MeV} } )^{1/2} ]. $ The
more quantitative studies presented in Refs.~\cite{lee,dawson} obtain
on the same basis the sneutrinos VEV bounds, $ v _\tau \le 12\
\text{GeV} ,\ v _e \le 2\ \text{MeV} .$ Another massive neutrino
eigenstate arises in the model of Hall and Suzuki~\cite{hallsuz} at
the one-loop level.  The comparison with experimental constraints
leads to the coupling constant bounds, $ \mu _{ue} < 1 \ \text{GeV}$
and $ { \mu_e \over \mu } \equiv { \mu_1\over \mu } < {1\over 150} . $

The unitary matrix transformation of the $ (L_i , \ H_d) $ superfields
which cancels away the bilinear RPV superpotential produces a variety
of lepton number violating terms in the scalar potential and the
trilinear Yukawa interactions. The transformed Lagrangian is examined
by Roy and Mukhopadhyaya~\cite{roybi} in the simplified case where one
only retains the bilinear interaction and VEV parameters of third
generation leptons, $ \mu_3 , \ v_3 , \ \mu ^2_{u3} = B_3 \mu_3
m_{\tilde G}$, while setting all the other RPV parameters to zero. The
mass spectrum of fermions and sfermions is determined after minimizing
the one-loop renormalized effective scalar potential.  An analysis of
the MSSM parameter space and the RPV parameters, subject to the
constraints from vacuum stability and the experimental $ \nu_\tau $
mass limit, shows that the allowed values for $ \mu_3 ,\ v_3 $ must
lie in a triangle shaped domain roughly bounded by the maximal values,
$ \vert \mu_3 \vert _{max} = 20 \ \text{GeV}, \ \vert v_3 \vert _{max} = 5 \
\text{GeV}$.  In addition to the normal RPV trilinear Yukawa
couplings, additional non-supersymmetric lepton number violating
trilinear Yukawa couplings with distinctive chirality structure are
produced in this scenario.  For illustration, we display a
representative sample of the corresponding lepton number violating
couplings, $ L _{EFF} = \rho_{i3i}\tilde e^ \star _{iL} \bar e^c_{3R}
\nu_{iL} + \o _{i3i}\tilde \nu _{iL} \bar e_{3R} e_{iL} +
\xi_{i3i}\tilde \nu ^ \star _{iL} \bar \nu^c_{3R} \nu_{iL} +
\O_{3ii}\tilde d^ \star _{iL} \bar e^c_{3R} u_{iL} + \L_{i3i} \tilde u
_{iL} \bar u_{iL} \nu ^c_{3R} + \cdots , $ where the coupling
constants, $ \rho_{i3i}, \ \o _{i3i}, \ \xi_{i3i},\ \O_{3ii},\
\L_{i3i} $, are fully predictable in terms of the free parameters, $
\mu _3,\ v_3,\ \ B_3 \mu _3 m_{\tilde G} $, and the regular MSSM input
parameters.  These coupling constants assume values of typical size
$O(10^{-3})$ or below.  There occurs lepton flavor changing charged
and neutral current couplings, produced through the gauge bosons
regular couplings, $ W^\pm \tchi ^\mp \tchi ^0, \ Z^0 \tchi ^0 \tchi
^0$, by the field mixing of neutralinos and charginos with the neutral
and charged charged leptons, respectively.  For an LSP neutralino, the
branching fractions for the neutral and charged two-body decay modes,
$ \tchi^0 \to \nu_\tau +Z^0, \ \tchi^0 \to \tau ^+ + W^-$, exhibit an
interesting dependence on the free parameters of the model.

It is important to keep in mind that the sneutrino VEVs, $v_i$, are
derived rather than free parameters depending on the soft
supersymmetry breaking interactions.  The extended bilinear R parity
violation option including the soft supersymmetry breaking
interactions generically features finite sneutrino VEVs.  The
phenomenological constraints on the coupling constant and VEV
parameters are usually variant under fields redefinitions, unlike the
physical observables.  This relatively common problem becomes
especially troublesome when one deals with several indistinguishable
fields.  Thus the fields gauge bases, diagonal with respect to the
gauge interactions, are defined up to unitary transformations of the
fields carrying the same gauge group representations. On the other
hand, the mass eigenstate bases for the quarks and leptons are
naturally singled out by the Yukawa interactions with the electroweak
Higgs bosons. In the presence of lepton number violating interactions,
the lack of quantum numbers distinguishing the down Higgs boson from
lepton doublets superfields enlarges the basis freedom to the unitary
transformations of the four vector fields, $L_\a = (H_d , \ L_i )$.
The redundancy in the RPV parameters for the leptons and sleptons can
be used to set a selected subset to zero.

The choice for the $L_i,\ H_d$ superfields characterized by vanishing
sneutrino VEVs, $v_i =0$, leads to a simplified block structure for
the neutralino and chargino mass matrices, $ M^{n} $ and $ M^{c}$,
which become explicitly independent of the trilinear RPV coupling
constants, $ \l _{ijk}$.  Based on this convenient field convention,
Bisset et al.,~\cite{bisset} perform a scan over the MSSM parameters,
$ \mu ,\ M_2$, and the bilinear parameters, $\mu _i$, subject to the
various experimental constraints including those currently available
for the neutrinos masses.  The allowed values for the effective
bilinear parameters are required to vary inside the narrow range,
$\hat \mu = (1 \ - \ 60) \ \text{GeV}$, for low $ \tan \b \simeq 2 $.
The allowed interval widens out with increasing $\tan \b \leq 45 $,
where it attains the wider range of variation, $\hat \mu = ( 10\ -\
1000 ) \ \text{GeV}$.  A comprehensive global analysis of the
admissible region of parameter space along with the derivation of
individual coupling constant bounds is attempted in the subsequent
work by Bisset et al.,~\cite{bisset00} for a large set of lepton
flavor changing processes induced at tree level by the bilinear RPV
interactions. A general discussion of the single VEV approach for the
$ L_\a = (H_d, \ L_i )$ system is presented by Kong~\cite{kong04}.


 A valuable insight into the experimental constraints is offered by
the intrinsic basis independent approach which is developed along
similar lines as that using the plaquette invariants for the CP
violation in the quarks Yukawa interactions.  Studies of the
constraints on RPV interactions have been presented in applications
dealing with the neutrino masses~\cite{banksnir}, thermal physics
effects~\cite{davidson0,davidson,davidson1}, Higgs boson
sector~\cite{davus00,groer01,grosshab03} and $Z$ gauge boson
observables~\cite{ferrandis}.  The tree level contributions from the
bilinear RPV interactions to the neutrino Majorana mass are found to
depend on the basis invariant angle variable, $\xi $, between the four
vectors $\mu _\a , \ v_\a $~\cite{banksnir} defined by, $ \sin ^2 \xi
= \vert \mu \times v \vert ^2 /(\hat \mu ^ 2 v_d^ 2 ) = 1 - (\mu \cdot
v ) ^ 2 /(\hat \mu ^ 2 v _d^ 2 ) = {1\over 2\hat \mu^2 v_d^2 } \sum _
{\a , \b } (\mu_\a v_\b -\mu_\b v_\a )^2 ,\ [\hat \mu ^2 =\sum _\a \mu
_ \a ^2 , \ v_d^2 =\sum _\a v_\a ^2 ]$.  The constraints set by the
neutrinos Majorana masses on the bilinear RPV parameters, $\mu _i , \
v_i$, are thus expressed by the geometric alignment condition between
the coupling constants and VEV parameters in the $L_\a = (L_i, \ H_d)
$ field space, described by the proportionality
relationship~\cite{banksnir}, $\mu_\a \propto v_\a $. This condition
is more quantitatively expressed by the basis independent bound, $
\sin \xi < O(( { m_{\nu _\tau } \over m_{\tilde Z } } )^\ud ) < 1.6
\times 10^{-2} ,\ [\hat \mu ^2 = \sum _\a \mu _\a ^2 ,\ v_d^2 = \sum
_\a v_\a ^2 ] . $ In the sector of charged higgsinos and leptons, the
above quoted constraint entails that the neutrinos and charged leptons
remain good electroweak doublets partners, thus leaving intact the
universality of charged current interactions.  The one-loop level
contributions to the neutrinos mass matrix, the Higgs boson mass
spectrum and the mass splittings between the neutrino-antisneutrino CP
even and odd eigenstates, $\tilde \nu _ {i+} , \ \tilde \nu _ {i-} $,
involve~\cite{groer01} other invariants constructed from the
supersymmetry breaking parameters, $ \tilde m^2 _{\a \b } , \ B_\a \mu
_\a $. The same basis independent approach is applied by Grossman and
Haber~\cite{grosshab03} in discussing the properties of the would-be
light majoron and its light scalar companion in general lepton number
violating supersymmetric models.

When general soft supersymmetry breaking bilinear terms, $ V _{soft} =
(B_\a m_{\tilde G } \mu _\a \tilde L_\a H_u + H. \ c. ) + \tilde m ^ 2
_{\a \b } \tilde L _\a^\dagger \tilde L _\b $, are present in the
scalar potential, the alignment condition, $\mu_\a \propto v_ \a $,
becomes equivalent to the two proportionality relations linking the
supersymmetric and non-supersymmetric parameters~\cite{banksnir}, $
B_\a \propto \mu _\a , \ \tilde m^2_{\a \b } \mu_\b \propto \mu _ \a
$. In order to ensure controllably small deviations with respect to
these two conditions, one needs to set non-trivial requirements on the
physics of the supersymmetry breaking sector.  Alternatively, one may
invoke some spontaneously broken horizontal
symmetry~\cite{borzum96,bhatta98} which accounts for the flavor
hierarchies in the leptons and neutrinos masses and mixing parameters.
Phenomenological studies of the bilinear and trilinear RPV 
contributions to the neutrino Majorana mass
matrix guided by horizontal  symmetry models  have been 
presented in several works of which two representative 
are given in Refs.~\cite{miralle00,gogola01}.

Since the running parameters, $ \mu_\a (Q)$, and the
sneutrinos VEVs, $v_\a $, renormalize in different ways, the alignment
condition, $ \mu_\a (Q) \propto v_\a $, is not invariant under the
renormalization group.  The misalignment effect induced by
renormalization is discussed within the supergravity framework for
supersymmetry breaking by Hempfling~\cite{hempf} and Nilles and
Polonsky~\cite{nilles97} in the case including the bilinear
interactions alone and by de Carlos and White~\cite{decarlosl} and
Nardi~\cite{nardi} in the more general case allowing for both bilinear
and trilinear interactions. Accounting for the renormalization group
evolution of coupling constants, the experimental limit on neutrino
masses leads to the following coupling constant bound~\cite{nardi}, $
\sum_i \vert \l ^{'} _{i33} \vert ^2 < [10^{-4} \ - \ 10^{-7}] $.  The
updated bound obtained recently by Allanach et al.,~\cite{allanach03}
$ \sum _i m_{\nu _i} < 0.71 \ \text{eV} \ \Longrightarrow \ \sum _i (3
\l '_{ijk} \l ^d _{jk} +\l _{ijk} \l ^e _{jk} ) < 2. \ \times 10^{-5}
$, translates into the strong single coupling constant bound, $ \vert
\l' _{333} \vert < 6. \ \times 10^{-6} $.

\subsection{Phenomenological constraints on bilinear R parity   
violation }
\label{secxxx1d}

The general effective Lagrangian for the majoron and its light scalar
companion, $J(x), \ \s (x)$, includes trilinear Yukawa couplings with
leptons of general form, $ L _{EFF} = \ud \bar \nu _i [ g^{\nu J}_{ij}
i\g _5 J (x) + g^{\nu \s }_{ij}\s (x) ] \nu _j + \bar e_i [ g^{e
J}_{ij} i\g _5 J (x) + g^{e \s }_{ij}\s (x) ] e _j$.  Bounding the
solar flux of majoron emission by the experimentally determined
luminosity at the center of the Sun and in the red giant
stars~\cite{palmoha} leads to the following bounds, $ \vert g^{eJ}_{11} \vert <
10^{-10},\ \vert g^{e J} _{11} \vert  < 10^{-12}$, respectively.  The astrophysical
constraints on the majoron couplings are reviewed in
Ref.~\cite{dearborn}.  The lepton number violating coupling constants
of majorons to neutrinos are required~\cite{abundance} to obey the
bound, $ \bar g^{ \nu J } \equiv \vert \sum _{ij} \vert (g ^{ \nu J}
_{ij})^{4} \vert ^{ 1/4 } < 1.5 \times 10 ^{-6 } $, based on the
observations from the supernova 1987A luminosity, and the bound, $
\bar g^{ \nu J } < O(10 ^{- 8 }) $, based on the cosmic baryon
asymmetry non-dilution by out of thermal equilibrium lepton number
violating reactions.

The experimental data from high energy colliders~\cite{gonzalez90}
also yield useful constraints on the couplings of leptons to the
majoron and its light scalar companion field.  The main representative
processes~\cite{bargerpakvasa} include: (i) Exotic semileptonic decay
modes of $K$ mesons into wrong sign leptons accompanied by emission of
a light scalar particle, $ K^+ \to e_i ^+ + \bar \nu +J ,\ K^+ \to
e_i^+ + \bar \nu +\s ;$ (ii) Neutrino-quark scattering with majoron
emission, $ \nu _ \mu + u \to e_i ^+ + d + J $; (iii) Majoron emission
in chargino decays, $\tchi ^ \pm \to \tau ^ \pm +J $; (iv) Lepton
flavor violating decays of charged leptons, $ e ^- _i \to e ^- _j +J
$; (v) Invisible decay modes of the Higgs boson, $ h \to J+J$.  Some
illustrative bounds on the $\nu \nu J $ and $ ee J$ coupling constants
inferred from these processes are given by~\cite{bargerpakvasa}: $
\vert g _{ee} ^ {\nu J }\vert ^2 < 1.8 \ 10 ^ {-4} , \ \vert g _{\mu
\mu } ^ {\nu J }\vert ^2 < 2.4 \ 10 ^ {-4} , \ [ K ^ + \to J + e_l ^ +
+ \bar \nu _l ] ; \ \vert g _{ee} ^ {\nu J }\vert ^2 < 4.5 \ 10 ^ {-5}
,\ [ (K ^ + \to e ^ + + \bar \nu _e )/ (K ^ + \to \mu ^ + + \bar \nu
_\mu )]; \ \vert g _{\mu \mu } ^ {\nu J } \vert ^2 < 2.5 \ 10 ^ {-2}
,\ [\nu + N \to J +e ^+ + \ \text{hadrons} ] . $

In the presence of sneutrinos VEVs, the neutrinos Majorana mass matrix
receives one-loop contributions~\cite{hallsuz,lee,dawson} from the
combined action of gauge and RPV interactions, $\tilde \nu (\bar e
\bar \chi ) $ and $ (\nu e ) \tilde e$.  The associated Feynman graph
is displayed by the diagram A.2 in Figure~\ref{figun}, with the
sneutrinos VEV conveniently represented as sneutrino tadpoles.  The
comparison with the experimental limits on neutrino masses leads to
the coupling constant bounds~\cite{dawson}, $\l _{imn} ({v_i \over 10
\ \text{ MeV} }) < ({\tilde m \over 250 \ \text{GeV} }) [ 10^2, \
1.5\times 10^4, \ 1.6\times 10^5 ] , \ [i=1,2,3] .$ Flavor
non-diagonal radiative two-body and three-body decay modes of leptons,
$ \mu ^\pm \to e ^\pm + \g $ and $\mu ^\pm \to e^\pm +e^++e^- $, as
well as two-body decay modes of neutralinos, $\tchi^0 \to \nu +\g , \
\tchi^0 \to \ e ^-+\pi ^ + $, can be initiated through one-loop Feyman
graphs of similar structure.

The mixing of neutralinos or charginos with the neutral or charged
leptons contribute to their three-body decay modes through the
processes represented by the Feynman diagram A.3 in
Figure~\ref{figun}.  The field mixing effect also induces, through the
neutral current $Z$-boson couplings to neutralinos or charginos pairs,
$ Z^0 \tilde \chi^0_l \tilde \chi^0_m, \ Z^0\tilde \chi^+_l \tilde
\chi^-_m ,$ the tree level lepton number violating two-body decay
modes, $ Z^0 \to \tilde \chi ^0 +\bar \nu_\tau ,\ \ Z ^0\to \tilde
\chi ^\pm +\tau ^\mp $. When energetically allowed, the former mode
can contribute to the $Z$-boson invisible decay width.  The predicted
branching fractions within the MSSM framework vary inside the
range~\cite{barbieri}, $ B (Z^0 \to f) \approx (3. \times 10^{-5} \ -
\ 3. \times 10^{-7}) $. The comparison by Campbell et
al.,~\cite{abundance} with the experimental limits leads to the
allowed intervals for the coupling constant upper bounds, $ (\mu _\tau
, \ v_\tau ) \in [0.1 \ - \ 1.]\ \text{GeV} $. The corresponding
branching fractions predictions in the model of Masiero and
Valle~\cite{masierovalle} are, $ B(Z ^0\to \tchi^\pm +\tau ^\mp ) <
6. \times 10^{-5}, \ B(Z^0 \to \tchi ^0 +\nu_ \tau ) < 10^{-4}.$ The
analogous $Z$-boson decay modes into slepton-antislepton pairs
decaying through the RPV interactions into four charged leptons, $ Z^0
\to e^- + \tau ^+ + e^+ +\tau ^- $, are examined by Brahm and
Hall~\cite{brahm}, based on a model for the bilinear RPV interactions
derived from the flipped $ SU(5)\times U(1)$ gauge unified theory.

The RPV bilinear interactions can affect the stability of the light
massive Majorana neutrinos by initiating the tree level three-body
decay modes, $ \nu ' \to 3 \nu $, and the one-loop level two-body
radiative decay modes, $ \nu '\to \nu +\g $, due to an anomalous $M1$
transition moment~\cite{roulet91,enqvist}.  In the model of Aulakh and
Mohapatra~\cite{aulak}, the breaking of supersymmetry produces a
trilinear coupling of the neutrino and majoron with the R parity odd
goldstino field, $\tilde G (x)$.  In the case of an ultralight
gravitino, this coupling can initiate the decay of a massive neutrino
by emission of a gravitino and majoron with the estimated partial
lifetime, $ \tau (\nu \to \tilde G +J) = (0.3 \ 10^3 s) ({ 100\text{
keV} / m_\nu })^3 ({\sqrt {F_S}/ 100 \ \text{GeV} })^4 $.  In the
model of Masiero and Valle~\cite{masierovalle}, the LSP neutralino
decays predominantly by emission of a neutrino-majoron pair, $ \tchi
^0 \to \nu + J$, or a neutrino-photon pair, $\tchi ^0\to \nu + \g $.
A sample of representative branching fraction predictions for the
$Z$-boson and $\tau $-lepton rare decays is given by: $ B(Z\to
\tchi^\pm +\tau ^\mp ) < 6. \times 10^{-5}, \ B(Z^0 \to \tchi ^0 +\nu_
\tau ) < 1.\times 10^{-4}, \ B(\tau ^\pm \to \mu ^\pm +J ) < 1. \times
10^{-3},\ \ B(\mu ^\pm \to e ^\pm + J) < 2.6 \times 10^{ -6},\ B(\tau
^\pm \to e ^\pm +J ) < 1.\times 10^{-4}.$ The neutralino decay modes
also have interesting astrophysical implications~\cite{berezinsky91}.

The neutralino-neutrino and chargino-lepton fields mixing produced by
finite bilinear RPV parameters, $\mu _i $ and $v_i$, produces a
variety of lepton flavor changing and lepton non-universal effects in
the neutral and charged current processes.  The approximate see-saw
type structure for the mass matrices~\cite{nowak96,faesskov98,joshi95}
entails a dependence of the RPV contributions on the misalignment
parameters, $\L _i= [ \mu <\tilde \nu _{iL} > - \mu _i <H_d^0>] $.
The induced flavor changing and non-universal corrections to the gauge
bosons couplings to fermions, $ Ze_i\bar e_j ,\ W ^-\nu _i \bar e_j $,
and to the semileptonic decays of mesons, are strongly correlated with
the contributions to the $\nu _\tau $ Majorana mass. For illustration,
the rare decay process, $ \mu ^-\to e^- +e^+ +e^-$, is found to set
the bound, $ m_{\nu _\tau } < 1 $ MeV.  The possibility of a
spontaneous CP symmetry violation initiated through the complex
sneutrino VEVs arises in a natural way, provided one allows for
generic soft supersymmetry breaking terms, as discussed by Joshipura
and Nowakowski~\cite{joshi95} and Nowakowski and
Pilaftsis~\cite{nowak96}.

The neutralinos-neutrinos and charginos-leptons mixing can initiate a
variety of lepton flavor changing or lepton number violating rare
processes at the tree level (semileptonic decays of mesons, leptonic
three-body decays of leptons, $\mu ^--e^-$ and $ Mu \to \overline{Mu}
$ conversion) and at the loop level (radiative decays of leptons, $\mu
^--e^-$ conversion in nuclei, electric dipole moments, anomalous
magnetic moments).  A systematic discussion is presented by Frank and
Huitu~\cite{frankhu01}, from which we quote the following
representative bounds involving the charginos mixing matrix, $\calv $,
and the CKM matrix: $ g^2 (\tilde u_{2L} ^{-2} + \tilde u_{3L} ^{-2} )
\calv ^\star _{31} \calv _{31} V_{cb} \leq 4.3 \times 10^{-4} , \
[b\to s + e^+ + e^-]; \ g^2 \calv ^\star _{41} \calv _{31} \tilde
\nu_{iL} ^{-1} \leq 6.6 \times 10^{-7} , \ [\mu ^-\to e^- + e^+ +
e^-]; \ \Re ( \calv ^\star _{31} \calv ^\star _{32} ) \leq 2.5 \times
10^{-5} , [\mu ^- \to e^- +\g ]$.  It is of interest to note that the
major fraction of the constraints on mixing matrix elements apply
independently of the sfermions masses.

The bilinear RPV interactions lead naturally to an hierarchical
neutrino mass spectrum in satisfactory agreement with the measurements
of atmospheric and solar neutrino oscillation data.  These results
have a direct impact on the lepton flavor changing radiative lepton
decay reactions.  Thus the possibility of getting observably large
rates for the decay mode~\cite{carval02}, $\mu ^- \to e ^- +\g $, is
favored by the atmospheric data but ruled out by the solar data.

The Higgs boson and sneutrino fields mixing by the soft supersymmetry
breaking RPV interactions has observable implications on the Higgs
sector mass spectrum and the Higgs boson production rates at the high
energy colliders.  It may also modify the Higgs boson decay modes and
add non standard Higgs decay modes, $h\to \nu + \tchi ^0 ,\ h\to \tau
^- + \tchi ^+ $, which become relevant in certain regions of the
parameter space. The studies of the neutral Higgs sector have focused
on the bilinear interactions alone~\cite{campos95,diaz98} or in
combination with the trilinear
interactions~\cite{davus00,changfeng00}.  The analysis by Davidson et
al.,~\cite{davus00} considers a basis independent description in a one
generation toy model.  The study by De Campos et al.,~\cite{campos02}
discusses the mass spectrum of the MSSM with bilinear R parity
violation in the scenario characterized by the conformal anomaly
mediated supersymmetry breaking.

Taking note of the sneutrino fields variance under linear unitary
transformations of the $L_\a = ( H_d , \ L_i) $ fields and of the fact
that certain phenomenological constraints involve directly the
sneutrinos VEVs, makes it important to quantitatively assess the
nature and size of the basis dependence of bounds on the relevant VEV
parameters, $v_\a $.  The study by Ferrandis~\cite{ferrandis} examines
this issue by working with a single finite VEV associated to the
tau-sneutrino, $<\tilde \nu _3> = v_3 /\sqrt 2 $, while restricting to
the third generation bilinear and trilinear coupling constants, $\mu
_3,\ \l '_{333}$.  The generalized case with three finite sneutrino
VEVs, $v_i$, is feasible.  Defining the three different consistent
choices of $ L_\a $ bases, $I,\ II, \ III$, by means of the
characteristic conditions, $I:\ \l ^{'I}_{333}=0; \ II:\ \mu
^{'II}_{3}=0; \ III:\ v_3 ^{III} =0$, one proceeds~\cite{ferrandis} by
first determining the coupling constant bound following from some
relevant physical constraint in the bases $ I$ or $II$ and next
translating this into a bound on the sneutrino VEV in the
corresponding basis, $ v_3 ^{I} =0$ or $ v_3 ^{II} =0$.  The strongest
condition on the sneutrinos VEVs is found to arise from the one-loop
correction to the decay width, $ Z \to b + \bar b $.  Starting from
the bound on $\l '_ {333} $ in the basis $III$, one finds upon
transforming to the basis $I$ the following bound on the sneutrinos
VEVs, $\sum _i ( v _i ^ 2 ) ^ \ud < 5. \ \text{GeV} \ \tilde \nu ^ \ud
.$

\section{Constraints on RPV  couplings from 
fundamental tests of Standard Model}
\label{secxxx2}

Our objective in the present section is to review the constraints on R
parity violation derived from the processes associated with
fundamental tests of the SM. First, we consider the high precision
measurements for the electroweak charged and neutral current
observables. The main topics include: (1) the generation universality
of the leptons and quarks interactions; (2) the neutrino scattering on
leptons and hadrons; (3) the fermion-antifermion pair production
reactions; (4) the $Z$-boson pole observables; and (5) the atomic
physics parity violation observables.  Unless otherwise stated, the
quoted results are obtained by employing SM predictions which include
both the tree and one-loop level contributions.  The initial
comparison with experiment by Barger et al.,~\cite{bargerg} was based
on the data available up to year 1989, using the tree level SM
predictions.  The extended study, performed by Ledroit and
Sajot~\cite{ledroit}, uses the current experimental results up to year
1998 and the improved determinations for the charged and neutral
current coupling constants including one-loop renormalization
contributions~\cite{langarad,erler98}.

Three other subjects lying at the interface between high precision
observables and R parity symmetry breaking are discussed in the
succeeding subsections. The first subject concerns the constraints
inferred from measurements of the anomalous magnetic dipole moments of
leptons.  The second subject addresses the possible correlations
between CP violation and R parity violation physics. The discussion
focuses on the RPV contributions to the polarization observables in
hadronic decays, the asymmetries in $ B $ mesons decay rates, and the
electric dipole moments of leptons and neutron.  The third subject
deals with the impact of R parity violation on the neutrino masses,
mixing and instability.

\subsection{Charged current interactions}
\label{secxxx2a}

\subsubsection{\bf Universality of lepton and quark charged current  
interactions}


The universality property of the quarks and leptons and of the quarks
generations or leptons generations is an automatic consequence of the
gauge nature of the electroweak interactions. Deviations from
universality are introduced for quarks through the unitary matrix
transformations connecting the flavor and mass bases.  The Fermi
coupling constant, $G_{F} $, is experimentally accessed by comparing
the measured muon beta decay width with the theoretical value
including the one-loop radiative corrections. The experimental
determination of the hadronic sector Fermi coupling constant is found
by comparing the body of super-allowed nuclear beta decay Fermi
transitions with the theoretical predictions including the radiative
and relativistic Coulomb effects~\cite{sirlin}.

The RPV amplitudes involving the local four fermion couplings of
leptons and quarks pairs are described by the sfermion exchange tree
level graphs depicted by the Feynman diagrams B.1-4 of
Figure~\ref{figun}.  The combined contributions with those from the
weak charged current gauge interactions are described by the effective
Lagrangian
\begin{eqnarray} 
L_{EFF} &=& -\bigg [{ g _2^2 \over 2m_W^2} \d_{i'j} \d_{j'i} -
{\l_{ijk} \l^\star _{i'j'k} \over 2m^2_{\tilde e_{kR}} } \bigg ] (\bar
\nu_{i'L}\g_\mu e_{jL} )(\bar e_{j'L} \g^\mu \nu_{iL}) \cr &-& \bigg [
{g _2^2 \over 2m_W^2} V_{j'j} +{\l'_{ijk} \l^{'\star } _{i'j'k} \over
2m^2_{\tilde d_{kR}} } \bigg ] (\bar e_{i'L}\g_\mu \nu_{iL} )(\bar
u_{j'L} \g^\mu d_{jL}) - {\l_{ijk} \l_{i'jk'} ^\star \over 2
m^2_{\tilde e_{jL}} } (\bar e_{kR} \g^\mu e_{k'R}) (\bar \nu_{i'L}
\g_\mu \nu _{iL}) + H. \ c. , \label{eqx2}
\end{eqnarray} 	
where the derivation of the above result, with the Dirac spinor
fermion fields understood as anticommuting variables, makes use of the
Fierz-Michel identities.

\subsubsection{\bf Fermi coupling constant redefinition}   

The four fermion local coupling responsible for the three-body muon
decay leads to the identification of a modified $\mu ^-$ decay
constant~\cite{bargerg}, $ {G_{F} \over \sqrt 2} = {g_2^2\over 8m_W^2}
(1+r_{12k}(\tilde e_{kR})), \quad [r_{ijk} (\tilde e_{kR})=
{m_W^2\over g_2^2 m^2_{\tilde e_{kR}} } \vert \l_{ijk}\vert^2 ] $
where $ r_{ijk} $ designates a convenient auxiliary parameter.
A significant part of the RPV contributions is absorbed into the
physical Fermi weak interactions coupling constant, which is
multiplicatively related to the SM coupling constant as, $ G_{F} =
G_{F} ^{SM} (1+r_{12k}(\tilde e_{kR}))$.  The ensuing redefinition of
$G_{F} $ represents an ubiquitous effect in the formulas expressing
deviations from universality in the charged and neutral current
observables.  A simple minded comparison of the experimental muon
decay width with the corrected coupling constant $G_F$, leads to the
coupling constant bound~\cite{bargerg}, $\vert \l _{12k}\vert
=(0.14\pm 0.05 ) \ \tilde e_{kR} $. The stronger bound, $\vert \l
_{12k} \vert < 0.060 \ \tilde e_{kR} $, is found upon using the
updated experimental information on the charged current parameters and
the $W$-boson mass, $ m_W$.

A quantitative discussion of the Fermi coupling constant redefinition,
$G_{F} \to G_{F} / (1+r_{12k}(\tilde e_{kR}))$, requires the
consideration of the field theory renormalization corrections to the
SM gauge sector parameters~\cite{dimohall,bargerg}.  The quantum level
one-loop corrected natural relations linking the renormalized input
parameters $\a ,\ G_{F} ,\ m_Z $ to the weak (Weinberg) angle and
$W$-boson mass parameters, have been derived in two popular
renormalization schemes: the modified minimal subtraction off-shell
regularization scheme (labeled by ${\overline {MS} } $) and the
on-shell regularization scheme~\cite{langarad}.  The two relations of
interest in the discussion of RPV corrections to $G_F$ are given by
\begin{eqnarray}
\text{Off-shell}\ \overline {MS}:\ m_W^2 &= & { \pi \a (1+r_{12k}
(\tilde e_{kR} ))\over \sqrt 2 G_{F} \sin^2\t_W (m_Z)\vert_ {\overline
{MS} } (1-\D r (m_Z) \vert_{\overline {MS} } )} , \cr
\text{On-shell}:\ \sin^2 \t_W &\equiv & 1-{ m_W ^2 \over m_Z ^2} = {
\pi \a (1+r_{12k} (\tilde e_{kR} ))\over \sqrt 2 G_{F} m_W^2 (1-\D r
)} , \label{eqx5}
\end{eqnarray}
where the quantities labeled by $\overline{MS}$ refer to the modified
minimal subtraction scheme and those without label refer to the
on-shell renormalization scheme.  The first off-shell scheme relation
can be regarded as a prediction for the $W$-boson mass $m_W$ using the
coupling constants $ G_F, \ \a $ and the weak interaction angle
parameter, $ \sin^2\t_W (m_Z)\vert_ {\overline {MS} }$, as inputs.
The second on-shell scheme relation can be regarded as a prediction
for the $W$-boson mass or weak interaction angle, assumed to be linked
together to all orders of perturbation theory by the natural relation,
$\sin ^2 \t _W = 1- m_W ^2/m_Z ^2$. The coupling constants $ G_F, \ \a
$ as treated as experimental inputs.  The corresponding auxiliary
parameters $\D r (m_Z) \vert _{\overline {MS} },\ \D r, $ are
renormalization scheme dependent functions which can be calculated in
terms of the basic input parameters and the various SM particles
masses.

The above quoted relations are used in fits to high precision
observables to evaluate $m_W$ or $\sin ^2 \t _W $ in terms of the
input parameters, based on the fact that the $W$-boson mass and weak
angle parameters are determined with poorer accuracy from experimental
measurements.  Instead, we propose here to determine the RPV coupling
constants $\l _{12k}$ from the quantum corrected relations by using
the values of the gauge sector parameters obtained in the SM fits to
the high precision experimental data.
The four input and derived parameters common to both relations are set
as~\cite{erler98}, $ \a ^{-1} = 137.035 ,\ G_F =1.16639 \times 10^{-5}
\ \ \text{GeV} ^{-2} , \ m_Z = (91.1867 \pm 0.0020) \ \text{GeV} , \
m_W= (80.405 \pm 0.089) \ \text{GeV}, $ based on the measurements of
the quantum Hall effect, muon decay width, $Z$-boson pole observables
and collider physics data, respectively.  Note that the alternate
experimental determination of the $W$-boson mass yields the value,
$m_W= (80.427 \pm 0.075) $ \ \text{GeV}. The fitted value of the weak
angle in the off-shell $\overline {MS} $ relation is given by, $\sin
^2 \t _W (m_Z) \vert _{\overline {MS} } = (0.23124 \pm 0.00017) $. The
weak angle in the on-shell scheme is fixed in terms of the $W$-boson
mass as: $\sin ^2 \t _W = 1- m_W ^2/m_Z ^2$.  The auxiliary parameters
are set at the values~\cite{marcianolanga}, $\D r = (0.0349 \pm 0.0019
\pm 0.0007 ),\ \D \hat r_W \equiv \D r (m_Z) \vert _{\overline {MS} }
= (0.0706 \pm 0.0011) $.
The comparison in the off-shell scheme case leads to a negative value
for $\vert \l _{12k}\vert ^2 $, therefore ruling out the coupling
constant $\l _{12k}$.  Taking into account the uncertainties on the
input parameters still leaves us with the possibility of inferring
finite coupling constant bounds.  The error on $ m_W$ dominates by far
all the other, and the $1\s $ level calculation leads to the coupling
constant bound, $\vert \l _{12k} \vert < 0.038 \ \tilde e_{kR}$.  The
comparison using the on-shell scheme also rules out the coupling
constant $\l _{12k}$, but yields the $1\s $ level coupling constant
bound, $\vert \l _{12k} \vert <0.046 \ \tilde e_{kR}$.  To illustrate
the importance of the uncertainties on the $W$-boson mass in the
on-shell scheme context, we consider the alternative prescription
using the experimental value for the on-shell renormalized weak angle,
$ \sin ^2 \t _W =(0.2260 \pm 0.0039) $, as determined from $\nu -N$
deep inelastic scattering data, while evaluating the $W$-boson mass
from the relation $ m_W ^2 = m_Z ^2 (1-\sin ^2 \t _W )$.  This
comparison yields the definite values, $ \vert \l _{12k}\vert = 0.081
\ \tilde e_{kR}$.  We conclude from the present discussion that the
constraint on $\l _{12k}$ set by the Fermi coupling constant
redefinition depends sensitively on the input value of the $W$-boson
mass $m_W$.  A useful consistency check on predictions is provided by
the simultaneous use of the off-shell and on-shell regularization
scheme relations.


The RPV contribution to the quark sector Fermi coupling constant,
$G_F$, is identified as the correction to the Cabibbo mixing angle
measured through the familiar up-quark three-body decay mode. In the
presence of R parity violation, the physical CKM matrix element can be
expressed as a function of the corresponding SM matrix element,
denoted as $V_{ud} ^{SM} $, by the relation~\cite{bargerg}: \bea &&
\vert V_{ud}\vert^2 = {\vert V^{SM} _{ud} +r'_{11k}(\tilde d_{kR})
\vert ^2 \over \vert 1+r_{12k}(\tilde e_{kR}) \vert ^2 } \simeq V^{SM}
_{ud} [1+ {2\over V^{SM} _{ud} } r'_{11k}(\tilde d_{kR}) - 2
r_{12k}(\tilde e_{kR}) ] ,
\label{eqvud} \eea 
where $V_{ud} $ designates the appropriate flavor mixing parameter of
light quarks to be compared with the experimental charged current
observable, $V_{ud} = G_V^{\D S =0} /G_{F} $. The corrections from $\l
'_{11k} $ and $\l _{12k} $ are seen to enter with opposite signs.
Under the single coupling constant dominance hypothesis, it follows
that only one of these two coupling constants can be non vanishing.
To extract a usable information, a helpful trick~\cite{bargerg}
consists in summing over the three quark generations for the
off-diagonal CKM matrix element $V_{ud_j} , \ [j=1,2,3]$ and invoking
the unitarity of the SM flavor mixing matrix, $ \sum _{j=1,2,3} \vert
V ^{SM}_{ud_j} \vert ^2 = 1 $.  The RPV contributions to $V_{ud_j}=
[V_{us} , \ V_{ub}],\ [j=2,3]$ are given by analogous formulas to that
for $ V_{ud} $ by effecting the replacement, $\vert \l '_{11k} \vert
^2 \to \vert \l _{11k} ^{'\star } \l '_{1jk} \vert ,\ [j=2,3]$, or
equivalently, $ r'_{11k} \to ( r'_{11k} r'_{1jk} )^\ud $. We recall
that the experimental information on $ V_{us} $ is accessed via the
strangeness changing decays $ K_{e3} ,\ K^+ \to \pi ^0 +e^++\nu $ and
that on $ V_{ub}$ via the charmless $B$ meson decays.  The unitarity
test for the CKM matrix is then expressed by the formula, \bea && \sum
_{j=1,2,3} \vert V_{ud_j}\vert^2 = \sum _{j=1,2,3} {\vert V^{SM}
_{ud_j} + (r'_{11k} (\tilde d_{kR})r'_{1jk} (\tilde d_{kR}) )^\ud
\vert ^2 \over \vert 1+r_{12k}(\tilde e_{kR})) \vert ^2 }, \cr &&
\simeq 1 - 2 r_{12k}(\tilde e_{kR}) + {2 V_{ud} } r'_{11k} (\tilde
d_{kR}) + {2 V_{us} } [[ r'_{11k} (\tilde d_{kR}) r'_{12k} (\tilde
d_{kR}) ] ^\ud + {2 V_{ub} } [ r'_{11k} (\tilde d_{kR}) r'_{13k}
(\tilde d_{kR}) ]^\ud . \eea At lowest order in the RPV corrections,
it is consistent to identify approximately the flavor mixing matrix
elements appearing in the right-hand side with the measured CKM matrix
elements, $V_{ud_j}^{SM} \simeq V_{u d_j} $.  This allowed us to
identify the quark flavor mixing parameters on the left-hand side with
the experimental values for the CKM matrix elements, implying the
approximate relation, $ \sum _{j=1,2,3} \vert V_{ud_j} ^{SM} \vert^2
\simeq 1$.

Only the RPV contribution to $V_{ud}$ would remain, upon invoking the
single coupling constant dominance hypothesis.  Using the experimental
values for the CKM matrix elements, giving $\sum _{j=1,2,3} \vert
V_{ud_j}\vert^2 = 0.9979 \pm 0.0021$, Barger et al.,~\cite{bargerg}
obtained the bound, $\vert \l _{12k} \vert <0.04 \ \tilde e_{kR} $ and
the $1\s $ level bound: $\vert \l ' _{11k} \vert <0.03 \ \tilde d_{kR}
$.  (The currently available fit to the quarks flavor mixing matrix
yields the updated value for the unitarity sum, $\sum _{j=1,2,3} \vert
V_{ud_j}\vert^2 = 0.9969 \pm 0.0022$, which is compatible with that
used above.)  It is also of interest to consider the contributions
involving the quadratic coupling constant products.  We have corrected
here the formula for the unitarity constraint used by Ledroit and
Sajot~\cite{ledroit} by including the dependence on the quadratic
coupling constant products in the off-diagonal matrix elements,
$V_{us}$ and $V_{ub}$.  Based on the same input information as used in
that work~\cite{ledroit}, we infer the following set of single and
quadratic coupling constant bounds: $\vert \l _{12k} \vert < 0.05 \
\tilde e_{kR}, \ \vert \l '_{11k} \vert < 0.02 \ \tilde d_{kR}, \
\vert \l ^{'\star }_{11k} \l '_{12k} \vert ^\ud < 0.04 \ \tilde
d_{kR}, \ \vert \l '_{11k} \l ^{'\star }_{13k} \vert ^\ud < 0.37 \
\tilde d_{kR} . $ To conclude, we note that a consideration of the
constraints on the unitarity sums, $\sum _{j=1}^3 \vert V _{cd_j}
\vert =1 $ or $\sum _{j=1}^3 \vert V _{cd_j} \vert =1 $, could be used
to set single coupling constant bounds for $ \l _{i2k} $ and $ \l
_{i3k} $.


\subsubsection{\bf Muon decay and  scattering processes} 

The related processes of muon leptonic three-body decay and
muon-lepton scattering provide sensitive probes of the Lorentz
covariant structure of the charged current interactions.  Precise
experimental data are available for the muon decay lifetime and the
energy and angular distributions of the emitted charged lepton.  The
deviations from the conventional $ V-A$ four fermion local couplings
for two pairs of neutrinos and charged leptons involve vector, scalar
and tensorial current couplings parameterized by the effective
Lagrangian, $ L _{EFF}= {4G_F\over \sqrt 2} \sum _{H,H'=(L,R)} \sum
_{F=S,V,T} g^F_{HH'} (\bar e _H \G _F \nu ) (\bar \nu \G _F \mu _{H'})
+ H. \ c. $, with the understanding that the $V$ coupling preserves
chirality and the $S,\ T$ couplings flip chirality.  The muon decay, $
\mu ^- \to e^- +\nu_\mu + \bar \nu _e $, double differential
distribution with respect to the emitted electron angle and energy, is
analyzed in terms of the Michel parameter, $\rho $, and three
additional parameters, $ \l , \ \xi , \ \delta $, given by known
functions of the parameters $ g^F_{HH'}$.  The measurements of
antineutrino-electron inelastic scattering cross sections, $\bar \nu_e
+ e^-\to \mu ^- + \bar \nu _\mu $, and of polarization observables in
the muon decay process, carried out in experiments at the Vancouver
TRIUMF laboratory and the Zurich PSI laboratory~\cite{gerber} aim at
sensitivities of order $10^{-4}$.  The review on muon physics by Kuno
and Okada~\cite{kuno01} covers both the high precision measurements of
the normal weak muon decay mode and the lepton flavor changing rare
decays.

 The tree level RPV contribution~\cite{cheung98} to the four lepton
interaction is part of the effective Lagrangian quoted in
eq.(\ref{eqx2}).
Focusing on the $ \G _S =1 $ Lorentz scalar vertex function, $g_{RR}
^S$,  Cheung and Zhang~\cite{cheung98} inferred by comparison with
the experimental limit, $\vert g_{RR} ^S \vert < 0.066 $, the
quadratic coupling constant bound: $ g_{RR} ^S= -{\sqrt 2 \over 4 G_F}
{ \vert \l _{131}\l ^\star _{232} \vert \over m_{\tilde \tau _L} ^2 }
\ \ \Longrightarrow \ \vert \l ^\star _{232} \l _{131} \vert <
2.2\times 10^{-2} \ \tilde \tau_L ^2 .$ Although the above bound is
weaker than that deduced by combining the individual coupling constant
bounds from the branching ratio, $ R_\tau $, namely, $ \l^{\star}
_{13k} \l_{23k} < 4.9 \times 10^{-3} \ \tilde e_{kR} ^2 $, it
nevertheless presents the advantage of being a more robust bound free
of unexpected cancellation effects.  The analogous comparison for the
other neutrino flavors in lepton number violating decays yields
similar bounds on the quadratic coupling constant products, $ \l
^{\star} _{ij1} \l _{2j2} / m^2 _{ \tilde \nu_{jL} } $.  The $ \tilde
e_{kR}$ exchange amplitude contributes to the vectorial $ \G _V =\g ^\mu $
four fermion coupling, $ g_{LL} ^V = -\sqrt 2 \vert \l_{12k}
\vert ^2 /(8 G_F m^2 _{ \tilde e_{kR} } ) $. The corresponding RPV
contribution is overwhelmed by the charged current $W$-boson exchange
contribution of same Lorentz structure, so that the current
experimental sensitivity does not warrant a comparison with the data
in this case.

\subsubsection{\bf  Rare semileptonic decays of heavy quark hadrons}

The transition amplitudes for the flavor changing neutral and charged
$D$ mesons semileptonic decay processes, $ D ^ + \to \bar K ^ 0 +e_i ^
+ + \nu _i ,\ D^+\to \bar K^ {0\star } +e_i^++\nu_i ,\ D^0 \to K^-
+e_i^++\nu_i , \ [e_i= e, \ \mu ;\ \nu_i = \nu_e,\ \nu_\mu ] $ have a
Lorentz structure involving pairs of form factors. For the purpose of
testing the lepton universality, one may circumvent the theoretical
uncertainties on these form factors by considering the ratio of decay
widths into electron and muon, respectively.  The ratios of decay
rates, $ R_{D ^ +} = \G ( D ^ + \to \bar K ^ 0 + \mu ^ + + \nu _\mu )
/ \G ( D ^ + \to \bar K ^ 0 + e ^ + + \nu _e ) $, are considered,
along with the similar ratios associated to the above two other decay
modes, $R_{D ^+} ^ \star , \ R_{D ^ 0} $.  The phase space dependence
on the final lepton mass is taken into account by setting the SM
predictions at the values: $(R_{D^+})^{SM} = (R_{D^+}^{\star } ) ^{
SM} = (R_{D^0})^{SM} = 1/1.03$.  The experimental data for the $D$
mesons branching fractions gives for these ratios~\cite{bhattachoud}:
$R_{D ^ +} = 1.06 {+ 0.48\choose -0.34} ,\ R_{D ^ +} ^ \star = (0.94
\pm 0.16) ,\ R_{D ^ 0} = (0.84 \pm 0.12) .$ For completeness, we also
quote the experimental values found by Ledroit and
Sajot~\cite{ledroit}: $ R_{D ^ +} = (1.2 \pm 0.6) , \ R_{D ^ +} ^
\star = (1.09 \pm 0.17) , \ R_{D ^ 0} = (0.933 \pm 0.085) $.  The RPV
effective Lagrangian for the process, $ D^+ (c\bar d) \to \bar K^0
(s\bar d) + \mu ^+ +\bar \nu _\mu $, as abstracted from
eq.(\ref{eqx2}), and the associated contributions to the $\mu ^+ /
e^+$ ratios of the $D$ mesons partial widths, read
as~\cite{altarelli}, \bea && L_{EFF} = {\vert \l ' _{22k} \vert
^2\over m^2 _{\tilde d_{kR} } } (\bar e ^c _{2R} u_{2L} ) (\bar d
_{2L} \nu ^c _{2R}) + H. \ c. \ \Longrightarrow \ { R_{D^+} \over
(R_{D^+} ) ^{SM} } = { R_{D^+}^\star \over (R^{\star } ) ^ {SM} _{D^+}
} = { R_{D^0} \over (R_{D^0}) ^{SM} } = {\vert 1+r'_{22k}(\tilde d_{k
R}) \vert ^2 \over \vert 1+r ' _{12k}(\tilde d_{kR} ) \vert ^2 }. \eea
The comparison with the experimental values leads to the single
coupling constant bounds:
\begin{eqnarray}
\vert \l '_{12k}\vert _{k=1,3} &<& 0.44 \ (0.34)\ \tilde d_{kR} , \
\vert \l '_{22k}\vert < 0.61 \ (0.39)\ \tilde d_{kR}, \ [R_{D^+}: \
D^+\to \bar K^0 ] ; \cr \vert \l '_{12k}\vert _{k=1,3} &<& 0.23 \
(0.29)\ \tilde d_{kR} , \ \vert \l '_{22k}\vert < 0.38\ (0.18) \
\tilde d_{kR}, \ [R^\star _{D^+}: \ D^+\to \bar K^{0 \star } ];\cr
\vert \l '_{12k} \vert _{k=1,3}\vert &<& 0.27 \ (0.34)\ \tilde d_{kR}
, \ \vert \l '_{22k}\vert < 0.21 \ \tilde d_{kR}, \ [R_{D^0}: \ D^0\to
K^-].
\label{eqx4b}
\end{eqnarray}
To illustrate the sensitivity with respect to the experimental
uncertainties, we have quoted first the $2 \s $ bounds obtained by
Ledroit and Sajot~\cite{ledroit} and next, by the numbers inside
parentheses, the $1 \s $ bounds obtained by Bhattacharyya and
Choudhury~\cite{bhattachoud}.

The flavor changing semileptonic inclusive decays of $B$ mesons, $ B^-
\to X_q +\tau ^-+\bar \nu _\tau $, acquire tree level contributions
from the gauge and RPV interactions. The latter corrections are
controlled by a single coupling constant if one uses the
representation of the RPV superpotential incorporating the up-quarks
flavor mixing.  The complete effective Lagrangian for the transition
amplitude to charmed mesons can be expressed as,
$ L_{EFF} = -V_{cb} [{4 G_F \over \sqrt 2} + { \vert \l ^{'} _{33k}
\vert ^2 \over 2 m^2_{\tilde d_{kR} } } ] (\bar c_L \g_\mu b_L) (\bar
\tau _L \g^\mu \nu_{\tau L} ) + H. \ c.  $ The comparison with the
experimental limit is performed in two recent works: Grossman et
al.,~\cite{grossman} find the coupling constant bound, $\vert \l
'_{33k}\vert < 0.12 \ \tilde d_{kR} $, whereas Erler et
al.,~\cite{erler} find, $\vert \l'_{33k} \vert < 0.32 \ \tilde d_{kR}
$. The mismatch between these two predictions reflects the model
dependence associated with the hadronic physics matrix element.

The RPV b-quark decay mode~\cite{raychaud96}, $ b \to c +e ^- + \tilde
\g $, controlled by the coupling constant $ \l '_{123}$, can initiate
the $B$ meson exotic decay mode, $B ^0 \to D ^+ + e^-+ \tilde \g $, in
the case of light photinos with masses in the range, $ m_{\tilde \g }
< 3 $ GeV.  The kinematic configuration of final states for this decay
channel is easily discernible from that characterizing the normal
charged current decay channel, $B ^0 \to D ^+ + e_i ^- +\bar \nu _i $.

The strangeness changing $\D S=1$ semileptonic three-body decay modes
of hyperons, $ \L \to p + e_i^- + \bar \nu_i , \ [e_i=e, \mu ] $ give
access to useful information on quadratic products of the $\l ' $
interactions. From the comparison with the measured decay widths,
Tahir et al.,~\cite{tahir} obtain the following $2\s $ level coupling
constant bounds,
\begin{eqnarray}
 \vert \l ^{'\star} _{11k} \l '_{12k} \vert &<& [1.3 \times 10^{-1}\
\tilde d^2_{kR}, \ 5.3 \times 10^{-3} \ \tilde d^2_{kR}] , \ [ \L \to
p+e^-+\bar \nu _e , \ \L \to p+\mu ^-+\bar \nu _\mu ] \cr \vert \l
^{'\star} _{11k} \l '_{12k} \vert &<& [ 8.5 \times 10^{-2}\ \tilde
d^2_{kR},\ 1.6 \times 10^{-2} \ \tilde d^2_{kR}] ,\ [\S ^- \to
n+e^-+\bar \nu _e , \ \S ^- \to n +\mu ^-+\bar \nu _\mu ] \cr \vert \l
^{'\star} _{11k} \l '_{12k} \vert &<& [1.2 \times 10^{-1} \ \tilde
d^2_{kR},\ 5.0 \times 10^{-2} \ \tilde d^2_{kR}],\ [\Xi ^- \to
\L+e^-+\bar \nu _e , \ \Xi ^- \to \L+\mu ^-+\bar \nu _\mu ]
\label{eqx4b1}
\end{eqnarray}
where, as indicated above, the two entries for each quadratic coupling
constant product type refer to the cases involving an electron and
muon in the final state.

\subsubsection{\bf Tests of $ e- \mu  - \tau $ lepton universality
in leptons and hadrons decays}

The tau-lepton three-body leptonic decays, $\tau^- \to e_j^- +\nu_\tau
+\bar \nu _j , \ [e_j= e, \ \mu ] $, provide useful probes of the
charged currents lepton universality.  Model independent tests can be
sought by comparing the ratios of decay widths for electron and muon
emission.  The $\tilde e_R$ tree level exchange RPV contributions to
the ratios of $\tau $ or $\mu $ leptons decay widths to final electron
and muon are given by~\cite{bargerg},
\begin{eqnarray}   &&
R_\tau \equiv {\G (\tau ^- \to e^- +\bar \nu_e +\nu_\tau ) \over \G
(\tau ^- \to \mu ^- +\bar \nu_\mu +\nu_\tau ) } = R_\tau ^{SM} [1+2
(r_{13k}(\tilde e_{kR}) -r_{23k} (\tilde e_{kR}))], \cr &&
R_{\tau \mu } \equiv {\G (\tau ^- \to \mu ^- +\bar \nu_\mu +\nu _\tau
) \over \G (\mu ^- \to e ^- +\bar \nu _e +\nu _\mu ) } = R_{\tau \mu }
^{SM} [1+2 (r_{23k}(\tilde e_{kR}) -r_{12k}(\tilde e_{kR}))].
\label{eq5p}
\end{eqnarray}
The comparison with the experimental results for these ratios,
$R_{\tau \mu } /R^{SM}_{\tau \mu } = (0.9987 \pm 0.0073) ,\ R_{\tau }
/R^{SM}_{\tau } = (1.0007 \pm 0.007) $, yields at the $2\s $ level the
coupling constant bounds~\cite{ledroit}, $ \vert \l_{13k} \vert < 7.\
\times 10^{-2} \ \tilde e_{kR}, \ \vert \l_{23k} \vert < 7.\ \times
10^{-2} \ \tilde e_{kR}\ [R_\tau ] ; \quad \vert \l_{23k} \vert < 7.\
\times 10^{-2} \ \tilde e_{kR}, \ \vert \l_{12k}\vert < 7.\ \times
10^{-2}\ \tilde e_{kR}\ [R_{\tau \mu }] . $

The RPV interactions also contribute to the local four fermion
operators built from quark and lepton pairs, $(\bar q q)(\bar l l) $,
which compete with the weak charged current contribution to the
familiar two-body leptonic weak decay modes of the charged $\pi
$-mesons, $ \pi^- \to \mu ^-+\bar \nu_\mu $, controlled by the pion
decay coupling constant, $F_\pi $.  A useful constraint on the
deviations to lepton universality, free from the experimental
uncertainties on $F_\pi $,
is found by considering the ratios of decay rates for electron and
muon emission~\cite{bargerg},
\begin{eqnarray} &&
R_\pi \equiv {\G (\pi ^-\to e^- +\bar \nu_e ) \over \G (\pi ^-\to \mu
^- +\bar \nu_\mu ) }= R_\pi ^{SM} [1+{2\over V_{ud}} (r'_{11k}(\tilde
d_{kR}) -r'_{21k}(\tilde d_{kR}))].
\label{eq4}
\end{eqnarray} 
The comparison between the experimental measurements and the SM result
for the ratios of partial decay widths, $R_{\pi } /R^{SM}_{\pi } =
1.230 \pm 0.004$, leads to the single coupling constant bounds, $
\vert \l '_{21k} \vert < 6. \times 10^{-2} \ \tilde d_{kR}, \ \vert \l
'_{11k}\vert < 3. \times 10^{-2} \ \tilde d_{kR} . $ Another useful
test of the lepton universality~\cite{bhattachoud}, free from the
uncertain hadronic physics inputs, is offered by the ratio between the
crossing symmetry related reactions of charged pion and tau-lepton
two-body decays,
\begin{eqnarray} R_{\tau \pi } &=& { \G (\tau ^- \to \pi ^- + \nu
_\tau ) \over \G (\pi ^-\to \mu ^- +\bar \nu_\mu ) } = R_{\tau \pi }
^{SM} { \vert V_{ud} + r'_{31k} (\tilde d_{kR}) \vert ^2 \over \vert
V_{ud} + r'_{21k} (\tilde d_{kR}) \vert ^2 } .
\label{eqx6}  \end{eqnarray} 
The comparison with experimental results for the ratio of decay
widths, $R_{\tau \pi } /R^{SM}_{\tau \pi } = 1.014 \pm 0.018 $, leads
to the single coupling constant bounds~\cite{ledroit}, $ \vert \l
'_{31k} \vert < 0.12\ \tilde d_{kR} ,\ \vert \l '_{21k} \vert < 0.08\
\tilde d_{kR} . $ The independent study by Kim et al.,~\cite{kim}
yields the bound: $ \vert \l '_{31k}\vert < 0.16 \ \tilde d_{kR}$.

The two-body leptonic weak decays of charged $D$ mesons, $ D^+_s
(c\bar s) \to e ^+ _i +\nu _j $, provide a useful probe of the lepton
universality through the comparison of decay rates for different final
lepton flavors. The RPV contributions for the $\tau $ and $\mu $
lepton emission modes predict a deviation from unity for the
corresponding ratio of branching fractions given by \bea && R_ {D_s}
(\tau \mu ) \equiv {B(D_s ^+\to \tau ^+ +\nu_\tau ) \over B(D_s ^+ \to
\mu ^+ +\nu_\mu ) } = { \vert V_{cs} + r'_{32k} (\tilde d_{kR}) \vert
^2 \over \vert V_{cs} + r'_{22k}(\tilde d_{kR}) \vert ^2 } , \eea
where we have suppressed the ratio of phase space factors, $ m_i ^ 2
(1-m_i ^ 2 /m_{D_s} ^ 2) ^2 $ for the emitted charged lepton mass, $
m_i $, in the reactions, $ D_s ^+ \to e^+ _i + \nu _ i $.  The
comparison with the experimental result yields the following set of
individual coupling constant bounds~\cite{ledroit}: $ R_ {D_s} (\tau
\mu )= (7.4\pm 3.7) \times 10^{-2} /(8.8\pm 3.9) \times 10^{-3} \ \
\Longrightarrow \ \vert \l '_{22k}\vert < 0.65 \ \tilde d_{kR},\ \vert
\l '_{32k}\vert < 0.52\ \tilde d_{kR}. $

\subsection{Neutral current  interactions} 
\label{secxxx2b}

\subsubsection{\bf Neutrino-matter elastic and deep inelastic scattering}


The neutrino elastic scattering cross section measurements with
leptonic targets are collected for $\nu_\mu $ and $ \bar \nu_\mu $
beams from the CHARM II Collaboration~\cite{charm} at Cern. The
neutrino deep inelastic scattering cross section measurements with
hadronic targets are collected by the CDHS and CHARM Collaboration at
Cern and the CCFR Collaboration at Fermilab~\cite{ccfr}.  The
experimental information~\cite{panman,perrierlanga} is described in
terms of the ratios of reaction rates for neutrino and antineutrino
beams or ratios of neutral current (NC) to charged current (CC)
contributions.  At energies below $m_Z$, the neutral current couplings
are described by the effective Lagrangian,
\begin{eqnarray} &&
L _{EFF}=- {4G_{F} \over \sqrt 2} \bar \nu_L\g_\mu \nu_L \bigg [
\sum_{f=e, \mu } g_L^{\nu f } \bar f_L\g^\mu f_L+ g_R^{\nu f } \bar
f_R\g^\mu f_R + \sum_{f=u,d} \e _L( f ) \bar f_L\g^\mu f_L + \e _R (f)
\bar f_R\g^\mu f_R \bigg ] . \label{eq7}
\end{eqnarray}
The RPV contributions to neutrino elastic scattering arise at tree
level order in terms of the Feynman diagrams C.1-2 in
Figure~\ref{figun}. The combined SM and RPV contributions to the
$Z$-boson vertex functions are described by the
formulas~\cite{bargerg}:
\begin{eqnarray} \bullet \quad && 
g^{\nu e} _L=(-\ud +x_W)(1-r_{12k}(\tilde e_{kR}) )-r_{12k}(\tilde
e_{kR}), \quad g^{\nu e} _R=x_W(1-r_{12k}(\tilde e_{kR}) )+
r_{211}(\tilde e_{1L}) +r_{231}(\tilde e_{3L}), \cr \bullet \quad &&
\e _L(d)=(-\ud +{1\over 3} x_W)(1-r_{12k} (\tilde e_{kR})
)-r'_{21k}(\tilde d_{kR}), \quad \e _R (d) ={x_W\over
3}(1-r_{12k}(\tilde e_{kR}) )+r'_{2j1}(\tilde d_{jL}), \cr \bullet
\quad && \e _L (u) = (\ud - { 2 \over 3 } x_ W) (1-r_{12k}(\tilde
e_{kR}) ) , \ \e _R (u) =- { 2 \over 3 } x_ W (1-r_{12k}(\tilde
e_{kR}) ) , \label{eqs7} \end{eqnarray} where $ x_W =\sin ^2 \t _W $,
the factors $ (1-r_{12k}(\tilde e_{kR}) )$ account for the Fermi
coupling constant redefinition and we note the absence of direct RPV
corrections for up-quarks.  The comparison with the CHARM II
Collaboration~\cite{charm} experimental results for the ratio of $\nu
_\mu $ to $\bar \nu _\mu $ cross sections, $ R_{\nu _\mu \bar \nu _\mu
}$, yields the coupling constant bounds~\cite{ledroit}: $ \vert
\l_{12k} \vert < 0.14 \ \tilde e_{kR}, \ \vert \l_{231} \vert < 0.11 \
\tilde \tau _{L}, \ \vert \l_{121}\vert < 0.13 \ \tilde e_{1L}. $ A
direct experimental determination of the vertex parameters, $ \e
_{L,R} ({q}) $ is also available through fits to the data from the
CDHS and CCFR Collaborations~\cite{ccfr}.  The comparison using the
tree and one-loop level SM contributions leads to the coupling
constant bounds~\cite{ledroit}: $ \vert \l_{12k}\vert < 0.13 \ \tilde
e_{kR}, \ \vert \l'_{21k}\vert < 0.15 \ \tilde d_{kR}, \ \vert
\l'_{2j1} \vert < 0.18 \ \tilde d_{jL}. $

The elastic scattering of $\nu_\mu $ and $\bar \nu_\mu $ on a proton
target also provide useful, accurate measurements of the weak
interaction angle~\cite{mannlanga,bargerz}.  Despite the high
sensitivity of the experimental data on $ \sin \t_W$, the
uncertainties from the axial nucleon form factor and the strange quark
partonic component in the nucleon wave function have precluded, so
far, a model independent analysis of the constraints on the RPV
interactions.

\subsubsection{\bf Fermion-antifermion pair production and 
$Z$-boson pole observables}
\label{confcc} 

\begin{figure} [ht]
\begin{center}
\leavevmode
\caption{\footnotesize \it Feynman diagrams {\bf A.1-A.3, B1.1-B.4, C.1-C.4, D.1-D.3}  
describing the perturbation theory
contributions to a selected set of processes initiated by the RPV
interactions.  The matter particle lines are represented by oriented,
labeled lines which propagate in time from left to right.  (This is a
useful convention which dispenses us from drawing the orientation
arrows.)  Scalar, fermion and vector particles are drawn as dashed,
continuous and wiggly lines.  A black dot for a particle ending in
vacuum signifies a field tadpole or VEV. A black dot or cross
insertion on a scalar particle line stands for a $ L-R$ chirality
mixing mass term.  A cross insertion on a fermion line stands for a
chirality flip or field mixing mass term.  We only display a
representative diagram for each given physical process, omitting the
other diagrams with different time orderings.  {\bf Diagrams} A: Majorana neutrino mass and field mixing terms at
tree level (A.1) and one-loop level (A.2).  Neutralino three-body
decay via $\tchi - \nu $ mixing (A.3).  
{\bf Diagrams} B: Interactions
of charged current type in decays of leptons and quarks (B.1-2) and in
semileptonic and leptonic decays of mesons (B.3-4).  
{\bf Diagrams} C:
Interactions of neutral current type in neutrino $\nu_\mu $-lepton
scattering (C.1-2) and lepton pair annihilation into fermion pairs
(C.3-4).  
{\bf Diagrams} D: Amplitudes at one-loop order for the
decays of $Z$ gauge boson into fermion pairs. }   
\vskip 1 cm \centerline
{\epsfig{figure=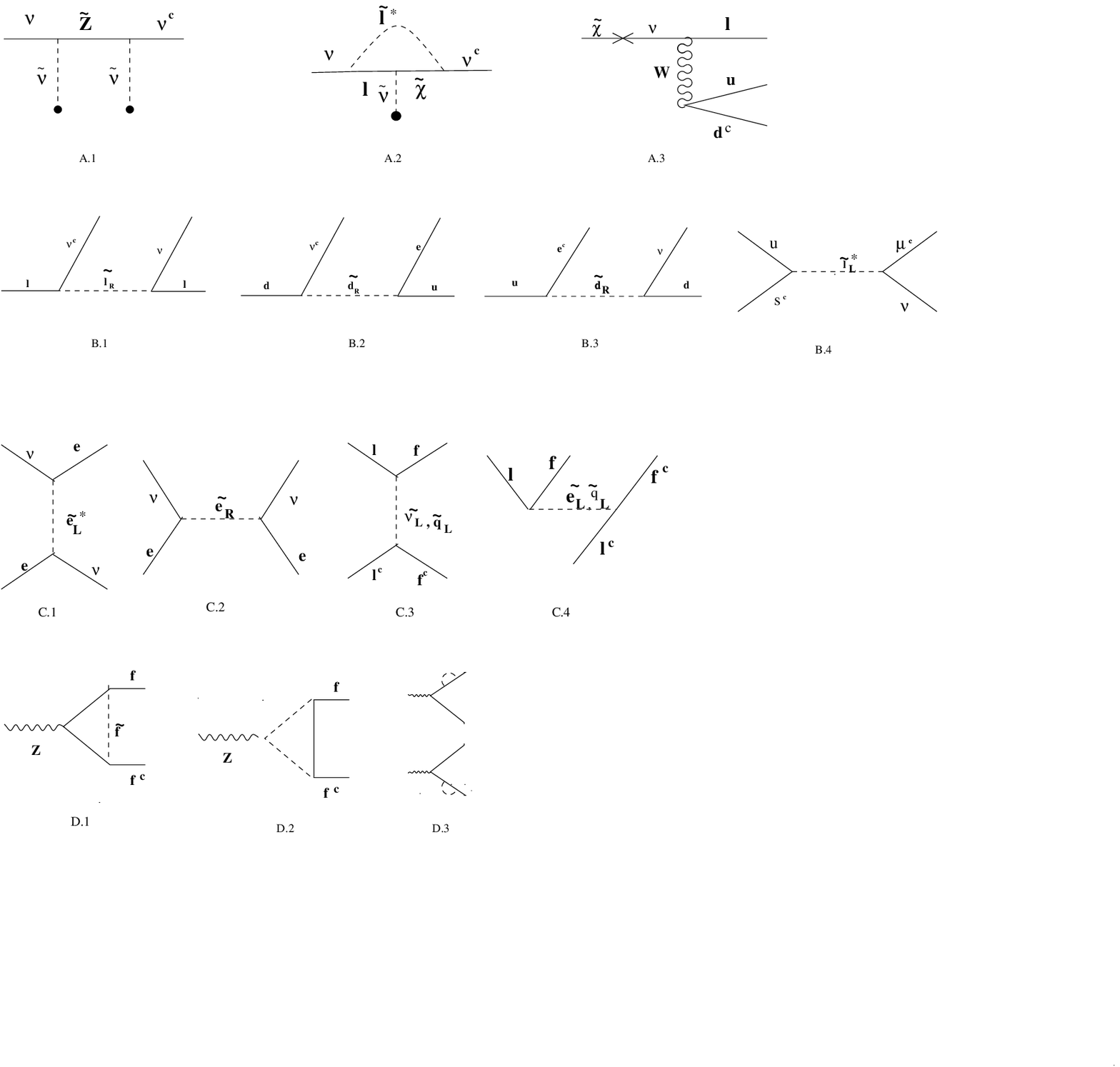,height=24cm,width=19cm} }
\label{figun}
\end{center}
\end{figure}

The experimental measurements of the fermion pair production reactions
at the leptonic colliders, $e^++e^-\to f +\bar f, \ [f=l,q] $ are
available over a wide range of initial energies~\cite{panman}.  The
neutral current couplings are encoded in the low energy effective
Lagrangian, $ L _{EFF} = - {4G_{F} \over \sqrt 2}\sum_{f=l, q} [g_L^e
\bar e_L \g ^\mu e_L + g_R^e \bar e_R \g ^\mu e_R ] [g_L^f \bar f_L \g
^\mu f_L + g_R^f \bar f_R \g ^\mu f_R ] $.
The measurements for $Z$-boson resonance production, $e^++e^-\to Z^0
\to f +\bar f$, collected by the experimental collaborations at the
LEP and SLD colliders, provide a wealth of high precision experimental
data.  The $Z$-boson pole observables include the total resonance
formation and hadronic cross sections, $\s _Z ,\ \s _{had} $, the
partial decay widths into fermion pairs, $\G_f , \ [f=l, h, c,b] $ the
leptonic and b-quark decays branching ratios, $R_l^Z , \ R_b ^Z $, the
forward-backward rate asymmetry, $ A^f _{FB}$, and the polarization
asymmetry for final and initial fermions, $ \calp _f ,\ A_{LR} $.  For
illustration, we write the formulas expressing the $Z$-boson pole
forward-backward asymmetry in the production of fermion-antifermion
pairs as a function of the neutral current vertex parameters, $
A^{l,q} _{FB} = {3\over 4} \cala ^e \cala ^{l,q} , \ \cala ^f=
{2g_V^fg_A^f\over \vert g_V^{f}\vert ^2 + \vert f_A^{f}\vert ^2 }
={\vert g_L^{f}\vert ^2 - \vert g_R ^{f}\vert ^2 \over \vert
g_L^{f2}\vert +\vert g_R ^{f}\vert ^2 } , $ where, $ g_{L,R} = (g_V
\pm g_A )/ 2 , \ (\cala ^f)_{SM} = -T_{3L}(f),\ [f=l,q].$ The tree
level RPV contributions, represented by the Feynman diagrams C.3-4 in
Figure~\ref{figun}, are described by the simple general formula, $
{(\cala ^e \cala ^f)^{SM} / \cala ^e \cala ^f } =\vert 1 + r_{ijk}
(\tilde f) \vert ^2, $ where $\tilde f$ is some appropriate
superpartner. The combined gauge and RPV contributions to the vertex
functions, $\cala ^f $, and the coupling constant bounds deduced in
the comparison with $Z$-boson pole experimental data are summarized by
the following formulas~\cite{bargerg,ledroit}:
\begin{eqnarray}
\cala ^e \cala ^e &=& {1\over 4} -\ud r_{ijk}(\tilde \nu_{iL} )\ \
 \Longrightarrow \ [ \vert \l_{ijk} \vert < 0.37 \ \tilde \nu _{iL}; \
 (ijk)= (121) (131)] \cr \cala ^e \cala ^\mu &=&{1\over 4} -\ud
 r_{ijk}(\tilde \nu_{iL} ) \ \ \Longrightarrow \ [ \vert \l_{ijk}
 \vert < 0.25 \ \tilde \nu _{iL}; (ijk)= (122), (132), (121), (321) ]
 \cr \cala ^e \cala ^\tau &=& {1\over 4} -\ud r_{ijk}(\tilde \nu_{iL}
 ) \ \ \Longrightarrow \ [\vert \l_{ijk}\vert < 0.11 \ \tilde \nu
 _{iL}; (ijk)= (213), (313), (131), (231) ] \cr \cala ^e \cala
 ^{u_{j=2} } &=& \cala ^e \cala ^{c} =-{1\over 4} -\ud r'_{1jk}(\tilde
 d_{kR} ) \ \Longrightarrow \ [\vert \l '_{12k} \vert < 0.21 \ \tilde
 d_{kR} ] \cr \cala ^e \cala ^{d_{k=2,3 } } & = & \cala ^e \cala ^{s,
 b} = {1\over 4} -\ud r'_{1jk}(\tilde q_{jL} ) \ \Longrightarrow \
 [\vert \l ' _{1j2} \vert < 0.28 \ \tilde u_{jL} , \ \vert \l ' _{1j3}
 \vert < 0.18 \ \tilde q_{jL}] \label{eq9}
\end{eqnarray}
where we have listed the various coupling constant bounds in
correspondence with the relevant configurations of the quarks and
leptons generations.
The available high precision measurements for the partial widths of
$Z$-boson decays into fermion-antifermion pairs, $Z\to f +\bar f , \
[f=q,l] $ justify pursuing the examination of RPV contributions up to
the one-loop order.  The $Z$-boson decay amplitudes into
lepton-antilepton and $b -\bar b$ quark pairs, $R_l^Z, \ R_b ^Z $, are
described by the Feynman diagrams, propagating internal lines of
fermions and sfermions, displayed by the graphs D.1-3 in
Figure~\ref{figun}.  The corrections to these branching ratios are
expressed by the formulas:
\begin{eqnarray}  &&
\d R_l \equiv {R_l^Z - (R_l^Z )_{SM} } \simeq - (R_l^Z )_{SM} \D_l
+(R_l^Z )_{SM} (R_b^Z)_{SM} \D_b, \ [R_l ^Z= {\G _h^Z \over \G _l^Z }
,\quad R_b ^Z= {\G _b^Z \over \G _h^Z } ] \cr && \ \d R_b = {\G _b^Z
\over \G _h^Z } - (R_b^Z)_{SM} \simeq (R_b^Z)_{SM} \D_b (1- \D_b { (\G
_b ^Z)_{SM} \over (\G _h ^Z)_{SM} }), \ [\D_f ={ \G (Z ^0\to f+\bar f)
\over \G_{SM} (Z^0 \to f+\bar f) } -1].
\label{eqx20}
\end{eqnarray}
The study by Bhattacharyya et al.,~\cite{ellishar}, focuses on the
dominant contributions from one-loop diagrams with an internal top
quark.  The lepton-antilepton pair decay channel, $Z ^0 \to e_J + \bar
e_J$, involves the coupling constant products, $\vert \l ' _{J3k}
\vert ^2 $, and the down quark-antiquark pair decay channel, $Z ^0 \to
d_J + \bar d_{J}$, the coupling constant products, $\vert \l ' _{i3J}
\vert ^2 $.  The comparison with the LEP-I experimental results for
$R_l^Z, \ [l=e, \ \mu , \ \tau ]$, leads to the coupling constant
bounds~\cite{ellishar}, $ \vert \l '_{13k}\vert < 0.63, \ \vert \l
'_{23k}\vert < 0.56, \ \vert \l '_{33k}\vert < 0.45, $ at $ 2\s $
level, and $ \vert \l '_{13k}\vert < 0.51, \ \vert \l '_{23k}\vert <
0.44, \ \vert \l '_{33k}\vert < 0.26 $, at $ 1\s $ level. The
comparison between the quoted $ 2\s $ and $ 1\s $ bounds illustrates
the accuracy that might be gained by improving the statistics.

The $\l ''$ interactions contribute directly at one-loop level to the
$Z$-boson decay branching fractions into bottom quark-antiquark pairs,
$ R^Z_b $, in terms of similar Feynman diagrams to those discussed
above. An indirect contribution also arises from $\l ''$ interactions
to the leptonic branching fraction, $R ^Z_l $.  The comparison with
LEP-I data for $R ^Z_l$ leads at $1\s \ (2\s )$ level to the coupling
constant bounds~\cite{sridhar}, $[\vert \l ''_{313}\vert , \ \vert \l
''_{323}\vert ] < 0.97 \ (1.46) , \ \vert \l ''_{312}\vert < 0.96 \
(1.45) $.  The analogous comparison with the experimental $Z$-boson
decay branching fraction, $ R^Z_b $, yields the weaker coupling
constant bounds, $ \vert \l ''_{312}\vert < 4.28 $ at the $1\s $
level, and $[\vert \l''_{313}\vert , \ \vert \l ''_{323}\vert ] <
1.89, \ \vert \l ''_{312} \vert < 5.45$ at the $3\s $ level.

A partially global statistical fit to the $Z$ gauge boson pole
hadronic observables, with theoretical predictions including the
one-loop MSSM and RPV contributions, is presented by Lebedev et
al.,~\cite{lebedev99}. The strongest constraints arise from the
$b$-quark asymmetry parameters, $ \cala ^b , \ A_{FB} ^b$.  The
comparison indicates that certain RPV coupling constants have
contributions of opposite sign to that required by experiment,
implying that these coupling constants must be set to zero.  As a
result, the set of coupling constants $ \l '_{i31}, \ \l '_{i32}, \ \l
''_{321} $ is ruled out at the $1 \s $ level and the set of coupling
constants $\l '_{i33}, \ \l ''_{331} $ is ruled out at at the $2 \s $
level.

The $Z$-boson decay modes into flavor non-diagonal configurations of
fermion-antifermion pairs, $ Z ^0 \to f_J+\bar f_{J'} ,\ [J\ne J']$,
with emitted leptons, down-quarks or up-quarks pairs, provide a useful
probe of the flavor changing effects at the high energy colliders.
The RPV contributions to the flavor off-diagonal $Z$-boson decay
 branching fractions, $ B_{JJ'} (f) \equiv {\G (Z\to f_J+\bar f_{J'}
 )+\G (Z\to f_{J'} +\bar f_J ) \over \G (Z \to \ \text{all}) } $,
 arise at one-loop order through the Feynman diagrams propagating
 pairs of fermion and sfermion internal lines which are displayed by
 the graphs D.1-3 in Figure~\ref{figun}.  A detailed discussion of the
 one-loop amplitudes is provided in Refs.~\cite{chemtobm}.  The
 leptonic $Z$-boson decay modes, $Z\to e_J +\bar e_{J'}, \ [ J\ne J'
 ]$ are examined by Anwar-Mughal et al.,~\cite{anwar}, by restricting
 consideration to the predominant loop contributions propagating top
 quarks.  The current experimental limits~\cite{wolftrippe} on the
 off-diagonal lepton pairs rates set bounds on several quadratic
 coupling constant products, $\l ' _{ijk} \l ^{'\star } _{i'j'k'} $,
 involving one or more third generation indices.  The deduced bounds
 are in general compatible with the existing ones.  We quote some
 illustrative quadratic coupling constant bounds: \bea \sum_k \vert \l
 ^{'\star} _{13k} \l '_{23k} \vert &<& 6.5 \ \times 10^{-2}, \ [ Z^0
 \to e ^\pm + \mu ^\mp ]; \quad \sum_k \vert \l ^{'\star} _{33k} \l
 '_{13k} \vert < 2.0 \ \times 10^{-1}, \ [ Z^0 \to e ^\pm +\tau ^\mp
 ]; \cr \sum_k \vert \l ^{'\star} _{23k} \l '_{33k} \vert &<& 1.6 \
 \times 10^{-1}, \ [ Z^0 \to \mu ^\pm + \tau ^\mp ] .\eea A systematic
 study of the RPV contributions to the $Z$-boson pole and off-pole
 flavor changing decay modes into quarks and leptons
 fermion-antifermion pairs is reported by Chemtob and
 Moreau~\cite{chemtobm}. The predicted branching fractions for the
 flavor off-diagonal lepton-antilepton pair decay channels along with
 the inferred quadratic coupling constant bounds are given by \bea &&
 B_{JJ'} (l) \simeq 4. \times 10^{-9} \ \bigg ( { \vert \l_{ijJ} \l
 _{ijJ'}^\star \vert \over 0.01 }\bigg )^2\ \Longrightarrow \ \vert
 \l_{ijJ}\l_{ijJ'}^\star \vert < [0.46,\ 1.1, \ 1.4 ] ,\cr &&
B_{JJ'} (l) \simeq 1.17 \times 10^{-7} \ \bigg ({ \vert \l ^{'\star }
 _{Jjk} \l'_{J'jk} \vert \over 0.01 }\bigg )^2 \ \Longrightarrow \
 \vert \l ^{'\star } _{Jjk} \l'_{J'jk} \vert < [0.38\times 10^{-1} ,\
 0.91\times 10^{-1} , \ 1.2\times 10^{-1} ] , \eea
holding for all fixed choices of the family indices, $(i, \ j)$ or
$(j,\ k)$, where the three entries are in correspondence with the
values of the family indices $[(J \ J')= (12),\ (23),\ (13)]$.  The
stronger bounds found for the $\l '$ coupling constants are due to the
extra color factor $N_c$ in the transition amplitudes and to the
possibility of accommodating an internal top quark line in the loop
contributions.

The joint effect on the flavor off-diagonal $Z$-boson decays, $Z \to
f_J + \bar f_{J'}$, of the trilinear coupling $\l '_{ijk} $ and the
bilinear supersymmetry breaking parameter, $\mu_{u3} ^2 = B_3 \mu _3
m_{\tilde G}$, is discussed by Atwood et al.,~\cite{atwood02}.  The
relevant auxiliary parameter, $B_3 \mu _3 /B _\mu \mu \simeq B_3
m_{\tilde G} \mu _3 /(m_A ^2 \cot \b ) $, controls the field mixing of
sleptons with Higgs bosons and can be monitored through its
contribution to the $ \nu _\tau $ mass. With the assignment, $\l
'_{332} = O(1)$, the predicted flavor changing branching fractions,
$B( Z ^0 \to b + \bar s) = O(10 ^{-6})$, significantly exceed those
quoted just above for the trilinear couplings alone.  The RPV physics
contributions are found to compete favorably with those expected from
the other new physics options.

\subsubsection{\bf Atomic physics parity violation}

The atomic physics parity violation (APV) experiments measure the
polarization vector optical rotation for the circularly polarized
light ray emitted in the atomic transitions, $ \vert nS > \to \vert n'
S >$.  The configuration mixing of atomic states is induced by the
parity violating quark-electron contact interaction Hamiltonian, $
H_{PNC} = {G_{F} Q_W \over 2 \sqrt 2 m_e } \sum _i \d ^3 (\vec x_i)
\vec \s _i \cdot \vec p_i , \ [Q_W= -2\big ( (A+Z) C_1(u)+(2A-Z)
C_1(d)\big ) ], $ with the flavor diagonal vertex functions, $ C_{1}
(q),\ C_{2} (q)$, described by the four fermion effective Lagrangian,
$ L _{EFF} = {G_{F} \over \sqrt 2} \sum_{q=u,d} C_1(q) (\bar e \g_\mu
\g_5 e) (\bar q \g^\mu q) +C_2(q) (\bar e \g_\mu e) (\bar q \g^\mu
\g_5 q). $ The reference experimental case is furnished by the $^
{133} _{55}\text{Cs} $ atomic transition~\cite{wood,woodlanga}, while the
experimental results for atomic physics transitions in $^{204} _{81}\text{Tl}
$ atom are already available and those for the $ ^1_1 H $ atom are
currently under study.  The combined contributions from gauge and RPV
interactions are described by the formulas~\cite{bargerg}:
\begin{eqnarray} C_1(u)&=&(-\ud +{4\over 3} x_W )(1-r_{12k}(\tilde
e_{kR})) -r'_{11k}(\tilde d_{kR}), \quad C_2(u)=(-\ud +2x_W
)(1-r_{12k}(\tilde e_{kR}))-r'_{11k}(\tilde d_{kR}); \cr C_1(d)&=&(\ud
-{2\over 3} x_W )(1-r_{12k}(\tilde e_{kR})) +r'_{1j1}(\tilde
q_{jL}),\quad C_2(d)=(\ud -2x_W )(1-r_{12k}(\tilde e_{kR}))
-r'_{1j1}(\tilde q_{jL}).
\label{eq10}
\end{eqnarray}
The comparison between the experimental data for the $^{133} _{55} \text{Cs}
$ atom and the SM loop improved predictions~\cite{erler98}, yields the
single coupling constant bounds~\cite{ledroit}: $ C_1(u):\ \vert
\l_{12k} \vert < 0.48 \ \tilde e_{kR}, \ \vert \l '_{11k} \vert < 0.28
\ \tilde d_{kR} ; \ C_1(d):\ \vert \l_{12k}\vert < 0.35 \ \tilde
e_{kR}, \ \vert \l '_{1j1}\vert < 0.26 \ \tilde q_{jL}. $ Focusing
instead on the discrepancies in the weak charge observable, $\d
(Q_W)\equiv Q_W^{exp}-Q_W^{SM}$, the comparison with experiment for
the $ _{55} ^{133}\text{Cs} $ and $_{81}^{204}\text{Tl} $ atoms can be expressed as:
\bea \d (Q_W(\text{Cs})) &=& 0.710 \pm (0.25)_{stat} \pm (0.80)_{syst} = -2
[72.07 r_{12k}(\tilde e_{kR}) +376 r'_{11k}(\tilde d_{kR}) -422r
'_{1j1}(\tilde q_{jL}) ], \cr \d (Q_W(\text{Tl})) &=& 1.9 \pm (1.2)_{stat}
\pm (3.4)_{syst} = -2 [116.89 r_{12k}(\tilde e_{kR}) +570
r'_{11k}(\tilde d_{kR}) -654r '_{1j1}(\tilde q_{jL}) ] , \eea where,
as explicitly indicated, the statistical and systematic errors appear
in the first and second positions.  Applying the single coupling
constant dominance hypothesis to the various additive contributions in
the $ ^{133} _{55} \text{Cs atom}$ case leads to the single coupling
constant bounds, $\vert \l_{12k} \vert < 0.15 \ \tilde e_{kR}, \ \vert
\l '_{11k}\vert < 0.06 \ \tilde d_{kR}, \ \vert \l '_{1j1} \vert <
0.04\ \tilde q_{jL}. $

One useful purpose of the global analysis of experimental data based
on the independent local four fermion operators, $ qqll$, is in
testing the consistency of the single coupling constant dominance
hypothesis. The initial analysis by Barger et al.,~\cite{bargerz} has
been updated by Barger and Cheung~\cite{barger00} in order to include
the new experimental measurement of the APV effect in the cesium atom,
$ \d Q _W (\text{Cs}) = (1.03 \pm 0.44)$, which deviates from the SM
prediction at the $ 2.3 \ \s $ level.  The Lorentz structure used for
the quarks and leptons interactions is the general one deduced from
the exchange of scalar leptoquarks, $[ S_0 ^L , \ S_0^ R, \ S_1^L, \
S_\ud ^L , \ S_\ud ^R ] $, which couple with the fermion field
subsets, $[ (l_L, q_L), \ (e_R, u_R,d_R), \ (l_L, q_L) , \ (l_L,
u_R,d_R), \ (e_R, q_L) ]$, respectively, with the lower suffix on $
S_I$ specifying the weak interactions $SU(2)_L$ group $ T_L$
representation label.  The fits to the resulting local couplings are
found to rule out the heavy scalar leptoquark, $ S_0^L$, which is the
counterpart of $ \tilde d _R$ in the RPV operator, $LQD^c $.  It also
appears that the favored leptoquark type, $ S_\ud ^L$, has no
counterpart in the supersymmetry framework with R parity violation.

\subsubsection{\bf Electron scattering polarization observables} 

The high precision observables~\cite{marcianolanga} associated with
the interference terms in the $ \g \ - \ Z ^0$ gauge boson exchange
amplitudes for the elastic and inelastic scattering of longitudinally
polarized electrons on proton and nuclear targets, $ e ^- _{L,R} +
N\to e^- +N,\ e^- _{L,R} + N\to e^- +X$, provide useful constraints on
new physics.  We discuss here the experimental constraints on RPV
couplings by making use of the values for the vertex parameters,
$C_{1,2} (u),\ C_{1,2} (d)$, inferred in the global analysis of four
fermion contact couplings by Barger et al.,~\cite{bargerz}.

We start with the longitudinal polarization asymmetry for the initial
state lepton in the elastic scattering on scalar $(J^P= 0^+)$ nuclear
targets, $ e^- _{L,R}+ A(0^+) \to e ^- + A(0^+)$. Fitting the RPV
contribution to the discrepancy between the SM prediction and the
experimental measurement from the Bates accelerator
experiment~\cite{souder,souderlanga}, $ e^-_{L,R} + ^{12}~C $, yields
the following single coupling constant bounds: \bea && \d [C_1(u)
+C_1(d)] = (0.137 \pm 0.033) -(0.1522\pm 0.0004) = -0.0152 \pm 0.033
\cr && \Longrightarrow \ \vert \l_{12k} \vert < 2.55 \times 10^{-1}\
\tilde e_{kR} , \ \vert \l '_{11k} \vert < 1.0 \times 10^{-1}\ \tilde
d_{kR} ,\ \vert \l '_{1j1} \vert _{1\s } < 1.1\times 10^{-1}\ \tilde
q_{jL},\eea where we have appended the index $1\s $ to signal a $ 1\s
$ level bound.  We discuss next the polarization asymmetry in the SLAC
experiment~\cite{prescott} for deep inelastic electron scattering
cross section on deuteron, $e^- +d \to e^- +X$.  The comparison with
the RPV contributions for the two relevant linear combinations of
vertex functions~\cite{bargerz}, $\d [2C_1(u) -C_1 (d) ] =- (0.22\pm
0.26)$ and $\d [2C_2(u) -C_2 (d) ]= (0.77\pm 1.23) $ yields the
following two sets of coupling constant bounds, \bea \bullet \quad &&
\vert \l '_{11k} \vert < 2.9 \times 10^{-1}\ \tilde d_{kR}, \ \vert \l
'_{1j1} \vert < 3.8 \times 10^{-1}\ \tilde q_{jL} , \ \vert \l_{12k}
\vert _{1\s } < 2.0 \times 10^{-1}\ \tilde e_{kR}; \cr \bullet \quad
&& \vert \l_{12k} \vert < 2.0 \ \tilde e_{kR}, \ \vert \l '_{1j1}
\vert < 7.1 \times 10^{-1}\ \tilde q_{jL} ,\ \vert \l '_{11k} \vert
_{1\s } < 3.9 \times 10^{-1}\ \tilde d_{kR}, \eea where the suffix
$1\s $ is used to signal $1\s $ level bounds.  For the initial state
electron polarization asymmetry in the quasi-elastic scattering
reaction, $ e^- _{L,R} + ^9Be \to N+X $, measured in the Mainz
accelerator experiment~\cite{mainz}, the comparison of the discrepancy
with respect to the SM prediction~\cite{bargerz}, ${\cal
{A}}_{Mainz}^{exp} - [2.73 C_1(u) -0.65 C_1(d) +2.19 C_2(u) -2.03
C_2(d)] = [-0.94\pm 0.19 ]-[-0.875\pm 0.014]=-0.065 \pm 0.19 $, with
the RPV contribution yields the single coupling constant bound, $
\vert \l '_{11k} \vert < 0.93\times 10^{-1}\ \tilde d_{kR}, \ \vert
\l_{12k} \vert _{1\s } < 3.0\times 10^{-1}\ \tilde e_{kR}, \ \vert \l
'_{1j1} \vert _{1\s } < 2.4 \times 10^{-1}\ \tilde q_{jL}.  $ As can
be seen, the bounds inferred from electron scattering experiments are
significantly weaker than those derived earlier from the atomic
physics parity violation experiments.


\subsubsection{\bf Anomalous  magnetic dipole moments}

The anomalous electromagnetic current magnetic dipole ($M1$) moment of
a spin-$\ud $ system is represented by the tensorial coupling in the
Lorentz covariant decomposition of the electromagnetic current matrix
element.  For a spin $\ud $ Dirac particle, described by the spinor
field $\psi _B (x)$, the anomalous moment $a_B $, in units of the
corresponding particle magneton, $ e/ 2 m_B$, is introduced through
the effective Lagrangian, $ L_{EFF} = - {i e \over 2 m_B} a_B \bar
\psi _B (x) \s_{\mu \nu } F^{\mu \nu }(x) \psi _B(x) , \quad [a_B =
\mu_ B ^\g -1 = {g_B \over 2} -1 ,\ B = e,\ \mu ,\ n] $ where $\mu_ B
^\g $ and $ g_B $ denote the total magnetic moment and gyro-magnetic
ratio parameters.

The anomalous $M1$ moments provide sensitive tests of new
physics~\cite{marciano,bern91}. The theoretical predictions include
the SM contributions corrected by higher-loop order electroweak
radiative corrections and hadronic radiative corrections.  Detailed
calculations of the MSSM loop corrections are also currently
available~\cite{carlos98,giudice97}.  The recently reported
measurement of the muon anomalous magnetic moment~\cite{brown01}
exhibits a $2.5\ \s $ deviation from the SM
prediction~\cite{davier98}, $\d a_\mu = (a_\mu )_{EXP} - (a_\mu )_{SM}
= 42.6 \ (16.5) \times 10^{-10} $.  Although the exact size of the
radiative hadronic contributions, included within $(a_\mu )_{SM}$,
remains an unsettled problem, the comparison using the existing
calculations~\cite{hadronmu98} indicates that the theoretical
uncertainties on this correction do not affect significantly the
quoted discrepancy.  The long-standing sign problem~\cite{hayakino01}
in the non-perturbative pion-pole contributions to the muon anomalous
magnetic moment, associated with the light-by-light scattering effect,
has been resolved in recent studies by Knecht et al.,~\cite{knecht01}.
Accounting for the correct sign brings the above deviation relative to
the SM down to the more modest $1.6\s $ level.

We discuss now the constraints inferred from the leptons anomalous
$M1$ moments.  The early study by Frank and Hamidian~\cite{hamidian}
obtained relatively insignificant coupling constant bounds. The
recently reported measurement of the muon anomalous magnetic
moment~\cite{brown01} has stimulated two detailed
studies~\cite{kimk01,adhir01} focused on the muon moment. The study by
Kim et al.,~\cite{kimk01} uses the so-called effective supersymmetry
framework in which the first and second generations sfermions assume
large masses, $m _{\tilde f _{lH} } = O(20)\ \text{TeV} , \ [l=1,2]$
while the third generation sfermions retain the standard $ O(100) $
GeV assigned to the supersymmetry breaking scale.  We note
parenthetically that under such a working hypothesis the RPV coupling
constant bounds for superpartner indices associated to the first two
generations would significantly relax. Since the first and second
generations sfermions effectively decouple, the calculation of loop
corrections may be restricted to the third generation sfermions only.
While the regular MSSM contributions from the third generation
superpartners to the muon $g_\mu -2$ moment are entirely negligible in
this framework, this is not the case for the RPV contributions.  The
one-loop contributions to the fermion $f_J$ anomalous $M1$ moment,
$a_{f_J} $, enter in two types, depending on whether the required
chirality flip between the external fermions takes place on the
external or internal fermion lines. The former contribution comes with
the mass factor, $ m_{f_J} $, and the latter with the mass factor, $
m_{f_J} \tilde m^{f2}_{LR} / m _{\tilde f }^2 $.  For the muon moment,
the external line chirality flip contributions of first type turn out
to be predominant.  The contributions from the $\l $
interactions~\cite{kimk01} are suppressed by cancellation effects
between the various one-loop terms associated with the intermediate
sleptons $ \tilde \nu _\tau ,\ \tilde \tau _{L} ,\ \tilde \tau _{R} $
which enter with the coupling constant factors, $ \vert \l _{32k}\vert
^2 ,\ \vert \l _{3j2} \vert ^2 ,\ \vert \l _{i23} \vert ^2 $,
respectively.  For the $\l '$ interactions, the dominant contributions
come from the one-loop diagrams with intermediate lines $ t^c,\ \tilde
b_R$, proportional to $\vert \l '_{233}\vert ^2 .$ The bounds inferred
under the single coupling constant dominance hypothesis are given by,
$ [\vert \l _{32k}\vert , \ \vert \l _{3j2}\vert ,\ \vert \l
_{i23}\vert ,\ \vert \l '_{2j3}\vert ] < 0.52\ ({m_{\tilde f } \over
100 \ \text{GeV}}) . $ Alternatively, if one focused solely on the
coupling constant, $\l _{322}$, based on the observation that this is
the least constrained of all the coupling constants involved, then the
inferred value of the RPV contribution to the anomalous moment, $
(a_\mu )_{RPV} \simeq 34.9 \times 10^{-10} ({100 \ \ \text{GeV}\over
\tilde m} )^2 \vert \l _{322} \vert ^2 , $ is seen by comparison with
the experimental result to be compatible with the perturbative
unitarity bound on $\l _{322}$.

It is of interest to consider a joint discussion of the RPV effects in
the muon magnetic moment and neutrino masses, based on the observation
that these observables are controlled by the same sets of coupling
constants.  The one-loop contributions to the neutrino masses involve,
however, chirality flip mass insertion terms for both fermion and
sfermion internal lines unlike those for the neutrino moments.
Adhikari and Rajasekaran~\cite{adhir01} observe that in order to get
RPV contributions to the muon anomalous magnetic moment and the
neutrino mass of the size required by current experiments, $a _\mu =
O(10^{-9}),\ m_{\nu_\mu } = O(1) $ \ \text{eV}, it is necessary to
suppress the one-loop contribution to the neutrino mass. This can be
achieved by postulating reduced values for the chirality flip slepton
mass parameters, $(\tilde m^{e2} _{LR})_{ij} $, or a mass degeneracy
with respect to the first two generations, $(\tilde m^{e2} _{LR})_{11}
\simeq (\tilde m^{e2} _{LR})_{22}$.  A natural way to satisfy the
latter constraints is by postulating the leptonic discrete symmetry
$D_H$ previously proposed by Babu and Mohapatra~\cite{babun1} in the
context of RPV contributions to the neutrinos Majorana masses and
electromagnetic $M1$ transition moments. The corresponding symmetry
group acts as a permutation symmetry on the lepton and muon fields, $
D_H:\ L_e \to L_\mu , L_\mu \to - L_e,\ e^c\to \mu ^c,\ \mu ^c \to
-e^c$, while leaving all other fields invariant.  The greatly reduced
number of surviving lepton number violating coupling constants, $\l
_{131}, \ \l _{232},\ \l _{123}, \ \l '_{3jk}$, is of help in
satisfying the experimental constraint from the muon anomalous moment.

\subsection{Combined  violation  of CP  and R parity symmetries}
\label{secxxx2c}

We discuss in the present subsection applications lying at the
interface of R parity and CP/T symmetry violation.  The possibility
that both symmetries are simultaneously broken has been recognized at
an early date~\cite{masiero}.  In recent years an increased attention
has been given to the possibility of embedding CP-odd complex phases
within the RPV coupling
constants~\cite{grossmanworah,kaplan,abel,guetta,jang,chen,shalom,shalom2,handoko,bhatta99}.
The proliferating number of supersymmetric and non-supersymmetric
couplings provides for a large variety of options.  The existing
studies deal with observables for the neutral $ K - \bar K $ system;
the electric dipole moment (EDM) of leptons and quarks; the CP-odd
rate asymmetries in the $Z$-boson or $B$ mesons decay channels or in
the production of fermion-antifermion or sfermion-antisfermion pairs
at the high energy colliders.



\subsubsection{\bf Neutral $K - \bar K $ mesons system}

The indirect CP violating observable, $\e $, representing the $K^0 -
\bar K^0 $ field mixing, with strangeness number change $ \D S =2 $,
is to date the single clearest experimental evidence of CP violation.
A convenient formula of use in evaluating the $\e $ physical parameter
is given by: $ \e \simeq \e _K = { e^{i\pi /4} {\Im } (M_{12}) \over
2\sqrt 2 {\Re } (M_{12}) } , \ [ M_{12} = -<K \vert L \vert \bar K >]
$, where $M_{12}$ denotes the off-diagonal matrix element of the
Hamiltonian mass operator in the $K ^0-\bar K^0$ flavor basis, such
that the splitting of $K_L\ - \ K_S$ mass eigenstates reads as, $ 2
{\Re } (M_{12}) \simeq \Delta m_K = m_{K_L} -m_{K_S}$.  The present
experimental value is determined via the $ K_{L,S} $ mesons hadronic
decay modes as~\cite{pdg00}, $ \e \approx \eta_{00}= (2.285 \pm 0.019)
\ \times \ 10^{-3} $.

\begin{figure} [ht]
\begin{center}
\leavevmode
\caption{\footnotesize \it Feynman diagrams {\bf E.1-E.10}    describing the perturbation theory
contributions to a selected set of processes initiated by the RPV
interactions. {\bf Diagrams} E:
Amplitudes $ \D S=2$ at one-loop contributing to the mixing of neutral
$ K \bar K $ mesons from the $ \l ' $ and $ \l ''$ interactions
(E.1-E.3) and from $ \l ''$ and gauge interactions (E.4).  Amplitudes
$ \D S=2$ for the decay of neutral $ K $ mesons from $ \l ' $
interactions at tree level (E.5-E.6) and one-loop level (E.7) and from
$ \l '' $ interactions at tree level (E.8).  Amplitude $ \D S=1$ from
$ \l '' $ interactions at one-loop level (E.9).  Amplitude for the B
meson rare decay channel, $ B^+ \to \bar K^0 + K^+$ (E.10). }   
\vskip 2 cm
\centerline {\epsfig{figure=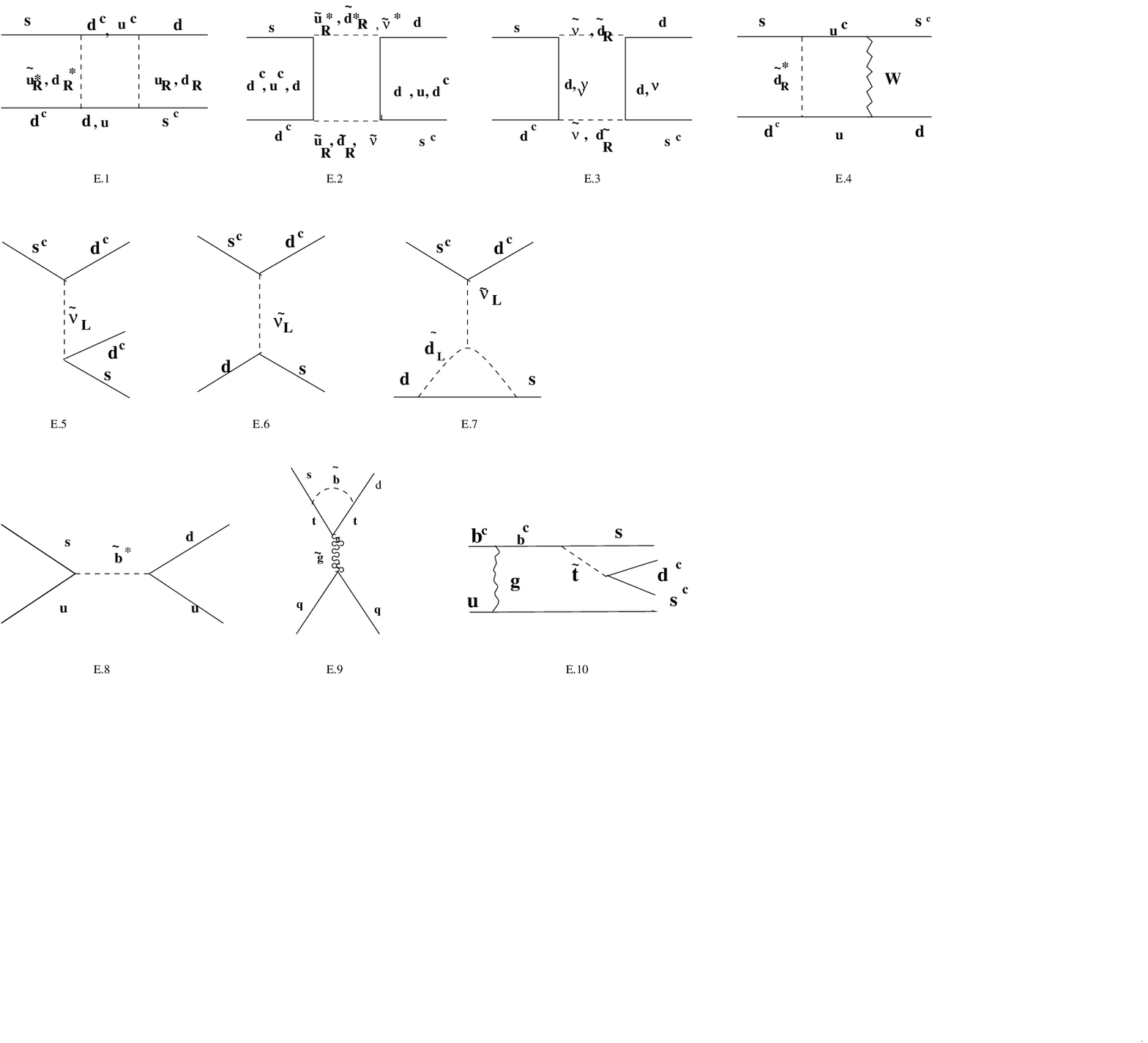,height=24cm,width=19cm} }
\label{fig2}
\end{center}
\end{figure}

The possibility of obtaining a finite contribution to $\e $ by
embedding a CP-odd complex phase within the $\l''_{ijk}$ coupling
constants, was initially envisaged by Barbieri and
Masiero~\cite{masiero}.  The relevant one-loop box diagrams involve
either pure RPV interactions or mixed gauge and RPV interactions, as
represented schematically by the Feynman graphs E.1-E.3 and E.4,
respectively, in Figure~\ref{fig2}.  Both types of Feynman diagrams
propagate a pair of internal fermionic quark lines, say, $t \ - \
\tilde b $ or $ c \ - \ \tilde b $, with a pair of squarks lines in
the first case and a pair of squark and gauge boson lines, $\tilde q\
- \ W $ in the second case.  The effective Lagrangian in the limit of
large scalar superpartners masses, $ m_{\tilde q} >> m_W , \ m_{\tilde
q} >> m_t$, is described by the approximate formula~\cite{masiero}, $
L_{EFF} = {\a V_{i2} V^\star _{1i} \over 4 \pi \sin ^2 \t _W } { \l
^{''\star } _{i13} \l ^{''} _{i32} \over m^2 _{\tilde q} } {m_c ^2
\over m_W ^2} \ln ({m_W ^2 \over m_{u_i} ^2 }) [(\bar s_{\a L } d_{\a
R}) (\bar s_{\b R} d_{\b L })- (\bar s_{\a L } d_{\b R}) (\bar s_{\b
R} d_{\a L})]
+ { \vert \l '' _{332} \l ^{''\star } _{331} \vert ^2 \over (4\pi )^2
m_{\tilde q}^2 } (\bar s_{\a R} \g_\mu d_{\a R} ) (\bar s_{\b R}
\g^\mu d_{\b R} ) + H. \ c. , $ where the first term is associated
with the mixed scalar-gauge exchange case and the second term with the
pure scalar exchange case.  The mixed gauge-scalar term with the
factor, $(m_{u_i} /m_W )^2 $, is dominated by the charm and top quark
exchange contributions and the pure scalar term by the top quark
exchange contribution.
The real and imaginary parts of the relevant RPV quadratic coupling
constant products~\cite{masiero} are expressed by the following
relations: \bea \bullet \quad &&
\l ''_{232} \l _{213}^{''\star } \simeq 0.33 \times 10^{-3} ({m_
{\tilde q } \over 100 \ \text{GeV} } )^2 { <K \vert L^{ \D S=2} \vert
\bar K >_{RPV} \over <K \vert L^{ \D S=2} \vert \bar K >_{SM} } ,\cr
&& \l ''_{332} \l _{331}^{''\star } < 0.3 \ \times 10^{-4} \ ({m_
{\tilde q } \over 100 \ \text{GeV} } )^2 {<K \vert L^{ \D S=2} \vert
\bar K >_{RPV} \over <K \vert L^{ \D S=2} \vert \bar K >_{SM} }; \cr
\bullet \quad &&
\l ''_{332} \l _{331}^{''\star } \simeq 0.71 \times 10^{-3} ({m_
{\tilde q } \over 100 \ \text{GeV} } ) { <K \vert L^{ \D S=2} \vert
\bar K >_{RPV}\over <K \vert L^{ \D S=2} \vert \bar K >_{SM} } , \eea
where the matrix elements for the strangeness changing effective
Lagrangian, $ L^{ \D S=2}$, represent the contributions to the $\bar K
^0 \to K ^0$ transition amplitudes from RPV and SM gauge interactions.
The results in the first entry above refer to mixed scalar-gauge
contributions from charm and top quark exchanges, respectively, and
those in the second entry to pure scalar top quark exchange
contributions. We have quoted updated values for the top quark
exchange contributions taking into account the present measured value
of the top quark mass and upper bounds on the flavor mixing matrix
elements, $V_{ts} < 0.05,\ V_{td} < 0.015$.

The direct CP violation physical parameter, $\e '$, is accessed
through measurements of the $ \D S =1 $ transition amplitudes for the
rare decay modes, $K_L ^0 \to \pi ^+ +\pi ^-,\ K_L ^0 \to \pi ^0+\pi
^0 $.  A convenient definition for $\e '$, of use for calculational
purposes, is given in the convention $\Im (A_2) =0$ by, $ \e ' \simeq
- {1 \over 20 \sqrt 2 } e ^{i ({\pi \over 2} + \d _2 -\d _0 )} \xi , \
[\xi ={ \Im (A_0)\over \Re (A_0) } , \ A_I = <(\pi \pi )_{I} \vert L
^{\Delta S =1} \vert K _L ^0 > ], $ where $\d _I ,\ [I=1,2]$ denote
the strong interactions phase shifts in the isospin coupled two-pion
states.  The dominant SM contribution from the $W$-boson penguin
diagram is given approximately as, $\vert \e '\vert \simeq 1.9 \times
10^{-5} \ \sin \delta $, where $\delta $ is the complex CP-odd phase
in the CKM matrix. The current experimental determinations of the
ratio of direct to indirect $CP$ violation~\cite{na48ktev} yield the
result, $\e '/\e = (21.2 \pm 4.6) \times 10^{-4}$.

The $ \l ' $ interactions contribute to the $\D S=1$ CP violating
amplitude at tree and one-loop levels through the Feynman diagrams E.6
and E.2-E.3 in Figure~\ref{fig2}, respectively. The $ \l ''$
interactions contribute at tree level in terms of the Feynman diagram
E.8 in Figure~\ref{fig2} and at one-loop level in terms of the gluon
exchange penguin Feynman diagram E.9 in Figure~\ref{fig2}, associated
to the vertex coupling, $ \bar d s g $. The associated contributions
in the limit of massive scalar superpartners are represented by the
effective Lagrangian~\cite{masiero}: $ L_{EFF} ^{\D S=1}= { \l ''
_{123} \l ^{''\star } _{131} \over 2 m^2 _{\tilde b _R } } [ (\bar
s_{\a R} \g^\mu d_{\a R} ) (\bar u_{\b R} \g_\mu u_{\b R})-(\bar s_{\a
R} \g^\mu u_{\a R} ) (\bar u_{\b R} \g_\mu d_{\b R} ) ]
+ {\a _s \over 4 \pi } { \l '' _{313} \l ^{''\star } _{323} \over 8
m^2 _{\tilde q} } \ln ({m_{\tilde q} ^2 \over m_t ^2}) (\bar s_R \l ^a
\g _\mu d_R)(\bar q_L \l ^a \g ^\mu q_L) + H. \ c. , $ where the
second term involves a gluon penguin diagram summed over the light
quark fields, $q (x)$, with $\l ^a$ denoting the $SU(3)_c$ color group
generators.  Thanks to the logarithm enhancement factor and the
possibility to select the generation configuration involving third
generation quarks, the one-loop gluon penguin diagram amplitude has
some chance to compete with the tree level diagram given by the first
amplitude. The fact that different generational configurations of the
RPV coupling constants contribute to the mixing and direct $CP$
violation parameters, $\e $ and $\e '$, shows explicitly that the $\l
''$ interactions have the ability to account for both milli-weak and
super-weak $CP$ violation, associated with the predictions, $\e '/\e =
O(10^{-3})$ and $\e '/\e = 0 $, respectively.  The above RPV
contributions to $\e '$ lead to the following two relations involving
the imaginary parts of quadratic coupling constant products, \bea &&
\vert \e ' \vert \simeq [1.\times 10^{-1} \ \Im ( \l '' _{123} \l ^{
''\star } _{113} ) \tilde q^{-2} , \ 1.\times 10^{+2}\ \Im ( \l ''
_{313} \l ^{'' \star } _{323} ) \tilde q ^{-2} ] \cr &&
\Longrightarrow \ \Im ( \l '' _{123} \l ^{ ''\star } _{113} )=
O(10^{-5} ) \ \tilde q^2 ,\ \Im ( \l '' _{313} \l ^{''\star } _{323} )
= O(10^{-8} ) \ \tilde q^2 ,\eea where we have used, for illustrative
purposes, the estimate for the experimental value, $\e ' =
O(10^{-6})$.  The penguin diagram mechanism is clearly providing the
strongest constraint.

The attractive possibility of introducing the CP-odd complex phase
through the soft supersymmetry breaking RPV trilinear couplings of
sfermions, $ V_{soft} = A^{\l '} _{ijk} \l'_{ijk} \tilde L_i \tilde
Q_j \tilde D_k^c$, is examined by Abel~\cite{abel}.  The contribution
to the $\D S=2$ CP violating amplitude is represented at the tree
level by the Feynman diagrams E.5 and E.6 in Figure~\ref{fig2}.  The
resulting imaginary part of the $K^0-\bar K^0$ mass shift parameter is
described by the approximate formula, $ \e \simeq 1.\times 10^{-2} \
\Im ( A ^{\l '} _{i11}\l '_{i11} - A ^{\l '} _{i12} \l ' _{i12} ) $.
The comparison with the experimental value of $ \e $ indicates that
significant cancellations ought to take place in this mechanism
between the two flavor non-diagonal configurations associated with the
soft parameter coupling constants, $A^{\l '} _{i11} , \ A ^{\l '}
_{i12} $. The $ \D S =1 $ four fermion coupling, $ (\bar s_R d_L)
(\bar d_R s_L)$, receives an RPV contribution from the one-loop level
sneutrino type penguin diagram depicted by the Feynman diagram E.7 in
Figure~\ref{fig2}.  The resulting contribution to the ratio of direct
to indirect CP violation observables is described as~\cite{abel}, \bea
&& \vert { \e ' \over \e } \vert \propto \vert { \Im (A _{i21} ^{\l '}
\l ' _{i21} +A _{i11} ^{\l '\star } \l ' _{i11} ) \l'_{i11}\l'_{i21}
\over \Im (A _{i21} ^{\l '} \l ' _{i21} +A _{i12} ^{\l '\star } \l '
_{i12} ) \l'_{i12}\l'_{i21} } \vert \simeq 10^{-7} \ { {\l '} _{i11}
\over {\l '} _{i12} } .  \eea Using the quadratic coupling constant
product bound, $ \vert {\l '} _{i12} {\l '} _{i21} \vert < 10^{-9}\
\tilde \nu_i ^2 $, one derives from the above result the lower bound,
$ \vert { \e ' \over \e } \vert > 10^2 \ {\l '} _{i11} {\l '} _{i21}
$.  Depending on the ratio of RPV coupling constants, $ {\l '} _{i11}/
{\l '} _{i12}$, one can simulate with the RPV interactions both a
super-weak or milli-weak scenarios.

\subsubsection{\bf Asymmetries  in hadron decay rates and polarization
observables}

The polarization observables in the $ K$ meson three-body semileptonic
and radiative decay modes, $K^+\to \pi ^0 +\nu _\mu +\mu^+ , \quad
(K_{\mu 3}) $ and $\ K^+\to \mu^+ +\nu +\g , \quad (K_{\mu 2\g } ) $
provide sensitive probes of $ T $ and $CP $ violations.  The muon
transverse polarization, $ P_T $, in the radiative $ K_{\mu 2\g } $
decay mode, is described by the final state muon polarization vector
$\vec s_\mu $ component orthogonal to the $\mu -\g $ production plane,
$ P_T ( K_{\mu 2\g } ) \simeq \vec s_\mu \cdot (\vec p_\mu \times \vec
p _\g ) $. The physical parameter, $ P_T $, is especially interesting
since the SM contribution vanishes identically, while the contribution
from final state interactions is expected to remain below the $
O(10^{-3})$ level.  The polarization of the emitted muons in both the
radiative and non-radiative decay modes should be experimentally
measured with an accuracy of $ O(10^{-3}) $ in ongoing (KEK
accelerator) and planned (BNL accelerator) experiments.

The RPV contribution to the subprocess, $ \bar s +u \to \mu ^+ +\nu $,
involves the tree level exchange of $ \tilde e_L$.  Using the current
bounds from $\mu ^\pm \to e^\pm +\g $ and $K^+ \to \pi^+ +\nu +\bar
\nu $, which are discussed below, Chen et al.,~\cite{chen} obtain the
estimate for the relevant product of coupling constants, $\Im (\l
^\star _{2i2} \l ' _{i12} )/ m^2_{\tilde e_{iL} } \simeq 10 ^{-4} $.
The ensuing prediction for the transverse muon polarization, $ P_T(
K_{\mu 2\g }) < 10 ^{-2} $, lies close to the presently attained
experimental sensitivity. A bound of same size is also found for the
transverse polarization in the non-radiative process, $ P_T( K_{\mu 3
})$.

The spin-independent CP violating decay rate asymmetries in the
$Z$-boson partial decays into fermion-antifermion pairs of different
flavors, $ Z \to f_J + \bar f_{J'} $, are represented by the
normalized differences of rates for the CP mirror conjugate reactions,
\be {\cal A}_{JJ'} (f) ={B (Z^0 \to f_J + \bar f_{J'} ) -B (Z^0 \to
f_{J'} + \bar f_{J} ) \over B (Z^0 \to f_J + \bar f_{J'} ) + B (Z^0
\to f_{J'} + \bar f_{J} ) } , \ [J\ne J' ; \ f=u, \ d, \ e ] .\ee
Finite contributions to the flavor decay asymmetries can arise by
embedding a CP-odd complex phase, $\psi $, within the RPV coupling
constants.  The relevant CP-odd combinations of parameters, invariant
under a rephasing of the quarks or leptons matter fields, are given by
products of four distinct RPV coupling constants.  The relevant
dependence on quartic products arises through the interference terms
between different configurations of the internal lines in the one-loop
RPV transition amplitudes~\cite{chemtobm}.  The asymmetries are
proportional to the imaginary parts of ratios of quadratic products of
the RPV coupling constants.  Should the RPV coupling constants exhibit
generational hierarchies, one would expect large enhancement or
suppression of the flavor rate asymmetries, depending on the flavor
configurations of the emitted fermions.  The predicted rate
asymmetries in the emission of charged leptons, down-quarks and
up-quarks, are of magnitude~\cite{chemtobm}, ${\cal A}_{JJ'} (f)
\simeq (10^{-1}\ -\ 10^{-3}) \times { \vert \l ^{'\star } _{iJk } \l
'_{iJ'k} \vert \over \vert \l ^{'\star } _{1Jk'} \l '_{1J'k'}\vert }
\sin \psi , \ [f=u,d,e] , $ where $\psi $ denotes a generic CP-odd
complex phase arising in the corresponding ratio of coupling constant
products.

\subsubsection{\bf Electric dipole moments}

The electric dipole moments of the spin $\ud $ leptons and baryons, $
d^\g _A$, are identified by the electromagnetic coupling effective
Lagrangian, $ L _{EFF} = -{i\over 2} d^\g _A \bar \psi _A (x) \s _{\mu
\nu }\g_5 F^{\mu \nu } (x) \psi _A (x), \ [A =n, \ p, \ e, \mu ] $.
Finite contributions to $ d^\g _A$ arise in perturbation theory from
short distance mechanisms involving one or higher loop order Feynman
diagrams.  The electric dipole moment of baryons can be directly
initiated through light quarks electric dipole moments, $d _q ^\g , \
[q=u,d,s]$ with the approximate non-relativistic quark model estimate
for the neutron electric dipole moment given by the approximate
formula, $d^\g _n= {4 \over 3} d^\g _d - {1 \over 3} d^\g _u$.  So
far, the discussion of RPV effects has been largely restricted to the
quarks electric dipole moments. A variety of competitive contributions
to the baryons electric dipole moments may, however, be induced
through higher dimensional operators in the quarks and gluon
fields~\cite{edm,chang,chemtob}, involving the dimension $5$ quarks
color electric dipole moment operators and the dimension $6, 8$
gluonic operators of third and quartic order in the gluon field
strength, $ GG\tilde G ,\ GGG\tilde G $.

Finite contributions to the electric dipole moments could in principle
arise from the CP-odd complex phases present in the regular soft
supersymmetry breaking interactions.  The RPV effects on the electric
dipole moment of leptons and quarks were originally
discussed~\cite{hamidian} in terms of one-loop mechanisms involving a
complex CP-odd phase in the trilinear scalar field interactions
parameter $A^d$. The corresponding Feynman diagram propagates a pair
of sfermion and fermion internal lines with chirality flip mass term
insertions.  The same one-loop mechanism combining complex RPV and
regular trilinear scalar interactions coupling constants, $\l '_{133}
=\vert \l'_{133}\vert e^{i\b } ,\ A^{u} = \vert A^{u} \vert e^{i\a
_{A} ^u }, \ A^{d} = \vert A ^{d} \vert e^{i\a _{A } ^d , \ } $, was
also considered in Ref.~\cite{adhi99}.  However, it appears that the
constraints on single RPV coupling constants inferred in both of these
works~\cite{hamidian,adhi99} are not valid ones, owing to an
inappropriate interpretation of the space-time structure of
interactions in applying the Feynman graph rules.

\begin{figure} [ht]
\begin{center}
\leavevmode
\caption{\footnotesize \it Feynman diagrams {\bf E.11-E.13, F.1-F.2, G.1-G.5,
H.1-H.2, I.1-I.4}     describing the perturbation theory
contributions to a selected set of processes initiated by the RPV
interactions.  {\bf Diagrams} E:
Amplitudes at two-loop order for the quark and electron electric
dipole moments from $ \l '' $ and $ \l '$ interactions (E.11 and
E.13). Amplitudes at one-loop order for the electron electric dipole
moment from $ \l $ interactions (E.12).  {\bf Diagrams} F: Amplitude
for the semileptonic K meson decay channel, $ K \to \pi + \nu +\bar
\nu $.  {\bf Diagrams} F: Amplitude
for the semileptonic K meson decay channel, $ K \to \pi + \nu +\bar
\nu $.  {\bf Diagrams} G: Tree level amplitude for the decay channels
of neutralinos or charginos (G.1) and sneutrinos (G.2).  One-loop
level amplitudes for the neutralino radiative decay channel, $\tchi ^0
\to \nu +\g $, induced by a sneutrino VEV (G.3), a neutralino-neutrino
mixing(G.4) and a trilinear interaction (G.5).  
{\bf Diagrams} H:
Amplitudes for at one-loop level for the radiative decays of charged
leptons involving the RPV interactions alone (H.1) and in combination
with the gauge interactions(H.2).  
{\bf Diagrams} I: Amplitude for the
process, $ d+d \to u+u+e+e$, initiating neutrinoless beta decay. The
various mechanisms correspond to scattering of sleptons mediated by
neutralino $t$-channel exchange (I.1), scattering of up-squarks
mediated by neutralino or gluino $t$-channel exchange (I.2),
production of a d-squark pair mediated by neutralino or gluino
$t$-channel exchange (I.3), scattering of down-squark with W-boson
mediated by neutrino $t$-channel exchange (I.4).     }   
\vskip 1 cm 
\centerline {\epsfig{figure=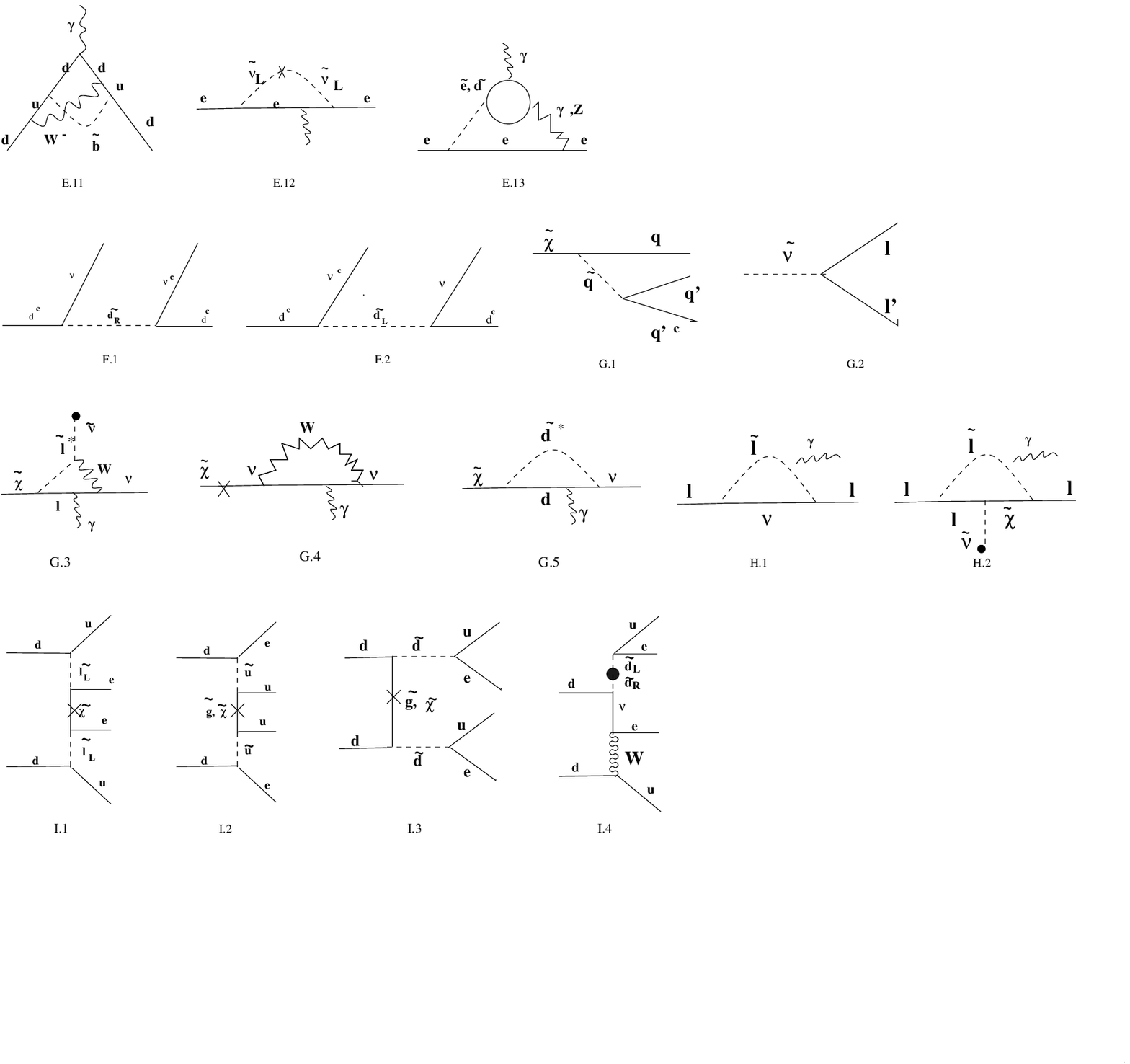,height=24cm,width=18cm} }
\label{fig3}
\end{center}
\end{figure}

The suspect character of the above mentioned analyses was first
realized in works by Godbole et al.,~\cite{godbole99} and Abel et
al.,~\cite{abel99}.  The fermion chirality selection rules imply, on
general grounds, that the RPV contributions to the electric dipole
moments of quarks and leptons are either absent or strongly suppressed
at one-loop order.  The reason is due to the chirality structure of
the bilinear and trilinear RPV interaction superpotential operators, $
LH_u, \ LLE^c , \ LQ D^c , \ U^c D^cD^c$. One readily sees that an $
\tilde e_R $ particle can never be emitted nor absorbed at an RPV
vertex.  In the one-loop diagram with $\tilde d ,\ \nu $ internal
lines, the chirality conservation also forbids the insertion on the
down-squark particle line of a $\tilde d_R - \tilde d_L $ chirality
flip mass term, unless this is accompanied by a neutrino Majorana mass
term insertion. The associated one-loop contributions to $d^\g _q$ are
thereby strongly suppressed by a factor proportional to $m_\nu $.  In
a similar way, the one-loop contribution with $\tilde \nu ,\ e $
internal lines to the electron moment due to the $\l $ interactions,
as depicted by the Feynman diagram E.12 in Figure~\ref{fig3}, is
non-vanishing only in the presence of a sneutrino Majorana mass term,
$ \tilde m^ M _{ij} \tilde \nu _{iL} \tilde \nu _{jL} + \ H. \ c.  $
The explicit calculations of one-loop Feynman diagrams, do show that
the $\l ''$ interactions do not contribute to the quarks moments,
while the $\l '$ and $\l $ interactions contribute only to the
d-quarks and charged leptons moments, respectively, by strongly
suppressed terms of form, $ d^\g _{d_k} \propto \l '_{ijk} \l ^{'\star
} _{i'kk'} (\tilde m ^{d2} _{LR}) _{k'j} (m_\nu ) _{i'i} m_{d_k} /( m
^2 _{\tilde d _{jL } } m ^2 _{\tilde d _{k'R } } ) $ and $ d^\g _{e_j}
\propto \l _{ijk} \l ^{\star } _{ij'j} (\tilde m ^{e2} _{LR}) _{j'k}
(m_\nu ) _{i'i} m_e /( m ^2 _{\tilde e _{jL } } m ^2 _{\tilde e _{kR }
}) .$ Only in the case of massive Majorana neutrinos can one find
finite one-loop order contributions to the $E1$ transition moments.
The chirality selection rules can also be used to determine the
dependence of the neutrino Majorana mass and electromagnetic $ M1 $
flavor diagonal or off-diagonal moments, or the charged fermions $ M1
$ transition moments, on the external fermions mass parameters.

At the two-loop order, several mechanisms involving the RPV
interactions, with mass insertion terms for the internal scalars and
fermions, can contribute to the leptons and quarks electric dipole
moments~\cite{masiero,godbole99,abel99}.  The various two-loop Feynman
diagrams involve the crossed or non-crossed exchanges of Higgs or
gauge bosons and sfermions lines along with fermion or boson particle
loops. These appear in four distinct classes which are
identified~\cite{changedm00} pictorially as rainbow, overlapping, tent
like and Barr-Zee types, respectively.  The graphs E.11 and E.13 in
Figure~\ref{fig3} display a sample of two-loop Feynman diagrams
contributing to the down-quark and electron electric dipole
moments. (The graph E.13 corresponds to what we term as the Barr-Zee
diagram.)  While the dependence on the RPV coupling constants is
generically of quartic order, a quadratic order dependence may also be
possible provided this is accompanied by a quadratic product of the
quark flavor mixing CKM matrix elements.  The various quartic products
involve, as expected, appropriately rephase invariant configurations
of the quark family indices.

The baryon number violating $ \l ''$ interactions contribute to the
electric dipole moments of quarks at two-loop order in terms of
three-point Feynman diagrams involving the crossed exchange of gauge
bosons and sfermions.  The corresponding representative graph
involving $W $ and $ \tilde d $ exchange is depicted by the Feynman
diagram E.11 in Figure~\ref{fig3}.  The CP-odd phase is embedded as a
relative complex phase for a pair of distinct $ \l ''$ coupling
constants.  The study by Barbieri and Masiero~\cite{masiero} leads to
a contribution to the neutron electric dipole moment in the limit of
massive internal squark, $ d^\g_n \simeq {e\a V_{i1} V _{2i} ^\star
\over 4\pi ^3 \sin ^2\t_W } {m_{u_i} ^2 m_{s} \over m^2_{\tilde b _R }
m_W^2 }\Im (\l ''_{i13} \l _{i32}^{''\star } ) ,\ [u_i =c, t] $ where
the suppression effect is seen to arise through the combined effect of
the flavor mixing factors, the light quark mass factors in the
numerator and the squark mass factor in the denominator.  The
corresponding contribution is maximized by selecting the third
generation $\tilde b$ squark along with the charm or top quark
internal lines.  Under the double coupling constant dominance
hypothesis, one deduces the following formulas for the imaginary parts
of quadratic coupling constant products~\cite{masiero}: \bea && \Im
(\l ''_{213} \l _{232}^{''\star } ) = 10^{-4} ({d_n ^\g \over 10^{-31}
e \times \ \text{cm} }) \tilde q ^2, \ \Im (\l ''_{312} \l
_{332}^{''\star } ) < 10^{-7} ({d_n ^\g \over 10^{-31} e \times \
\text{cm} }) \tilde q ^2,\eea where we have updated the result for the
internal top quark contribution by accounting for the measured top
quark mass and flavor mixing matrix elements.

The $\l ,\ \l ' $ interactions contribute to the electron and quark
electric dipole moments, $ d^\g _e$ and $ d ^\g _q $, through two-loop
diagrams with one fermion closed loop attached to the external line by
a pair of sfermion and gauge boson internal lines.  The comparison
with the experimental limits for the electron and neutron moments
yields the coupling constant bounds~\cite{godbole99,abel99,herczeg99}:
$ \Im (\l _{211} ^\star \l '_{233} ) < 5.  \times 10^{-6} \ \tilde \nu
^2 , \ \Im (\l _{1j1} ^\star \l '_{j33} ) < 6.  \times 10^{-7} \
\tilde \nu_j ^2 , \ [d ^\g _e]; \quad \Im (\l ^{'\star } _{i11} \l
^{'} _{i33} ) < 1.\times 10^{-4} \ \tilde m ^2 , \ \ [d ^\g _n] $.
The analogous comparison, based on the two-loop diagram with two
crossed sfermionic loops attached to the external line, yields bounds
on quartic coupling constant products~\cite{abel99}: \be{m_l \over
m_\tau }\Im (\l _{1mn} \l ^\star _{jln} \l ^\star _{iml} \l _ {ij1} )
< 10 ^ {-6 } ,\ \ {m_l \over m_t}\Im (\l ' _{1mn} \l ^{'\star } _{jln}
\l ^{ '\star } _{iml} \l ' _ {ij1} ) <3. \ 10 ^ {-6} ,\ee in
correspondence with the electron and quark electric dipole moments,
respectively.  The comparison by Abel et al.,~\cite{abel99} for the
neutrino electric dipole moment, using the experimental limit, $\vert
d _ \g ^ {\nu }\vert \leq 2.5 \times 10^{-22} e \ \text {cm} $, yields
the coupling constant bounds, $ \Im (\l _{i21} \l ^\star _{i' 12} ) <
5.\times 10^{-2} ,\ \Im (\l ' _{i32} \l ^ {'\star } _{i'23} ) < 2.4
\times 10^{-3}$.

The indirect contributions to the neutron electric dipole moment
involving purely hadronic operators can compete with the above
considered direct contributions.  Chang et al.,~\cite{changedm00} find
that the Barr-Zee type two-loop diagram inducing the quarks color
electric dipole moment operator yields a significant contribution to
$d_n ^\g $ which sets the following coupling constant bounds, $\Im (\l
^{'\star } _{i33} \l '_{i11} ) < 1.2 \times 10^{-5},\ \Im (\l ^{\star
} _{i33} \l '_{i11} ) < 33.  \times 10^{-5}.$ One interesting
exception to the absence of one-loop effects occurs upon considering
the combined contributions from bilinear and trilinear interactions to
the off-diagonal squark mass squared submatrix linking the left and
right chirality squarks.  As pointed out by Keum and
Kong~\cite{kongedm00}, the extra complex matrix element terms, $ \d
(\tilde m^{d2} _{LR})_{jk} = - v_u (\mu ^\star _i \l ' _{ijk} /\sqrt 2
) $, have the ability to induce a down quark electric dipole moment
through the familiar gluino-squark exchange one-loop diagram.  The
comparison with the experimental limit on the neutron electric dipole
moment leads to the coupling constant bound, $\Im (\mu ^\star _i \l
'_{i11} ) < 1. \times 10^{-6}.$

The chirality selection rules forbidding the one-loop contributions
from trilinear RPV couplings are not in force if one combines together
the bilinear and trilinear couplings or if one includes the soft
supersymmetry breaking RPV bilinear couplings~\cite{choi01}. The
reason is due to the field mixing terms of leptons with gauginos
induced by the bilinear couplings.  Choi et al.,~\cite{choi01}
consider the choice of field basis characterized by vanishing
sneutrinos VEVs and assume CP-odd complex phases of order unity. The
comparison with the experimental limits on the electron and quark
electric dipole moments leads to the following typical bounds: $ d _e
^\g :\ \vert \l _{1j1} \vert < 1.3 \times 10^{-2} \times \ ( { (\sum
_i \mu ^2_i ) ^\ud \over \vert \mu _j \vert } ) \ (\tchi ) ^ \ud
\tilde \nu ( {m_\nu \over 0.1 \ \text{eV} } ) ^ {- \ud } ;
\quad d _q ^\g :\ \Im ( {\l '_{ij1} V ^\dagger _{j3} (B \mu _i
m_{\tilde G} ) ^\star \over m ^2 _{\tilde e _j } } ) < 2.5 \times 10
^{-6} \ ({0.01 \over V _{31}}) m _{\tilde e _j } . $

The atoms electric dipole moments are described by the local four
fermion effective Lagrangian for the electron-nucleon system,
parameterized by the space and time parity reversal violating
couplings~\cite{Khrip97}, $ L _{EFF}= -{G_F\over \sqrt 2} [C ^p_{S}
(\bar e i\g_5 e)(\bar p p) +{i\over 2} \e ^{\a \b \g \delta } C^p_{T}
(\bar e \s _{\a \b } e) (\bar p \s _{\g \delta } p ) ] +
[p\leftrightarrow n] , $ involving two pairs of scalar and tensorial
vertex parameters, $C ^{p} _{S},\ C^{p}_{T} $ and $C ^{n} _{S},\
C^{n}_{T} $ for the proton and neutron.  The comparison with the
currently available experimental limits for the $ ^{133} \text{Cs} $ or $ \
^{205} \text{Tl}  $ atoms~\cite{Khrip97}, using the RPV contributions of
two-loop order, yields the bounds~\cite{herczeg99}: $ \Im (\l _{1j1}
^\star \l '_{j11} ) < 1.7 \times 10^{-8} \ \tilde \nu_j ^2 , \
[j=2,3]. $

\subsubsection{\bf Asymmetries  in $B$ meson hadronic decays}

The hadronic decay modes of the flavor eigenstates of neutral $B -
\bar B $ mesons provide sensitive tests of the SM and useful probes of
new physics~\cite{carter80,dunietz86}. A comprehensive discussion is
provided in the textbook by Bigi and Sanda~\cite{bigi00}.  A large
number of $B$ meson decay modes have been experimentally observed of
which an illustrative representative sample is given by, $ B_d ^0 \to
J/\Psi +\rho ^0, \ B_d ^0 \to J/\Psi + K_S , \ B_d ^0 \to D^\pm +\pi
^\mp , \ B_d ^0 \to K^\pm +\pi ^\mp $. The SM amplitudes are
determined, in general, at tree level by the quark subprocesses, $
b\to \bar q + q+ q' , \ [q =u,d,c ] $.  The $B$ meson decay modes,
$B^0 \to K^0+\bar K^0, \ B^0 \to \phi ^0+ \pi^0 , \ \ B^0 \to \phi +
K_S ^0$, are of special interest since the underlying quark
subprocesses, $ \bar b \to \bar d +d+\bar d, \ \bar b \to \bar d +d
+\bar s, \ \bar b \to \bar s +s+\bar s$, receive SM contributions only
beyond the tree level.

The decay rate asymmetries in the $B$ mesons nonleptonic decays into
CP eigenstates final states yield a direct useful information on the
CKM matrix~\cite{grossmanworah}. In recent years several anomalies
between the SM predictions and experiment have been identified for $B$
meson decays into light flavored mesons.  Thus the CP-odd rate
asymmetries in the $ B^0 \to \pi ^+ +\pi ^-$ modes are found to
exhibit significantly high branching fractions in comparison to the SM
predictions, while those in the decay modes, $ B^ {\pm ,0} \to \eta '
+ K ^ {\pm ,0} , \ B^ {\pm ,0} \to \eta + {K ^ { \pm ,0}} ^ \star $,
differ from SM predictions by factors of $ 2\ -\ 3$.  An
incompatibility is also observed between the mixing type CP-odd
asymmetry in the decay, $ B^0\to \phi ^0+ K_S ^0$, and that in the
decay mode, $ B ^0\to J/\Psi + K_S ^0$, which is used as a standard in
extracting the flavor unitarity triangle angle parameter, $\sin \b $.
The theoretical description of the above decay processes relies on
robust methods and inputs~\cite{ali98}.

 The time dependent structure of the CP-odd decay rate asymmetry in
the quantum oscillating system of $B^0-\bar B^0$ mesons is described
by the formula:
\begin{eqnarray} &&  {\G (B^0(t)\to f(CP) )
-\G (\bar B^0(t)) \to f(CP) \over \G (B^0(t)\to f(CP) ) +\G (\bar
B^0(t) \to f(CP) ) } = {(1-\vert r_{f(CP)} \vert ^2 )\over (1+\vert
r_{f(CP)} \vert ^2 ) } \cos (\D mt) - { 2\Im r_{f(CP)} \over (1+\vert
r_{f(CP)} \vert ^2 ) } \sin( \D m t ), \cr && [\D m \equiv m_{B_H} -
m_{B_L},\ r_{f(CP)} \equiv \l = {q \over p} {\bar A \over A } ,\
{q\over p} = { 1-\e _B \over 1+\e _B } = e^{-2i\phi_M} ,\ {\bar A
\over A } = {\bar A (\bar B^0\to f(CP) ) \over A (B^0 \to f(CP) ) }
=e^{-2i\phi_D} ]
\label{eqx9p} \end{eqnarray} where the oscillation frequency depends
on the mass splitting parameter between the heavy and light mass
eigenstates, $ \D m $, and the oscillation amplitude, $ r_{f(CP)} $,
is given by a product of two factors associated with indirect and
direct CP violation, respectively, with the factor, $q/p$, including
the flavor eigenstates complex mass mixing, and the factor, $
e^{-2i\phi_D} $, the ratio of CP conjugate transition amplitudes.  The
$ B ^0-\bar B^0$ field mixing, with $\D b =2$ selection rule, arises
in the SM at one-loop level, while the $ \D b =1$ nonleptonic decay
transition amplitudes for the bottom or anti-bottom quarks, $ b \to
d_i+ \bar q'+ q'', \ [q' , q'' = (u, c, d, s)]$, arise from either
tree level or one-loop penguin type diagrams.  Under the assumption
that the SM and RPV interactions contribute additively to the mixing
and decay amplitudes, the total mixing and decay contributions,
distinguished by the labels $ M, \ D$, can be parameterized by two
pairs of modulus and complex phase parameters, $ r_X, \ \t _X , \ [X=
M, D]$ defined in terms of the off-diagonal matrix element of the mass
operator and transition amplitude as, $ M_{12}= M_{12} ^{ SM} (1+ r_M
e^ {i\t_M } ), \ A = A_{SM} (1 +r_D e ^{i\t _D } ) ,\ \bar A = \bar
A_{SM} (1 +r_D e ^{- i\t _D } ) . $ The ratio of transition
amplitudes, $ \bar A /A $, becomes a pure complex phase in the
simplified cases where the final state CP-even strong interactions
phase is the same for all the additive contributions to the decay
amplitudes.

The RPV tree level contributions to the $ B ^0\ - \ \bar B ^0 $ mixing
amplitudes are described by the effective Lagrangian, \be L_{EFF} =
-\ud \l' _{ijk} \l ^{'\star } _{ij'k'} \bigg [{1 \over m_{\tilde \nu
_iL} ^2 } (\bar d_{\a kR} \g ^\mu d_{\b k'R}) (\bar d_{\b j'L} \g _\mu
d_{\a jL})+ {1 \over m_{\tilde e_iL} ^2 } (\bar d_{\a kR} \g ^\mu
d_{\b k'R}) (\bar u_{\b j'L} \g _\mu u_{\a jL})\bigg ] +\ H.\ c. \ee
The associated contribution to the physical mixing parameter, $r_M$,
can be expressed as~\cite{guetta}: $ r_M \simeq 10^ 8 \ \vert \l ^{'
\star } _{i13} \l '_{ i31} \vert \tilde \nu ^{ -2}$.

We proceed now to the RPV contributions in the decay amplitudes.  The
B meson decay transitions into pure CP eigenstates occur by $ \tilde
e_L $ and $\tilde d_R$ exchange in the modes, $ B ^0 \to J/\Psi + K_S
^0 ,\ (b\to \bar c cs); \ B ^0 \to D^0 + \pi ^0, \ (b\to u \bar c d);
B ^0 \to \pi ^0 + \pi ^0, \ \pi ^0 + \rho ^0,\ (b\to u\bar u d); B ^0
\to K_S ^0 +\pi ^0 , \ B ^0 \to K_S ^0+\rho ^0, \ (b\to u \bar u \bar
s); $ and by $ \tilde \nu _L $ exchange in the modes, $ B ^0 \to K_S
^0+K_S ^0 , \ B ^0 \to \phi ^0 +\pi ^0 , \ (b\to s\bar s d); B ^0 \to
\phi ^0 + K_S ^0 \ (b\to s s \bar s). $ Detailed studies of the
induced contributions to the CP-odd asymmetry parameter $ r_D$ in the
various decay channels are presented in Refs.~\cite{kaplan,jang}.  The
existing bounds on the $\l ' $ and $\l '' $ coupling constants
involved in the $B$ meson decays yield useful constraints on the
asymmetry parameters, $ r_D$. The RPV contributions exhibit different
patterns~\cite{kaplan} depending on whether one considers heavy mesons
decay channels, such as, $ B ^0 \to J/\Psi + K_S ^0, \ B ^0 \to D^+
+D^- $, or light mesons decay channels, such as, $ B^0\to \phi ^0+K_S
^0, \ B^0\to \phi ^0+\pi ^0, B^0\to K_S ^0+K_S ^0 $.  The latter decay
modes generally yield stronger signals, $ (1+ r_D )\simeq \vert
A_{RPV} /A_{SM} \vert >> 1$.  The study undertaken by Jang and
Lee~\cite{jang} accounts for the CP-odd contributions from mass mixing
and uses updated values for the Wilson coefficients of the allowed
four fermion operators.  The typical size of the predicted asymmetry
parameters in various decays varies in the range $ (10^{-3}\ - \
10^{-4})$ for the $\l ' $ interactions, and in the wider range
$(10^{0}\ -\ 10^{-1})$ for the $\l '' $ interactions.

The joint study of the two neutral $B$ meson decay modes, $ B ^0 \to
\phi ^0 + K_S ^0 $ and $ B ^0 \to J/\Psi + K_S ^0$, is of special
interest in searches of new physics, as noted by
Guetta~\cite{guetta}. The SM predictions yield equal decay phases,
$\phi_D$, for these two decay modes, up to a small controllable
theoretical uncertainty on the phases difference, $\D \phi = \vert
\phi (B_d \to \phi ^0+ K_S ^0) - (B_d \to J/\Psi +K_S ^0 ) \vert <
O(10^{-1} ),$ where the complex phase $ \phi $ is related to the decay
amplitude parameters as, $ \phi = \tan ^{-1} [ r_D \sin \t _D /(1 +
r_D \cos \t _D ) ].$ The fact that the experimental sensitivity should
also soon reach the $ O(10^{-1} ) $ level, motivates one to envisage a
comparison with the RPV contributions to these parameters.  The
reaction $ B \to \phi + K_S ^0 $ is governed by the subprocess $ b\to
s +\bar s +s$ and is assigned a transition amplitude from the combined
gauge and RPV interactions of form: \be A (B_d ^0 \to \phi ^0 + K_S
^0)= \bigg [-{G_F \over \sqrt 2 } V^\star _{ts} V_{tb} C_W + {\l ^{
'\star } _{i23} \l ' _{i22} + \l '_{i32} \l ^{ '\star }_{ i22} \over 8
N_c m^2 _{\tilde \nu _{iL} } } \bigg ] <K_S ^0 \phi ^0\vert (\bar s\g
^\mu s )(\bar s \g ^\mu b) \vert B ^0_d >,\ee where the Wilson
coefficient in the SM prediction is evaluated as, $C_W (m_W) \simeq
2. \times 10^{-2} $.  With a similar formula applying to the second
decay mode, related by the substitution $s \to c$, one can express the
RPV contributions to the auxiliary physical parameters
as~\cite{guetta}, $ r_D (B_d^0 \to \phi ^0 +K_S ^0 ) \simeq 8. \times
10^ 2 \vert \l ^{ '\star } _{i32} \l ' _{i22} + \l '_{i23} \l ^{
'\star }_{ i22} \vert ({m_W \over m_ {\tilde \nu _i } })^2 , \ \ r_D
(B_d \to J/\Psi + K_S ^0 ) \simeq 2. \times 10^ 2 \vert \l ^{ '\star }
_{i23} \l ' _{i22} \vert ({m_W \over m_ {\tilde e _{iL} } })^2 .$
Using the existing bounds on the relevant coupling constants, yields
the encouraging estimate for the phases difference~\cite{guetta}, $ \D
\phi \simeq O(1) $.

The neutral or charged $B $ meson decays into non-pure CP channels are
also of use in inferring experimental constraints on new physics.  The
CP-odd decay asymmetries for neutral $B$ mesons are given by the
differences between the decay rates for the CP conjugate transitions,
$ B^0 \to f ,\ \bar B^0 \to \bar f $. Interesting
signals~\cite{kaplan} from the RPV contributions are also expected for
the charged $B $ meson decays CP-odd asymmetries for the decay
reactions, $ B^+ \to f ,\ B^- \to \bar f $.  Representative examples
are given by $ \ B_d^+ \to J/\Psi +K^+ , \ B_d^+ \to \pi^++\pi^0 $.

For the charged $B$ meson decay mode, $ B _d^\pm \to \pi^\pm +K ^0 $,
the comparison of the tree level RPV contribution to the transition
amplitude~\cite{bhatta99}, $ A_{RPV} \approx [ \l '_{i13} \l ^{'\star
} _{i12} /m^2 _{\tilde \nu _{iL} } ] ( b \bar s) (d\bar d)$, with the
measured branching fraction yields the useful $1\s $
bound~\cite{bhatta99}, $\vert \l '_{i13} \l ^{'\star } _{i12} \vert <
5.7\times 10^{-3}$.  The CP-odd rate asymmetry in this decay mode is
the more interesting to the extent that the SM contribution there is
expected not to exceed $ 40 \% $, whereas the RPV
contribution~\cite{bhatta99} can easily enhance the asymmetry by
nearly $100\% $.

Several recent studies~\cite{asym01,asym02} have attempted to resolve
the various observed anomalies in the $B$ mesons decays by invoking
the tree level contributions from the RPV interactions.  Kundu and
Mitra~\cite{asym02} consider the $\l '_{ijk}$ interactions tree level
contribution to the subprocess, $ b\to s + s+ \bar s $, controlling
the $B$ meson decay modes into strange mesons. The scan over the MSSM
parameter space, including the experimental constraints from $B$ meson
decays, leads to the following inequality, $ 1.3\times 10^{-3} \leq
\vert \l' _{i32} \l ^{'\star } _{i22} \vert \leq 2.3 \times 10^{-3} $.
Focusing on the measured anomalies in the branching fraction and
CP-odd asymmetry for the decay mode, $ B^0 \to \pi ^+ +\pi ^-$,
Bhattacharyya et al.,~\cite{asym01} find that a satisfactory
explanation can be sought by considering the tree level RPV
contribution provided this obeys the quadratic coupling constant
product bound, $ \vert \l' _{i11} \l ^{'\star } _{i13} \vert \leq 2.5
\times 10^{-3} $.
 
The observation of decay modes for the bottom quark baryons has a good
potential of discovery for the new physics.  Experimental measurements
are already available for the hyperon semileptonic and nonleptonic
decay modes, $\L _b \to \L + e_i ^- +\bar \nu +X ,\ \L_b \to \L +
J/\Psi $.  More experimental data is anticipated to come in the future
from the $B$ meson factories.  Motivated by this perspective,
Mohanta~\cite{mohanta00} has examined the tree level RPV contribution
to the transition amplitude of the two-body process, $\L _b \to p +
\pi ^-$, which involves quadratic products of the coupling constants,
$ \l ' _{ijk} ,\ [(ijk)= (211), \ (213),\ (311),\ (313) ]$.  The decay
rate asymmetry, $a_{CP}\propto \G (\L _b \to p + \pi ^- ) -\G (\bar \L
_b \to \bar p + \pi ^+ ) $, and the polarization asymmetry observable,
$\a \propto \Re (S ^\star P) $, controlled by the interference term
between the S-wave and P-wave parity violating and conserving
amplitudes, give access to two useful CP-odd observables.  The current
bounds on the above coupling constants are found to lead to a strongly
suppressed rate asymmetry but a strongly enhanced polarization
asymmetry.

\subsection{Neutrino masses,  mixing and  instability}
\label{neutx1}

The experimental observations of the neutrinos mass and mixing
parameters can be grouped in three main classes: (1) The laboratory
experiments accessing the light neutrino masses through the weak
decays of hadrons and nuclei and the effective electron neutrino
Majorana mass parameter, $ <m_{\nu_e }> = \sum _j (V^{'\dagger } )
^2_{1j} m _{\nu _j} $, measured in the $ \b \b _{0\nu } $ neutrinoless
nuclear double beta decay reactions; (2) The neutrinos flavor
oscillation experiments with nuclear reactors, solar~\cite{skamio},
atmospheric~\cite{fukudas01,ahmad01} and particle
accelerators~\cite{lsnd}, yielding information on the neutrinos
pairwise mass differences and mixing angle
parameters~\cite{bahcall98}; (3) The cosmological constraints bounding
the mass density of neutrino cosmic relics by the Universe critical
density and by the requirement of producing the observed anisotropies
in the density fluctuations responsible for the galactic structure.

We discuss in the present subsection four main topics relating to: (i)
Neutrino Majorana masses, (ii) Neutrino electromagnetic dipole
moments,(iii) Sneutrino Majorana masses, (iv) Models of charged
leptons and neutrino mass matrices.  The last topic has recently
witnessed a surge of interest motivated by the aim to extract from the
oscillation measurements an information on the generational structure
of the neutrinos mass matrix. A variety of future promising projects
under development use high energy neutrino telescopes~\cite{bonn01}
and dedicated detectors.  The phenomenological aspects of neutrino
masses and interactions are discussed in the review by Bilenky et
al.,~\cite{bill99}.

\subsubsection{\bf Neutrinos Majorana mass  matrix}

The Majorana mass matrix of the electroweak doublet neutrinos,
described by the effective Lagrangian, $L_{EFF}= - \ud (m_{\nu })_{ij}
\bar \nu ^c_{iR} \nu _{jL} + H.\ c.$, acquires contributions from the
lepton number violating $\l '$ or $\l $ interactions at the one-loop
order. The one-loop Feynman graphs involve exchange of
fermion-sfermion pairs, $d - \tilde d_{kH}$ or $l - \tilde l_{kH}, \
[H= L,R]$.  The chirality selection rules require the insertion of
fermionic and bosonic mass terms for the internal fermion and
sfermion, with the latter involving the mass matrices for down-squarks
and charged leptons, $(\tilde m^{d2} _{LR})_{ij} ,\ (\tilde m^{e2}
_{LR})_{ij} $.
The following useful approximate formulas for the one-loop RPV
contributions to the neutrino mass matrix~\cite{dimohall,godbole} are
found by neglecting the mixing between sfermions of same charge and
chirality, $ (\tilde d_{jL} ,\ \tilde d_{k' R}) $ and $ (\tilde e_{jL}
,\ \tilde e_{k' R})$, and taking the limit where the $L, R$ internal
sfermions masses are much larger than the corresponding internal
fermions masses, $ m _{\tilde d_{jL} } \sim m _{\tilde d_{k' R}} >>
m^d _k $ and $ m _{\tilde e_{jL} } \sim m _{\tilde e_{k' R}} >> m^e _k
$: \begin{eqnarray} && ( m_{\nu })_{i'i} \simeq { N_c \over 8\pi^2} \l
^{' } _{ijk} \l ^{' } _{i'kk'} { (\tilde m^{d2} _{LR} )_{jk'} m_{d _k}
\over m_{\tilde d_{jL} } ^2 } + { 1 \over 8\pi^2} \l _{ijk} \l
_{i'kk'} { (\tilde m^{e2} _{LR} )_{jk'} m^e_{k} \over m_{\tilde e_{jL}
} ^2 },
\label{eqx12}
\end{eqnarray}
The internal fermions mass terms in the numerators clearly indicate
that the largest contributions are those arising from the $\l ' $
interactions with internal line quarks belonging to the third or
second generations. Comparing with the experimental limit on the
$\nu_e$ mass inferred from $\b \b _{ 0\nu } $ neutrinoless double beta
decay, $m_{\nu _e} < 5 \ \text{ eV} $, and using the estimate for the
soft breaking flavor diagonal parameters, $ (\tilde m^{d2} _{LR} )_{j}
\approx 0.1 \times m_{\tilde d_j } m_{d_j } $, Godbole et
al.,~\cite{godbole} deduced the coupling constant bounds, $\vert \l
'_{133} <\vert 3.5\times 10^{-3} , \ \vert \l '_{122}\vert < 7.\times
10^{-2} .$ To exhibit the dependence on the various input parameters,
we quote the same bounds in a more explicit way, \be \vert \l '_{133}
\vert < 1.73 \times 10^{-4} ({m_{\nu _i} \over 5\ \text{eV} } )^\ud
({m_ b \over 4.44 \ \text{GeV} } )^\ud ({ m_{\tilde d_{jL} } ^2\over
\tilde m^{d2} _{LR} } ) ^\ud , \quad \vert \l '_{122} \vert < 8.8
\times 10^{-4} ({m_{\nu _i} \over 5 \ \text{ eV} } )^\ud ({m_s \over
0.170 \ \text{GeV} } )^\ud ({ m_{\tilde d_ {jL} } ^2\over \tilde
m^{d2} _{LR} })^\ud . \label{eqgodb1} \ee A similar comparison of the
RPV contributions to neutrino diagonal mass matrix elements with the
experimental limits on $\nu_\mu $ and $ \nu_\tau $ masses yields the
bounds: $\vert \l '_{233}\vert < 0.63 \ ( m_ {\tilde d} /\tilde
m^d_{LR}) , \ \vert \l '_{333} \vert < 7.6 \ (m_ {\tilde d} /\tilde
m^d_{LR}) .$ The coupling constant bounds inferred for the first and
second families, $ \l'_{111},\ \l'_{ 112} , \ \l'_{121}$, are
uninterestingly weak ones.

\subsubsection{\bf Neutrino electromagnetic dipole  transition moments} 

The anomalous dipole moments of neutrinos~\cite{palmoha} are
represented by the Pauli-Dirac flavor diagonal, static and chirality
changing operators ($\nu _L \to \nu _R = \nu ^c_L$) and the Majorana
flavor non-diagonal, dynamic and lepton number violating operators
($\nu _{iL} \to \nu ^c_{jL}, \ [i\ne j]$).  The experimental
constraints from neutrino oscillations set the following typical
bounds on the transition moment and mass splittings for neutrino
pairs~\cite{palmoha,voloshyn86}: $ (\mu _{\nu _e \nu _X} ) \leq (0.1 \
- 1.)  \times 10^{-10} \mu _B ,\ \vert \D _{\nu _e \nu _X} \vert =
\vert m_{\nu _e} ^2 - m_{\nu _X} ^2 \vert < 10 ^{-7} \ \text{eV} ^2 .$
The observed burst of neutrino flux from the supernova SN 1987A place
a severe bound on the flavor diagonal Dirac anomalous magnetic
moment~\cite{palmoha}, $ \mu ^D_{\nu _J} < (0.1 \ - 1.) \times
10^{-12} \ \mu _B .$ The size of the quoted experimental bound on the
moment is seen to be much lower than the value obtained for the
one-loop contribution to the neutrino Dirac static magnetic
moment~\cite{fujik80} in the SM with right chirality neutrinos, $ \mu
_{\nu _J } ^D \simeq { 3 e G_F \over 8\sqrt 2 \pi ^2} m_{\nu _J
}\approx 3.1 \times 10^{-19} \mu_B ({ m_{\nu _J }\over 1 \ \text{ eV}
}) $.

Tight correlations between the contributions to the neutrinos moments
and masses are present in most new physics models, with the
experimental constraints being generally stronger on masses than on
moments.  In the case of Dirac neutrinos, the possibility to suppress
their Dirac masses without touching their moments led
Voloshin~\cite{voloshyn88} to postulate an $SU(2)_\nu $
particle-antiparticle type symmetry under which the
neutrino-antineutrino field pairs, $ (\nu _i, \ \nu _i ^c) $,
transform as doublets.  Using the representation of the Dirac
electromagnetic dipole moment and mass operators, $\bar \nu _{iR } \s_
{\rho \s } \nu _{iL}= - \nu ^{cT}_{iL} C ^\dagger \s_ {\rho \s } \nu
_{iL} , \ \bar \nu _{iR }\nu _{iL} = - \nu ^{cT} _{iL} C ^\dagger \nu
_{iL}$, where $ C$ denotes the Dirac algebra charge conjugation
matrix,
one finds that the moment and mass operators are symmetric and
antisymmetric under the substitution, $\nu _{iL} \leftrightarrow \nu
^c_{iL} $, hence transform as singlet and triplet representations
under $SU(2)_\nu $, respectively.  Thus only the neutrino moment
receives a contribution in the unbroken $SU(2)_\nu $ symmetry
limit. The discrete symmetry subgroup acting as, $ D_\nu : \nu \to \nu
^c,\ \nu ^c\to - \nu $, might do as well and is obviously to be
preferred since the $SU(2)_\nu $ symmetry suffers a spontaneous
breaking at the electroweak scale.

In the strict MSSM without right handed neutrinos, Babu and
Mohapatra~\cite{babun} proposed an alternative attractive suppression
mechanism based on the horizontal symmetry combining the discrete
group, $ D _H :\ \nu _e \to \nu _\mu ,\ \nu _\mu \to -\nu _ e $, with
the abelian lepton number group, $U(1)_{L_e-L_\mu } $.  Considering
the neutrinos moment and mass operators, one finds that the $L_e-L_\mu
$ symmetry forbids the flavor diagonal Majorana mass operators, $ \bar
\nu _e ^c \nu _e ,\ \bar \nu _\mu ^c \nu _\mu $, while the $D_H $
symmetry forbids the off-diagonal mass operator, $ \bar \nu _e ^c \nu
_\mu ,$ but allows the off-diagonal moment operator, $ \bar \nu _e ^c
\s _{\rho \s } \nu _\mu $, which are even and odd under $D_H$,
respectively.  In the context of broken R parity symmetry, Babu and
Mohapatra~\cite{babun1} considered the lepton number symmetry group, $
U(1)_{L_e-L_\mu } $, in combination with the horizontal discrete
symmetry group, $D _H \subset SU(2)_H $, acting on the electroweak
doublet and singlet lepton fields of the first two generations as, $ D
_H:\ \quad (L_e , e^c) \to (L_\mu , \mu ^c) , \quad (L_\mu , \mu ^c)
\to ( -L_e , - e ^c), $ with all other fields left invariant.  The
$D_H$ symmetry restricts the allowed RPV interactions to the lepton
number violating subset, $\mu _\tau ,\ \l_{131} = \l_{232},\ \l_{123}
$ and $ \l' _{3jk}$, with a similar restriction for the soft
supersymmetry breaking interactions.  The one-loop diagrams with
sleptons and leptons internal lines yield a vanishing contribution to
the flavor off-diagonal, $ \nu _e \to \nu _\mu $, neutrino mass but a
finite contribution to the $M1$ transition moment, as described by the
formulas, \bea && m_{\nu _e \nu _\mu } =0,\ \mu _{\nu _e \nu _\mu }
\simeq {e m_\tau \sin( 2 \t ) \l _{123} ^\star \l _{131} \over 8\pi
^2} \bigg ( {1 \over m_{\tilde e_1} ^2 } [\ln {m_{\tilde e_1} ^2 \over
m_\tau ^2 } -1] - (\tilde e_1 \leftrightarrow \tilde e_2 )\bigg ),\eea
where the $\tilde e$ and $\tilde \mu $ systems are described by
$2\times 2$ mass matrices, with pairs of mass eigenvalues and mixing
angles denoted as, $ [m_{\tilde e_I} ,\ \t_ {\tilde e}] ,\ [ m_{\tilde
\mu _I} ,\ \t_{\tilde \mu }], \ [I=1,2]$.  We have set, for
simplicity, $ \t \equiv \t_ {\tilde e} = \t_{\tilde \mu } $.  The
predicted neutrino transition moment~\cite{babun1} lies within the
experimentally interesting range, $\mu _{\nu_e \nu _\mu } \approx \
(10^{-11} - 10^{-10}) \mu_B $.  However, the necessary breaking of the
$D_H$ symmetry, which is required to reproduce the finite $e - \mu $
mass splitting, generates additional radiative contributions to the
off-diagonal neutrino mass term, $m _{\nu_e \nu _\mu } $.  Including
the soft supersymmetry breaking mass terms for smuons and
selectrons~\cite{babun1}, $ V_{soft} = \d m^2_{\tilde \mu _L} \tilde L
^\dagger _{2L} \tilde L _{2L} + \d m^2_{\tilde \mu _R} \tilde e ^\star
_{2R} \tilde e _{2R} $, along with the one-loop contribution with
$\tchi - \tilde l $ internal lines to the charged leptons mass
splitting, $ m_e- m_\mu $, one obtains the following additional
contributions to the neutrino masses, \bea && \d _1 (m _{\nu _e \nu
_\mu }) \simeq {\l ^\star _{123}\l _{131} \over (4\pi )^2} m_\tau \sin
^2 (2\t _{\tilde e} ) \bigg [ -{\d \tilde m^2_{1} \over m^2_{\tilde
e_1} } {\d \tilde m^2_{1} \over m^2_{\tilde e_2} -m^2_{\tilde e_1} }
\ln {m^2_{\tilde e_1}\over m^2_{\tilde e_2} }- (1\leftrightarrow
2)\bigg ] , \cr && \d _2 (m _{\nu _e \nu _\mu }) \simeq {\l ^\star
_{123} \l _{131} \over (4\pi )^2} \sin( 2 \t _{\tilde \tau })
(m_e-m_\mu ), \eea where $\t _{\tilde \tau }$ designates the mixing
angle in the $ \tilde \tau _L - \tilde \tau _R$ system.  The leptons
mass splitting, $ m_e -m_\mu $, and the neutrino mass bound, $m_{\nu_e
} =m_{\nu_\mu } < 10 \ \text{ eV} $, can both be satisfied but at the
cost of a fine tuning in adjusting the flavor dependent mass
differences, $\d \tilde m_1 ^2 = m ^2_{\tilde e_1 } - m ^2_{\tilde
\mu_1 } $ and $\d \tilde m_2 ^2 = m ^2_{\tilde e_2 } - m ^2 _{\tilde
\mu _2 } $.  The off-diagonal neutrino mass contribution in the second
entry, $\d _2 (m _{\nu _e \nu _\mu }) $, can be bounded by adjusting
the mixing angle, $\t _{\tilde \tau }$.

With a diagonal structure for the chirality flip mass matrix for the
sfermions, $(\tilde m ^{e2} _{LR})_{ij } = \tilde m _0 m ^e_i \d_{ij}
, \ [\tilde m _0 \simeq A m_{3/2} ] $, the natural size of the RPV
contributions to the $ M1$ transition moment is~\cite{barbnum1}, $\mu
_{\nu _e \nu _\mu } \simeq 10 ^{-13} \mu _B$.  The dominant one-loop
diagrams with the $\l $ interactions involve $\tau -\tilde \mu $ and
$\tau - \tilde e $ internal lines.  Under the hypothesis of a dominant
pair of RPV coupling constants, one finds mutually proportional
contributions to the off-diagonal neutrino mass and magnetic moment
matrices expressed by the relation, $ \vert {\d (\mu_{\nu_e \nu_\mu })
\over \mu_B } \vert \simeq \d (m _{\nu_e \nu_\mu } ) {8m_e \over
m_{\tilde e } ^2 } [\ln {m_{\tilde e } \over m_f} -1] , $
which implies that the bounds on the neutrino moment and mass would be
simultaneously saturated.  The cosmological limit on the sum over
light neutrino masses yields a quadratic coupling constant bound,
which entails in turn a severe bound on the neutrino transition
moment, $ m_{\nu_e \nu_\mu } < 10 \ \text{ eV} \ \Longrightarrow \
\vert \l^{\star} _{123} \l _{232} \vert < 4. \times 10^{-4} ( m
_{\tilde f } / 100 \ \text{GeV} ) ^2 \ \Longrightarrow \ (\mu_{\nu_e
\nu_\mu } /\mu_B ) < (0.2 - 0.4) \times 10^{-13} \ (m_{\tilde f } /
100 \ \text{GeV} ) ^{-2}. $

The flavor structure of the chirality mixing scalar mass matrix
$(\tilde m ^{e2}_{LR})_{ij} $ is a crucial input in determining the
size of RPV contributions to the neutrinos masses and moments.
To explore the impact of departures from the flavor diagonal ansatz,
Barbieri et al.,~\cite{barbnum1} used guidance from a postulated
horizontal $SU(2)_H$ symmetry.  The relationship, $m _{\nu_e \nu_\mu }
\propto (\tilde m ^{e2} _{LR})_{11}-(\tilde m ^{e2} _{LR})_{22} $,
shows that a hierarchy in the chirality flip sfermions mass matrix
needs to be imposed only between the diagonal and off-diagonal entries
in order to satisfy the inequality, $(\tilde m ^{e2} _{LR})_{11}
\approx (\tilde m ^{e2} _{LR})_{22} >> \vert (\tilde m ^{e2}
_{LR})_{11}-(\tilde m ^{e2} _{LR})_{22} \vert $.  However, invoking an
horizontal symmetry faces the problem that even a small breaking of
the symmetry, consistent with the allowed flavor mixing of the scalar
superpartners, could lead to unacceptable radiative contributions to
the $e-\mu $ fermion mass splitting.  If the symmetry breaking is
represented by soft slepton mass terms, the required suppression of
radiative corrections cannot be achieved without some fine tuning of
different input parameters.  The solution advocated by Barbieri et
al.,~\cite{barbnum1} consists in assuming that the $ SU(2)_H$ symmetry
breaking affects the Yukawa interactions sector only, so that the
$e-\mu $ mass splitting is accounted for directly in terms of the tree
level parameters.  Although the symmetry is exact in the supersymmetry
breaking sector, there still occur one-loop contributions to $m_e
-m_\mu $, involving the one-loop Feynman diagram with $ \tilde e -
\tchi ^0$ exchange, which, however, may be appropriately bounded by
requiring the bound, $(\tilde m ^{e2}_{LR})_{11} < 100 \ \text{GeV} ^2
$.  The resulting RPV contribution to the neutrino moment is enhanced
in this case to the value, $ \mu _{\nu _e \nu _\mu } \sim O(10^{-12})
\mu _B$.

Analogous contributions to the neutrinos masses and moments arise from
the $\l '_{ijk}$ interactions.  The information on the neutrino mass
matrix elements accessed from the experimental data on neutrino flavor
oscillation can be used to fix the values of the $\l '_{ijk}$ coupling
constants and hence deduce predictions for the neutrino
moments. Bhattacharyya et al.,~\cite{bhaklap} consider two main
scenarios characterized by large and small mass splittings between the
down-squarks of different generations relative to the supersymmetry
breaking scale.  In the large mass splitting case, $ m_ {\tilde d_j}
>> m_ {\tilde d_k} $, corresponding to the situation considered by
Barbieri~\cite{barbnum1}, the one-loop neutrino moment has a
logarithmic dependence on mass involving the term, $[ \ln (m_{\tilde
d_k} / m_{d_j} ) -1 ]$.  The alternative case of nearly degenerate
down-squarks, $ m_ {\tilde d_j} \simeq m_ {\tilde d_k} $, also
presents a logarithmic enhancement factor but involving now the
fermion masses, $ \ln (m_{d_j} / m_{d_k}) $.  The various combinations
of quadratic coupling constant products, $\l ' _{ijk} \l ^{'\star }
_{i'j'k'} $, compatible with neutrino mass matrix elements in the
range, $ m_{\nu _i \nu _j} = [0.01 - 2.5 ]$ \ \text{eV}, are found to
lead to contributions to the magnetic moments falling in the range, $
\mu _{\nu _i \nu _j}/\mu _B = [10^{-15} - 10^{-16}] $.

For the neutrino $M1$ moment, the direct comparison by Abel et
al.,~\cite{abel99} of the one-loop contribution with the experimental
limit, $\vert \mu _{\nu _e} \vert \leq 1.5 \times 10^{-10} \mu _B$,
gives the coupling constant bounds, $ \vert \l _{121} \l ^\star _{212}
\vert < 0.58 ,\ \vert \l ' _{123} \l ^ {'\star } _{232} \vert < 0.030
$.

The neutrino instability due to the radiative decay mode, $\nu ' \to
\nu +\g $, $M1$ initiated by the electromagnetic transition moment has
an important impact on the possibility of the light neutrino cosmic
relic to qualify as a hot dark matter candidate.  The correlation
between the neutrinos mass and lifetime is described roughly by the
relations, $m _\nu \simeq 28. \ \text{eV}, \ \tau _\nu > 0.9 \ \times
10 ^{23} \ \text{s} $.  The possibility that the RPV one-loop
contributions can account for the requisite small enough mass and long
lifetime is examined by Roulet and Tommasini~\cite{roulet91}.  The
approximate formula for the radiative decay lifetime, $ {\tau (\nu '
\to \nu +\g ) \over 0.9 \ \times 10 ^ {23} \ \text{ s}} \simeq ( {m
_{\nu } \over 28 \ \text{eV} }) ^ {-3} ( {\mu _{\nu \nu ' } \over 10 ^
{-14} \mu _B }) ^ {-2} \ , $ in light of the above discussed results
for the neutrinos transition mass and moment, shows that the
correlation between $ m_\nu $ and $ \tau _\nu $ can be satisfied by
setting the relevant product of RPV coupling constants at the value, $
\vert \l _{232} \l ^\star _{233} \vert \simeq 2.3 \times 10 ^ {-4} $,
while allowing arbitrary values for the corresponding ratio, $ \vert
\l _{232} /\l ^\star _{233} \vert $. The latter ratio \ may either be
$ O(1)$ or exhibit a large hierarchy.

\subsubsection {\bf Neutrino  propagation in matter} 

The option of a broken R parity symmetry offers an attractive
resolution of the solar neutrino flux deficit problem avoiding
altogether the need of introducing massive neutrinos.  This
possibility rests on the presence of finite tree level contributions
to the flavor diagonal and and off-diagonal scattering amplitudes of
neutrinos on leptons and quarks, $\nu_\a + f \to \nu_{\b } +f , \ [f=
u, \ d, \ l;\quad (\a , \b ) = e ,\ \mu , \ \tau ] $ which may
significantly affect the time evolution of the neutrinos fields in
their travel from the Sun core to the Earth.  Scenarios using these
contributions alone or in combination with the familiar flavor
oscillations and the resonant enhancement MSW
(Mikheyev-Smirnov-Wolfenstein) effect~\cite{msw7886} are discussed by
several authors~\cite{barger91}.  The effective four fermion local
couplings describing the neutrinos-fermions elastic scattering
amplitudes along with the defining formulas for the auxiliary physical
parameters, $ \e _f,\ \e ' _f $, associated to the ratios of neutral
to charged current contributions are described by the formulas, \bea
&& L _{EFF}= - G_{F} \sqrt 2 \sum _{d, u , e } (\bar \nu_{x L } \g_\mu
\nu_{ x'L} ) [ g_{V xx'} ^{\nu f} ( \bar f\g^\mu f) + g_{A xx'} ^{\nu
f} ( \bar f\g^\mu \g _5 f)] + H. \ c. , \cr && \bigg [\e _f ={1 \over
A^{W} } A^{NC} (\nu _e f ) = { g^{\nu f} _{Vex } } ,\ \e ' _f = {1
\over A^{W} } [A^{NC} (\nu _x f ) - A^{NC} (\nu _e f)] = [g^{\nu f}
_{V xx } - g^{\nu f} _ {V ee }] \bigg ] .\eea The neutrino field time
evolution in the two-flavor case, $ \nu _e \to \nu _x $, is described
by the coupled linear differential equations, \bea && i { \dh \over
\dh t } \pmatrix{ \nu_e (t) \cr \nu_x (t) } = \pmatrix{ M_{ee} &
M_{ex} \cr M_{xe}& M_{xx} } \pmatrix{\nu_e \cr \nu_x } , \cr &&
[M_{ee}= 0, \ M_{ex} = M_{xe} = {\D _{ex} \sin (2\t _{ex}) \over 4
E_\nu } +B , \ M_{xx} = {\D _{ex} \cos (2\t _{ex}) \over 2 E_\nu } +C
, \cr && [\D _{ex} = \vert m _{\nu _e } ^2 - m _{\nu _x } ^2 \vert ,\
B = \sqrt 2 G_F (\e_d n_d +\e_u n_u + \e_e n_e ), \ C = \sqrt 2 G_F
(\e ' _d n_d +\e '_u n_u + (\e ' _e -1) n_e )], \eea

where $ E_\nu $ denotes the neutrino energy, $ \D _{ex},\ \t _{ex} $
the difference of mass squared and mixing angle, the matrix elements
$M_{ex}, \ M_{xx} $ denote the flavor changing (off-diagonal) and
conserving (diagonal) contributions to the forward (vector and axial
vector) scattering amplitudes of neutrinos with matter particles in
the Sun, and $ n_f , \ [f= d,\ u ,\ e] $ denote the number densities.
The two additive contributions, $ M_{ex}$ and $ M_{xx} $, in the
transition amplitudes are associated with neutrino oscillation and
neutral current interactions with quarks and leptons. The auxiliary
parameters $\e _f , \ \e ' _f , \ [f= d,\ u ,\ e] $ parameterize the
ratios of neutral current to charged current scattering amplitudes of
neutrinos with matter particles.

The RPV contributions for the two pairwise combinations of neutrino
flavors, $ \nu _e -\nu _\mu , \ \nu _e -\nu _\tau$, are studied in
several works~\cite{barger91}.  For illustration, we quote
representative formulas for the scattering amplitudes on down-quarks
and charged leptons in the mixed $\nu_e- \nu_\tau $ case, \bea \bullet
\quad && \e _d = {1\over 4\sqrt 2 G_F} [{\l '_{3j1} \l ^{'\star }
_{1j1} \over m_{\tilde d_{jL} }^2} - {\l '_{11k} \l ^{'\star } _{31k}
\over m_{\tilde d_{kR} }^2} ], \ \e _e = {1\over 4\sqrt 2 G_F}
{\l_{1j1} \l ^{'\star } _{3j1} \over m_{\tilde e_{jL} }^2},\cr \bullet
\quad && \e ' _d = {1\over 4\sqrt 2 G_F} { \vert \l ^{'2}_{3j1}\vert
^2 - \vert \l ^{'2} _{1j1} \vert ^2 \over m_{\tilde d_{jL} }^2 } + {
\vert \l ^{'2}_{11k}\vert ^2 - \vert \l ^{'2} _{31k} \vert ^2 \over
m_{\tilde d_{kR} }^2},\ \quad \e ' _e = {1\over 4\sqrt 2 G_F} { \vert
\l ^{2}_{331}\vert ^2 - \vert \l ^{2} _{1j1} \vert ^2 \over m_{\tilde
e_{jL} }^2} . \eea The down-quark auxiliary parameters are constrained
through the known constraints on the RPV coupling constants from the
leptons rare decay modes, $ \mu ^\pm \to e ^\pm + \g, \ \tau ^\pm \to
\rho ^0 + e^\pm $, with bounds of order, $[\vert \e_e \vert,\ \vert
\e_d \vert ] < O(10^{-2})\ - \ O(10^{-5}), \ [\vert \e '_e \vert,\
\vert \e ' _d \vert ] < O(10^{-2}) $.  The solutions for $\e _d, \ \e
' _d $ reproducing the mean neutrino counting rates in the $ ^{37} Cl
$ and Kamiokande-II experiments are studied by Barger et
al.,~\cite{barger91} in a variety of scenarios involving either
massless neutrinos, or massive neutrinos with flavor off-diagonal
effects, or also massive neutrinos with flavor diagonal and
off-diagonal effects.  The solar neutrino deficit may be explained on
the basis of matter enhancement by neutral current flavor diagonal and
off-diagonal contributions involving the down quarks.  For instance,
the scenario with massless neutrinos selects annular regions for the
auxiliary parameters, $\e _d \simeq [0.01, \ 0.1], \ \e ' _d \simeq
0.6$.  In the scenario involving a mass degenerate pair of neutrinos,
$\D _{ex} =0$, the experimental data can be fitted with auxiliary
parameters lying in the ranges, $ \e _{[d, u] } \approx (10^{-3}\ -\
10^{-2}) , \ \e '_{[d, u]} \approx \ (0.5 -\ 0.8) $.  One solution is
found which favors a neutrino resonance crossing in the interval of
density ratios, $n_n/n_e \in [\ud , \ {1\over 6} ] $, the suppression
effect depending only on the neutrino type and not its energy. The
fact that the $\nu _e \to \nu _\tau $ transition is favored over the $
\nu _e \to \nu _\mu $ transition shows how the broken R parity
symmetry option may lead to a solution to the solar problem radically
different from conventional one.

Several recent works~\cite{adhi02,dreimoreau02} examine the extent to
which the RPV contributions to the neutrino scattering processes in
the sun might offer an acceptable solution to the combined data for
the solar and atmospheric neutrino experiments.  The study by Dreiner
and Moreau~\cite{dreimoreau02} for the complete three-flavor case,
taking into account the energy spectrum measured by the Sudbury (SNO)
collaboration of the recoil electrons emitted in the $^8B $ beta
decay, reaches the conclusion that the MSW solution for the solar
neutrino flux is the preferred one.

\subsubsection{\bf  Karmen anomalous process}

The time anomaly distribution in the neutrino beam experiment,
reported by the Karmen collaboration~\cite{karmen1,karmen2} at the
Rutherford Laboratory, has stimulated strong interest towards an
interpretation of the observed anomaly on the basis of a broken R
parity symmetry.  The anomaly resides in the observation of an
anomalous time structure for the neutrinos produced from the decay of
the stopped pions beam. The signal can be explained in terms of the
unusual two-body decay reaction of charged pi-mesons, $\pi ^+ \to \mu
^+ +X ^0$, with emission of a neutral fermion of mass, $ m_X = m_ {\pi
^+} - m_ {\mu ^+} -5 \ \text{keV} = 33.9 $ \text{MeV}.  The anomalous
structure of the time profile requires that the fermion $X$ decays
while traversing the detector with a rate, $B( \pi ^+ \to \mu ^+ + X
^0) \G (X ^0 \to e ^+ +e^- + \nu ) = 2.6 \ \times 10 ^{-11} s ^{-1} $.
The experimental data exhibits a correlation between the neutral
fermion particle decay lifetime, $ \tau _X $, and its production rate
by pion decay, which is illustrated by the typical values, $ B(\pi ^+
\to \mu ^+ +X) ={\G (\pi ^+ \to \mu ^+ +X)\over \G (\pi ^+ \to \mu ^+
+\nu _\mu )} \simeq 1.2 \times 10^{-8} ,
\quad \tau _X > 0.3 \ \mu \text{s} . $

Attractive interpretations of the Karmen experiment observations have
been proposed on the basis of a broken R parity symmetry, with the
light fermion $X$ identified with a light (photino or zino) neutralino
$ X= \tchi ^0 = (\tilde \g ,\ \tilde z)$.  In the bilinear option, one
must require the bound~\cite{nowak96}, $\mu < 10 $ MeV.  In the
trilinear option, Choudhury and Sarkar~\cite{choudsak96} considered
the case of a light neutralino which decays by neutrino emission
through the two-body radiative decay mode, $\tchi ^0 \to \nu +\g
$. The production and decay amplitudes are both controlled by the
single RPV coupling constant $ \l '_{211}$.  This interpretation may
not be compatible with the recent experimental data~\cite{karmen2},
since the proposed mechanism entails the observation of a so far
unseen mono-energetic photon at an energy, $ E_\g \sim 17 $ \
\text{MeV}.

The recent study by Choudhury et al.,~\cite{choud99} distinguishes the
production process, controlled at the tree level by the operator, $ \l
'_{211} L_2 Q_1D^c_1$, from the decay process, involving the
three-body decay mode $ \tchi ^0 \to e^+ +e^- +\nu _j$, controlled at
the tree level by the operators, $\l _{1j1} E_1\nu _j E^c_1$.  The
consistency with the experimental observations for the correlation
between $B(\pi ^+ \to \mu ^+ +\tchi ^0) $ and $\tau (\tchi ^0 \to \g
+\nu _\mu )$ selects a domain in the plane of the coupling constants $
(\l '_{211} , \ \l _{1j1}) $ which is bounded approximately by the
upper and lower bounds~\cite{choud99}: $ \vert \l '_{211}\vert < 3.0
\times 10^{-5} \ \tilde f^2 , \ \vert \l _{1j1} \vert > 7.4 \times
10^{-4} \ \tilde f^2 $.  The light neutralino postulated in the above
discussed proposals~\cite{choudsak96,choud99} poses certain
compatibility problems with the cosmological constraints associated
with the $\tchi ^0$ relic abundance and their possible large
production rate in supernovas.  These problems can, however, be evaded
if one takes the photino instability into account.

\subsubsection{\bf Sneutrino Majorana masses}

Alongside with the familiar Hermitian scalar mass parameters of
superpartners, $\tilde m^2 _{ij} $, the supersymmetry breaking
interactions of sneutrinos and antisneutrinos~\cite{hirschosc,haber1}
may also allow for Majorana type scalar mass parameters, $\tilde
m^2_{Mij} $. These parameters enter the effective quadratic Lagrangian
for the complex sneutrinos and antisneutrinos fields as, $ L_{EFF}=- [
\ud (\tilde m^2_M)_{ij} \tilde \nu_i \tilde \nu_j + H. \ c. ]  -
\tilde m^2_{ij} \tilde {\bar \nu_i } \tilde \nu_j .$ Like the
fermionic neutrinos Majorana mass terms, $ L_{EFF}= -\ud m _{\nu ij}
\bar \nu ^c_{iR} \nu_{jL} +\ H.\ c. $, the bosonic sneutrinos Majorana
mass terms violate lepton number by two units, $ \D L (\tilde m_M) =\D
L (m_\nu )= - 2. $ In the case of a conserved CP symmetry, the
electroweak Higgs bosons, sneutrinos and antisneutrinos split into
CP-even and CP-odd eigenstates, $ h^0, \ H^0, \ \tilde \nu _{i+}$ and
$ A^0, \ \tilde Z^0, \ \tilde \nu _{i-}$, where $ Z^0_L $ represents
the electroweak Goldstone boson mode absorbed as the $Z$-boson
longitudinal spin component and the sneutrino CP eigenstate fields are
identified with the real and imaginary parts of the complex sneutrino
fields, $ \tilde \nu _{i+} = \sqrt 2 \Re (\tilde \nu_i ) = \tilde
\nu_{1i} , \ \tilde \nu _{i-} = \sqrt 2 \Im (\tilde \nu_i )= \tilde
\nu_{2i}, \quad [ (\tilde \nu _i, \ \tilde {\bar \nu _i } = \tilde \nu
^\star _i ) = {1 \over \sqrt 2 } (\tilde \nu_{1i} \pm i \tilde
\nu_{2i}) ]. $ The CP-even and CP-odd sneutrinos, $ \tilde \nu _{i\pm
} $, mix in different ways with the neutral Higgs bosons fields $ [
H^0, \ h ^0] $ and $ [ A^0, \ Z^0_L ] $ of same quantum numbers.  The
Majorana mass terms affect the mixing patterns in the CP-even and
CP-odd scalars sectors by lifting the mass degeneracies between the
sneutrinos-antisneutrinos mass eigenstates which would otherwise
assemble into three generations of pairwise degenerate massive pairs.
In the single generation case, for instance, the mass splitting
assumes the simple form,
$ m^2 _{\tilde \nu _{+} } - m^2 _{\tilde \nu _{-} } = (\tilde m^2 +
\vert \tilde m _M \vert ^2 ) - (\tilde m^2 - \vert \tilde m _M \vert
^2 ) = 2\ \vert \tilde m _M \vert ^2 . $

The RPV contributions to the Majorana sneutrino and neutrino masses
feature strong correlations, in the sense that the Majorana neutrino
mass terms may radiatively induce Majorana sneutrino masses at
one-loop level and vice-versa~\cite{haber2}.  The radiative generation
mechanism may be used with advantage to explain the generational
hierarchies in the neutrinos masses or the sneutrinos-antisneutrinos
mass splittings.  Tight correlations also exist between the neutrino
and sneutrino Majorana masses and the $\b \b _{0 \nu }$ double beta
decay amplitude, which take place on a pairwise basis and at variable
loop levels.  The calculated corrections to the neutrino and sneutrino
Majorana masses and the neutrinoless double beta decay rate, $ m_\nu
,\ \tilde m_M ,\ R(\b \b )_{0 \nu }$, are found to obey empirical
linear relations of form~\cite{klapdoretal}, $ z_i = \sum _j A_{ ij }
z_j + B_i , \ [ z_i = ( m_\nu ,\ \tilde m_M ,\ R ( \b \b _ { 0 \nu } )
)] $ where the calculable quantities, $ A_{ ij } , \ B_i $, are
accessed by evaluating higher order loop diagrams associated to the
various observables.  Thus finite contributions to $\b \b _{0 \nu }$
arise from $ m_\nu $ at tree level and from $\tilde m_M$ at one-loop
level, while finite contributions to $m_\nu $ and $ \tilde m_M$ arise
from $\b \b _{0 \nu }$ at $4 $ and $ 5$ loop levels, respectively, and
from $ \tilde m_M $ and $ m_\nu $ at one-loop level, respectively.
The resulting contribution from the sneutrinos Majorana mass terms is
found by Hirsch et al.,~\cite{klapdoretal} to impose the bounds, $
\tilde m_M < 2. \ \text{GeV} \ \tilde m^{3/2} $ or $ 11. \ \text{GeV}
\ \tilde m^{7/2} $, for models involving neutralinos dominated by a
bino component ($ \tilde B $) or a higgsino component ($ \tilde H$),
respectively.  The one-loop contribution to the neutrino mass induced
by the Majorana sneutrino mass term, $ \tilde m_M \tilde \nu \tilde
\nu $, requires the inequality, $ \tilde m_M < (60 \ - \ 125 ) \
\text{MeV} \ ({m_\nu \over 1\ \text{eV} })^\ud $.

In the bilinear option of broken R parity symmetry, the tree level
contributions to the sneutrinos-antisneutrinos mass splittings are
controlled by the misalignment between the four vector parameters, $
B_\a \mu _a $ and $ v_\a $~\cite{haber2}.  This condition parallels
the parameters alignment condition, $ \mu _a \propto v_\a $, needed to
suppress the tree level contributions to the neutrinos masses.  The
supersymmetric and non-supersymmetric auxiliary alignment parameters
involving the linear combinations, $ \xi = {v_i \over v_d} - {\mu _i
\over \mu } , \ \eta _i= {v_i \over v_d} - {\mu _i B_i \over \mu B} $,
respectively, would clearly coincide in the case of a generation
universal supersymmetry breaking.

The structure of the scalar potential simplifies considerably in the
field basis choice, $v_i =0 , \ [i=1,2,3]$ corresponding to the
identification, $ L_0 = H_d$.  Using this field convention, Grossman
and Haber~\cite{haber2} consider the three lepton generations one at a
time while neglecting the intergenerational mixing between sneutrinos.
The various $\tilde \nu _i - \tilde \nu _i ^\star $ pairs are found,
in general, to split in mass due to the contributions controlled by
the soft supersymmetry breaking parameters, $ B_i \mu _i m _{\tilde G}
= [ \tilde m ^2 _{di} + \mu _i \mu \cot \b ]$.  The independent tree
level mass splittings for fixed generations are described by the
approximate formula, valid to leading order in the parameter $B_i$,
\begin{eqnarray}  &&
\D m ^ 2 _{\tilde \nu _i } \equiv \tilde m ^2 _{\tilde \nu _{i+} } -
\tilde m ^2 _{\tilde \nu _{i-} } = { 4 (B_i m_{\tilde G} \mu_i m_Z )^2
m^2 _{\tilde \nu_i \tilde \nu_i ^ \star } \sin ^2 \b \over \prod_{ H_I
}
( m^2 _{\tilde \nu_i \tilde \nu_i ^ \star } - m^2 _{H_I} ) } \simeq {
(B_i\mu_i m_{\tilde G})^2 \sin ^2 \b \over m_Z ^2} ,\cr && [m^2
_{\tilde \nu_i \tilde \nu_j ^ \star }= (\tilde m^2_{L} )_{ij}
+\mu_i\mu_j -{ g_2^2+g_1^{2} \over 8} (v_u^2 -v_d^2) \d _{ij} ,\ H_I =
(H^0,\ h^0,\ A^0)]
\label{eqx19} \end{eqnarray}
yielding contributions to the sneutrinos-antisneutrinos mass
splittings of order, $ \D m _{\tilde \nu } << 1\ \text{GeV} $.  Using
the approximate estimate for the neutrino mass, $m_\nu \simeq m_Z \cos
^2 \b \sin ^2 \xi $, one can express the characteristic ratio, $ r_\nu
\equiv {\D m _{\tilde \nu} \over m_\nu} $, of the sneutrinos mass
splitting to neutrino mass as, $ r_\nu \simeq {(B_i \mu _i m_{\tilde
G} )^2 \tan ^2 \b \over m_Z^2 \mu_ i ^2} $.
The parameter $r_\nu $ assumes natural values, $O(1)$, although larger
values lying below $O(10^3) $ can also be accommodated.  This pattern
for $r_\nu $ contrasts with that found in the supersymmetric see-saw
model where the expected values cover the wider range~\cite{haber1}, $
10 ^{-3} < r_\nu < 10^ 3.  $ Contributions to the
sneutrino-antisneutrino mass splittings in all generations also arise
at the one-loop level.

The basis independent formalism for the bilinear RPV interactions
contributions to the sneutrino-antisneutrino mass splittings is
developed in a later work by Grossman and Haber~\cite{groer01}, using
a perturbative procedure valid for small RPV couplings.  The relevant
supersymmetry breaking invariant parameters involve the mass matrix
for the $\tilde \nu _ \a = (H ^0_d,\ \tilde \nu _i) $ neutral scalar
fields, $m^2 _{\tilde \nu \tilde \nu ^\star } $, and the four vector,
$c _ \a = (m^2 _{\tilde \nu \tilde \nu ^\star } )_{\a \b } b_\b / b ^2
$, through the following set of scalar products, $ v^2 , \
b^2 , v \cdot b, \ v \cdot c, \ b \cdot c ,\ 
[b_\a = B_\a \mu _\a m_ {\tilde G},\ b^2
=\sum _\a b^2 _\a ]$  and traces, $Trace ( m^{2p} _{\tilde
\nu \tilde \nu ^\star } ) = (\vert b \vert \tan \b )^p + \sum _i m^
{2p}_{\tilde \nu _i } $.  The fact that the mass splittings vanish when
the four vectors, $ v_\a , \ b_\a $, become parallel, follows in this
approach from the explicit dependence of $ \D m ^2 _{\tilde \nu _i} $
on the generalized cross products, $ (b \times v ) , \ (b \times c) $.

The sneutrinos-antisneutrinos mixing has important implications on the
collider physics tests of R parity symmetry
violation~\cite{hirschosc,haber1,haber2}.  In the regime of large
sneutrino mass splitting, $ \D m_{\tilde \nu } > 1\ \text{GeV} $, the
production of sneutrino-antisneutrino pairs tagged by the leptonic
decays, $ \tilde \nu \to e^\pm +\tchi ^\mp $, can be used to
reconstruct the sneutrinos masses~\cite{haber1}. The resonant
sneutrino or antisneutrino formation~\cite{feng98} in $ e^+e^-$ or
$q\bar q$ collisions may initiate interesting signals.  The off-shell
sneutrino and antisneutrino exchange processes can also be tested in
the high energy leptonic colliders~\cite{shalom,haber2} and hadronic
colliders~\cite{shalom,haber2} through the fermion-antifermion pair
production reactions, $e^+ + e^- \to [ \tilde \nu , \ \tilde {\bar \nu
}] \to f +\bar f $.  The study by Bar-Shalom et al.,~\cite{shalom} of
tau-antitau lepton pair production yields bounds on coupling constant
products of form, $ \vert \l^{\star} _{232} \l ' _{311} \vert < 3. \
\times 10^{-3} , \ \vert \l^{\star} _{232} \l ' _{322} \vert < 1.1 \
\times 10^{-2}$.

The regime of small sneutrino mass splitting, $ \D m_{\tilde \nu } <<
1\ \text{GeV} $, is favorable for the observation of
sneutrino-antisneutrino oscillations, provided the oscillation time is
shorter than the sneutrinos lifetime~\cite{haber1}, $ x_\nu >1,\ [
x_\nu = { \D m _{\tilde \nu }\over \G_{\tilde \nu } }]$, and the
branching fractions into the leptonic tagging decay modes, $\tilde \nu
\to e^\pm +\tchi ^\mp $, are appreciable.  In the presence of CP
violation, the $\tilde \nu - \tilde {\bar \nu } $ oscillations in the
sneutrino resonance formation reactions, $ e^++e^- \to ( \tilde \nu ,
\tilde \nu ^\star ) \to \tau ^+ +\tau ^- , \ p+\bar p \to ( \tilde \nu
, \tilde \nu ^\star ) \to \tau ^++ \tau ^- + X $, may exhibit finite
CP-odd or CP-even double spin correlation observables associated with
the spin polarization of the $ \tau ^+ - \tau ^-$ final state
pair~\cite{shalom2}.  Using the bilinear RPV interactions with a
coupling constant fitted to the neutrino oscillation
data~\cite{chuncp02} gives one the ability to induce significant
contributions to the oscillation and CP violation observables of the
sneutrino-antisneutrino system which may be observed at the high
energy colliders.

\subsubsection{\bf Early  studies of neutrino masses and mixing}

The specific structure of the neutrinos mass matrices predicted on the
basis of broken R parity symmetry have attracted considerable interest
in recent years.  Natural hierarchies can be accommodated by combining
together contributions from tree and loop levels, possibly
complementing these by the contributions from non-renormalizable
operators~\cite{hempf,nilles97,enqvist}.  As an introduction to the
detailed discussion of this issue in the next subsection, we review
here the constraints deduced in early studies of the neutrinos and
charged leptons mass matrices.

One of the first studies to focus on the one-loop contributions to the
neutrinos mass matrix is due to Enqvist et al.,~\cite{enqvist}. Using
experimental data available in the early 1990 years, the mass
difference and mixing angle parameters in various two-flavor mixing
schemes are fitted by assuming the successive dominance of pairs of
lepton number violating coupling constants.  A combined fit to the
neutrino oscillation experimental data for the $ \nu_\mu - \nu_\tau $
pair and the Majorana mass inferred from $\b \b _ { 0 \nu } $,
determines the allowed regions in the planes associated to the product
and ratio of coupling constant pairs of the form: $ [\vert \l ^{'\star
} _{323} \l '_{333}\vert , \vert \l ^ \star _{322} \l '_{323}\vert ] <
O(10^{-6})\ -\ O(10^{-8} ),\ [\vert \l^{'\star } _{233}/\l'
_{333}\vert , \vert \l ^ \star _{323}/\l' _{322}\vert ] \simeq
[O(10^{-3}) \ - \ O(10^2 ) ]. $

The astrophysics constraints on unstable neutrinos, assuming a
predominant radiative two-body decay mode, $\nu '\to \nu +\g $, yield
several inequalities linking the neutrinos masses and radiative decay
lifetimes.  An illustrative form for the experimental limits is given
by the approximate inequalities, $ m_\nu < 100 \ \text{ eV} , \ \tau
_\nu (s)/ m_\nu (eV) > 10 ^ {23} $.  Useful bounds involving quadratic
products and ratios of RPV coupling constant pairs can be
inferred~\cite{enqvist} on this basis, of which a representative
sample is given by, $ \vert \l^{'\star }_{323} \l '_{333} \vert <
4. \times 10^{-6} ({\l^{'\star }_{323}\over \l '_{333} }) ^{2/3}
m_{\tilde q} ^{8/3} \tilde m ^{-1} , \quad m_\nu (eV) < 13. \
({\l^{'\star }_{323}\over \l '_{333} })^{-1/3} m_{\tilde q} ^{2/3},
\quad \tau_\nu (s) > 2.  \times 10^{24} \ ({\l^{'\star }_{323}\over \l
'_{333} })^{-1/3} . $

A general study of the bilinear RPV contributions to the neutrino mass
matrix has been presented by Joshipura and
Nowakowski~\cite{joshinowak95}.  The initial motivation stemmed from
the observation that the coupling constant values consistent with the
baryon asymmetry non-erasure constraint, $(\mu _i, \ v_i ) =
O(10^{-6})\ \text{GeV} $, yielded neutrino mass splittings, $\D _{ex}
\sim [10^{-6} \ - \ 10^{-10} ] \ \text{eV} ^2$, roughly reproducing
the vacuum solution value for solar neutrino oscillation.  An
exploration of the model parameter space also indicates a
compatibility of the baryon asymmetry non-erasure constraints with a
large neutrino mixing angle.

\subsection{Developments  initiated  by recent neutrino oscillation data} 

The phenomenon of neutrino flavor oscillations has a natural
interpretation in terms of finite mass splitting and mixing angle
parameters between the three light neutrinos.  The recently
accumulated experimental evidence has provided useful information on
the transformation matrices linking the leptons flavor and mass field
bases, which has considerably narrowed the choice of solutions for the
charged lepton and neutrino mass matrices.  Motivated by the new
information leading to solutions for mass differences and mixing
angles in different neutrino pairs, several studies have attempted to
examine the implications of these results on the RPV interactions.
The comparison of the neutrino Majorana mass matrix with experimental
data yields then definite values, rather than bounds, for the coupling
constants.

The existing studies have sought to reproduce the solutions for the
neutrinos mass and mixing parameters by focusing on the RPV bilinear
interactions
only~\cite{joshi95,romao97,vissani96,kapnel,romao99,hirschvalle00,hirsle02,hirsle03,chun299,bi99,benakli97},
on the trilinear interactions
only~\cite{roybi,adhi99,rakshit,kong99,drees98,clav99}, or on a
combination of both~\cite{chun99,bednya98,joshi99,biti99}.  The
literature on this issue is quite extensive and will only be discussed
here in a global fashion.  For lack of space, our discussions will
emphasize the main issues without reviewing the fine points raised in
the various works.

\subsubsection{\bf Bilinear interactions} 


The key property of bilinear R parity violation lies in the ability to
explain several hierarchies in the masses and mixing angles by
invoking the tree and loop level contributions.  The experimental
constraints are not restricted to the bilinear couplings themselves,
since the adequate suppression of the misalignment parameters
controlling the tree level contributions, $ \L _i = \mu v_i -v_d \mu_i
<< 1 $, puts demands on the supersymmetry breaking effects responsible
for the generation of sneutrino VEVs, $v_i$.

The bilinear RPV contributions to the neutrinos mass matrix satisfy
two attractive general properties~\cite{joshinowak95,kapnel}: (i) The
leptons flavor mixing matrix is automatically CP conserving; (ii) It
depends on two real mixing angles only, $ \sin \t _1 = {\mu_1 / (\mu
_1 ^2 +\mu _2^2 )^\ud }, \ \sin \t _2 = {(\mu_1 ^2 + \mu_2^2 )^\ud /
(\mu _1 ^2 +\mu _2^2 + \mu _3^2 )^\ud } .$ These remarkable properties
can be traced to the existence of a residual $U(1)$ symmetry
suppressing the mass terms for the linear combination of neutrinos
fields~\cite{kapnel}, $ \cos \t_1 \nu_e (x) -\sin \t_1 \nu_\mu (x)$.
A predictive model motivated by the gauge mediated supersymmetry
breaking is proposed by Kaplan and Nelson~\cite{kapnel}.  Naturally
suppressed sneutrinos VEVs, $ v_i << \mu_i$, are achieved by assuming
generation universal soft supersymmetry breaking parameters.  All
three solutions to the solar neutrino data (small and large angle MSW
and just-so vacuum propagation) can be reproduced by fitting the two
available mixing angles.
Two other hierarchically ordered contributions to the mass matrix can
arise in the model.  The heaviest neutrino, $\nu_\tau (x) \propto \mu
_i \nu_i (x)$, acquires in the limit, $\mu _i << \mu $, the tree level
mass, $ m_{\nu_\tau } \simeq (g _2^2 /4 M_2+ g _1^{2}/4 M_1 ) v_{ \tau
}^2 , \ [\sin ^2 \xi =v_{ \tau }^2/ v_d ^2] $.  The one-loop Feynman
diagrams, with internal lines $\tau - \tilde \tau $ or $ b - \tilde b$
and zero, one or two sneutrino tadpole insertions, generate smaller
Majorana masses for the other neutrinos.  The approximate
relationship, $ m_{\nu _\mu } / m_{\nu _\tau } \sim O(\l _\tau ^4) $,
links the neutrino mass hierarchy to the existence of a large $\tan \b
$ parameter.

The study by Rom\~ ao et al~\cite{romao99} includes the neutrino
oscillations data among the set of experimental constraints upon
performing a general scan over the MSSM parameter space. The search
allows for finite parameters, $\mu_i $, with radiatively generated
sneutrinos VEVs.  The $\nu _\mu -\nu _\tau $ oscillation solution to
the atmospheric muon data requires the corresponding alignment
parameters to satisfy the conditions, $ [\L _\mu , \L _\tau ] \simeq
(0.03 - 0.25) \ \text{GeV}, \ \L _\mu /\L _\tau \simeq O(1) $.
Similar constraints are also obtained for the $\nu _e -\nu _\mu $
matter MSW and vacuum oscillation solutions, associated with large and
small mass difference and mixing angle parameters, respectively.
Other studies of the neutrinos masses and mixings with similar focus
are presented in Ref.~\cite{bi99}.

The constraints from the solar and atmospheric neutrino oscillations
data become particularly severe when one incorporates the one-loop
contributions to the mixing and mass terms. Comprehensive studies
within the bilinear R parity violation option of the
neutralino-neutrino field mixing up to one loop order are presented in
the studies by Hirsch et al.,~\cite{hirschvalle00,hirsle02,hirsle03}.
The predicted strongly reduced neutrino effective
mass~\cite{hirschvalle00}, $<m_{\nu _e }> \simeq 0.01 $ eV, would yield
unobservably small contributions to the $\b \b _{0\nu }$ reaction
rates.  The complete gauge invariant treatment~\cite{hirsle02}
confirms that the atmospheric data mass scale with maximal mixing
arises at tree level while the solar data mass scale arises at loop
level. A bimaximal mixing is only possible if one relaxes the
universal generational boundary conditions on supersymmetry breaking.
Useful approximate analytic formulas are provided in
Ref.~\cite{hirsle03}.

We consider next the issue of sterile neutrinos.  Chun~\cite{chun299}
considers a type I axion model framed within the gauge mediated
supersymmetry breaking approach.  The axino component of the axion
chiral superfield, $\Phi $, acts as a sterile neutrino which mixes
with the left-handed neutrinos.  The postulated K\"ahler potential, $
K = C ^\dagger _I C _I + \Phi ^\dagger \Phi + \sum _I {x_I \over F_a}
( \Phi ^\dagger + \Phi ) C ^\dagger _I C _I$, includes a
non-renormalizable trilinear term which couples $\Phi $ to the
observable and hidden sectors chiral superfields, $ C_I$, where $F_a$
denotes the $U(1)_{PQ}$ symmetry scale and the parameters $x_I$ are
determined by the Peccei-Quinn symmetry charges of the fields $C_I$.
A small axino mass term is assumed to arise via the see-saw mechanism
involving a hidden sector field, $ m_{\tilde aI} \simeq x_I F_I /F_a \
\Longrightarrow \ m_{\tilde a} \simeq m_{\tilde aI} ^2 / M_I \sim x_I
M_I ^3 / F_a ^2$, where $\sqrt {F_I} =M_I $ denotes an intermediate
mass scale which can take values in the range, $ 10^{2}\ -\ 10^{6} $
GeV.  The trilinear K\"ahler potential term coupling the axion with
neutrinos, $\Phi \nu _i ^\dagger \nu _i $, contributes to the neutrino
mass via the bilinear RPV interactions induced by the neutrinos
auxiliary field components, $F_{\nu _i} \simeq \mu _i v \sin \b /\sqrt
2 $. The resulting axino-neutrino mass mixing term, $m_{\tilde a \nu
_i }\simeq {x_{\nu _i} F_{\nu _i} \over F_a } \simeq {x_ {\nu _i} \mu
_i v \sin \b \over \sqrt 2 F_a } \simeq ({\mu _i \sin \b \over 0.6 \
\text{MeV} }) \ ({10 ^ {12} \ \ \text{GeV} \over F_a }) \ 10 ^ {-4} \
\text{eV} , $ explain the solar or atmospheric neutrino flux deficits
in terms of an oscillation to a sterile neutrino.  The solar neutrino
constraints set the bounds, $ \mu_{1} /\mu \simeq 10^{-5}, \ \mu_{2}
/\mu \simeq 10^{-6} $, for the favored value, $F_a = 10^{12} $ GeV,
while the atmospheric neutrino constraints are compatible with the
bound, $ \mu_{2} /\mu \simeq 10^{-5} $ for $ F_a = 10^{10} \
\text{GeV} $.

An alternative candidate for a sterile neutrino can be sought among
the modulino fields, corresponding to the fermionic superpartners of
the moduli scalar fields generically present in the superstring
compactification models. A trilinear superpotential induced by the
gravitational interactions and coupling a moduli chiral superfield $T$
with a lepton and a Higgs boson~\cite{benakli97}, $ W = \l _\a T LH_u
, \ [\l_a = m_\a / M_ \star ]$, can produce a neutrino-modulino mass
mixing term, $m_a 
<\tilde T> $.  Other mechanisms involving light modulino fields are
discussed by Benakli and Smirnov~\cite{benakli97}.

\subsubsection{\bf Trilinear  interactions} 

Constraining tests of the trilinear RPV interactions are obtained by
fitting the one-loop contributions to the neutrinos Majorana mass
matrix to the experimental mass matrix inferred from the neutrino
oscillation solutions.  Rakshit et al.,~\cite{rakshit} reconstruct the
neutrino mass matrix from the one-loop contributions of the $ \l ' $
and $ \l $ interactions by assuming a flavor diagonal chirality mixing
mass term for the internal sfermions, $ (\tilde m ^{f 2} _{LR})_{ij} =
m_f \tilde m _0 \d _{ij}.$ Choosing an input value for the lowest
neutrino mass eigenvalue allows one to infer values rather than bounds
for the large number of relevant single and quadratic coupling
constant products.  Some representative predictions are: $\vert \l
_{111} ^ {'}\vert ^2 = (9.7 \times 10^{ -4} \ -\ 1.1 \times 10^{ -2})
, \ \vert \l_{122} \vert ^2 = (6.6\times 10^{ -6}\ - \ 7.2\times 10^{
-5}) ,$ where the intervals of variation are associated with the
interval of values assumed for the input neutrino mass, $ m_{\nu _1}
\in [ 0. \ - \ 0.1 ] \ \text{ eV} $.  An analogous analysis aimed at
the solar and atmospheric solutions, invoking an approximate flavor
symmetry for the RPV interactions, is presented by Kong~\cite{kong99}.
The comparison in the $\nu _\mu -\nu _\tau $ case yields the
representative coupling constant predictions, $\l '_{233} \simeq \l
'_{333 } \simeq 10^{-5} $.  Adhikari and Omanovic~\cite{adhi99}
present results for the subset of coupling constants, $ \l _{133} , \
\l _{ 233} , \ \l _{333}, \ \l'_{133}, \ \l'_{233}, \ \l'_{232}, \
\l'_{132}$, fitted to the solar, LSND and atmospheric data as well as
the $ \b \b _{0\nu } $ data.

A significant improvement in the predictive power is made possible by
postulating an horizontal discrete symmetry which limits the number of
unknown RPV coupling constants~\cite{drees98,clav99}.  Clavelli and
Frampton~\cite{clav99} discuss a search for the allowed domain in the
nine coupling constants $ \l _{ijk}$ parameter space solving for the
neutrinos oscillation parameters, the $\b\b _{0\nu } $ data and
astrophysical constraints, assuming that a small subset of the
coupling constants are predominant. The solutions for the solar MSW
and atmospheric parameters can be reproduced by allowing for a pair of
finite coupling constants lying in the ranges, $ 0< \l _{131} < 0.1 $
and $ 0< \l _{121} < 0.1$, while maintaining all the other coupling
constants at much smaller values of order, $ 10^{-3} \ - \ 10^{-4}$.
Drees et al.,~\cite{drees98} consider an ansatz for the neutrinos
Majorana mass matrix involving a maximal $\nu _\mu \ - \ \nu _\tau $
mixing and a weakly mixed $ \nu _e $.  The RPV one-loop contributions,
$ (m_\nu )_{ii'} \propto \l ' _{ijk} \l ' _{i'kj} m_{d_j} m_{d_k} /
m_{\tilde f}, $ can reproduce the observations in the case of a
hierarchy free generational structure for the coupling constants,
provided one assumes the presence of suitable texture zero entries in
the neutrinos mass matrix.  This prescription is motivated by the
possibility that some horizontal discrete symmetry would set certain
matrix entries to zero while allowing the non-vanishing entries to be
of same magnitude.  The preferred horizontal symmetry breaking
direction in the quarks generation space is set by the strange
quark. A fit assuming $\l ' _{ 133} = 0 $ and $ \l ' = \l ' _{ 233}
\simeq \l ' _{ 233} , $ with the overall mass scale set by the $b$
quark mass $ m_b $, selects the coupling constant value $\l ' \simeq
7. \ \times 10 ^{-5 }. $ If the mass scale were rather set by the
strange quark mass, the corresponding coupling constant value would
get enhanced to $\l ' \simeq 2.5 \ \times 10 ^{-3 } .$ With a
postulated baryon number conserving $Z_3 $ symmetry, the possible
choices of symmetry charges lead inevitably to an explicit breaking of
the flavor symmetry by the regular Yukawa interactions.  The preferred
case, associated with the strange quark mass breaking, yields a
strange quark Yukawa coupling constant $\l^d _{22} $ of the same order
as the fitted value of $ \l ' $.

The one-loop contributions to neutrino masses might get accidentally
suppressed, as in the case of small chirality flip scalars mass
parameters, $ \tilde m^2 _{LR} / \tilde m _0 ^2 << 1$, or in the case
of finely tuned cancellations between different coupling constants
related by symmetries.  Motivated by these considerations, Borzumati
and Lee~\cite{borzu02} examine the effective two-loop contributions to
the neutrino masses from the one-loop Feynman diagrams with $\tilde
\nu - \tchi ^0 $ exchange, including the one-loop renomalization
corrections to the intermediate sneutrinos-antisneutrinos mass
splitting, $\tilde \nu _i - \tilde \nu _i ^\star $. The two-loop
effect involves the soft supersymmetry breaking trilinear RPV coupling
constants, $ A^\l _{ijk},\ A^{\l '}_{ijk} $, along with the
supersymmetric coupling constants, $\l _{ijk},\ \l ' _{ijk},$ in
flavor configurations distinct from those entering the one-loop
effect.  The combined one- and two-loop contributions yield robust
bounds on the coupling constants, $\l _{i33} , \ \l ' _{i33} $, along
with useful bounds on the soft supersymmetry breaking RPV triscalar
coupling constants, $ A ^ \l _{i33}, \ A ^ {\l '} _{i33}.$

\subsubsection{\bf Combined bilinear and trilinear interactions}

Exploiting the large ratio between the tree and one-loop contributions
to the neutrino masses, $ m ^{tree} _\nu /m ^{loop} _\nu \sim 10 ^2$,
gives one the ability to account for interesting structures of the
neutrinos mass matrix.  The resulting fits to the oscillation
solutions are strongly influenced by the mode of supersymmetry
breaking. Using a supergravity framework for the supersymmetry
breaking, Chun et al.,~\cite{chun99} consider the set of trilinear
coupling constants, $ \l '_{i33} , \ \l _{i33}$, while allowing for
the bilinear coupling constants, $\mu _i$, to be radiatively induced
via the renormalization group scale evolution.  Different pairwise
combinations of solutions for the oscillations parameters (solar,
atmospheric data, LSND data) are explored corresponding to the
hierarchical or degenerate patterns of the neutrino mass matrix.  In
order to reproduce the experimental results from the solar and
atmospheric neutrinos data combination, with three active neutrinos
and no sterile neutrino, one must require a large $\nu _\mu - \nu
_\tau $ mass hierarchy, $ m_{ \nu _\mu } / m_{\nu _\tau } = \chi
\approx 7 - 40 , $ and a large mixing angle, $ \sin ^2 (2 \t ) >
0.82$.  The renormalization group evolution produces sneutrino VEVs,
$<\tilde \nu _i> \simeq { v_d \over 8 \pi ^2 } (a_i \l '_{i33} \l _b +
b_i \l_{i33} \l _b) \ln (M_X/m_Z) $, with $a_i,\ b_i $ being
calculable constants.  The coupling constants must obey bounds of
form, $ (\vert \l '_{i33}\vert, \ \vert \l_{i33}\vert  ) 
< [O(10^{-4}) - O(10^{-5})] $.
Cancellations between tree and loop contributions occur in the regime
of small $\tan \b $ parameter.  The comparison with the neutrino
oscillation solutions favors a large triscalar parameter, $ A $, and
small electroweak gaugino mass parameters, $ M_{1}$ and $ M_{2} .$

General fits to the neutrino oscillation data, using the combined tree
and one-loop level contributions, are attempted in several
works~\cite{abada99,haugg00,davilosa00,haugg01} with the purpose to
determining the allowed ranges for the various RPV coupling constants.
Haug et al.,~\cite{haugg00} compare the combined tree and one-loop
level contributions from the bilinear interactions and the trilinear
interactions, $ \l _{i33}, \ \l '_{i33}$,  with
a phenomenological light neutrinos mass matrix with
entries representing upper limits inferred from the oscillation
experiments
and the $ \b \b _{0\nu } $ measurements.  
The subsequent work by Haug et al.,~\cite{haugg01}
examines the compatibility of the LSND data
with the three light neutrino generations scenario.  Using an extended
analysis of the oscillation data accounting for the constraints from
$\nu_e + e $ inclusive scattering reactions on the neutrino CP phases
in the effective neutrino mass, $ <m_{\nu_e }> = \sum _i V^ {' \dagger
2} _{ei} m_{\nu _i} e ^{i \l  _i} $, it is found that the effective
mass obtained in the fit including the LSND data is an order of
magnitude larger than that obtained in the fit excluding it.  The
latest study by Abada et al.,~\cite{abadabha02} includes the recent
SNO experimental data.  Chun et al.,~\cite{chunpark02} focus
specifically on the bilarge neutrino mass matrix ansatz accommodating
two maximal mixing angles, $[\t _{12}, \t _{23}] \simeq \pi /4 $. This
can be satisfactorily accounted for by allowing for a mild fine tuning
on the non-universal soft supersymmetry breaking parameters, as
described by the parameters differences, $B-B_i, \ \tilde m^2_{H_d} -
\tilde m^2_{L_i} $.  The fits to neutrino masses by Grossman and
Rakshit~\cite{grossrak03} using the tree and loop order contributions
with generic RPV supersymmetry breaking parameters also accommodates
all existing data with a mild fine tuning of parameters.

The renormalization corrections to the trilinear RPV interactions may
generate finite contributions to the soft parameters in the scalar
potential which may in turn induce non-vanishing sneutrino VEVs.  For
the one-loop contributions from trilinear interactions to neutrino
masses, Joshipura et al.,~\cite{joshi99} find that an account of the
feed-back effect of the sneutrino VEVs amplifies the constraints on
the corresponding trilinear coupling constants.  Within the
renormalization group supergravity approach to supersymmetry breaking,
the coupling constant bound imposed by the $\nu _\tau $ mass limit,
$\l '_{133} < O(10^{-3}) $, is thus strengthened to $\l '_{133} <
O(10^{-5}) $ in the calculation accounting for the sneutrinos VEV.
The values of the lepton number violating coupling constants fitted to
the neutrino mass matrix vary in the range, $ ( \l , \ \l ' ) =
[10^{-3} \ \tilde m \ - \ 10^{-5}\ \tilde m ]$.  These results lend
hope to the prospect that some manifestation of a lepton number
violation might be observed in the future at the high energy
colliders.

Having in hand fitted values for the lepton number violating trilinear
and bilinear coupling constants, $\l _{ijk} $ and $ \mu _i$, gives one
the ability to make detailed predictions on related phenomenas.  By
combining the information gleaned from neutrino physics with the
stringent quadratic coupling constant bounds available from nucleon
decay, one can thus infer strong single bounds on the baryon number
violating coupling constants~\cite{biti99}, $\l '' _{ijk} < O(10^{-9})
$, irrespective of the generation configurations. Important
implications also hold on the collider physics.  On side of the
three-body RPV decay modes of LSP neutralinos into three leptons, the
finite field mixing of neutrinos with neutralinos from the bilinear
interactions also induces the two-body decay modes, $\tchi ^0 \to \mu
^\pm +W ^\mp , \ \tchi ^0\to \tau ^\pm + W ^\mp $.  Based on fits to
the neutrino oscillation solutions, Choi et al.,~\cite{biti99}
present, as a function of the MSSM parameters, predictions for the
various branching fractions and decay lengths associated to the
two-body decay modes, while Chun and Lee~\cite{biti99} study the
expected lepton flavor asymmetries in decay rates.  As demonstrated by
Datta et al.,~\cite{bi99}, an observation of the neutralino LSP
two-body decays might be accessed at the Fermilab Tevatron through
searches of characteristic like-sign dimuon or ditau signal
events. With similar motivations in mind, towards linking the neutrino
data to collider physics tests, recent works have focused on the
multilepton signals~\cite{das03} and the LSP decay
modes~\cite{hirrod03}.  The RPV coupling constant values, $\l'_{233}
\sim \l'_{333} = O(10^{-4})$, as deduced in fits to the atmospheric
neutrino oscillation data, can be tested~\cite{bartia02} at the
Tevatron collider by searching for the observable signals of $ l +
bbb$ jets in the RPV decays of the produced pairs of charginos and/or
neutralinos, $\tchi ^+ \tchi ^0,\ \tchi ^0 \tchi ^0$.  The decay
lengths of LSP sleptons, $\tilde e_{iR}$, can be used to discriminate
between the bilinear and trilinear R parity breaking options.  The
typical predictions for the branching fractions from trilinear
interactions are~\cite{bartl03}, $ B(\tilde e_i \to e_j + \nu _k ) <
0.5 $, while the typical ones from the bilinear interactions are, $
B(\tilde e_1 \to e + \nu ) \simeq 1$.

The broken R parity symmetry may have an important impact on the
future experimental projects using collimated high energy neutrino
beams or detector telescopes for extraterrestrial ultrahigh energy
neutrinos.  The RPV flavor non-diagonal couplings of neutrinos with
matter fermions can compete with the neutrino flavor oscillation
effects in inducing the flavor changing processes, $\nu _i + N \to e_j
+X,\ [i\ne j]$ in experiments using the neutrino and antineutrino
beams at the neutrino factories. The contributions to the $\tau ^-$
production rates in the reactions, $\nu _\mu +d \to \tau ^- + u,\ \nu
_\mu +\bar u \to \tau ^- + \bar d,$ involving the coupling constant
product~\cite{dattagan01}, $ \l ^{'\star }_{213} \l '_{313} / m^2
_{\tilde b_R } $, are found to dominate over those  originating 
from the flavor neutrino oscillation mechanism.  
The deep inelastic neutrino-nucleon
and antineutrino-nucleon inclusive scattering reactions may be used
for the purpose of detecting the neutrinos from extraterrestrial
sources of high energy, $E_\nu \in [10 \ \text{GeV} ,\ 1 \ \text{TeV}]
$, and ultrahigh energy neutrinos, $E_\nu \in [10 ^{15}\ \text{eV} ,\
10 ^{21} \ \text{eV}] $.  The RPV contributions to the charged and
neutral current reactions, $\nu _i+ N \to e_i^- +X,\ \nu _i + N \to
\nu _i +X,$ involving the squark resonant formation, can yield
observably large enhancements of the predicted SM
rates~\cite{cargg98}.

\section{Tests of trilinear RPV  interactions  
in scattering and rare decay processes}
\label{secxxx3}

The rare decay and scattering processes involving the leptons and
hadrons provide a rich source of experimental constraints on the new
physics.  The hadron and/or lepton flavor changing reactions are
largest in number, while the lepton and/or baryon number
non-conserving reactions are those furnishing the strongest
constraints.  The hadronic structure physics plays an important r\^ole
in the analysis of low energy processes involving hadrons, as can be
appreciated by consulting the textbook by Donoghue et
al.,~\cite{donog92} and the lecture notes by Buras~\cite{hadron}.

To review the results obtained for R parity symmetry violation, we
have organized the discussion into four subsections, where we discuss
in succession the hadron flavor violating processes, the lepton flavor
violating processes, the lepton number violating processes and the
baryon number violating processes.

\subsection{Hadron flavor changing processes}
\label{secxxx3a}

The experimental constraints from flavor changing neutral current
processes are automatically satisfied in the SM by virtue of the GIM
suppression mechanism~\cite{gim70,glashow}.  For the MSSM, another
source of flavor changing neutral current contributions arises through
the non-universality with respect to the quarks and leptons
generations of the soft supersymmetry breaking interactions.  With the
RPV interactions, several new flavor changing sources become
available.

\subsubsection{\bf Mixing and decay of neutral mesons}

The mass difference and mixing observables for the strange, charmed
and beauty (bottom) quark neutral mesons, $K \ -\ \bar K, \ [\D S=2] ;
\ D\ -\ \bar D, \ [\D c =2] ; \ B \ -\ \bar B, \ [\D b =2] $ are
described by the real and imaginary parts of the amplitudes for the
flavor off-diagonal quark subprocesses, $ s +\bar d \to \bar s+ d, \ c
+\bar u \to \bar c +u , \ d +\bar b \to \bar d+ b $.  The tree level
RPV contributions due to sneutrino $ t$-channel exchange are described
by the effective Lagrangian, $ L_{EFF}= -\sum _{j,k} F '_{jkkj} (\bar
d_{kR}d_{jL}) (\bar d_{kL} d_{jR}) + H. \ c.  ,\ [ F '_{abcd} = \sum_i
{ \vert \l ' _{iab} \l _{icd}^{'\star } \vert \over m ^2 _{\tilde
\nu_i } } ] .$ The comparison with experimental data for the neutral
mesons mass splittings yields the following strong bounds on the
auxiliary parameters~\cite{roy}: $F'_{1221} < 4.5\times 10^{-9}, \ [
K-\bar K];\
\quad F'_{1331}< 3.3\times 10^{-8}, [B-\bar B] $.  Under the double
coupling constant dominance hypothesis, these results translate into
the representative quadratic coupling constant bounds, $ \vert \l
^{'\star} _{i21} \l'_{i12} \vert < 4.5\times 10^{-9}\ \tilde \nu_{iL
}^{2} , \ [K- \bar K]; \ \vert \l ^{'\star} _{i31} \l'_{i13}\vert <
3.3\times 10^{-8} \ \tilde \nu_{iL }^{2},\ [B- \bar B] .$

At the one-loop level, the RPV interactions contribute at orders $\l
^{'4} $ or $\l ^{''4}$, via $ s$-channel and $ t$-channel exchanges of
pairs of sfermions, as displayed by the box diagrams E.1-E.3 in
Figure~\ref{fig2}.  There are also mixed RPV and gauge contributions
of order $\l ^{'2} g^2 $ or $\l ^{''2}g^2$, represented by box
diagrams propagating a sfermion and a charged $W$-boson or Higgs
boson, as displayed by diagram E.4 in Figure~\ref{fig2}. The pure and
gauge mixed RPV box diagrams contributions have an approximate
quadratic and linear dependence on the sfermions mass parameter,
respectively.  Several studies have been devoted to the $ K ^0-\bar
K^0 $ system.  The $\l '' $ interactions were first considered by
Barbieri and Masiero~\cite{masiero}.  The coupling constant bounds
found in the updated study by Carlson el al.,~\cite{carlson} read as,
$ \vert \l ^{'' \star } _{332} \l ''_{331}\vert < \ \text{min} \ [6.
\times 10^{-4}\ { m_{\tilde t} \over m_W }, \quad 3. \times 10^{-4} \
({ m_{\tilde t} \over m_W } )^2],\ \vert \l^{'' \star }_{232} \l
''_{231}\vert < \text{min}\ [ 6. \times 10^{-4} { m_ {\tilde c} \over
m_W } , \ 2.  \times 10^{-4}\ ({ m_{\tilde t} \over m_W } ) ^2 ]. $
The quadratic coupling constant bounds obtained by de Carlos and
White~\cite{decarlosq} are given by, $ \vert \l ^{''\star } _{213}
\l''_{223}\vert < 2. \times 10^{-2} \ \tilde q^2, \ \vert \l^{''\star
} _{213} \l''_{323} \vert < 4. \times 10^{-3} \ \tilde q^2, \ \vert
\l^{''\star } _{313} \l''_{223} \vert < 8. \times 10^{-2} \ \tilde
q^2, \ \vert \l^{''\star } _{313} \l''_{323} \vert < 4. \times 10^{-2}
\tilde q^2.$ Including the contributions to $\D m_K $ from the Yukawa
and mixed Yukawa-gauge interactions along with the QCD corrections,
Slavich~\cite{slavich00} obtains the bound, $\vert \l^{''\star }
_{313} \l''_{323} \vert < 3.3 \times 10^{-2} $.  Similar studies have
been devoted to the mixing of the heavy quarks neutral mesons.  The
bounds inferred for the $ B\ -\ \bar B$ system are weaker than those
quoted above for the $ K \ -\ \bar K$ system. For the $ D\ -\ \bar D$
system, Carlson et al.,~\cite{carlson} obtain the coupling constant
bound, $ \vert \l ^{''\star } _{232} \l ''_{132} \vert < 3.1 \times
10^{-3} ({ m_{\tilde s} \over m_W} )^2$.

Bhattacharyya and Raychaudhuri~\cite{bhattaray} note that the dominant
contribution from the mixed RPV and gauge interactions box diagrams to
the $\D m_K, \ \D m_B$ mass differences arises from the exchange of $
(q, \ \tilde l)$ and $ (H^\pm ,\ W^\pm ) $ pairs with the top quark
and transversely polarized $ W^\pm $.  Under the double coupling
constant dominance hypothesis, one obtains several useful quadratic
bounds of which we quote some representative cases, $ \vert \l
^{'\star } _{i31} \l '_{i32} \vert < 7.7 \times 10^{-4}, \ \vert \l
^{'\star } _{131} \l '_{122}\vert < 1. \times 10^{-4}, \ [ K - \bar
K]; \quad \vert \l ^{'\star }_{i31} \l '_{133}\vert <1.3 \times
10^{-3}, \ [ B -\bar B]. $ The competition between tree and loop level
RPV contributions to the $ K-\bar K$ mixing is reflected by a strong
basis dependence of the associated quadratic coupling constant
bounds. Upon considering the two extreme choices of basis for the
quarks superfields defined by the choices for the CKM quarks flavor
mixing matrix, $ V= V_L^{d\dagger } $ and $ V= V^u_L$, Huitu et
al.,~\cite{huituang99} find that the tree level mechanism dominates in
the first choice and the loop level mechanism in the second choice.

A finite contribution to the $K-\bar K$ mass difference can arise from
a single dominant coupling constant~\cite{agashe} if one accounts for
the quark flavor mixing.  Starting from the current field basis
description of the RPV superpotential, one obtains the flavor changing
$\D S =2$ effective Lagrangian in the form, $ L_{EFF}= -{\vert
\l'_{ijk}\vert ^4\over 128\pi^2} V_{j2} V^\star _{j1} ({1\over
m^2_{\tilde \nu_i} } +{1\over m^2_{\tilde d_{kR} } } ) (\bar d_L\g_\mu
s_L) (\bar d_L\g_\mu s_L) + H.\ c.  $
Several single coupling constant bounds are obtained by Agashe and
Graesser~\cite{agashe} in the case involving a small Higgs bosons VEVs
ratio parameter, $\tan \b =1$, of which we reproduce a representative
subset, $ \vert \l '_{imk} \vert < 0.11 \ [m_{\tilde \nu_{iL} } ^{-2}
+ m_{\tilde d_{kR} } ^{-2} ]^{-1/4} , \ [K - \bar K]; \quad \vert \l
'_{ijk} \vert < 0.16 \ [m_{\tilde e_{iL} } ^{-2} + m_{\tilde d_{kR} }
^{-2} ]^{-1/4},\ [D-\bar D]; \ \vert \l '_{i3k} \vert < 1.1 \
[m_{\tilde \nu_{iL} } ^{-2} + m_{\tilde d_{kR} } ^{-2} ]^{-1/4} , \
[B_d -\bar B_d] . $


\subsubsection{\bf  Rare leptonic decays  of hadrons}

The leptonic and semileptonic rare decay modes of the strange and
 beauty flavored mesons provide useful probes of new physics owing to
 the availability of detailed SM predictions~\cite{litt93}.  We review
 in the present subsection the leptonic two-body decay channels
 involving charged lepton-antilepton pairs or charged lepton-neutrino
 pairs, $ M ^0\to e^-_i+ e ^+ _j , \ M ^- \to e^- _i +\bar \nu _j, \ [
 M= K^0 _{L} , \ K^\pm , \ B ^{0} _{[d, s]}, \ B ^{\pm } _{[d, s]} ] $
 leaving the discussion of the semileptonic decay channels, $ M\to M
 '+ e_i +\bar e_j , \ M\to M ' +\nu _i + \bar \nu _j, \ M\to M '+
 e^-_i + \bar \nu _j ,\ [M' = \pi , \ X_q]$ to the next subsection.
 There is a nice complementarity between the above quoted leptonic and
 semileptonic decays.  Both processes are described by the same quark
 subprocesses, $ d_k + \bar d_l \to e_i +\bar e_j$, and the same
 hadron flavor changing operators, while differing only in the initial
 and final hadronic states. For illustration, the processes, $K^- (s
 \bar u) \to \pi ^- (d\bar u) +e^++e^-,\ K ^0(s\bar d) \to e^++e^-$,
 are determined by the hadronic matrix elements, $<\pi ^- \vert \bar s
 \g_\mu d \vert K^-> $ and $<0 \vert \bar s \g_\mu d \vert K^0> $,
 respectively.  The general effective Lagrangian, for a fixed
 configuration of the generation indices, involves up to ten
 independent four fermion local operators suitably selected from the
 Lorentz covariant decomposition in Fermi invariants, $ L_{EFF} =
 \sum_{\G _\a = S,V,T,A,P} [C^q _\a ( \bar q \G _\a Q)( \bar l \G _\a
 l) + C^{ 'q} _\a ( \bar q \G _\a Q)( \bar l \G _\a \g _5 l)], \ [Q=
 b,\ c, \ s; \ q = u, \ d , \ s ] $.  For the leptonic decays only
 three independent operators survive, which can be chosen
 as~\cite{nardiguetta97,grossman2}, $ C_P (\g _5 )(\g _5),\ C'_P (\g
 _5 )(1),\ C_A (\g_\mu \g _5 )(\g_\mu \g _5)$.
 
The SM contributions to the flavor changing processes, $ d_k + \bar
d_l \to e_i +\bar e_j , \ [k\ne l,\ i\ne j]$ arise from one-loop box
and penguin diagrams.  For a final state configuration dominated by
left-chirality leptons, the invariance under the electroweak group, $
SU(2)_L $, can be invoked to relate the transition amplitudes with
lepton-antilepton and lepton-neutrino pairs, as illustrated by the
leptonic decays, $ M ^0\to e_i + \bar e_j $ and $ M^- \to e_i ^- +\bar
\nu_j $.  The predicted SM branching fractions for the hadron and/or
lepton flavor changing decays generally fall short of the experimental
results.
The current experimental data reveal several cases in which the number
of observed events exceed the SM predictions by one to a few orders of
magnitude.  Focusing on these reactions may be used with profit to set
bounds on quadratic products of the RPV coupling constants.  The RPV
contributions to the subprocesses, $ d_k + \bar d_l \to e_i +\bar
e_j$, arise at tree level through $\tilde \nu_L , \ \tilde u_L ,\
\tilde d_{kR},\ \tilde d_{kL} $ exchanges, with the transition
amplitudes represented by the effective Lagrangian, \bea && L_{EFF} =
{\l'_{ijk} \l ^\star _{ij'k'} \over m^2_{\tilde \nu_{iL}}} (\bar
d_{kR}d_{jL})(\bar e_{j'L}e_{k'R}) - {\l'_{ijk} \l ^{'\star } _{i'jk'}
\over 2 m^2_{\tilde u_{jL}}} (\bar d_{kR} \g ^\mu d_{k'R})(\bar
e_{i'L}\g _\mu e_{iL}) \cr && + { \l ^{'\star } _{ijk} \l
'_{i'j'k}\over 2 m^2 _{\tilde d_{k R} } } (\bar d_{jL} \g ^\mu d_{j'L}
) (\bar \nu_{iL} \g^\mu \nu_{i'L} ) - {\l ^{'\star } _{ijk} \l
'_{i'jk'}\over 2 m^2 _{\tilde d_{jL} } } (\bar d_{kR} \g ^\mu d_{k'R}
) (\bar \nu_{i'L} \g^\mu \nu_{iL} ) + \ H. \ c. \label{eqxx5} \eea
The long-lived and short-lived neutral mesons two-body decays into
charged lepton-antilepton pairs, $ K ^0_{L,S} \to e _i + \bar e_j $,
are described by the effective Lagrangian~\cite{roy},
\begin{eqnarray}  &&
L_{EFF} = \bigg [\ud (\bar e_{iL} e_{jR}) (\bar d_R s_L \mp \bar s_R
d_L ) { {\cal B}_{ij} \choose {\cal D}_{ij} } + H.\ c.  \bigg ] -{1
\over 4} (\bar e_{iL} \g ^\mu e_{jL}) \ (\bar s_R \g _\mu d_R \mp \bar
d_R \g _\mu s_R) { {\cal A}_{ij} \choose {\cal C}_{ij} } ,\cr && [
({\cal A}_{ij} , \ {\cal C}_{ij} ) = \sum_{n, p} \tilde u_{nL}^{-2}
V_{np} ( \l ^{'\star } _{ip1} \l ' _{jn2} \mp \l ^{'\star } _{ip2} \l
' _{jn1} ), \ ({\cal B}_{ij} ,{\cal D}_{ij}) = \sum_{n} \tilde
\nu_{nL}^{-2} \l ^{\star } _{nij} (\l ' _{n12} \mp \l ' _{n21} )].
\end{eqnarray}

The patterns of relative signs in the quadratic coupling constant
forms, $[{\cal A }_{ij} , \ {\cal B}_{ij} ] $ and $ [{\cal C}_{ij},\
{\cal D}_{ij} ]$, associated with the $ K_L$ and $ K_S$ mesons
couplings, respectively, demonstrate the complementarity between the $
K_L $ and $K_S$ mesons decays, as observed by Choudhury and
Roy~\cite{roy}.  The comparison with the experimental limits for the
decay branching fractions yields the bounds, \bea \bullet \quad &&
{\cal B}_{11} < 2.5 \times 10^{-8} , \ \quad [K_L ^0 \to e ^- + e ^+]
\cr \bullet \quad && \vert {\cal D}_{22} \vert ^2 +0.099 \Re ({\cal
D}_{22}) ^2 +0.1 \Re ({\cal D}_{22} {\cal C}^\star _{22} )+0.0025\vert
{\cal C}_{22} \vert ^2 < 3.1 \times 10^{-9}, \ [ K_S ^0 \to \mu ^- +
\mu ^+ ]\eea along with other bounds associated with the lepton
generation off-diagonal decay modes.  Several derived quadratic
coupling constant bounds can be inferred from these results by
invoking the double coupling constant dominance hypothesis, of which
we quote below a representative sample involving the first and second
generation leptons,
\begin{eqnarray} 
[ \vert \l ^\star _{121} \l '_{212} \vert , \vert \l ^\star _{121} \l
'_{221} \vert ] & <& 2.5 \times 10^{-8} \ \tilde \nu _{2L} ^2,\ [
\vert \l ^\star _{131} \l '_{312} \vert , \ \vert \l ^\star _{131} \l
'_{321} \vert ] < 2.5 \times 10^{-8} \ \tilde \nu _{3L} ^2, \ [K_L ^0
\to e^+ + e^-]; \cr [\vert \l ^\star _{122} \l '_{112} \vert , \vert
\l ^\star _{122} \l '_{121}\vert ]& <& 3.8 \times 10^{-7} \ \tilde \nu
_{2L} ^2 ,\ [\vert \l ^\star _{232} \l '_{312} \vert , \ \vert \l
^\star _{232} \l '_{312} \vert ] < 3.8 \times 10^{-7} \ \tilde \nu
_{3L} ^2 ,\ [K_L ^0\to \mu^+ + \mu^-]; \cr [\vert \l ^\star _{122} \l
'_{212} \vert , \ \vert \l ^\star _{122} \l '_{221} \vert ] & <& 2.3
10^{-8} \ \tilde \nu _{2L} ^2, \ [ \vert \l ^\star _{132} \l '_{312}
\vert , \ \vert \l ^\star _{132} \l '_{321} \vert ] < 2.3 \times
10^{-8} \ \tilde \nu _{3L}^2, \ [K_L ^0\to e^\pm + \mu^ \mp ].
\label{eqx11b}
\end{eqnarray}  

The corresponding leptonic decays of the light quark pseudoscalar
mesons yield weaker constraints.  The bounds obtained by Kim et
al.,~\cite{kim} for the neutral $\pi $-meson flavor off-diagonal and
diagonal two-body leptonic decays are given by, $ \sum_i \vert
\l'_{i11} \l ^{ \star } _{i12} \pm \l '_{i11} \l ^{ \star } _{i21}
\vert 0 < 0.14 \ \tilde \nu_{iL} ^2 , \ [ \pi^0 \to e^\pm + \mu^ \mp ]
$.  The analogous $J/\Psi $ meson decay modes, $ J/\Psi \to e_i ^\pm
+e_j ^ \mp $, lead to insignificant bounds~\cite{kim}.

Significant one-loop box diagram contributions to the leptonic decays
also arise from pure RPV and mixed RPV gauge interactions.  The study
by Bhattacharyya and Raychaudhuri~\cite{bhattaray} for the $\l '$
interactions yields the quadratic coupling constant bounds, $ \vert \l
^{\star }_{1j1} \l '_{1j2} \vert < 8.6 \times 10^{-5} ,\ [K_L ^0 \to
e^+ + e^-]; \quad \vert \l ^{\star } _{2j1} \l '_{2j2} \vert < 5.8
\times 10^{-6} ,\ [K_L ^0\to \mu ^+ + \mu ^-]. $

The heavy flavored mesons decays which have attracted wide interest in
view of the wide variety of the accessible experimental information.
For the neutral $ B$ meson decay modes, Jang et al.,~\cite{jangkim}
obtain several bounds which improve on previously obtained bounds in
all cases. A representative sample of their results reads as,
\begin{eqnarray}  &&
[ \vert \l ^\star _{121} \l'_{213} \vert , \vert \l ^\star _{121}
\l'_{231} \vert ,\ \vert \l ^\star _{131} \l'_{313} \vert ,\ \vert \l
^\star _{131} \l'_{331} \vert ] < 4.6 \times 10^{-5} \ \tilde m^2 , \
[ B_d ^0\to e^+ + e^- ]; \cr && [ \vert \l ^\star _{122} \l'_{113}
\vert , \ \vert \l ^\star _{122} \vert \l'_{131}, \ \vert \l ^\star
_{232} \l'_{313} \vert , \ \vert \l ^\star _{232} \l'_{331} \vert ] <
2.4 \ 10^{-5} \ \tilde m^2 , \ [ B_d ^0\to \mu ^+ + \mu ^- ]; \cr && [
\vert \l ^\star _{121} \l'_{113} \vert , \ \vert \l ^\star _{122}
\l'_{213} \vert , \ \vert \l ^\star _{132} \l'_{331} \vert , \ \vert
\l ^\star _{231} \l'_{331} \vert ] < 4.5 \times 10^{-5} \ \tilde m^2 ,
\ [B_d ^0\to e^\pm +\mu^\mp ]; \cr && [ \vert \l ^\star _{131}
\l'_{131} \vert ,\ \vert \l ^\star _{123} \l'_{213} \vert ,\ \vert \l
^\star _{133} \l'_{331} \vert ,\ \vert \l ^\star _{231} \l'_{231}
\vert ]< 4.9 \times 10^{-4} \ \tilde m^2 , \ [ B_d ^0 \to e^\pm + \tau
^\mp ]; \cr && [ \vert \l ^\star _{123} \l'_{131} \vert , \ \vert \l
^\star _{232} \l'_{213} \vert , \ \vert \l ^\star _{232} \l'_{231}
\vert , \ \vert \l ^\star _{233} \l'_{331} \vert ] < 6. \times 10^{-4}
\ \tilde m^2 , \ [ B_d ^0 \to \mu ^\pm +\tau ^\mp ].
\label{eqx12b}
\end{eqnarray}
For the charged $ B$ meson decays, Erler et al.,~\cite{erler} infer
the quadratic coupling constant bound, $ \vert \l ^\star _{131}
\l'_{333} \vert < 0.075\ \tilde e_{3L}^2, \ [ B^- \to e ^- +\bar \nu ]
$.  Focusing on the purely leptonic two-body decays of charged D
mesons, $D^{\pm } \to e ^{\pm } +\nu , \ D_s ^{\pm } \to e ^{\pm }
+\nu $, and the corresponding ones for the B mesons, $ B^{\pm }_u, \
B^{\pm }_c $, Akeroyd and Recksiegel~\cite{akeroyd03,aker02} find that
the tree level contributions from the RPV $\l $ and $ \l '$
interactions can lead to large enhancements of the associated rates as
predicted on the basis of the SM.

Improved quadratic coupling constant bounds can be deduced by
considering ratios of rates for the mesons leptonic decays reactions,
$\pi ^-\to e ^- +\bar \nu ,\ K ^-\to e ^- +\bar \nu ,\ B ^-\to e ^-
+\bar \nu , $ in different lepton flavors, thus avoiding the
consideration of poorly determined hadronic physics parameters. We
complement the bounds quoted above by displaying some representative
bounds obtained for the charged mesons leptonic
decays~\cite{dreipole02}, $ \vert \l ^{' \star} _{i11} \l _{3i1} \vert
< 3.4 \times 10^{-6} \tilde e_{iL} ^2 , \ [\pi ^- \to e_l ^- + \bar
\nu _i ] ; \ \vert \l ^{' \star} _{i12} \l _{3i2} \vert < 1.3 \times
10^{-3} \tilde e_{iL} ^2 , \ [K ^- \to e_l ^- + \bar \nu _i ] ; \
\vert \l ^{' \star} _{i13} \l _{3i2} \vert < 7. \times 10^{-4} \tilde
e_{iL} ^2 , \ [B ^- \to e_l ^- + \bar \nu _i ] $, and for the neutral
mesons leptonic decays~\cite{dreipole02}, $ \vert \l ^{' \star} _{311}
\l _{312} \vert < 3. \times 10^{-3} \tilde \nu _{iL} ^2 , \ [\pi ^0
\to \mu ^- + e ^+ ] ; \ \vert \l ^{' \star} _{i21} \l _{i12} \vert <
6. \times 10^{-9} \tilde \nu_{iL} ^2 , \ [K ^0 \to \mu ^- + e ^+ ] ; \
\vert \l ^{' \star} _{i23} \l _{i12} \vert < 7. \times 10^{-5} \tilde
\nu_{iL} ^2 , \ [B ^0 \to \mu ^- + e ^+] $.  Another analysis aimed at
the $K$ mesons decays is presented by Belyaev et al.,~\cite{belya00}.
We quote a sample of their quadratic coupling constant bounds for the
semileptonic three-body decays, $ \vert \l ^{' \star} _{i12} \l _{i21}
\vert < 4.5 \times 10^{-6} \tilde \nu _{iL} ^2 , \ [K ^+ \to \pi ^+ +
\mu ^- + e ^+ ] ; \ \vert \l ^{' \star} _{i22} \l ' _{i12} \vert < 1.2
\times 10^{-5} \tilde \nu _{iL} ^2 , \ [K ^0 \to \mu ^- + \mu ^+ ] $.
Of special interest in this context~\cite{belya00} is the decay mode
involving the emission of a pair of like-signs leptons, $ K ^+ \to \pi
^ - + e_i ^+ + e_j ^+ $, which arises through tree and one-loop
contributions sensitively depending on the $\tilde b_L - \tilde b_R$
mixing.

Useful constraints involving the tau leptons can be inferred from the
experimental data on $ B$ mesons decays gathered at the high energy
leptonic and hadronic colliders.
The signals for the $ B_{d}, \ B_{s} $ mesons leptonic and
semileptonic decay modes, $ B ^- \to \tau ^- +\bar \nu _\tau ,\quad B
^0 \to \tau ^+ + \tau ^- $ and $ B^-\to \tau ^- + \bar \nu _ \tau +X,
\ B ^0 \to \tau ^+ +\tau ^- + X$, can be selected by searching for $
B$ meson final states characterized by the $ \tau $-leptons cascade
decay and a large missing energy.
From the failure to observe the corresponding signals on a sample of
LEP collider data, Grossman et al.,~\cite{grossman2} infer the
branching ratio limits, $ B(B ^0_{[d, \ s]} \to \tau ^+ +\tau ^- ) <
[1.5 \%, \ 5. \% ] ,\ B(B^0_d \to X +\tau ^+ +\tau ^- ) < 0.5 \% ]
$. Although these limits exceed the SM prediction by nearly four order
of magnitudes, they can still set useful constraints on the RPV
contributions involving the $\l ^\star \l '$ coupling constant
products.  The above limits deduced from experimental data yield the
quadratic coupling constant bounds, \bea [ \vert \l '_{i23} \l ^\star
_{i33} \vert , \ \vert \l '_{i32} \l ^\star _{i33} \vert ] &<& 1.2
\times 10 ^{ -2} \ \tilde e ^2_{iL} ,\ [B^0_s \to \tau ^- +\tau ^+ +X
_s , \ B^0_s \to \tau ^- +\tau ^+ + X ] ; \cr [ \vert \l '_{i13} \l
^\star _{i33} \vert ,\ \vert \l '_{i31} \l ^\star _{i33}] \vert &<&
0.67 \times 10 ^{ -2} \ \tilde e_{iL} ^2 , \ [B^0_d \to \tau ^- +\tau
^+, \ B^0_d \to \tau ^- +\tau ^+ + X ].  \eea The steady improvement
of the statistics in measurements of $B$ meson decays at the
asymmetric $ B $ meson factories should strongly stimulate the
development of dedicated studies of flavor changing effects in $B$
mesons decays.

A predictive study of the lepton flavor changing $B$ mesons leptonic
decay modes, $B ^0\to \tau ^+ + \mu ^- ,\ B ^0\to \tau ^+ + \mu ^- +X
$, is developed~\cite{guet99} by using the existing coupling constant
bounds and invoking $U(1)$ flavor symmetry models.  The tree level RPV
contributions involving the coupling constant products, $ \l ^{'\star
} _{ij3} \l _{i32}/ m^2 _{\tilde \nu _i } ,\ \l ^{'\star } _{i3j} \l
_{i23} / m^2 _{\tilde \nu _i } ,\ \l ^{'\star } _{2j3} \l _{3jk}/ m^2
_{\tilde q _k } ,$ yields branching fractions, $ O(10 ^{-9} ) \ - \
O(10 ^ {-7} ) $, which are not particularly suppressed relative to the
flavor diagonal ones, $ B ^0 \to \tau ^+ \tau ^- $.

\subsubsection{\bf Rare semileptonic decays of hadrons}

The RPV contributions to the semileptonic decay modes of neutral and
charged mesons, $ K_{S,L}^0 \to \pi ^0 + e^-_i + e ^ +_j , \ K^\pm \to
\pi ^\pm + e^- _i +e ^ + _j $, exhibit interesting correlations with
the corresponding leptonic decay modes discussed in the above
paragraphs. These reactions, along with the three-body semileptonic
rare decays of $K$ mesons with neutrino-antineutrino pair emission,
$K^{+,0}\to \pi^{+,0}+\nu +\bar \nu $, represent hallmarks for SM
tests and new physics searches.  For charged mesons, the dependence on
the hadronic wave function factor can be removed by considering the
branching ratio to the $ K_{e3}$ decay, $ { \G (K^+\to \pi^++\nu +\bar
\nu ) / \G (K^+\to \pi^0+e^++\nu ) } $.  Aside from the information on
the flavor mixing matrix element, $ V_{td}$, the interest in these
decay rates stems from the high sensitivity of the experimental
measurements and the tight control on the theoretical uncertainties
for the SM input parameters and the long distance hadronic physics
contributions~\cite{litt93,bigi,litt98}.  Nevertheless, the inference
of constraints on new physics is becoming more ambiguous as the
experimental sensitivity~\cite{adler97} is currently reaching the same
$O(10 ^{-11})$ level as that of the theoretical uncertainties.

The RPV interactions contribute at tree level to the process $K^+\to
\pi^++\nu_i +\bar \nu_j $ through the quark subprocess, $ d^c \to \nu
+\nu ^c+ d^c$, involving a $\tilde d_{kR}$ or a $ \tilde d_{kL}$
exchange, as displayed by diagrams F.1-2 in Figure~\ref{fig3}.  The
four fermion coupling is represented by a term in the effective
Lagrangian quoted in eq.(\ref{eqxx5}).
The comparison by Choudhury and Roy~\cite{roy}, for the experimental
decay rate summed over neutrino generations, gives a bound on the
quadratic coupling constant form, $ \sum_{ij} \vert {\cal E }_{ij}
\vert ^2 < 2.3 \times 10^{-9},\ [{\cal E }_{ii' }= \sum _k (\tilde
d^{-2} _{kR} \l ^{'\star }_{i2k} \l '_{i' 1k} - \tilde d^{-2} _{jL} \l
^{'\star }_{ij1} \l '_{i' j2}) ] $. Under the double coupling constant
dominance hypothesis, one can infer useful quadratic coupling constant
bounds by selecting different configurations of generation indices in
the inequality, ${\cal E }_{ij} < 4.8 \times 10^{-5} \ \tilde m^2$.  A
representative sample of the strongest bounds is given by~\cite{roy},
$ \vert \l ^{'\star }_{i1k} \l '_{j2k} \vert < 4.8 \times 10^{-5} \
\tilde d_{kR}^2,\quad \vert \l ^{'\star }_{ij2} \l '_{i' j1} \vert <
4.8 \times 10^{-5} \ \tilde d_{jL}^2 $.

Starting from the current basis field representation of the RPV
interactions, one may transform to the mass basis fields by
substituting for the down-quark fields in the effective interaction, $
\bar d_k d_{k'} \to \bar d'_k d'_{k'} \simeq V_{k'1} V^\star_{k2} \bar
s d +\cdots $.  The resulting effective Lagrangian, along with the
individual coupling constant bounds~\cite{agashe}, derived under the
single coupling constant dominance hypothesis, are given by, $
L_{EFF}=-{\vert \l '_{ijk}\vert^2 \over 2m^2_{\tilde d_{kR} } } V_{j1}
V^\dagger _{2j} (\bar s_L \g_\mu d_L)( \bar \nu_{iL} \g_\mu \nu_{iL})
+ H. \ c. \ \Longrightarrow \ \vert \l '_{imk}\vert < 1.2 \times
10^{-2} \ {\tilde d_{kR}} , \ \vert \l '_{i3k}\vert < 0.52\ {\tilde
d_{kR}} . $

The available experimental limit on the inclusive semileptonic
three-body decay modes of neutral $B$ mesons, $ B(B^0 \to X + \nu
+\bar \nu ) < 7.7 \times 10^{-4} $, lies at an order of magnitude
above the SM prediction.  The comparison with the tree level RPV
contribution from the quark subprocess, $ b \to s +\nu +\bar \nu $,
leads to the quadratic coupling constant bounds~\cite{grossman}: $
\vert \l ^{'\star }_{ijk}\l'_{i' 3k} \vert< 1.1 \times 10^{-3}\
{\tilde d_{kR} }^2, \quad \vert \l ^{'\star }_{ijk } \l ' _{i'j3}\vert
< 1.1 \times 10^{-3}\ {\tilde d_{jL} }^2 , \ [B ^0\to X_q +\nu +\bar
\nu ]. $

\subsubsection{\bf Rare nonleptonic  decays of
light and heavy quark hadrons}

The hadronic rare decays of $B$ mesons involving a change of quark
flavor provide a useful information on the $\l _{ijk} ,\ \l ' _{ijk} $
interactions.  The measurements by the BaBar Collaboration of decay
modes, $B ^0 \to \phi + \pi ^0,\ B^0 \to \phi + \phi $, feature a
strong suppression with respect to the SM prediction.  The comparison
by Bar-Shalom et al.,~\cite{eilam03} with the RPV tree level
contribution yields the useful bounds on coupling constant products, $
\vert \l '' _{i23} \l ^{''\star } _{i21} \vert < 6.  \times 10^{-5} \
\tilde u_{iR} ^2 ,\ [\vert \l ' _{i32} \l ^{'\star } _{i12} \vert ,\
\vert \l ' _{i23} \l ^{'\star } _{i21} \vert ] < 4.\times 10^{-4} \
\tilde u_{iR} ^2 $.  For the radiative decay mode, $\bar B_d^0 \to
\phi +\g $, Li et al.,~\cite{li03} obtain useful bounds on the
coupling constant products, $\l '' _{i23} \l ^{''\star } _{i12} , \ \l
' _{i32} \l ^{'\star } _{i12} $.  The decay modes of charmed quark
mesons are also actively searched at the $B$ meson factories.


Carlson et al.,~\cite{carlson} examine the contributions to the
charged $B$ meson decay flavor changing processes, $B^+\to \bar K^0
+K^+,\ B^- \to K^0 +K^- ,\ B^+\to K^0 +\pi ^+ $, based
on the tree level Feynman diagram depicted by graph E.10 in
Figure~\ref{fig2}.  The intermediate gluon line is spacelike and
propagates far off the mass shell, which validates the perturbative
QCD treatment of the transition amplitude for this exclusive process
employing the methods of perturbative QCD light-cone
physics~\cite{brodsky}. The initial state one-gluon exchange amplitude
is described by the dominant quark-antiquark wave function component
of the $B$ meson, $ B^+ \to \bar b +u$.  The comparison to the
experimental limits on the two-body $B$ meson hadronic decays
yields the coupling constant bounds~\cite{carlson}, \be \vert \l
^{''\star } _{i32} \l ^{''\star } _{i21}\vert < 5. \times 10^{-3} ({
m_{\tilde q_i } \over m_W} )^2 , \ [ B^+\to \bar K^0 +K^+]; \ \ \vert
\l ^{''\star }_{i31} \l ''_{i21}\vert < 4.1 \times 10^{-3} ({
m_{\tilde q_i } \over m_W} )^2 , \ [ B^+\to K^0 +\pi ^+] .\ee The
alternative description of the above discussed
processes~\cite{carlson}, using the heavy quark symmetry approach,
yields similar coupling constant bounds differing by $15 \% $ from the
above quoted bounds.

\subsubsection{\bf Rare top quark decay modes}
\label{secxxx3atop}

The rare decay modes of the top quark can be used both to identify its
actual production in the high energy collider reactions and to infer
useful constraints on new physics.  An early study by Dreiner and
Phillips~\cite{phillips91} proposed to search for the top quark by
means of the multilepton final state signals initiated at the high
energy colliders by the RPV cascade decays.  Motivated by the
experimental observation of top-antitop quark pair production at the
Fermilab Tevatron collider reaction, $p+\bar p \to t+\bar t$, several
authors~\cite{agashe,erler} have recently examined the constraints
imposed by the RPV top quark decay modes.  The SM final states
initiated in the production reaction, $p+\bar p \to t+\bar t$, by the
top quark semileptonic weak decay mode, $ t \to b +W ^+$, are
characterized by three main signatures, involving a dilepton pair, a
single lepton accompanied with hadronic jets, and purely hadronic
jets, respectively.  In the presence of RPV decay channels for the top
quark, the branching fractions, $ R_B (x)$, associated to a fixed
final state, $X$, are modified as, \bea R_B (x) &=& {B( t\bar t \to X
) \over B( t\bar t \to X )_{SM} } = (1-x)^2 + \d R_B (x) = {1\over
(1+R_t)^2 } + \d R_B (x) , \cr [ x &=& {R_t\over 1+R_t} , \ R_t = {\G
_{RPV}(t\to b+ \tilde \tau ) \over \G (t \to \ \text{all} ) } ]
\label{eqtop1} \eea where the parameter $x$ includes the RPV
contributions to the decay branching fraction, while the contributions
from the additional multijet events are included in $\d R_B (x) $.
For the dilepton events, one has obviously, $\d R_B (x)=0 $.  The RPV
contributions to the branching fractions increments, $\d R_B (x)$,
depend on the auxiliary ratio parameter, $ x$, and the $b$-quark
tagging efficiencies, $\e _{m, n}$. Note that the $b$-quark jets
identification is affected by the tagging efficiency parameters, $\e
_{m, n}$, defined as the probabilities to correctly identify $m \
b$-quark jets out of a total number $n$ of jets.  The top quark
two-body decay channel with slepton emission, $t\to \tilde e^+_i + d_k
, $ is initiated by the $\l '_{i3k}$ interactions with a rate, $ \G
(t\to \tilde e^+_i+d_k) \simeq {\vert \l '_{i3k} \vert ^2 k ^2 \over
16 \pi } (1- ({m_{\tilde l_i} \over m_t })^2)^2 $, where $k \approx
{m_t / 2}$ denotes the final state center of mass momentum.  An
identical formula holds for the RPV down-squark emission reaction, $t
\to \tilde d_i + e^+_k $.

As an initial test, one may attempt to compare the calculated RPV
contribution to $ R_B(x) $ with the ratio of the experimental total
production cross section~\cite{cdf}, $ \s (t\bar t)_{exp} = 6.8 {+ 3.6
\choose -2.4 } \ \text{pb},$ to the corresponding QCD
prediction~\cite{catani}, $\s (t\bar t) _{QCD} = 5.52 {+ 0.07 \choose
-0.45 } \ \text{pb} $.  From the experimental uncertainties on the
partial decay channels, one infers the coupling constant
bound~\cite{erler}, $\vert \l '_{333}\vert < 1.3 $, at the $ 2 \s $
level, using $m_{\tilde \tau _L} =100 \ \text{GeV} $.  A more direct
test is provided by the experimental data for the total $ t\bar t$
production rate.  Comparing the average rates for the D0 and CDF
collaboration experiments~\cite{cdf} with the QCD prediction of Catani
et al.,~\cite{catani} based on the formula, \be {\s (p +\bar p \to t +
\bar t )_{QCD}\over \s (p +\bar p \to t + \bar t )_{exp} } = [1+ {\G
(t \to b + \tilde \tau ^+ ) \over \G (t \to b + W^+) }] ^{2} = (1+0.70
\vert \l ^{'} _{332}\vert ^2 )^{2}, \ee yields the $2 \s $
bound~\cite{ledroit}: $ \vert \l ' _{l3n} \vert < 0.55$.  For the
purely hadronic RPV decay channels, a similar comparison aimed at the
ratio, $ R_B (x) = {\s_{t\bar t} ^{exp} / \s_{t\bar t} ^ {QCD}} \simeq
(1-x)^2$, gives the coupling constant bounds, $\vert \l ''_{3jk} \vert
< 1.25$.

We discuss now the studies specialized to specific final states.  In
the case of a neutralino LSP, the leptonic events associated to the
top quark cascade decays, $ t \to \tilde e^+_i+ d_k ,\ \tilde e^+_i
\to \tchi^0 +e ^+ _i \to [e ^+_i + \nu _ j + b+ \bar d_k , \ e ^+_i +
\bar \nu _ i +\bar b + d_k] $, are governed by the same set of
coupling constants, $\l '_{i3k} $.  These reactions can influence the
final states events by initiating deviations with respect to the $e \
- \ \mu $ lepton universality, through a large hierarchy in the ratio
of coupling constants, $ \l '_{13k}/\l '_{23k} $, and, for $ k=3$,
multijet events with a surplus of $b-$quarks.  The comparison of
branching fraction ratios for single-$e$ to single-$\mu $ events, $
B(t\bar t \to e +jets) / B(t\bar t \to \mu +jets) $, with the
corresponding experimental ratio of events with one charged lepton and
two $b $-quark jets, $ N(e+jets)/ N(\mu +jets) = 1 {+a\choose -b} $,
gives the coupling constant bounds~\cite{agashe}, $\vert
\l'_{13n}\vert < 0.41, \ [n=1,2] $.

 A different line of reasoning is followed by Barger et
al.,~\cite{fp2}, by invoking the possibility that the $\l '_{333}$
initiated decay mode, $ t\to b + \tilde \tau ^+$,
may not be mistaken from the regular $2l + 4 $ jets signal.  The RPV
channels would then reduce the SM prediction for $t-\bar t$ event
rates by the correction factor, $ x \simeq R_t = {\G (t \to b +\tilde
\tau ^+ ) \over \G (t \to b + W ^+) } \simeq 1.12 \ \vert \l ^{'}
_{333}\vert ^2(1-{m_{\tilde \tau _L} ^2 \over m_t ^2 } )^2 $, yielding
with the input, $ m_{\tilde \tau } \simeq m_W$, the prediction $R_t
\simeq 0.70 \vert \l ^{'} _{333}\vert ^2 $.  This result clearly shows
that coupling constant values, $ \l ' = O(1)$, are required in order
for the competition to be effective.  Using the fixed point value for
the coupling constant, $\l '_{333} \simeq 0.9$, the RPV mode would
deplete the SM signal by the factor, $ (1+ R_t) ^{-2} \simeq (1 +0.70
\vert \l '_{333} \vert ^2 ) ^{-2} \simeq 0.4 $.
A similar analysis applies for the $\l ''_{323}$ interaction which
initiates the hadronic two-body decay channels, $ t\to \bar b +\tilde
{\bar s} , $ cascading to $5$ jets final states.  The expected RPV
correction factor to the SM signal from the decay modes, $ t \to \bar
b +\tilde s ^\star ,\ t \to \bar s+ \tilde b ^\star $, reads as, $ (1+
R_t) ^{-2} \simeq (1 +0.16 \vert \l ''_{323} \vert ^2 ) ^{-2} \simeq
0.75 $. Both of the above correction factors go in the wrong direction
in comparison with the observed trends, since the experimental rates
are found to exceed the SM predictions for all the channels.  The
attendant conflict with the experimental observations can be avoided,
however, either by ruling out the relevant coupling constant or by
closing the relevant decay channels, as by assuming that the sleptons
and squarks are heavier than the top quark.

In the bilinear R parity violation option with finite sneutrino VEVs,
the additional couplings between the charged Higgs bosons and sleptons
arising from the fields mixing can contribute to the top quark decay
modes, $ t \to \tilde \tau ^+ + b , \ t \to \tau ^+ +\tilde b $, with
rates increasing with $\tan \b $.  The analysis by Navarro et
al.,~\cite{navarro} of the MSSM predictions
subject to the usual constraints on the mass spectrum, the neutrino
mass bound, $ m_{\nu _\tau } < (18 \ \text{ MeV}\ -\ 1 \ \text{ eV} )
$, and the allowed intervals for the top quark decay branching
fractions, $[ B(t\to \tilde \tau ^+ + b ) , \ B(t \to \tau ^+ +\tilde
b ) ] \sim (0.025\ - \ 0.10) $, selects the domain of RPV parameters,
$ v_3 ,\ \mu_3$, bounded by the intervals, $ 0< \vert v_3 \vert < 20 \
\text{GeV} ,\ 0< \vert \mu_3 \vert < 100 \ \text{GeV} . $

The experimental sensitivity on the top quark decay branching
fractions attained with the CDF and D0 collaboration detectors
currently reaches the orders of magnitude, $ 10^{-3} $ or $ 10^{-4} $.
An improved experimental sensitivity is expected with the Cern LHC
collider detectors.  Motivated by this prospect, Yang et
al.,~\cite{yang98} examine the rare two-body decay channels involving
the emission of a final state vector boson, $ t\to c + V , \ [V= Z,
\g, g] $.  The one-loop RPV amplitudes are related by means of
crossing to the fermion pair production amplitudes, $ V \to \bar t +
c$. The predicted branching fractions for emission of a photon,
$Z$-boson or gluon are given by, $ B( t\to c +[Z, \g, g] ) = [0.36
\times 10^{-4}, \ 0.09 \times 10^{-5}, \ 0.16 \times 10^{-3} ] \ (
\sum_{j\ne k} \vert \l ^{''\star}_{2jk} \l ''_{3jk}\vert )^2 ,$
respectively.  The resulting coupling constant bounds are not very
strong ones but are still complementary to the other existing bounds.
The RPV one-loop contributions to the flavor changing top quark decay
into the Higgs boson, $ t\to c+h$, are found by Eilam et
al.,~\cite{eilam01} to reach branching fraction values, $ O(10^{-5})$,
when one uses the current bounds on the $\l' , \ \l ''$ trilinear
coupling constants.  The detection of the top quark
decays~\cite{han00}, $ t \to \tilde \tau + b , \ t \to \tau + b +
\tchi ^0 $, at the run II of the Tevatron collider can set the useful
$ 2 \s $ level bound, $ \vert \l ' _{333} \vert < (0.38 \ -\ 0.24)$.

\subsubsection{\bf Radiative decays of hadrons}

The recent measurements by the CLEO and ALEPH collaborations of the
neutral $B_s$ meson radiative decays, $ B^0_s \to K^0 + \g ,\ B^0_s
\to K^{0\star } + \g $, have aroused a wide interest. These hadron
flavor changing reactions, along with the analogous $B$ mesons
inclusive decay reactions, $ B_s \to X_s + X $, take place in the SM
through one-loop contributions to the subprocesses, $b \to s + \g $
and $b \to s + g $, respectively.  Extensive discussions of the MSSM
corrections have also been developed to these reactions which offer
sensitive probes of new physics.

The so-called indirect RPV contributions produced by renormalization
group corrections are examined by De Carlos and White~\cite{decarlosq}
within the supergravity approach to supersymmetry breaking.  The
radiative loop contributions can enhance the rate for $b \to s +\g $
by nearly an order of magnitude if one uses values for the relevant
coupling constants at the unification scale, $ [\l' _{121}(M_X) , \
\l' _{131}(M_X)] = 0.05, \ [\l'' _{112} (M_X), \ \l' _{113}(M_X) ] =
0.1 $. The direct type one-loop RPV contributions yield the quadratic
coupling constant bounds, $ \vert \l ^{'\star }_{i2k} \l '_{i3k}\vert
< 0.09 \ [{2 \over \tilde \nu_{iL}^{2} } - {1 \over \tilde d_{iR}^{2}}
] ^{-1}, \ \vert \l ^{'\star }_{ij2} \l '_{ij3}\vert < 0.03 \ [{2
\over \tilde e_{iL}^{2}} - {1 \over \tilde d_{jL}^{2} } ] ^{-1}, \
\vert \l ^{''\star }_{i2k} \l ''_{i3k}\vert < 0.16\ \tilde
q_{iR}^{2}. $

 The small but finite discrepancy with respect to the SM prediction
featured by the experimental rate measurements of the radiative $B$
meson inclusive decay reaction, $ B^0 \to X_s +\g $, may be used to
constrain the new physics contributions.  The resulting quadratic
coupling constant bounds for the baryon number violating coupling
constants turn out, however, to be rather weak~\cite{chakra01}, $\vert
\l ^{''\star } _{3j2} \l'' _{3j3} \vert < 0.35 $.
  
The comprehensive discussion of the $b\to s + \g $ transition
amplitude makes use of the renormalization group approach to the
relevant dimension $6$ effective Lagrangian, $ L_{EFF} = \sum _i C_i(Q
) O_i(Q ) $, where $Q$ is the running momentum. For illustration, we
recall the definitions of the first two local operators, $ O_1 = (\bar
s_{\a L }\g ^\mu b_{\a L }) (\bar q_{\b L } \g _\mu q_{\b L }) ,\ O_2
= (\bar s_{\a L }\g ^\mu b_{\b L }) (\bar q_{\b L } \g _\mu q_{\a L
}). $ The various Wilson coefficients, $C_i(Q ) $, are mixed together
by renormalization effects.  The general formalism is adapted to the
broken R parity symmetry case by Besmer and Steffen~\cite{besmer00}.
In the presence of both gauge and RPV interactions, the total basis of
independent operators, $O_i$, is enlarged from the 8 allowed operators
in the SM case to a total of 28 operators.  In the conventional
formalism, one chooses some input boundary conditions for the various
parameters at the unification scale, $Q = M_X$, and runs the momentum
scale evolution from the gauge bosons and heavy quarks masses down to
the low energy hadronic mass scale.  As it turns out, the impact of
the RPV interactions cannot be easily assessed because of the strong
sensitivity of results to the supersymmetric partners spectrum.
Nevertheless, the contributions are found to become significant for
values of the trilinear coupling constants, $ [\l_{ijk}, \l '_{ijk},
\l ''_{ijk} ] > O(10^{-1}) $.

\subsection{Lepton  flavor changing processes}
\label{secxxx3b}

\subsubsection{\bf Radiative decays of charged leptons}


The radiative two-body lepton decays are among the simplest processes
probing the lepton flavor changing effects. The contributions from the
SM loop corrections are well understood while extensive studies have
attempted to assess the impact of the MSSM radiative
corrections~\cite{masierofcnc,hagelin94,gabbiani96}.

The muon radiative decay reaction, $\mu ^ \pm \to e ^\pm +\g $, has
focused most attention because of the greater experimental sensitivity
accessible in the experimental measurements.  An illustrative example
of the one-loop trilinear RPV contribution to the leptons radiative
decays is displayed by the Feynman diagram H.1 in Figure~\ref{fig3}.
In the presence of sneutrino VEVs the mechanism illustrated by the
Feynman diagram H.2 in Figure~\ref{fig3} can take place through the
combined bilinear RPV interactions and the gauge interactions.  The
trilinear RPV transition amplitude~\cite{decarlosl}
is described schematically by the formula, $ A \propto { (\l^{\star}
\l ) m_\mu \over (4\pi )^2 m^2_{\tilde f} } f(m^2_f/ m^2_{\tilde f}
)$, with $f,\ \tilde f $ referring to the fermion and sfermion
internal lines.  The comparison with the experimental limits results
in the coupling constant bounds~\cite{decarlosl}, $ \l^{\star} _{31n}
\l_{32n} < 4.6 \times 10^{-4} \ \tilde e_L^2,\ \l^{\star} _{lm1}
\l_{lm2} < 2.3 \times 10^{-4} \ \tilde e_L^2 . $ A systematic
discussion of the trilinear $ \l $ and $\l '$ interactions one-loop
contributions to the muon radiative decay, $\mu ^\pm \to e^\pm +\g $,
is presented by Chaichian and Huitu~\cite{chaichian}.  A
representative sample of the strongest quadratic coupling constant
products is given by, \bea \bullet \quad && \vert \l^{\star} _{121} \l
_{121} \vert < 0.57 \times 10^{-4} ,\ \vert \l^{\star} _{131} \l
_{131}\vert < 0.57 \times 10^{-4} ,\ \vert \l^{\star} _{23k} \l _{131}
\vert < 1.1\times 10^{-4} ; \cr \bullet \quad && \vert \l ^{'\star }
_{2mk} \l '_{1mk}\vert < 4.5 \times 10^{-4} , \ \vert \l ^{'\star }
_{23n} \l '_{13n}\vert < 7.7 \times 10^{-3} ,\ \vert \l ^{'\star }
_{233} \l '_{133}\vert < 1.0 \times 10^{-2} , \eea for $ k=1,2,3 $ and
$ m,n =1,2.$ The study by Masiero~\cite{masierotau} of the tau-lepton
decay reaction, $\tau ^- \to e^- + \g $, yields the coupling constant
bound, $\vert \l ^{'\star } _{1jk} \l ' _{3jk} \vert < 1.2 \ \times
10^{-2}. $

The lepton flavor changing decay modes with two-photon emission, $e ^-
_i \to e ^- _j + \g + \g $, are recently discussed by Gemintern et
al.,~\cite{gemin03}.  Along with the reducible type one-loop Feynman
graphs, the RPV transition amplitude includes the contribution from
the irreducible type graphs described by the $t$-channel exchange of
sneutrinos which decay as, $\tilde \nu \to \g +\g $, through the
familiar triangle loop diagram.  The reaction rates, estimated by
using the existing bounds for the relevant coupling constant products,
happen to lie well within the current range for an experimental
observability. With a reasonable estimate for the expected
experimental sensitivity on the branching fraction, $ B(\mu ^-\to e ^-
+ \g + \g ) < 10^{-14} $, one finds the following quadratic coupling
constant bounds, $[\vert \l _{122} \l ^\star _{233}\vert , \ \vert \l
_{121} \l ^\star _{133}\vert ] < 1.8 \ \times 10^{-3},\ \vert \l
_{122} \l ^{'\star } _{211} \vert < 8.8 \ \times 10^{-4},\ \vert \l
_{131} \l ^{'\star } _{i33}\vert < 1.4 \ \times 10^{-2}.$

\subsubsection{\bf Three-body decays of leptons}

The three-body decay modes of charged leptons, $e^\pm _m\to e^\pm_i
+e^-_j+e^+_k$, offer promising probes for the observation of lepton
flavor number non-conservation effects compatible with a conserved
total lepton number, $ L= \sum _i L_i$.  A large set of flavor
configurations for the trilinear coupling constants can be accessed
through the various final states, by contrast to the real photon
emission reaction discussed in the preceding subsection. The RPV tree
level contributions are described by the effective
Lagrangian~\cite{roy},
$ L _{EFF} = [ F_{mijk} (\bar e_{iR}e_{mL}) (\bar e_{jL} e_{kR})
+F_{kjim} (\bar e_{iL}e_{mR}) (\bar e_{jR} e_{kL}) ] +(i\to j) + H. \
c. , $
where the relevant generation indices in the auxiliary parameters, $ F
_{abcd}=\sum_i m_{\tilde \nu_i }^{-2} \l _{iab} \l _{icd}^\star $, run
over the sets: $ (ab)= [13, 23, 31, 32], \ (cd)= [11, 12, 21, 22]$.
Barring the possibility of accidental cancellations between different
operators, the comparison with experimental decay rates can be used to
deduce bounds on quadratic coupling constant forms.  A representative
sample of the strongest bounds reads as~\cite{roy}, \bea &&\vert
F_{1112}\vert ^2 + \vert F_{2111}\vert ^2 < 4.3 \times 10^{-13},\ [\mu
\to 3e] ; \quad \vert F_{1113}\vert ^2 + \vert F_{3111}\vert ^2< 3.1
\times 10^{-5}, \ [\tau \to 3e]; \cr && \vert F_{2223} \vert ^2 +
\vert F_{3232}\vert ^2 < 4.1 \times 10^{-5}, \ [\tau \to 3\mu ]. \eea
Under the double coupling constant dominance hypothesis, one derives
several quadratic coupling constant bounds of which we quote a
representative sample, \bea && [\vert \l^{\star} _{i11} \l _{i12}\vert
, \ \vert \l^{\star} _{i21} \l _{i12} \vert ] < 6.5 \times 10^{-7}
\tilde \nu_{i}^2, \ [\vert \l^{\star} _{i11} \l _{i13}\vert , \ \vert
\l^{\star} _{i31} \l _{i13}\vert ] < 5.5 \times 10^{-3} \ \tilde
\nu_{i}^2, \cr && \vert \l^{\star} _{i22} \l _{i23} \vert < 6.4 \times
10^{-3} \ \tilde \nu_{i}^2, \ \vert \l^{\star} _{i32} \l _{i32}\vert <
6.4 \times 10^{-3} \ \tilde \nu_{i}^2.\eea

\subsubsection{\bf Charged lepton conversion}

The reaction of muonium-antimuonium atom
conversion~\cite{ponte58,feinberg}, $Mu (\mu^+e^-)\to \overline{Mu} (
\mu^-e^+)$, is a valuable experimental probe of the lepton flavor
changing transition, $\mu ^\pm \leftrightarrow e^\pm $, obeying the
selection rules for the lepton flavor numbers, $\D L_\mu = - \D L_e =
-2$. The interest in the $ Mu \leftrightarrow \overline{Mu} $
oscillation was initially motivated by the possible realization of
lepton number conservation by the less restrictive multiplicative
$Z_2$ discrete symmetry~\cite{feinberg0}.  The $Mu \to \overline{Mu}$
transition is forbidden in the SM unless one includes massive Majorana
neutrinos which initiate the process at the one-loop order by the box
diagram involving the exchange of a neutrino pair.

The initial discussion of $Mu \to \overline{Mu} $ oscillation by
Feinberg and Weinberg~\cite{feinberg} considered two cases associated
with ensembles of muonium atoms $(\mu^+e^-)$, prepared in a gaseous
phase or embedded in a crystal, sitting in the hyperfine states $1S_{F
=0, 1}$ or in higher excited states.  A characteristic experimental
signature for the transition is the observation of the fast electrons
emitted during the decay of the produced muons.  The
muonium-antimuonium system is treated as a two-dimensional quantum
mechanical system whose time evolution is governed by a $2\times 2$
mass matrix, $M$, with the $Mu \to \overline{Mu}$ transition initiated
by the off-diagonal mass parameters, $ M_{12} = M_{21} = 2 \d m $.
The binding effects from an external electric or magnetic field
contribute with opposite signs to the two diagonal matrix elements
associated with the CP mirror conjugate muonium and antimuonium
systems.  Due to the extreme smallness of the non-diagonal matrix
elements relative to the diagonal ones, $ \d m << [ M_{11} ,\ M_{22}]
$, even small environmental effects can drastically affect the
conversion of muonium into antimuonium. The conditions set on the
observation of unquenched transitions are discussed in
Ref.~\cite{feinberg} and the current experimental status is discussed
in Ref.~\cite{hou96}.

The RPV contribution is described by the effective
Lagrangian~\cite{kim,halprin,moha92}, \be L_{EFF} (Mu \to
\overline{Mu})= {4G_{Mu \to \overline{Mu}} \over \sqrt 2} (\bar \mu_L
\g ^\mu e_L ) (\bar \mu_L \g_\mu e_L) + H. \ c.  , \quad [ {G_{Mu \to
\overline{Mu}} \over \sqrt 2} = -{\vert \l _{i 21} \l _{i12}
^\star\vert \over 4 m^2_{\tilde \nu _{iL} } } ] \ee where the coupling
constant parameter, $G_{Mu \to \overline{Mu}}$, is related to the mass
shift, $\d m $, by the formula, $ \d m = {16G_{Mu \to \overline{Mu}}
\over \sqrt 2 \pi a ^3 } \simeq {G_{Mu \to \overline{Mu}} \over G_F }
\ (2.1 \times 10^{-12} \ \text{eV}) , \ [ a= (m_e e^2)^{-1}] $, with
$a$ designating the atomic Bohr radius.  The time integrated
transition probability to observe an $\vert \overline{Mu}>$ state,
starting at an initial time $t=0$ with an $\vert Mu > $ state, is
described as: $ P(Mu \to \overline{Mu}) = {\d m ^2 / [ 2(\d m ^2 + \D
E ^2 + {1\over 4} \G _\mu ^2 ) ] } \simeq 2 \vert \d m \vert ^2/ \G
_\mu ^2 $, where $\G _\mu $ designates the muon decay width.  The
comparison with the experimental limit on the time integrated
transition probability, $ P _{exp} (Mu \to \overline{Mu}) < 2.1\times
10^{-9} \ \Longrightarrow G_{Mu \to \overline{Mu}} \le 9.6\times
10^{-3} G_F $, leads to the quadratic coupling constant
bound~\cite{kim}: $\vert \l_{i21} \l^\star _{i12} \vert < 6.3 \times
10^{-3}\ {\tilde \nu_{iL} } ^2 $.

The lepton flavor changing process for the exotic wrong flavor muon
decay mode, $ \mu ^+ \to e^+ + \bar \nu_e + \nu_\mu $~\cite{halprin},
is closely related to the $Mu \to \overline{Mu} $ oscillation process.
The transition amplitude is described by the analogous effective
Lagrangian, \be L_{EFF} = {4G_{\mu }^{(e)} \over \sqrt 2} (\bar \mu_R
\g ^\mu e_R ) (\bar \nu_{\mu L } \g_\mu \nu _{eL}) + H. \ c. , \quad [
{G_{\mu }^{(e)} \over \sqrt 2} = {\l_{i12} \l ^\star _{i21} \over
m^2_{\tilde e_{iL} } }] .\ee Using the quadratic bound, $\vert
\l_{312} \l^\star _{321} \vert < (2. - 3.) \times 10^{-3}$, inferred
from the existing single coupling constant bounds, one obtains the
prediction for the effective vertex parameter~\cite{halprin}, $G_{\mu
}^{(e)} /G_F< 2.  \times 10^{-2} \tilde \nu_\tau ^{-2},$ which lies
close to the experimental sensitivity that can be currently attained
for this reaction.

The nuclear physics conversion process, $ \mu ^- +N \to e ^- +N $,
where the muons are transformed into electrons following their nuclear
capture from an atomic orbit, is a sensitive probe of new
physics~\cite{moevogel}. The branching fraction of the conversion
reaction relative to the dominant nuclear muon capture reaction,
$ R^A_{\mu e} = \G (\mu ^- +A \to e ^- +A ' ) / \G (\mu ^- +A \to
\nu_\mu +A ') $, should be measured for the nuclei $ ^{48} _{22}
\text{Ti} $ and $ ^{179} _{79}\text{Au} $ with high experimental
sensitivity by the SINDRUM II experiment at the PSI-Zurich accelerator
and the MECO experiment at the BNL accelerator.  The RPV contributions
to the nuclear conversion reactions involve specific quadratic
coupling constant forms in the $\l '$ coupling constants. For the $
^{48} _{22} \text{Ti} $ target nucleus, the bound found by Kim et
al.,\cite{kim} is given by: \be [\sum_j \vert \l ^{'\star }_{2j1} \l
'_{1j1} \vert {\tilde u_{jL} }^{-2} -2 \sum_i (\vert \l ^{'\star
}_{i11} \l '_{i12} \vert {\tilde \nu _{iL} }^{-2} \pm \vert \l
^{'\star }_{i11} \l '_{i21}\vert {\tilde \nu_{iL} }^{-2} ) -\sum_k
{70\over 14} \vert \l ^{'\star } _{21k} \l '_{11k}\vert {\tilde d_{kR}
} ^{-2} ] < 1.6 \times 10^{-7} .\ee Using the pair coupling constant
dominance hypothesis, one may single out in turn the various quadratic
product terms in the above quadratic form.  The representative sample
of bounds which are least exposed to cancellations is given by, $\vert
\l ^{'\star }_{2j1} \l '_{1j1} \vert < 1.6 \times 10^{-7} {\tilde
u_{jL} }^{2} , \ \vert \l ^{'\star }_{11k} \l '_{21k} \vert < 3.2
\times 10^{-8} {\tilde d_{kR} } ^{2}.$

A closely related observable is furnished by the radiative conversion
reaction, $ \mu ^- +A \to e^- +\g +A $, corresponding to the coherent
photon emission nuclear reaction with the target nucleus remaining in
its ground state.  The electromagnetic current vertex, $ \g \mu e $,
is parameterized in the familiar way in terms of electric and magnetic
anapole and dipole form factors, $ E_0, \ M_0,$ and $ E_1, \ M_1 $.
Upon comparing the relative merits of the radiative and non-radiative
conversion amplitudes, a delicate balance between advantages and
disadvantages must be taken into account.  Three main arguments are in
favor of the radiative conversion reaction~\cite{huitu98}.  Firstly,
the penalty from the extra electromagnetic coupling constant factor is
compensated by the coherent photon coupling which brings an extra
power of the nuclear electric charge $Z$.  Secondly, the price in
allowing for an electromagnetic $\mu e \g $ coupling at the one-loop
level, is compensated by the presence of a large logarithm from the
loop amplitude, $ \ln ( m_f/m_{\tilde f } )$.  Thirdly, the wider
freedom in choosing the internal lines allows the possibility of
inferring bounds on configurations of the indices involving the heavy
quarks and leptons generations.  The RPV one-loop level contributions
yield bounds on the quadratic coupling constant
products~\cite{huitu98}, $ \l^{\star} _{ij2} \l '_{ij1},\ \l^{\star}
_{2jk} \l '_{1jk}, \ \l ^{'\star }_{2jk} \l '_{1j'k'} $ which are
generically weaker by one to two orders of magnitude than those
inferred from the tree level contributions~\cite{kim}. Two
exceptionally strong bounds referring to higher generation
configurations read as, $\vert \l^{\star} _{23k} \l '_{13k}\vert <
O(10^{-5}) ,\ \vert \l ^{'\star }_{23k} \l '_{13k} \vert < 8.7 \times
10^{-5}$. Taken together, the ordinary and photonic conversion
mechanisms provide some of the strongest bounds on the $ \ \l ^{'\star
} \l '$ products, thus making the conversion reaction a very promising
case study for future experimental improvements.

A systematic study of the normal and radiative $\mu ^-\to e^-$
conversion processes using a refined description of the hadronic and
nuclear structure, notably by accounting for the strange quark sea
component, is presented by Faessler et al.,~\cite{faesslermu}.  The
predictions include the lepton-chargino mixing contributions from the
bilinear interactions as well as the trilinear interactions loop
contributions.  The bounds deduced from the current limit of
non-observation of the $\mu ^-\to e^-$ conversion in the $
^{48}\text{Ti} $ nucleus are generically stronger than the existing
ones.  We quote below a representative sample of coupling constant
bounds inferred from a comparison of the experimental limits with
predictions obtained for the radiative and non-radiative reactions,
\bea \bullet \quad && \mu ^- \to e^- :\ \vert \l ^{'\star }_{2mk} \l
'_{1m'k}\vert < O(10^{-8}), \ \vert \l ^{'\star } _{23k} \l '_{1m'k}
\vert < [ O(10^{-5}) \ - \ O(10^{-6}) ] , \cr && \vert \l ^{'\star
}_{i 11} \l _{imn} \vert < O(10^{-9}), \ \vert \l ^{'\star }_{i22} \l
_{imn} \vert < O(10^{-9}),\ \vert v_1 v_2 \vert < (80 \ \text{ MeV}
)^2 \tilde m^2 \ \hat B ,\cr && \vert \mu_1 \mu_2\vert < (80 \ \text{
MeV} )^2 \tilde m^2\ \hat B , \ [\vert v_1 \mu_2 \vert , \ \vert v_2
\mu_1 \vert ] < (80 \ \text{MeV} )^2 \tilde m^2 \ \hat B , \cr \bullet
\quad && \mu ^- \to e^- +\g :\ \vert \l^{\star} _{1j1} \l _{1j2} \vert
< O(10^{-6}), \ \vert \l^{\star} _{2jk} \l _{13k} \vert < O(10^{-6}),
\eea where $\hat B = { ( (R^{\mu e}) _{exp} / 7. \times 10^{-13} )^\ud
} ] $ denotes the square root of the lepton number conversion reaction
branching fraction scaled by the value of the corresponding current
experimental limit.

The constraints on a broken R parity symmetry from the $(\mu ^-, e^-)
$ nuclear conversion reaction are examined in several other recent
works~\cite{gouvea01}.  A new mechanism combining the bilinear and
trilinear interactions in one-loop diagrams with neutralino and
sfermion internal lines is discussed by Cheung and
Kong~\cite{kong01}. The deduced bounds involve products of bilinear
and trilinear coupling constants of generic form, $ [\vert {\mu _i
^\star \over \mu } \l _{i21} \vert , \ \vert {\mu _i ^\star \over \mu
} \l _{i12} \vert ] < 1.5 \times 10^{-7}. $ Kosmas et
al.,~\cite{kosmas01} examine the one-loop mechanism for the
subprocess, $ \mu ^- + q \to e ^- + q$, described by the $t$-channel
exchange of a sneutrino emitted at the vertex, $\mu ^-\to e^- +\tilde
\nu $, decaying through the $b$-quark triangle diagram into a pair of
gluons, $\tilde \nu \to g + g $, which are subsequently absorbed by
the nucleon valence quark.  The comparison with the experimental limit
for the nucleus, $ ^{197} \text{Au}$, yields the quadratic coupling
constant bounds, $[\vert \l ^\star _{121} \l '_{123} \vert ,\ \vert \l
^\star _{212} \l '_{233} \vert , \vert \l ^\star _{312} \l '_{333}
\vert ,\ \vert \l ^\star _{321} \l '_{333} \vert ] < 8.5 \times
10^{-7} $.

\subsubsection{\bf Semileptonic two-body decays of $\tau $- lepton}

A valuable source of information on lepton flavor violation effects is
offered by the $\tau $ lepton semileptonic decay modes.  Of special
interest are the two-body decay processes into pseudoscalar and vector
mesons, which are observed experimentally for various flavor
configurations of the emitted leptons and mesons, $ \tau ^- \to e_i^-+
P^0 ,\ \tau ^- \to e_i^-+ V^0 , \ [e_i =e, \mu ;\ P^0 =\pi ^0, \eta ,
K^0 ; \ V=\rho^0, \omega , K^\star ]. $ The RPV interactions
contribute to these processes via sneutrinos or squarks tree level
exchange.
We quote a representative sample of the strongest bounds deduced by
Kim et al.,~\cite{kim} from a comparison with the experimental limits
on the decay rates,
\begin{eqnarray} \bullet \quad    
[\vert \l _{i31} \l ^{ '\star } _{i11}\vert ,\ \vert \l _{ i13} \l ^{
'\star } _{i11}\vert ]& < & 6.4 \times 10^{-2} \ \tilde \nu_{iL}^2, \
[\tau^- \to e^- +\pi ^0 ]; \cr [\vert \l _{i31} \l ^{' \star }
_{i12}\vert , \ \vert \l _{ i13} \l ^{ ' \star } _{i12}\vert ] &<& 8.5
\times 10^{-2} \ \tilde \nu ^2_{iL} , \ [\tau^- \to e^- +K ^0 ]; \cr
\bullet \ [\vert \l _{i32} \l ^{' \star } _{i11}\vert , \ \vert \l _{
i23} \l ^{' \star } _{i11\vert }] & < & 3.6 \times 10^{-2} \tilde
\nu_{iL} ^2, \ [\tau^- \to \mu ^- +\pi ^0 ]; \cr [\vert \l _{i32} \l
^{' \star } _{i12}\vert ,\ \vert \l _{ i23} \l ^{' \star } _{i12}\vert
] &<& 7.6 \times 10^{-2} \ \tilde \nu_{iL} ^2, \ [\tau^- \to \mu ^- +K
^0 ]; \cr \bullet \ [\vert \l _{i31} \l ^{' \star }_{i11}\vert , \vert
\l _{ i13} \l ^{' \star }_{i11}\vert ] & < & 4.5 \times 10^{-3} \
\tilde \nu_{iL} ^2, \ [\vert \l _{i31} \l ^{' \star } _{i22},\ \vert
\l _{i13} \l ^{ ' \star } _{i22}\vert ] < 4.5 \times 10^{-2} \ \tilde
\nu_{iL} ^2 , \ [\tau^- \to e^- +\eta ^0 ]; \cr [\vert \l _{i32} \l
^{' \star } _{i11}\vert , \ \vert \l _{i23} \l ^{ ' \star }
_{i11}\vert ] &<& 4.8 \times 10^{-3} \ \tilde \nu_{iL} ^2 , \ [\vert
\l _{i32} \l ^{' \star } _{i22}\vert , \ \vert \l _{i32} \l ^{ ' \star
} _{i22}\vert ] < 4.8 \times 10^{-2} \ \tilde \nu _{iL} ^2 , \ [\tau^-
\to \mu ^- +\eta ^0 ]; \cr \vert V^\dagger _{jp} V_{pj'} \l ' _{3j1}
\l ^{' \star } _{1j'1}\vert & < & 3.5 \times 10^{-3} \ \tilde u_{p L}
^2 ,\ \vert V^\dagger _{j1} V_{1j'} \l ' _{3jp} \l ^{' \star } _{1j'p}
\vert < 3.5 \times 10^{-3} \ \tilde d_{p R} ^2 , \ [\tau^- \to e^-
+\rho ^0 ].
\label{eqaconv1}
\end{eqnarray}
 Several other coupling constant bounds arise from the decay processes
involving $e_i = e, \mu $ leptons and $ K^{\star 0}, \o $ vector
mesons~\cite{kim}.  The closely related $\tau $ semileptonic decay
modes with neutrino emission~\cite{bhattachoud} leads to the $1\s $
bound, $\vert \l '_{31k}\vert < 0.16 \ \tilde d_{kR}, \ [ \tau^-\to
\pi^- +\nu _\tau ]$.

\subsection{Lepton number non-conserving processes }
\label{secxxx3c}

\subsubsection{\bf Double beta decay reactions}

The neutrinoless double nuclear beta decay reaction, $\b \b _{0\nu }
$, takes place by the nucleon level process, $ n +n \to p+p + e^- +
e^-$, initiated through the lepton number violating subprocess, $ d+d
\to u +u +e+e$.  These rare reactions bear close similarities with the
lepton number conserving double beta decay $ \b \b _{2\nu }$
reactions, $ n +n \to p+p + e^- + e^- +\bar \nu _e +\bar \nu _e $. The
relevant experimental situation corresponds to a parent nucleus whose
beta decay channel, $ A_Z \to A_{Z+1} +\b ^- $, is energetically
closed, but which is allowed to decay to the daughter nucleus, $A
_{Z+2} $, via the two-step beta decay process involving virtual
transitions to the neighboring nucleus, $A_Z \to A _{Z+1} +\b ^- \to A
_{Z+2} +\b ^- +\b ^- $.  The double beta decay processes probe the
nuclear structure through the nuclear ground state matrix elements of
two-current correlation functions.  Only the regular $ \b \b _{2\nu }$
reactions have been experimentally observed so far, while active
searches are currently pursued for the neutrinoless $ \b \b _{0\nu }$
reactions, which offer sensitive probes of new physics.  On side of
the supersymmetric RPV contributions, the other promising mechanisms
involve the exchange of light or heavy Majorana neutrinos.  A
comprehensive discussion of the double beta decay reactions and of the
related reactions, associated with the electron-positron and
muon-positron conversions and double electron nuclear capture is
presented in the review by Vergados~\cite{verga86} for the case
without emission of neutrinos, $e ^- + A_Z \to e^+ + A_{Z-2} , \ \mu
^-+ A_Z \to e^++ A_{Z-2} ,\ e^-+ e^- + A_Z \to A_{Z-2} $, and also the
case with neutrino pair emission.  An updated discussion from a modern
perspective of the neutrinoless double beta decay reactions is
presented by Vergados~\cite{verga01}.  The transition amplitudes for
the $ \b \b _{0\nu }$ and two-lepton processes naturally separate into
a long range part, associated to neutrino and photon exchange,
respectively, and a short range part associated to massive particles,
$Z$-boson and $W^- - W^+$ pair exchange, respectively.  The $ \b \b
_{0\nu }$ and two-lepton processes give access to effective neutrino
masses defined by the weighted sums over neutrino masses and lepton
flavor mixing matrix elements, $\b \b _{0\nu }:\ <m_ {\nu _e} > \equiv
<m_\nu > _{e e} = \sum _j V^{' \dagger } _{e j} V^{' \dagger } _{ej}
e^{i\l _j } ,\ (\mu ^-, e^-) : \ <m_\nu > _{\mu e }= \sum _j V^{'
\dagger } _{\mu j} V^{' \dagger } _{ej} e^{i\l _j } ,\ (\mu ^-, e^+) :
\ <m_\nu > _{\mu e ^+ } = \sum _j V^{'T} _{\mu j} V^{'T}_{ej} e^{-i\l
_j } ,$ where $\l _j$ denote the CP intrinsic phases of the neutrino
mass eigenstates.

Having identified the quark level amplitudes, one must first convert
the description from the quark level to the nucleon level, by
expressing the two-nucleon transition amplitudes for the process, $ n
+n \to p+p + e^- + e^-$, in terms of non-relativistic nuclear two-body
operators of zero or finite ranges.  To evaluate the transition
amplitude, one must then perform~\cite{verga86,haxton86} the
summations over the nuclear intermediate states or use a closure
approximation over the nuclear states.  In the former case, the
nuclear matrix elements can be evaluated by truncating the sums over a
basis of shell model states or of excited states of the random phase
approximation (RPA) ground state.  In the latter case, the nuclear
transition amplitude is described by the diagonal ground state matrix
element of two-body operators of generic form~\cite{ejiri00}, $< A_{Z}
(0 _f ^+ ) \vert \sum_{i\ne j} h (E, \vec r_{ij} ) \tau ^- _i \tau ^+
_j \vert A_{Z} (0_i ^+ )>,\ < A_{Z} (0 _f ^+ ) \vert \sum_{i\ne j} h
'(E ,\vec r_{ij} )\tau ^- _i\tau ^+ _j\vec \s _i \cdot \vec \s_j \vert
A_{Z} (0_i ^+ )> $, with form factors, $ h, \ h'$, depending on the
emitted electrons energies, $E$, and the nucleons coordinates,
multiplied by the nucleons spin and isospin operators.  A useful
formalism providing an easy access to the nuclear matrix elements of
the short range local operators and the long range neutrino exchange
operators of $\b \b _{0\nu } $ encountered in the various particle
physics models is presented by P\"as et al.,~\cite{pasko01}.  A
general review of the subject is presented by Faessler and
Simkovic~\cite{faessler99}.  While the nuclear structure effects may
strongly suppress the calculated transition amplitudes for $ \b \b
_{2\nu }$, they have a less drastic effect on the $ \b \b _{0\nu }$
amplitudes.  The sensitivity to various existing mechanisms is
discussed by Simkovic and Faessler~\cite{sim02}.  The nuclear physics
formalism of the $(\mu ^-, e^-) $ conversion reactions in nuclei,
$R_{\mu e^-}$ is presented in Ref.~\cite{kosm97} and that for the
exotic double charge exchange $(\mu ^-, e^+) $ conversion reactions in
nuclei in Ref.~\cite{divari02}.

The experimental measurements of the double beta decay nuclear
reactions, $(Z,N)\to (Z+2,N-2)+e^-+e^-$, are performed for even-even
heavy nuclei, with a representative sample of the nuclear transitions
given by, $^{48} \text{Ca} \to \  ^{48} \text{Ti} ,\ ^{76} \text{Ge} \to
 \   ^{76} \text{Se} ,\ ^{82} \text{Se} \to  \   ^{82} \text{Kr} , \ ^{100}
\text{Mo} \to  \   ^{100} \text{Ru} ,\ ^{128} \text{Te} \to  \   
^{128}\text{Xe} $. The experimental setups use geochemical (Se, Zr, Te
nuclei) or radiochemical (U, Pu) techniques and employ dedicated
detectors with Ge semiconductor material, cryogenic, scintillation, or
beta ray tracking.  The most stringent experimental limits are those
obtained by the Moscow-Heidelberg
collaboration~\cite{baudis97,heidel01}, with the corresponding
experimental limits for the $^{76} \text{Ge} $ nucleus $ \b \b _{0\nu
} $ half-life given by, $T _ \ud > [1.1\ \times 10 ^ {25},\ 1.5\
\times 10 ^ {25}] $ y, respectively.  The future projects aim at a
half-life sensitivity of order, $ 10 ^ {26} $ y.  A summary of the
available experimental information along with a review of the future
experimental projects can be found in Ref.~\cite{expbb}.  A recent
review on double beta decay reactions is presented by Elliott and
Engel~\cite{elliott04}.  The typical experimental limits on the
two-lepton reaction rates currently attained in experiments at the
TRIUMF, PSI and BNL accelerators are: $ R ( \text{Au} (\mu ^-, e^-)) <
5. \ \times 10 ^{-13 } , \ R ( ^{27} \text{Al} (\mu ^-, e^+) ^{27}
\text{Na} ) < (4.6 \ - \ 4.2 ) \ \times 10 ^{-12 } $.

\subsubsection{\bf Neutrinoless double beta decay from broken R parity
violating interactions}

The RPV interactions have the ability to contribute, jointly with the
gauge interactions, direct tree level transition amplitudes to the $\b
\b _{0\nu }$ process, independently of the indirect contribution to
the familiar long range neutrino exchange amplitude produced through
the RPV neutrinos Majorana mass.  The $\b \b _{0\nu }$ amplitudes
naturally separate into long range light neutrino exchange and short
range massive gaugino exchange terms, the former being controlled by
the effective neutrino mass parameter, $<m_{\nu_e }> $.  A variety of
short range tree level mechanisms are available involving the short
range $t$-channel exchanges of massive gauginos.  The early
realization of this possibility by Mohapatra~\cite{moha} was followed
by a detailed study by Vergados~\cite{verga87} of the particle and
nuclear physics aspects.  The RPV amplitudes are represented by
specific dimension $9$ operators of scalar and tensor Lorentz
covariant structure arising from a variety of particle exchange
mechanisms.  The RPV short range mechanisms are found to compete
favorably with the indirect neutrino exchange mechanisms induced by
the RPV contribution to the Majorana neutrino mass, once the
corresponding contribution to the effective neutrino mass drops below
the exceedingly small value~\cite{verga87}, $< m_\nu > \simeq 0.05 $
eV.

The direct RPV contributions, displayed by the Feynman diagrams I.1-3
in Figure~\ref{fig3}, involve the sequential $t$-channel exchange of
a pair of sfermions and a gaugino, where the sfermion may be a slepton
$\tilde e_L $ or a squark $ \tilde u_{L},\ \tilde d_{R} $ and the
gaugino may be a neutralino or a gluino~\cite{moha,hirsch}.  The
associated transition amplitudes are described in the limit of large
gauginos masses by local six fermion couplings having the same
space-time structure but differing in the quarks and leptons
generational and gauge color structure.  The low energy effective
Lagrangian for diagram I.1 in Figure~\ref{fig3} has the general form,
$ L_{EFF}= {\l'_{ijk} \l ^{'\star }_{lmn} \over \tilde m ^4 m_\tchi }
(\bar u_{\a jL} d_{\a kR}) (\bar u_{\b mL} d_{\b nR}) (\bar e_{iL}
e^c_{lR}) + H. \ c.  $ For the dineutron system, the contribution from
photino, zino and gluino exchange is described by the approximate
formula derived initially by Mohapatra~\cite{moha},
\begin{eqnarray}  && L_{EFF} ^{\D L =2} = 2\vert \l^{ '} _{111}\vert ^2 (\bar
u_{\a L } d_{\a R }) (\bar u_{\b L } d_{\b R }) (\bar e_{L} e^c_{R})
 \bigg [ { g_2^2 \over \cos ^2 \t_W m _{\tilde Z} } \sum_{\tilde f =
 \tilde u_L, \tilde e_L, \tilde d_R } { 2 T_3^{L,R} - 2 Q(\tilde f)
 \sin ^2 \t _W \over m ^4_{\tilde f} } \cr && + { g_2^2 \sin ^2 \t_W
 \over 2 m _{\tilde \g} } \sum_{\tilde f = \tilde u_L, \tilde e_L,
 \tilde d^\star _R } {Q(\tilde f ) \over 2 m ^4_{\tilde f} } + {4
 g_3^2 \over 3 m _{\tilde g}} \sum_{\tilde f = \tilde u_L, \tilde d_R
 } {1 \over m ^4_{\tilde f} } \bigg ] + H. \ c. ,
\label{eq11} 
\end{eqnarray}
where $Q (\tilde f)$ denotes the electric charge.  To account for the
gauginos masses, one needs to substitute for the corresponding mass
factors, $ m ^{-1} _{\tilde V} \to m _{\tilde V} <{ q^2 \over q^2 + m
^2_{\tilde V}} > , \ [\tilde V = \tilde Z, \ \tilde \g , \ \tilde g
]$, where $q$ denotes the gauginos momentum and the averaging refers
to the nuclear Fermi motion.  The gluino exchange contribution is
obtained by performing a projection over the color singlet quark
bilinear operators, using a permutation reordering of the quark
spinors and a projection on the color singlet component, $ (\bar u_{\a
L} d_{\b R } ) (\bar u_{\b L} d_{\a R } ) \simeq {4\over 3} (\bar
u_{\a L} d_{\a R } ) (\bar u_{\b L} d_{\b R } ) $.  The Fermi type
nuclear matrix element for the $\b \b _{0\nu }$ process, $ M _{F-F} =
<A_{Z+2} (0^+) \vert (\bar u_{\a L} d_{\a R } ) (\bar u_{\b L} d_{\b R
} ) (\bar e_L e^c_R) \vert A_Z (0^+) >$, can be roughly evaluated by
using an approximate prescription~\cite{moha} which relates it to the
Gamow-Teller nuclear matrix element, $M_{GT-GT}$, for the
corresponding $\b \b _{2\nu }$ process. The proposed estimate reads
as, $M _{F-F} \sim 10^{-2} M_{GT-GT}$.  The comparison with the
experimental limit on the $\b \b _{0\nu }$ half-life leads to the
coupling constant bounds, $\vert \l '_{111}\vert <0.48\times 10^{-9/4}
\ \tilde f^2 \tilde g ^\ud, \quad \vert \l '_{111}\vert < 2.8 \times
10^{-9/4} \ \tilde f^2 \tchi ^\ud , $ in correspondence with the
gluino and neutralino exchange contributions, respectively.

The interplay between the long and short range amplitudes has been
studied in the bilinear broken R parity option by Hirsch and
Valle~\cite{hirval99}.  The indirect RPV contributions to the neutrino
effective mass, $<m_{\nu_e }> $, are found to be the dominant ones,
with the constraints from the limits on the $\b \b _{0\nu }$ rates
yielding the individual coupling constant bounds, $ [\vert <\tilde \nu _1>
\vert ,\ \vert \mu _1 \vert  ]= O(100) $ keV for small $\tan \b = O(1)$.  For variable
$\tan \b = 1 \ - \ 50 $, the limits from the combined fits to the data
are less restrictive and the following variation intervals, $ [\vert  <\tilde
\nu _1>\vert   ,\ \vert \mu _1\vert   ] = O(0.1) - O(1) $ MeV.

A significant progress in the treatment of the $ \b \b _{0\nu }$
reaction rates, especially with respect to the particle and nuclear
physics aspects, has been accomplished in recent years.  The various
tree level Feynman diagram contributions from $\tilde \chi , \tilde g,
\tilde q $, exchange can be grouped together, after a Fierz-Michel
reordering of the Dirac spinors, into scalar and tensorial Lorentz
covariant operators entering the effective Lagrangian
as~\cite{faessler99,faessler98,faessler97,wodecki},
\begin{eqnarray} && 
L_{EFF} ^{(\D L =2)}= {4 G_F ^2 \over m_p} (\bar e_L e^c_R ) [ \eta
_{PS} (\bar u_{\a L }d_{\a R} )^2 - {1\over 4} \eta _{T} (\bar u_{\a L
} \s _{\mu \nu } d_ {\a R} )^2] + H. \ c., \label{eqx14}
\end{eqnarray} 
where the auxiliary parameters gather additive contributions from the
gaugino and sfermion exchange amplitudes depending on the RPV and
gauge coupling constants and the particle masses, $\eta _{PS} =
\eta_{\tchi \tilde e} + \eta_{\tchi \tilde f}+\eta_{\tchi } +
\eta_{\tilde g } + 7 \eta '_{\tilde g } , \ \eta_{T} = \eta_{\tchi } -
\eta_{\tchi \tilde f} + \eta_{\tilde g } - \eta '_{\tilde g } $. We
quote, for illustration, the explicit expression for the gluino
exchange contribution, $\eta_{\tilde g } = {\pi \a _s \over 6 G_F^2 }
{\vert \l ^{'} _{111}\vert ^2 m_p \over m_{\tilde d_R } ^4 m_{\tilde g
} } [1 + ({ m_{\tilde d_R } \over m_{\tilde u_L } } ) ^4 ] $.  Note
that some disagreement regarding certain color factors seems to
persist~\cite{wodecki} between different published works. The
constraints from the $\b \b _{0\nu }$ process are very sensitive to
the inputs used for the soft superpartners spectrum.  The improved
limit on $\l '_{111}$ was first obtained by Faessler et
al.,~\cite{faessler98} using the pi-meson exchange mechanism of
neutrinoless double beta decay.  The comparison with experimental data
for the $^{76} Ge$ nucleus, yields the bound~\cite{hirsch,hirsch1},
$\vert \l'_{111} \vert < 3.3\times 10^{-4} {\tilde q} ^2 {\tilde
g}^{\ud } $, within the minimal supergravity framework, and the
alternative bound~\cite{wodecki,wodecki2}, $\vert \l'_{111} \vert <
3.2 \times 10^{-5} {\tilde q} ^2 {\tilde g}^{\ud } , $ within the
minimal model for the gauge mediated supersymmetry breaking
framework~\cite{dimothomas}.
The sensitivity to the superpartners mass spectrum of the bound on $
\l ' _{111}$ is examined by Uehara~\cite{ueharabb02} jointly with the
constraints deduced from the $ K-\bar K $ mass difference and the
reaction $ K ^+ \to \pi ^+ + \nu + \bar \nu $.  New limits have been
recently obtained~\cite{gozdz04} from a study of $\b \b _{0\nu }$
reaction rates in several medium and heavy nuclei using two nucleon
and pion exchange models.  The resulting bounds are: $\vert \l
'_{111}\vert < 2.75 \times 10^{-5} \ \tilde q ^2 \tilde g ^\ud ,\
\vert \l '_{111}\vert < 2.73 \times 10^{-3} \ \tilde e ^2 \tilde \chi
^\ud $, in the neutralino and gluino mediated supersymmetry breaking
cases, respectively.

An interesting competitive mechanism for the $\b \b _{0\nu }$
transition was proposed by Babu and Mohapatra~\cite{babu} using the
$t$-channel scalar-vector type exchange of a sfermion and a charged
$W$-boson linked together by an intermediate neutrino exchange. The
transition amplitude combines the amplitude for $ d+\bar \nu \to e ^-
+u$, induced by $\tilde d_L-\tilde d_R$ exchange, with the $ W ^ \pm $
charged current exchange amplitude for the process, $ \nu +d \to e ^-
+u$. The complete reaction chain, $d+d\to ( \tilde d \nu ) +(W^- u)\to
(u+e ^-) + (u+e ^-)$ is depicted by the Feynman diagram I.4 in
Figure~\ref{fig3}.  The corresponding gaugino-sfermion-neutrino
exchange diagram~\cite{babu} is similar to the SM neutrino exchange
diagram, except that no chirality flip mass insertion is needed for
the internal neutrino line in the present case.  The chirality flip in
the RPV amplitude is transferred to the exchanged down-squark scalar
particle which must be inserted with a $\tilde b_L -\tilde b_R$
chirality flip mass mixing term.  The effective neutrino mass
suppression factor, $<m_{\nu _e}>$, in the SM amplitude is replaced by
a neutrino propagator factor, ${ <m_{\nu _e}> \over q^2 } \to { 1\over
\g \cdot q }$, where the momentum transfer variable, $q$, may be
approximately treated by taking an average over the nucleons Fermi
motion inside the nucleus or by identifying $q$ with the nuclear Fermi
momentum, $ \vert \vec q \vert \simeq p_F \approx 100 \ \text{ MeV}$.
The squark exchange amplitude for the subprocess $ d \to \nu +e _i ^-
+u$ is described by the effective Lagrangian,
\begin{eqnarray}  &&
L_{EFF}= -{4 G_F \over \sqrt 2 } [ \e_1^{ee_i} (\bar d_R u_L) (\bar
\nu ^c_R e_{iL} ) + \e_2^{ee_i} (\bar d_R \nu_L) (\bar u ^c_R e_{iL} )
] + H. \ c. , \quad [e_i =e, \mu ] \cr && \quad [ \e_2^{ee} = { \l
_{131}^{'\star } \l '_{113} (\tilde m^{d2} _{LR} )_{33} \over 2\sqrt 2
G_F m^2_{\tilde b_L} m^2_{\tilde b_R} } , \ \e_2^{e\mu } = { \l
_{213}^{'\star } \l '_{131} (\tilde m^{ d2} _{LR} )_{33} \over 2\sqrt
2 G_F m^2_{\tilde b_L} m^2_{\tilde b_R} } , \ (\tilde m^{d2} _{LR}
)_{33}= m_b (m_{\tilde G } A_b +\mu \tan \beta ) ].
\label{eqx15}
\end{eqnarray}  where we have recorded in the second line above 
the dominant contribution to the vertex parameters $\e_2^{ee},\
\e_2^{e\mu }$ from the third generation $b$-squarks and displayed the
contribution to the chirality flip down-squark mass term produced at
the electroweak symmetry breaking.  Combining the above amplitude with
that of the charged current $W$-boson exchange subprocess, $ d \to
\bar \nu +e ^- +u$, leads to the effective Lagrangian for the $\b \b
_{0\nu } $ process, $ L_{EFF} ^{\D L=2} = - G_F^2 \e_2^{ee} (\bar u_L
\s _{\l \rho } d_R) (\bar u_L \g_\mu d_L) (\bar e_ L \g^ \mu {1\over
\g \cdot q } \s ^ {\l \rho } e^c_R ) + H. \ c. $ Updated predictions
for the Babu and Mohapatra mechanism~\cite{babu} are presented by
Hirsch et al.,~\cite{hirsch} with the resulting bounds for the third,
second and first down-quark generations given by, $\vert \l ^{'\star}
_{113}\l'_{131}\vert < 7.9 \times 10^{-8}, \ \vert \l ^{'\star}
_{112}\l'_{121}\vert < 2.3 \times 10^{-6}, \ \vert \l '_{111}\vert ^2
< 4.6 \times 10^{-5} $, respectively.

The contributions to the $\b \b _{0\nu }$ reaction rates initiated by
both bilinear and trilinear RPV interactions, including the soft
supersymmetry breaking interactions, are discussed in the review by
Faessler and Simkovic~\cite{faessler99}. The study embodies an
improved treatment of the hadronic structure, using a quark-nucleon
duality mapping for the scalar, pseudoscalar, vector and axial vector
$(S, P, V, A )$ current operators. The nuclear structure description
is also improved by evaluating the two-body nuclear matrix elements in
the nuclear RPA (random phase approximation) scheme.  The $t$-channel
sequential exchange of sfermion, neutralino and sfermion may be viewed
as a formal two-current effect, with the scalar type current induced
by the RPV interactions described schematically as, $\d L _{RPV} /\d
\tilde f _k (x) \simeq \l '_{ijk} \bar f_i (x) f '_j (x) $.  Since the
$t$-channel exchange of massive superparticles is well described in
terms of two-nucleon contact operators, one may evaluate reliably the
neutrinoless amplitude in a nuclear closure approximation.  The
neutrino mass term induced by the RPV bilinear interactions contribute
to the $\b \b _{0 \nu } $ transition in much the same way as in the
conventional $t$-channel neutrino exchange mechanism, with the massive
Majorana neutrino emitted and absorbed by two charged gauge current
couplings.  This is found to represent the predominant contribution.
The condition for a non-observation of the $\b \b _{0 \nu } $
transition in $ ^{76} \text{Ge} $, consistently with the experimental
limit on half-life, $ T_\ud (0^+ \to 0^+ ) < 1.1 \times 10^{25} \ y $,
leads to the following bounds on the first generation coupling
constant and VEVs parameters~\cite{faessler99,faessler98}: $ \vert
\mu_1 \vert < 470\ \text{ keV}, \ \vert <\tilde \nu_1 > \vert < 840\
\text{ keV}, \ \vert \mu_1 \l'_{111}\vert < 100 \ \text{ eV} , \ \vert
<\tilde \nu_1 > \l'_{111}\vert < 55 \ \text{ eV} ,\ \vert
\l'_{111}\vert < 1.3\times 10^{-4} \ \tilde q^2 \tilde g ^\ud .$ The
above quoted bounds on the bilinear coupling constants inferred from
the $\b \b _{0\nu }$ reaction rates are seen to be significantly
stronger than those deduced from the comparison with the neutrino
masses, $\mu _1 < 15. \ \text{GeV} , \ <\tilde \nu_1 > < 7. \
\text{GeV} $.

Motivated by the observation that the charged current initiated
neutrino exchange mechanism contributes at a significant level to the
$\b \b _{0\nu }$ amplitude, Hirsch~\cite{hirsch99} has focused on the
contributions from misalignment effects to the effective neutrino
mass, $<m_{\nu _e}>$.  The resulting bilinear RPV contributions are
proportional to the misalignment parameter, $ <m_{\nu _e}> \simeq 2g_2
^2 M_2 (v_e \mu -v_d \mu _e)^2 / Det' (M^n) $.  The comparison with
experimental limits on $\b \b _{0\nu }$ reaction rates leads then to
exclude the interval of values for the coupling constant and VEV
parameters~\cite{hirsch99}, $[\mu_e,\ v_e ] \in [ O(10^{-1}) \ - \ O(
1) ] \ \text{ MeV} $.  Even assuming a perfect alignment, $ \mu _e
\propto v_e$, at the classical level, yielding $<m_{\nu _e}> = 0$,
misalignment effects can arise at the one-loop level leading to finite
contributions to $<m_{\nu _e}>$.  The comparison with data yields the
bilinear coupling constant bound, $ \mu _e/\mu < 0.01$.

The atomic orbit capture reaction of positive muons, $\mu^- +A_{Z,N}
\to e^+ +A_{Z-2, N+2} $, with the $ \mu^- \to e^+ $ conversion taking
place in nuclei, is closely related~\cite{palmoha} to the $\b \b
_{0\nu }$ reaction.  The lepton number violation occurs through the
process, $\mu ^- + u \to d +\bar \nu $, mediated by $\tilde d_L-
\tilde d_R$ exchange. The associated amplitude is represented by the
same four fermion contact Lagrangian as that defined in
eq.(\ref{eqx15}) where the relevant vertex functions are given by the
off-diagonal parameters, $\e^{e\mu }_i,\ [ i=1,2]$.  The complete
effective Lagrangian is then obtained from that for $ \b \b _{0\nu } $
by substituting $ \e_2^{ee} \to \e_2^{e\mu }$.  The predicted
branching fraction, represented by the approximate formula for the
coupling constant product, $ \vert \l _{213}^{'\star } \l '_{131}
\vert \simeq 2.3 \times 10^{-2} \ [B(\mu ^- \to e^+)/ 10^{-12} ] $,
shows the extent to which an improvement of the current experimental
sensitivity could usefully constrain the RPV interactions.

\subsection{Baryon number non-conserving processes}
\label{secxxx3d}

The phenomenon of matter instability is a well documented subject
thanks to the extensive discussions developed in the context of grand
unification
theories~\cite{weinbergs,weinberg80,langacker,weinberg82,masiero1}.
With the $B, \ L $ non conservation in the broken R parity symmetry
case taking place at the supersymmetry breaking mass scale, one
disposes of particularily severe constraints on a large number of
couplings obtained by invoking perturbative Feynman diagram mechanisms
of increasing complexity with respect to the loop order and the number
of participating particles.  Recall that the
classification~\cite{weinbergs,wilczee,weinberg80} of gauge invariant
higher dimensional operators in the quarks, leptons, Higgs and gauge
bosons fields reveals that the dangerous operators that vehicle $ B $
and $ L$ number violation appear first in the SM at dimension $\cddd
=6$ with the selection rule, $\D B= \D L= -1 , \ [ \D B= B_f -B_i, \
\D L= L_f -L_i] $.  For orientation, we note that the selection rules
for the higher dimension dangerous operators are: $ \D B=- \D L= -1 $,
at dimension 7; $ \D B= \D L= -1 $ at dimension 8; $ \D B= {1\over 3}
\D L= -1 $ and $ \D B= -2, \ \D L=0 $ at dimension 9; $ \D B =-{1
\over 3} \D L =-1 $ at dimension 10; $\D B=+ {1\over 3} \D L= -1 $ at
dimension 11; and $ \D B=- {1\over 3} \D L= 1 $ at dimension 12.  So
far, most studies have restricted consideration to the dimension $
\cddd \leq 7$ operators.

\subsubsection{\bf Single nucleon decays}

The RPV contributions to single nucleon decay arise at the tree level
upon combining the $\l ' $ and $ \l ''$ interactions in the way
illustrated by the schematic formula, $ (L_lQ_mD_n^c)
(U_i^cD_j^cD_k^c) ^\dagger = (\nu _l d _m d_n^c - e_lu_md_n^c) ( u_i^c
d_j^c d_k^c)^\dagger . $ The contraction of down-quark superfields,
$d^{c\dagger} - d^c $, associated with the $\tilde d_{kR}$ squarks
$s$-channel exchange graph given by the diagram J.1 in
Figure~\ref{fig4}, contributes a $B-L$ conserving $B+L$ violating
amplitude.  The Dirac spinor representation of the effective
Lagrangian is given by,
\begin{eqnarray}  &&  L_{EFF} = 
{\l ^ { '' \star } _{ijk} \l '_{lmk} \over m^2_{\tilde d_{kR} } } \e ^
 {\a \b \g } [ (\bar u^c_{i \a L } \g^\mu d_{m \b L}) (\bar \nu
 ^c_{lR} \g _\mu d_{j \g R } ) + (\bar u^c _{i\a L}\g^\mu u_{m\b L
 })(\bar e^c_{lR}\g _\mu d_{j \g R} ) ] + H. \ c.
\label{eqx16} \end{eqnarray}

Following the review by Langacker~\cite{langacker}, one can express
the proton partial lifetime by means of the approximate formula, $\tau
(p\to f) = 1/ \G (p\to f) = \L _{EFF} ^2 a _{pf} , \ [ a _{pf} ^{-1} =
(5.52 \ - \ 0.35) \times 10^{29} \ \text{GeV} ^4 / \text{yr} ,\ \L
_{EFF} ^2 \simeq (\l '' \l ^{'\star } )/\tilde m^2 ]$, where the
auxiliary quantity, $a _{pf} $, includes the contributions from the
hadronic operator matrix element, radiative corrections and phase
space. The experimental bound, $\tau (p\to f) > 10^{32} $ yr, leads to
the quadratic coupling constant bounds, $\vert \l ^ { '' \star }
_{11k} \l ^{'\star } _{lmk}\vert < [10^{-25} \ - \ 10^{-27}] \ {\tilde
d_{kR} } ^2 , \ [ p \to \pi^0 +l^+, \ p \to \pi^+ +\bar \nu _l ; \ p
\to K^0 +l^+, \ p \to K^+ +\bar \nu _l ] $ where the cases $ m=1, 2$
are in correspondence with the decay channels involving the emission
of a pion and strange meson, respectively.  The alternative
contraction involving the opposite chirality down-quark superfields, $
d - d^{c \dagger }$, can contribute if accompanied by the insertion of
a mass term, $\tilde m_{LR} ^{d2}$, flipping the chirality of the
exchanged down squark.  The corresponding $\tilde d_L - \tilde d_R $
tree level exchange graph, displayed by the Feynman diagram J.2 in
Figure~\ref{fig4}, contributes a $ (B-L) $ violating, $(B+L)$
conserving amplitude, represented by the effective Lagrangian,
\begin{eqnarray} L_{EFF} =
\l^{''\star }_{ijk}\l ^{'} _{lmn} {(\tilde m^{d2}_{LR})_{jm}\over
m^2_{\tilde d_{jR}} m^2_{\tilde d_{mL}}} \e_{\a \b \g } (\bar u^c_{i
\a L} d_{k\g R} ) (\bar \nu_{lL} d_{n\b R})+ H. \ c.
\label{eq11p} \end{eqnarray}

Strictly speaking, one should assign the above quoted operator an
effective dimension, $\cddd =7$, in view of the proportionality of the
chirality flip squark mass term to the electroweak symmetry breaking
mass scale, $(\tilde m_{LR} ^{d2})_{ij} = M^d _{ij} (m_{\tilde G} A ^d
+\mu \tan \b ) $.  The $B-L$ violating operator initiates the single
nucleon two-body decay modes, $ p \to \pi ^ + + \nu _l , \ \ n \to \pi
^ 0+ \nu _l , \ n \to \pi ^ + + e^- _l$. The three-body proton decay
modes, $p \to \pi ^ + + \pi ^ 0+ \nu _l $, can occur through the
reaction scheme, $p\to (u+u) +d \to (u+u) +(u^c+d^c+ \nu _l) \to (u+
d^c)+(u+u^c) + \nu _e \to \pi ^+ + \pi ^0 +\nu _l.$ Assuming
tentatively a matrix element for the $B-L$ violating hadronic operator
of same size as that for the previous $B+L$ violating operator, yields
the following coupling constant bounds derived from the experimental
limit on the nucleon lifetime, $\vert \l '_{lj1} \l ^{''\star }_{1j1}
\vert <( 10^{-25} \ - \ 10^{-27}) \ \tilde d_{jL} ^2 \tilde d ^2_{jR}
{ (100 \ \text{GeV} )^2\over (\tilde m^{d 2} _{LR}) _{jj}} . $

The restriction to quadratic coupling constant bounds in
configurations involving the first two light quark generations can be
removed upon considering a one-loop level order dressing of the
initial tree level transition amplitudes~\cite{smirnov}.  The loop
diagrams are obtained from the tree level ones by adding a vertex
diagram dressing, $\tilde d u d$, or a box diagram dressing, $ u+ d\to
_{h^+}\to d + u \to_{\tilde d}\to \bar \nu +d$, where the internal
lines propagating in the loops are charged or neutral Higgs bosons or
winos.  The price that one must pay for the suppressed loop amplitudes
remains affordable thanks to the strong sensitivity of the
experimental limits on the single nucleon decay reactions.  The ratios
of loop to tree amplitudes, $\xi \equiv A_{\text{loop}} /
A_{\text{tree}} $, are determined by the contributions associated with
the CKM matrix, the Higgs matter coupling constants, $\l ^{u,d} =
m_{u,d} /v_{u,d} $, and the $( 4 \pi )^ 2$ loop factors.  The
generational structure in the combination of $B $ and $L $ violating
interactions, $\l '_{jki} \l ^{''\star }_{pmn} $, leads one to
distinguish two class of contributions depending on whether or not the
generation indices of $ d$ or $ d^c$ fields match between the $\l '
LQD^c $ and $\l '' U^c D^c D^c $ interaction operators.  The matching
case includes two sub-cases corresponding to the matching of indices
within the couplings $ d d^c $ and $ d^cd^c$, respectively, thus
contributing to the four fermion operators, $(\nu d) (ud) $ or $(\nu
^c d) (ud) $, respectively.  The no-matching case requires the
equality $ i=k$ and the condition that all of the indices in $\l
^{''\star }_{pmn}$ are different, which entails the presence of an
additional suppression factor in the numerator represented by the
external quarks momenta, $p_q$.  We display below results for three
illustrative examples~\cite{smirnov}, \bea \bullet &&
\xi \simeq {2 m_{u_p} m_b \over (4\pi )^2 v^2 } V_{13} V_{p1} \
\Longrightarrow \ \vert \l ^{'\star }_{i33} \l ''_{112} \vert <
{10^{-24} \over \xi } \approx 10^{-7} , \cr \bullet &&
\xi \simeq { 2 m_{d_n} m_{u_p} m_b \over (4\pi )^2 v^2 \tilde m }
V_{1n} V_{1p} \ \Longrightarrow \ [\vert \l ^{'\star }_{i21} \l
''_{123}\vert ,\ \vert \l ^{'\star }_{i32} \l ''_{112}\vert ,\ \vert
\l ^{'\star }_{i11} \l ''_{123}\vert ,\ \vert \l ^{'\star }_{2j1} \l
''_{113}\vert ] < 10^{-9} , \cr \bullet &&
\xi \simeq {2m_{u_p} m_b p_q \over (4\pi ) ^2 v^2 \tilde m } V_{13}
V_{pn} \tan \b
\ \Longrightarrow \ \vert \l ^{'\star }_{j33} \l ''_{112} \vert
<10^{-18} ,\ \vert \l ^{'\star }_{j22} \l ''_{113} \vert <10^{-9} ,\
\vert \l ^{'\star }_{3j1} \l ''_{112} \vert <10^{-9}, \eea where the
three successive entries correspond to the $(d^c d^c)$ matching, $(d
d^c) $ matching and no-matching cases, respectively, and the numerical
estimates are obtained by setting the superpartners masses at $ \tilde
m \approx O(1) \ \text{TeV}$. (Note that the coupling constant
conventions used by Smirnov and Vissani~\cite{smirnov} are related to
ours by the permutation of indices, $\l ''_{ikj} \to \l ''_{ijk} $.)
The consideration of loop dressing effects~\cite{smirnov} yields
useful bounds on the various possible combinations of coupling
constant pair products, $\vert \l '_{ijk} \l ^{'\star} _{pqr}\vert <
O(10^{-7}) \ -\ O(10^{-9}) $.  Stronger bounds of order of magnitude,
$\vert \l ^{'\star} \l ''\vert < O(10^{-11}) $, can even be deduced if
one takes the CKM flavor mixing of sfermions into account.

We discuss now other mechanisms for single nucleon decay involving the
combined action of different types of RPV interactions.  Bhattacharyya
and Pal~\cite{palbhatta} consider the tree level amplitude involving
the regular Yukawa interaction of quarks with the $H_u$ Higgs boson
combined with the trilinear and bilinear interactions, $\l '' _{ijk},\
\mu_i$, as illustrated by the Feynman diagram J.3 in
Figure~\ref{fig4}.  The associated $(B+L)$ conserving, $(B-L)$
violating effective Lagrangian, \be L _{EFF} = {\l ^{''\star }_{112}
\l^u _1 \mu_l \over m^2_{\tilde u_R} m_{\tilde H_u} } [(d^c s^c) ^
\dagger (u \nu ) +( d ^ c s ^ c) ^ \dagger (d l ) ] + \ H.\ c. \ee
with $\l^u _1$ designating the diagonal up-quark Yukawa coupling
constant, can initiate the proton decay modes, $p \to K^++\nu _l $ and
$ p\to K^+ + \pi ^++ l^-$.  Using the experimental limit on the proton
decay partial lifetime, $ \tau (p \to K^+ + \nu ) > 10^{32}\
\text{y}$, and assuming a higgsino mass, $ \mu \approx m _{\tilde H_u
} = 1 $ TeV, leads to the coupling constant product bound, $\vert \l
''_{112} \mu_l /\mu \vert < 10^{-23} \ \tilde u^2_{ R } \times
(m_{\tilde H} /\mu )$.  The one-loop Higgs boson or gaugino dressing
mechanism can be invoked to infer quadratic coupling constant bounds
for all possible generations.  The relevant Feynman diagram with four
interaction vertices, three of which involve the regular interactions
of quarks with Higgs bosons, is described by the effective Lagrangian,
\bea && L_{EFF}= {\l^u_i \l^d_j \over (4\pi )^2} V^{\star } _{i1}
V_{1j} \l '' _{ijk} \l_k ^d {\mu _l \mu \over m^2_{\tilde d _R} m
^2_{\tilde H} } (d^cs^c)(u\nu ) + \ H.\ c. \eea The resulting one-loop
level bounds vary inside the intervals~\cite{palbhatta}, $ \l
''_{ijk}{ \mu_l \over \mu } < (10^{-15}\ - \ 10^{-25}) \tilde d^2_R
({m _{ \tilde H } \over 100 }) ^2 $, for the configurations of
generation indices, $(jk)= (21, 31, 32),\ i=1,2,3 $, with $ \tilde H $
referring to a higgsino dominated neutralino mass eigenstate. For
definiteness, we quote a representative sample of bounds, $ [\vert \l
''_{321} {\mu_l \over \mu }\vert ,\ \vert \l ''_{331} { \mu_l \over
\mu }\vert ,\ \vert \l ''_{33} { \mu_l \over \mu } \vert ] <
[10^{-18}, \ 10^{-17}, \ 10^{-18} ] \ \tilde d_R ^2 \ ({m_{\tilde H}
\over \mu } ) ^2 . $

Chun and Lee~\cite{chunlee99} use the constraints from atmospheric
neutrino oscillation data on the bilinear RPV coupling constants, $\mu
_i $, jointly with the single nucleon decay constraints on the
quadratic coupling constant products, $\l ^{'\star } \l '' ,\ \l ''
\mu _l $, in order to derive individual bounds on the complete set of
baryon number violating trilinear coupling constants.  The resulting
individual coupling constant bounds are of form, $\vert \l ''_{ijk}
\vert < [10^{-10} \ - \ 10^{-19}] $.

The combination $\l'' \l ^\star $ of baryon and lepton number
violating R parity odd interactions can contribute to single nucleon
decay through the quark subprocess, $u ^c d^cs^c \to e ^\pm _i e ^\mp
_j \nu _k $, described by the tree level Feynman diagram involving the
sequential $t$-channel exchange of squarks, sleptons and
gauginos~\cite{long97}.  The corresponding dimension $\cddd =9$
effective Lagrangian, $ L_{EFF} = {g _2^2 \l ''_{112} \l^{\star}
_{ijk} \over m_{\tilde Z} m _{\tilde d_i} ^2 m _{\tilde e_j} ^2 }
(\bar d _1 u^c_1) (\bar e_i d_i ^c) (\bar e_k e_j) + H.\ c. $ is $(B -
L)$ conserving and $(B+ L)$ violating.  To illustrate the physical
relevance of this mechanism, Long and Pal~\cite{long97} consider the
extended gauge symmetry model in which the electroweak symmetry group
$SU(2)_L$ embedded into an $SU(3)_L $ group.  The resulting model with
gauge group, $SU(3)_c \times SU(3)_L \times U(1)_N$, has two
distinctive features: (1) The extra matter content dictated by the
anomaly cancellation constraints require the number of generations to
be a multiple of $3$; (2) The baryon number violating four fermion
coupling induced by the RPV down-squark exchange amplitude favors the
strangeness changing mode, $ p\to K^+ + \bar \nu $, as the dominant single
nucleon decay channel.  The R parity odd purely leptonic interactions,
$L_i L_j E^c_k$, arise in this model with the fully antisymmetric
lepton generation structure, $\l _{ijk} = \e _{ijk} \l $.  The
coupling, $ u^c d^c s^c \to \mu ^\pm e^\mp \nu _\tau $, resulting from
the combined $\l'' , \ \l ^\star $ interactions favors again the
strangeness changing four-body proton decay mode, $p \to K^+ +\mu ^\pm
+ e ^\mp + \bar \nu_\tau $.  The comparison with the experimental
bound for proton instability gives the quadratic product coupling
constant bounds, $\vert \l^{\star } \l ''_{112} \vert < 10^{-16}$.

The possibility of inducing single nucleon decay by the combined $\l $
and $ \l ''$ interactions applies, however, quite generally and is not
restricted to a specific model.  The relevant amplitudes
arise~\cite{bhatt99} through appropriate one-loop or two-loop diagrams
consisting of two tree or one-loop subdiagrams linked together by an
internal neutralino line.  The single proton decay modes, $ p\to K^+ +
e^\pm + \mu ^\mp +\bar \nu ,\ p\to K^+ +\nu + \nu + \bar \nu,\ p\to
K^+ + \bar \nu $, lead to the set of quadratic coupling constant
bounds lying in the range, $\vert \l
_{ijk} \l ^{''\star }_{112} \vert   < [O(10^{-21} ) \ - \ O(10^{-16}) ] $, 
while the analogous decay modes
with emission of $\pi ^+ $  lead for all the RPV quadratic
coupling constant combinations to bounds varying inside the range,
$\vert \l _{ijk}  \l  ^{''\star }_{i'j'k'} \vert < [O(10^{-12} ) \ -
\ O(10^{-3} ) ] $.

Carlson et al.,~\cite{carlson} examine the tree level mechanism
involving the sequential $s$-channel exchange of $\tilde d , \tilde
\chi^+ , \tilde e$, as represented by the Feynman diagram J.4 in
Figure~\ref{fig4}.  The corresponding amplitudes can initiate the
single proton decay mode, $ p\to e^ +_k + \nu_i + \nu_j $, obeying the
selection rule, $ \D B=- \D L = - 1$.  The comparison with the
experimental proton lifetime limit yields bounds for products
involving any $\l_{ijk} $ with the subset of coupling constants $ \l
'' _{i'j'k'} $ carrying at least one light first generation index.  A
representative sample of the strongest coupling constant bounds reads
as: $\vert \l^{\star } _{ijk} \l ''_{121} \vert < 10^{-14}, \ \vert
\l^{\star } _{ijk} \l ''_{131} \vert < 10^{-13}, \ \vert \l^{\star }
_{ijk} \l ''_{132} \vert < 10^{-12}, \ \vert \l^{\star } _{ijk} \l
''_{221} \vert < 10^{-13}, \ \vert \l^{\star } _{ijk} \l ''_{321}\vert
< 10^{-12}. $ If enough third generation fields are involved, the
predicted single nucleon rates can be maintained at an observable
level by considering the one-loop level dressing effects.  The
consideration of the wino one-loop dressing of the vertex $ d+ u \to
\tilde d ^\star $ yields amplitudes depending on quadratic products of
coupling constants with two or three heavy second or third generation
indices present in the coupling constants $\l '' _{i'j'k'}$. The
resulting bounds are, however, weaker: $\vert \l ^{\star } _{ijk} \l
''_{331} \vert < 10^{-3},\ \vert \l^{\star } _{ijk} \l ''_{332}\vert <
10^{-2}, \ \vert \l^{\star } _{ijk} \l ''_{231} \vert < 10^{-2} , \
\vert \l^{\star } _{ijk} \l ''_{232} \vert < 10^{-3}. $

The combined action of the $\l '$ and $ \l ''$ interactions can
initiate three-body single nucleon decays, $p \to K^0 + e^+ + \pi ^0$,
obeying the selection rule, $ \D B = + \D L =-1 $. The tree level
Feynman graph representing the reaction scheme, $ u \to d^c_j + \tilde
d_k ^\star \to d^c_j + (e^+ + u ^c_i)$, is displayed by diagram J.5 in
Figure~\ref{fig4}.  The comparison with the experimental limit on the
proton lifetime yields strong bounds for quadratic coupling constant
products comprising any $\l '_{ijk} $ with a $ \l '' _{i'j'k'} $
coupling constant carrying at least a single second or third
generation index, $\vert \l ^{'\star } _{ijk} \l ''_{l21}\vert <
10^{-9}, \ \vert \l ^{'\star } _{ijk} \l ''_{l31} \vert < 10^{-9} , \
\vert \l ^{'\star } _{ijk} \l ''_{l32} \vert < 10^{-9}, \ [l=1,2]. $ A
small set of coupling constant products, $\vert \l ^{'\star } _{33k}
\l ''_{221} \vert <10^{-1}, \ \vert \l ^{'\star }_{12k} \l
''_{231}\vert <10^{-2}$, still remains weakly constrained.  The
one-loop dressing effects allow one to extend the constraints to cases
involving several third generation quarks or leptons.  These effects
may be associated with the self-energy or mass renormalization
corrections of non-diagonal character with respect to the quark
generations, $ \tilde d \to \tilde b $, or with the box diagrams
corrections for the internal lines, $ W - \tilde u$.  The comparison
with the nucleon decay lifetime, in the cases involving two second or
third generation indices, give the coupling constant
bounds~\cite{carlson}: $ \vert \l ^{'\star } _{11n} \l ''_{332}\vert <
10^{-2}, \ \vert \l ^{'\star } _{12n} \l ''_{332}\vert < 10^{-2}, \
\vert \l ^{'\star } _{2mn} \l ''_{331} \vert < 10^{-3},\ \vert \l
^{'\star } _{2mn} \l ''_{332} \vert < 10^{-3},\ \vert \l ^{'\star }
_{213} \l ''_{331} \vert < 10^{-5},\ \vert \l ^{'\star } _{223} \l
''_{332} \vert < 10^{-5}. $
Nevertheless, there still remains $30$ weakly constrained $\l ' \l ''$
products out of the total number of $ 243$ products.  For instance,
one finds $\vert \l ^{'\star } _{lm3} \l ''_{33n}\vert < O(1)$. The
consideration of loop dressing mechanisms may possibly remove the
suppression effects and give improved bounds for all the $\l ' \l ''$
coupling constant products.

\subsubsection{\bf Exotic single nucleon decay channels and 
rare $B$ meson decay channels}

We discuss now the exotic type baryon number violating processes
controlled by the $\l ''$ interactions alone.  Single nucleon decays
can occur in this context only for final state channels involving the
emission of a supersymmetric particle.  The two-body proton decay into
a neutralino and strange meson, $ p \to \tilde \chi ^0 +K ^+ $,
arising from the tree level $ \tilde s$ -quark exchange process, is
energetically allowed neutralino masses inside the range, $ m_{\tilde
\chi } < m_p - m_K$.  The resulting coupling constant bound
is~\cite{chang96}, $\l ''_{112} < 10^{-15}$.

The existence of ultra-light gravitino or axino
particles~\cite{kimrep,rajagopal,bagger} is a characteristic feature
of the low energy gauge mediated supersymmetry breaking approach.
Single nucleon decays, involving the emission of a strange meson
accompanied by sufficiently light gravitino or axino, can be initiated
through a single dominant $\l ''$ coupling constant by making use of
the $\tilde s $ tree level exchange process, $u+ d \to \bar s +\tilde
G, \ u+ d \to \bar s +\tilde a $, depicted by the Feynman diagram J.13
in Figure~\ref{fig4}.  The comparison with the experimental limit on
the nucleon decay lifetime leads to the following coupling constant
bounds for the gravitino and axino emission cases~\cite{choit}: \bea
\vert \l '' _{112}\vert &<& 5.5 \times 10^{-15} \tilde m^2 ({m_{\tilde
G }\over 1 \ \text{ eV} }) , \quad [ p \to K^+ + \tilde G ] \cr \vert
\l '' _{112} \vert &<& 7.7 \times 10^{-15} C_ q ^{-1} \tilde m^2 ({F_a
\over 10^{10} \ \text{GeV} }) ,\quad [ p \to K^+ + \tilde a ] .\eea
For the axino emission case, $ F_a$ denotes the axion decay mass scale
and $C_q$ is a model-dependent parameter associated with the trilinear
couplings of quarks and leptons to the axion supermultiplet. In models
where the light quarks carry a $ U(1) _{PQ} $ charge, $C_q \simeq O(1)
$, while in models where they do not, $C_q \simeq [O(10^{-2}) -
O(10^{-3}) ]$, due to the need to generate the axion matter fields
Yukawa couplings at the one-loop level.  The loop dressing mechanism
for the single nucleon decay modes with gravitino or axino emission
provides useful bounds on the complete set of coupling constants, $\l
''_{ijk}$.  A representative sample of the corresponding one-loop
processes is depicted by the Feynman diagrams J.14-J.16 in
Figure~\ref{fig4}.  Choi et al.,~\cite{choil} consider the one-loop
triangle diagrams for the vertices, $ u d \tilde s , \ \bar d \tilde G
\tilde d , u \bar s \tilde d $ and box diagrams, involving the Higgs
boson, W boson, gaugino or squark boson internal lines. Accounting in
a qualitative way for the suppression factors in the amplitudes due to
the loops integral, the flavor changing effects and the superpartners
masses, yields the following intervals of variation for the coupling
constant bounds, applying to all possible generations~\cite{choil},
$\vert \l '' _{ijk} \vert < [10^{-7 }\ \tilde m^{2} \ - \ 10^{-9}
\tilde m^{3}] ({m_{\tilde G } \over 1 \ \text{ eV} }) , \quad \vert \l
'' _{ijk} \vert < [10^{-7 }\ \tilde m^{2} \ - \ 10^{-9} \ \tilde
m^{3}] ({F_a \over C_q \times 10^{10}\ \text{GeV} }) $, in
correspondence with the gravitino and the axino emission cases,
respectively.  A representative subset of the bounds for the gravitino
and axino particles emission reads as: \bea [{\vert \l '' _{112}\vert
\over \tilde m^2}, \ {\vert \l ''_{113}\vert \over \tilde m^3},\
{\vert \l '' _{323}\vert \over \tilde m^2 }] &<& [0.55 \times 10^{-16}
, \ 0.11 \times 10^{-10}, \ 0.55 \times 10^{-8} ] ({m_{\tilde G }
\over 1 \ \text{eV} }) , \cr [{\vert \l '' _{112}\vert \over \tilde
m^2},\ {\vert \l ''_{113}\vert \over \tilde m^3},\ {\vert \l ''
_{323}\vert \over \tilde m^2}] &<& [0.77 \times 10^{-16} , \ 0.25
\times 10^{-10}, \ 0.77 \times 10^{-8} ] \ ( {F_a \over C_q \ 10^{10}
\ \text{GeV} }) . \eea

The $B$ meson decay modes can provide potentially useful signals of
baryon number violation initiated by the $\l ''$ interactions, despite
the fact that the current experimental sensitivity to $B$ meson decays
is still well below that attained in the nucleon or nuclear decay
experiments.  Carlson et al.,~\cite{carlson} examine the mechanism for
$b$-quark decay with $\D B=-\D L = + 1$, represented by the one-gluon
exchange Feynman diagram J.6 in Figure~\ref{fig4}, which features the
important advantage of being calculable in perturbative QCD.  The
branching fractions of $ O(10^{-8})$ predicted for the decay
channels~\cite{carlson}, $ B^+ \to p +\bar \nu_\tau, \ B^+ \to n+\tau
^+ $, are not promising in terms of an experimental observability at
the $ B$ meson factories.  The tree level mechanism involving a
sequential $t$-channel exchange of $\tilde b ,\ \tilde u , \ \tilde
\chi^+ $, as displayed by diagram J.7 in Figure~\ref{fig4}, may
initiate $ \D B=2$ baryon number violating B-meson decay channels, $B
^0\to \L +\L $ or $ B^0 \to \S^++\S^-$.  The estimated order of
magnitude of the branching fractions, $O(10^{-8}) \times \vert \l
^{''\star } \l '' \vert $, also lies well below the sensitivity
attained at the existing $ B$ meson factories.  With the fluxes of $B$
meson secondary beams attained at the current asymmetric accelerators,
the sensitivity on the branching fractions of rare decay channels
currently lies at the level, $ O(10 ^ {-5}) $.  One may thus conclude
from the above results that the exploration of baryon number violation
through the $ B$ meson decays~\cite{thorndike} is unlikely to compete
with the conventional studies based on nucleon and nuclear decays.

\subsubsection{\bf Double baryon number violation  reactions} 

The neutron-antineutron transition, $n\to \bar n$, is described by the
$ \D B =2 $ effective Lagrangian bilinear in the neutron-antineutron
Dirac fields, \be L_{EFF}= -\pmatrix{\bar n & \bar n^c } \pmatrix{ m_n
+ V _n & \d m \cr \d m ^\star & m_n+ V _{\bar n} }\pmatrix{n \cr n^c
}, \ee where the off-diagonal matrix element, $\d m$, designates the
free neutron-antineutron mass splitting parameter and the complex
terms, $ V _n,\ V _{\bar n }$, in the diagonal matrix elements refer
to potential energy binding effects.  The new physics contributions to
$\d m$ are represented in the electroweak symmetry limit by $6$
independent $\cddd =9$ six quark operators involving the two
structures in terms of the chiral (Weyl spinor) quark fields, $ d_Rd_R
d_Ru_Rq_Lq_L,\ d_Rd_R q_Lq_Lq_Lq_L $.  Note that the allowance for
electroweak symmetry breaking, as would be appropriate for mechanisms
taking place close to the Fermi mass scale, introduces a total of $18$
independent six fermion operators~\cite{rao82}.  The quark model
evaluation of $n \to \bar n $ transition matrix elements for the six
quark operators is discussed by Pasupathy~\cite{pasupathy}, Misra and
Sarkar~\cite{misra83} and Rao and Schrock~\cite{rao82}.  The results
can be summarized by the schematic formula for the quark wave function
inside the nucleon: $ \vert \psi (0)\vert ^4 \simeq <\bar n \vert O
^i_{\cddd =9} \vert n> = {N^6 I _i(p)\over p^3 (4\pi )^2} \simeq (1. \
- \ 0.1) \times 10^{-5} \ \text{GeV}^6 $, where $ I _i(p) $ denote
calculable wave function overlap integrals, with $p$ standing for the
average momentum of the bound quarks and $N$ for the normalization
constant of the valence quark cavity mode.

The experimental searches for $n-\bar n$ oscillation make use of the
long flight path of the small velocity ultra-cold neutrons to enhance
the oscillation probability, $ d P _{n \to \bar n} /dt \sim (\d
m\times t ) ^ 2 $.  The experimental limit on the $n-\bar n $ free
oscillation time obtained by the Institute La\"ue-Langevin
collaboration~\cite{baldo94} stands at the value, $ \tau _ {osc}
\equiv 1/\d m > 0.86 \times 10^8$ s.  The oscillation period, $ \tau _
{osc} $, is also accessed through searches of nuclei disintegration
proceeding via two nucleon decay processes.  The experimental searches
at the Kamiokande and Fr\'ejus underground
laboratories~\cite{takberg86} yield limits on the lifetime of $^ {16}
\text{O} $ and $^{56} \text{Fe} $ nuclei which translate into the
upper limit on the oscillation time, $ \tau _ {osc} > (1-2) \times
10^8$ s.  The initial proposal of the $n-\bar n$ conversion process
was made by Kuz'min~\cite{kuzmin70}, but the wider interest in this
reaction was revived sometime later by Mohapatra and
Marshak~\cite{mohaak80} and other authors with the advent of grand
unified gauge theories.  Other related processes are the
hydrogen-antihydrogen atom oscillation~\cite{moha80}, $ H(p+e ^ -) \to
\bar H( \bar p+e ^ +) $, and the double nucleon decay
reaction~\cite{aruel82}, $p + p \to e ^ + + e ^ + $, obeying the
selection rules, $ \D B = \D L= -2$.  The prospects to improve the
experimental limit on $ \tau _ {osc} $ with observations of the free
neutron oscillation process are more hopeful, as appears from the
currently planned proposal at the ORNL reactor~\cite{kamysh97} which
aims at pushing down the experimental limit by two orders of
magnitude.

For neutrons immersed inside nuclear matter, the $n-\bar n$
oscillation has observable manifestations in terms of nuclei
disintegration initiated by the two nucleon decay reactions, $ A_Z \to
n + (A_Z -1) \to \bar n + (A_Z-1) \to X + (A_Z-2)^\star , \ [X= \pi, \
2\pi , \ 3\pi,\ \cdots ]$ where $X $ denotes the decay products from
nucleon-antineutron pair annihilation.  Using the qualitative relation
between the nuclear and two nucleon decay lifetimes, $ N \tau _{nucl}
\equiv \tau_{NN} \simeq \tau^2 _{osc} \G_a = \G_a /(\d m) ^2 $, where
$\G _a$ is the $ n - \bar n$ annihilation rate, one can express the
experimental limit from matter stability as, $\tau_{osc} > 10^{7.5}\
\text{s} $ or equivalently as, $ \d m \equiv \tau _{osc} ^{-1}<
10^{-31.5} \ \ \text{GeV}.$ The discussion of $n-\bar n$ conversion in
nuclei in terms of the two nucleon nuclear decay reactions hinges on
the relatively poorly understood $\bar n -N$ annihilation process. The
multibody decay channels are accounted for in an optical approximation
approach by adding a phenomenological imaginary part to the familiar
$\bar n -N$ meson exchange potential~\cite{dover92}.

The nuclear physics aspects of the $\D B=2$ nuclear disintegration
reactions are discussed in several precursor
works~\cite{riazuddin82,chet81}.
The currently standard approaches rests on the seminal work by
Chetyrkin et al.,~\cite{chet81}. This uses a perturbation theory
approach in which the initial $n\to \bar n$ oscillation transition,
$(A-1)_Z + n \to (A-1)_Z + \bar n$, is followed by $\bar n -N$
annihilation from the compound nuclear system, $\bar n + (A-1)_Z$.  A
simplified formula for the nuclear decay rate is obtained by making
use of the closure approximation.  The annihilation amplitude is
expanded on a basis of quasi-stationary nuclear states, $ \vert
((A-1)_Z + \bar n )_i>$, with complex excitation energies, $ \D M_i -i
\G _i /2 $, which can be identified with the differences in complex
potential energy experienced by the bound neutrons and antineutrons, $
\D M_i -{i\over 2} \G _i = V_n - V_{\bar n}.$ Under the assumption
that the spacings of levels and the differences of annihilation widths
are small and cluster around average values, $ M_i -M_j << \ <\D M > ,
\ \G_i - \G_j << \ <\G _a>$, one can use a closure approximation to
express the nuclear decay rate by the simplified formula, $\tau ^{-1}
_{nucl}= \G_{nucl} = {N \over \tau_{NN} } = N \G _{NN} = { N <\G_a>
(\d m ) ^2 / ( <\D M ^2> +{1\over 4} <\G_a^2> ) } \equiv {N (\d m ) ^2
\over T _R } , \ [ T _R^{-1} = {<\G_a> / ( <\D M ^2> +{1\over 4}
<\G_a^2>) } ] $.  Improved semiquantitative discussions based on the
above approach, but using distinct approximations, are reported by two
groups of authors~\cite{gal83,alberico}.  Dover et al.,~\cite{gal83}
define the requisite average antineutron width and mass shift
parameters by means of an optical potential description of the $\bar n
-(A-1) _Z $ system, while Alberico et al.,~\cite{alberico} evaluate
the sum over nuclear states within a shell model description, using
realistic inputs for the complex two-body potentials in the $N-N$ and
$ N-\bar n$ systems.  The results obtained by means of these two
calculational procedures differ at quantitative rather than
qualitative levels.  An alternative nonrelativistic diagram approach
advocated by Kondratyuk~\cite{kondra96}, which claims a strong
suppression of the $n-\bar n $ transition rate in nuclei relative to
that found in the above described standard approach, has been
dismissed as unrealistic~\cite{hufner98} by H\"ufner and
Kopeliovich~\cite{hufner98}. The latter authors reason on the basis of
a simplified quantum mechanical two-channel formalism in which the
diagonal elements of the mass matrix are described by nuclear density
dependent local complex optical potentials for the neutron and
antineutron folded with a spatially distributed $n -\bar n$
annihilation rate function.

\subsubsection{\bf Broken R parity contributions to neutron-antineutron  oscillation
and double nucleon decay}

\begin{figure} [ht]
\begin{center}
\leavevmode
\caption{\footnotesize \it Feynman diagrams {\bf J.1-J.16}     describing the perturbation theory
contributions to a selected set of processes initiated by the RPV
interactions. {\bf Diagrams} J:
Amplitude describing two-body nucleon decay channels from the combined
action of the $ \l ' $ and $\l ''$ interactions, $ \D B =\pm \D L =
-1$, (J.1-2), and of the bilinear and trilinear interactions, $ \D B =
\D L = -1$, (J.3).  Other multi-body nucleon decay channels from the
combined $\l '' \ \l $ interactions
(J.4) and $\l '' \ \l ' $ interactions (J.5).  $B$ meson decay
channels, $ B ^+\to n +e^+$, initiated by the $\l ''$ interactions
with single baryon number violation $(\D B =-\D L =-1)$ (J.6). Double
baryon number violation B meson decays, $ B ^0 \to \L + \L , \ B \to
\S ^+ + \S ^- ,\ (\D B =-2)$ (J.7).  Amplitudes for the quark
subprocess, $ u+d+d \to u^c +d^c +d^c $, initiating the $n\to \bar n$
transition (J.8-9, J.11) and the $n\to \bar \Xi $ transition (J.10).
Amplitude for the quark subprocess initiating the double nucleon decay
reactions, $ p+p\to K^++K^+,\ n+n\to K^0+K^0 $ (J.12).  Amplitudes for
the subprocesses initiating single nucleon decays with emission of
gravitino or axino at tree level (J.13) and the corresponding one-loop
level dressing diagrams (J.14 - J.16).     }   
\vskip  1 cm 
\centerline{\epsfig{figure=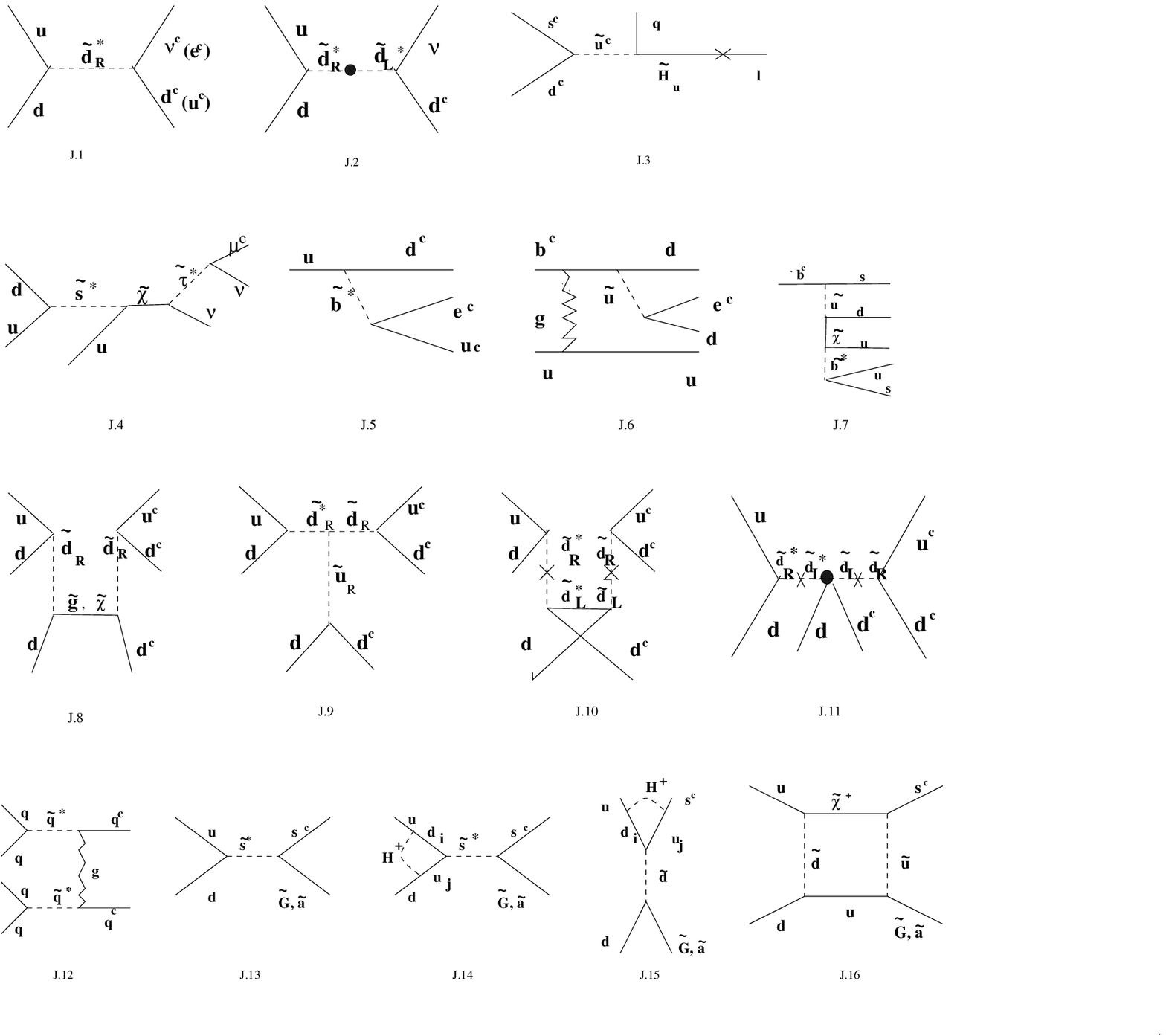,height=21cm,width=19cm} }
\label{fig4}
\end{center}
\end{figure}

Proceeding from the above brief survey of the double baryon number
violation phenomenology, we focus now on the constraints inferred from
experimental data on the RPV interactions.  Two distinct RPV tree
level mechanisms, displayed by the diagrams J.8-9 in
Figure~\ref{fig4}, were initially considered by
Zwirner~\cite{zwirner}.  The predominant contribution arises from two
$s$-channel exchange of d-squarks linked together by a gluino
exchange.  The $t$-channel gluino exchange Feynman diagram J.8 in
Figure~\ref{fig4} is described by the effective Lagrangian,
\begin{eqnarray} && L _{EFF} ^{\D B=2} = { C g_3^2  \over m_{\tilde g } }
 { (\tilde m^{d2}_{RR} )_{kK } \over m^2_{\tilde d_{kR} } m^2_{\tilde
d_{KR} } } { (\tilde m^{d2}_{RR} )_{K k' } \over m^2_{\tilde d_{k'R} }
m^2_{\tilde d_{K R} } } \l^{''\star }_{ijk}\l'' _{i'j'k'} (\bar u^c_{i
L} d_{jR} ) (\bar d^c_{K L} d_{KR}) (\bar u ^c _{i'L} d_{j'R}) + {
j\leftrightarrow k \choose j '\leftrightarrow k' } + H. \ c. ,
\label{eq12}
\end{eqnarray}
where the color indices have been suppressed and the various
combinatorial factors included in the constant overall factor $ C$.
The $n\to \bar n $ matrix element picks up the indices, $
i=j=K=i'=j'=1,\ k=k'=2 $, resulting in the amplitude, $ L _{EFF}
\simeq { C g_3^2 \vert \l'' _{112} \vert ^2 \over m_{\tilde g } }
\vert { (\tilde m^{d2}_{RR} )_{12} / m^2_{\tilde d_{1R} } m^2_{\tilde
d_{2R} } }\vert ^2 (\bar u^c_L d_R) (\bar d^c_L d_R) (\bar u ^c_L d_R
) $, whose contribution is strongly suppressed by the down-squark
off-diagonal mass matrix element factor, $ (\tilde m^{d2}_{RR})_
{12}$.  Comparing the approximate estimate for the oscillation period
with the current experimental limit on nuclear decay
yields the coupling constant bound~\cite{zwirner,halldimo,hinchliffe},
\bea && \tau _{osc} ^{-1} = { 16 \vert \l ^{''2} _{112} \vert g_3 ^2
\vert \psi (0)\vert ^4 \over 3 m_{\tilde q} ^4 m_{\tilde \chi } }
\vert { (\tilde m^{d2}_{RR})_{12} \over m^2 _{\tilde d _ {kR } } }
\vert ^2 \ \Longrightarrow \l ''_{11k} < [ 10^{-7} - 10^{-8} ] {10^8 \
s \over \tau_{osc} } ({\tilde m \over 100 \ \text{GeV} })^{5/2} \times
{(\tilde m^{d2}_{RR})_{12}\over m^2 _{\tilde d _ {kR } } } . \eea The
second mechanism proposed by Zwirner~\cite{zwirner} is represented by
diagram J.9 in Figure~\ref{fig4} in terms of an intermediate vertex
at which three squarks, emitted by the quark lines via the $ \l ''$
Yukawa interactions, jointly annihilate via the soft supersymmetry
breaking interaction, $ A_{ijk}^{\l ''} \l '' _{ijk} \tilde
U^c_i\tilde D^c_j \tilde D^c_k $.  Being of order $\l ^{''4} $ and
suppressed by the same flavor changing mass term as that mentioned
above, the contribution from this mechanism is expected to be
insignificant.

An important distinction between the $n-\bar n$ oscillation and two
nucleon decay processes arises in the context of broken R parity
symmetry.  As noted initially by Dimopoulos and Hall~\cite{halldimo},
one may bypass the strong suppression from the squarks flavor changing
mass factor by considering the nuclear decay channels involving pairs
of $ K $ mesons in the final states. The antisymmetric structure of
the $\l ''_{ijk}$ coupling constants imposes generation non-diagonal
configurations in the $ d^c _j $ quarks, which favor the strangeness
changing transition, $ n \ - \ \bar \Xi $, over the familiar
transition, $n \to \bar n$.  Motivated by this observation, Barbieri
and Masiero~\cite{masiero} considered the double strangeness changing
process, $ u +d + d \to u^c + s^c + s^c, $ displayed by the Feynman
diagram J.10 in Figure~\ref{fig4}, which contributes to the $ \D S
=2$ transition, $ n \to \bar \Xi $.  The corresponding effective
Lagrangian has the same formal structure as that in eq.(\ref{eq12}),
\bea && L _{EFF} ^{\D B=2} = { C ' g_3^2 \over m_{\tilde g } } {
(\tilde m^{d2}_{LR} )_{kK } \over m^2_{\tilde d_{kR} } m^2_{\tilde
d_{KL } } } { (\tilde m^{d2}_{LR} )_{K k' } \over m^2_{\tilde d_{k'R}
} m^2_{\tilde d_{K L } } } \l^{''\star }_{ijk}\l'' _{i'j'k'} (\bar
u^c_{iL} d_{jR} ) (\bar d^c_{KR} d_{KL}) (\bar u ^c _{i'L} d_{j'R}) +
{j\leftrightarrow k \choose j ' \leftrightarrow k' } + H. \ c. , \eea
but now includes the flavor changing mass factor from the $LR$
chirality flip down-squarks off-diagonal mass matrix element, $
(\tilde m^{d2}_{LR})_{12} $, which causes a weaker suppression than
the $RR$ down-squarks off-diagonal mass matrix element.  The $ n \ -\
\bar {\Xi } $ matrix element selects the indices, $ i=i'=k=k'=1,\
j=j'=K=2 $, resulting in the amplitude, $ L _{EFF} \simeq { C' g_3^2
\vert \l'' _{121} \vert ^2 \over m_{\tilde g } } \vert { (\tilde
m^{d2}_{LR} )_{12} \over m^2_{\tilde d_{1R} } m^2_{\tilde d_{2L} }
}\vert ^2 (\bar u^c_L s_R) (\bar d^c_R d_L) (\bar u ^c_R s_L ) $.  The
$ n \ -\ \bar {\Xi } $ oscillation initiates baryon and strangeness
number violating double nucleon decay processes with $\D B= \D S= -2$,
such as, $ p+p \to K^++K^+ $, or, $ n+n \to K^0+K^0 $, contributing to
nuclear decay modes with $K$ mesons in the final state, $ ^{16}
\text{O} \to \ ^{14}\text{C} + K^{+} + K^{+} ,\ ^{16} \text{O} \to \ 
^{14}\text{O} + K^{0} + K^{0} $.  The gain from flavor mixing is
frustrated by the energetically more suppressed off-shell transition,
$ n \to \bar \Xi $, due to the larger mass splitting.  On the other
hand, whether the $n-\bar \Xi $ system annihilation rate is similar or
very different from that of the $n-\bar n $ system is not clearly
understood so far.  The comparison with the experimental limit for the
$ ^{16}\text{O} $ nucleus lifetime, $ \tau_{nucl} \simeq (10^{-2} y) \
\vert \l ^{''}_{112}\vert ^{-4} (\tilde q ^4 \tilde g)^2 $, using the
tentative estimates for the average nuclear mass shift and
annihilation width parameters, $<\D M > \simeq \vert m_n - m_ {\bar
\Xi } \vert \le 400 \ \ \text{MeV},\ <\G _a > \simeq 200 \ \text{MeV}
$, leads to the coupling constant bound~\cite{masiero}, \bea && \vert
\l ''_{11k} \vert < 10^{-17/2}\ \tilde g ^{1/2} \tilde d_{kR} ^2
({\tau _{nucl} \over 10^{ 32} \ \text{y} } )^{-1/4}
({ 10^{-6} \ \text{GeV} ^{6} \over < \bar \Xi ^0 \vert ududss \vert n
> })^{1/2} .\eea

In view of the lack of solid information on the chirality flip
supersymmetry breaking mass parameters, Goity and Sher~\cite{goity}
question the above proposal to constrain the coupling constants $ \l
''_{11k}$ by using the nuclear decay limits. The point is that the two
nucleon decays might still be predominantly initiated through the $
n-\bar n $ oscillation, provided one finds a suitable mechanism for
the requisite flavor changing interaction.  The proposed alternative
mechanism~\cite{goity} involves the coupling constant $ \l ''_{113} $
and intermediate sbottom quarks, with contributions to the flavor
changing vertex, $ \tilde {\bar b } _L + d \to \tilde b_L + \bar d $,
initiated via a $ W ^\mp -\tilde W ^\pm $ box diagram.  The
corresponding Feynman graph, displayed by diagram J.11 in
Figure~\ref{fig4}, uses the chain of reactions, $ (u+d+ d_L ) \to
\tilde b_R ^\star +d_L\to \tilde b ^\star _L +d_L \to _{ q\tilde q}\to
d_L^c + \tilde b_L \to d_L^c+ b_R \to (d^c + u^c+ d^c)$, where the
intermediate amplitude for the process, $ \tilde b _L ^\star +d \to
\tilde b _L +d^c$, is described in terms of a $W$-boson and gaugino
exchange box diagram.  The choice of intermediate $\tilde b$-squarks
is the most favorable one for the purpose of maximizing the diagonal
mass matrix element factor, $ (\tilde m^{d2}_{LR})_{kk} /m^2_W) $.
The corresponding fully calculable transition amplitude for this
process is described by the effective Lagrangian, \bea && L _{EFF}
^{\D B=2} \simeq { 3 g^4_2 \vert \l ^{''} _{113}\vert ^2 (\tilde
m^{d2}_{LR})^2 _{33} m_{\tilde W} \over 8\pi ^2 m^4_{\tilde b_L}
m^4_{\tilde b_R} } J (m^2_{\tilde W} , m_W^2, M^{u2}_j, m^2 _{\tilde u
_{j'}} ) \xi_{jj'} (\bar u^cd)(\bar d^cd)(\bar u^cd) + H. \ c., \eea
involving a diagonal element of the chirality flip down-squarks mass
matrix, a loop momentum integral factor, $J $, and a product of CKM
matrix elements, $ \xi_{jj'} = V^\dagger _{bu_j} V _{u_jd} V^\dagger
_{bu_{j'} } V _{u_{j'}d} $.  The numerically evaluated amplitude leads
to a coupling constant bound lying inside the range, $ \vert
\l''_{131} \vert < [2. \times 10^{-3} \ - \ 1.\times 10^{-1} ] $, in
correspondence with squark masses varying inside the interval, $
\tilde m = (200 \ - \ 600) \ \text{GeV} $.  The analogous
bound~\cite{goity} on the coupling constant $\l''_{112} $ is weaker by
a factor, $ (\tilde m^{d2}_{LR}) _{22} / (\tilde m^{d2}_{LR}) _{33}
\sim m_s/m_b \simeq 4. \times 10^ {-2} $, and varies in the range, $
\vert \l''_{121}\vert < [5. \times 10^{-2} \ - \ 2.5 ] $.  Three
additional one-loop box diagrams involving the exchange of gaugino and
$W$ gauge boson and quark-squark pairs need to be included for a truly
complete description of the transition amplitudes, as noted by Chang
and Keung~\cite{chang96}.  A single diagram from this set turns out to
give the predominant contribution. The resulting bounds for the
associated coupling constants read: $ \vert \l ''_{321} \vert < [2.1
\times 10^{-3}, \ 1.5 \times 10^{-2}] ({m_s \over 200 \ \text{ MeV} }
)^{-2} , \ \vert \l ''_{331}\vert < [2.6 \times 10^{-3}, \ 2.  \times
10^{-2}] , $ where the numerical results in the two entries are in
correspondence with the values of the scalar superpartners mass
parameter $ \tilde m= [100, \ 200 ] \ \text{GeV} $.
 
The search for the dominant RPV nuclear decay channel can be addressed
from still another perspective by considering a calculation bypassing
the initial $ n-\bar n $ or $ n -\bar \Xi $ oscillation stage.  The
alternative approach proposed by Goity and Sher~\cite{goity} deals
directly with the hadronic level transition.  Although one may
envisage several two-nucleon decay channels involving the production
of strange baryons, $ N+ N \to K + \L , \ N+ N \to K + \S $, it is
reasonable to assume that the contribution from the purely mesonic $
\D B = \D S = -2 $ channels, $ N + N \to K+K$, dominates the total
cross section, $ \s ( N + N \to X) $.  The RPV transition amplitude is
represented at the quark level by the $s$-channel exchange, mediated
by a gluino $t$-channel exchange, of a pair of strange squarks
producing two pairs of quarks.  The corresponding reaction scheme, $
(q_iq_j) (q_lq_m)\to (\tilde {\bar q}_k \tilde {\bar q '}_m )
\to_{\tilde g}\to (\bar q_k \bar q_n) $, is represented by the Feynman
diagram J.12 in Figure~\ref{fig4}.  The amplitude for the specific
flavor configuration, $ (u +d ) + (d + u) \to \tilde s ^\star + \tilde
s ^\star \to s ^c + s^c $, is described, up to suitable permutation
reordering of the fermion spinor fields and saturation of the color
indices, by the effective Lagrangian, $ L _{EFF} = {16 g_3  ^2 \vert \l
^{''} _{112} \vert ^2 \over 3 m_{\tilde g} m ^4_{\tilde q} } \e _{\a
\b \g } \e _{\d \rho \s } [(\bar u ^c _{\a L} d _{\b R}) (\bar u ^c
_{\d L} d _{\rho R}) (\bar s ^c _{\g L} s _{\s R}) + \cdots ] + H. \
c. , $ where the ellipsis refers to the nine distinct permutations of
the spinor fields.  The nuclear decay cross section for the reaction,
$ ^{16}O \to \ ^{14}C+ K^{+ } +K^{+ }$, is evaluated within a Fermi gas
model.  Using the impulse approximation, one evaluates the total
nuclear decay rate as an incoherent sum over the $NN $ pair integrated
cross sections weighted by the nucleon momentum distributions.  The
momentum integral folds the elementary reaction cross section with the
nuclear momentum distributions of nucleon pairs.  The resulting bound
reads as~\cite{goity}, $\vert \l ''_{121}\vert < 10^{-15} {\cal
R}^{-5/2},$ in terms of the auxiliary mass ratio parameter, ${\cal R}
= {\tilde \L / (m_{\tilde g} m_{\tilde q}^4)^{1/5} } $, where the
parameter $\tilde \L $ describes a hadronic mass scale introduced
through a dimensional analysis estimate for the hadronic and nuclear
transition matrix elements.  We note, for the sake of comparison, that
the coupling constant bound obtained within the direct
approach~\cite{masiero}, $ \l ''_{121}< 5. \times 10^{-16} {\cal
R}^{-5/2}$, lies quite close to the present bound.  Varying the
auxiliary parameter ${\cal R}$ inside the interval, $(10^{-3} \ - \
10^{ -6} )$, one finds a coupling constant bound varying inside the
range, $\vert \l ''_{121}\vert < [10^{-7} \ - \ 10^{0}]$. In spite of
the strong dependence on the hadronic and nuclear structure inputs,
one can quote the preferred estimates for the coupling constant bounds
as~\cite{goity}: $ \vert \l ''_{121}\vert < 10^{-6}, \ \vert \l
''_{131} \vert < 10^{-3}, $ for the choice of superpartner mass $
\tilde m = 300 \ \text{GeV} $.


\section{Renormalization group scale evolution}
\label{secin2}

The quantum field theory renormalization effects have a strong impact
on the phenomenological properties of the MSSM. The importance of
tying together physics at widely different distance scales appears
most vividly within the supergravity framework for supersymmetry
breaking, although it is clearly not restricted to just that case.
The renormalization corrections from the gauge and Yukawa interactions
impose useful constraints on the RPV interactions, which, in turn, may
significantly affect the renormalization group structure of the MSSM.
In the present section, we discuss this two-sided connection between
broken R parity symmetry and the renormalization group by reviewing
the following four main issues,  dealing with the constraints from
perturbative unitarity, the infrared fixed points, the grand
unification of gauge interactions and the supersymmetry breaking
effects.  The discussion is developed within the supergravity
framework and is mainly focused on the RPV trilinear interactions.

\subsection{Perturbative unitarity constraints}

The condition that the renormalization group scale evolution is free
from ultraviolet divergences imposes perturbative unitarity or
triviality constraints on the RPV coupling constants. One of the
earliest studies, due to Brahmachari and Roy~\cite{brahmarchi}, dealt
with the baryon number violating interactions restricted to those
involving the maximal number of third generation particles.  The
derived bounds, $ [\l '' _{313} , \ \l '' _{323}] < 1.12$, depend
weakly on $\tan \b $ and increase smoothly with the input value for
the top quark mass with a divergent Landau pole appearing at, $ m_t
\approx 185 \ \text{GeV} $.  The unitary bounds for the other coupling
constants $\l '' _{ijk} $ are discussed by Goity and Sher~\cite{goity}
in a simplified approach using decoupled renormalization flow
equations for specific quadratic combinations of the coupling
constants.  Under the single coupling constant dominance hypothesis,
the resulting bounds are found to vary inside the range, $\l '' _{3jk}
< [1.10 \ - \ 1.25] $.  Owing to the repulsive nature of the Yukawa
interactions, one expects these bounds to strengthen as one relaxes
the single coupling constant dominance assumption.

The renormalization group formalism for the case including the RPV
interactions is discussed by Allanach et al.,~\cite{review99,allner}
in the minimal supergravity grand unification framework.  Useful
results for the anomalous dimensions valid up to two-loop order are
provided in this work.  The maximal RPV coupling constants compatible
with perturbative unitarity assume the updated values, $\l_{323}
(m_t)=0.93, \ \l' _{333} (m_t)=1.06, \ \l '' _{323} (m_t)=1.07 $, for
$\tan \b =5$, with a slow decrease in the coupling constant values
taking place upon increasing $\tan \b $.

\subsection{Infrared fixed points}

The infrared fixed points of the renormalization group flow equations,
$\dh G _I /\dh t = \b _I (G) , \ [t= -\log Q^2 ] $ for the set of
coupling constant parameters, $G _I$, are defined as the zero
solutions for the beta functions equations, $\b _I (G _{J \star } )
=0$. These correspond to values towards which the coupling constants
are focused or attracted in the infrared limit, $t \to \infty $. In
the SM or MSSM cases, one is often satisfied with the quasi-fixed
points solutions exhibiting the focusing property only for the ratios
of Yukawa coupling constants to the color gauge coupling
constant~\cite{jack98}.

Solutions for the Yukawa coupling constants featuring an infrared
quasi-fixed point behaviour continue to exist as one switches the RPV
interactions.  The non-trivial fixed points are of interest in that
they set absolute upper bounds on the coupling constants.  Requiring
the lower bound on the top quark mass, $ m_t > 150 \ \text{GeV} $,
excludes domains in the parameter space for $\l _t$ and the RPV
coupling constants~\cite{brahmarchi}.  Upon introducing the RPV
coupling constants one at a time, there does arise solutions with
simultaneous quasi-fixed points for $\l_t$ and/or $\l_b$ and for the
RPV coupling constants~\cite{fp2}. The fixed point values are $ (i)\
\l_t \simeq 0.94, \ \l ''_{323} \simeq 1.18 ; \quad (ii) \ \l_t \simeq
1.07, \ \l '_{333} \simeq 1.07; \quad (iii)\ \l_t \simeq 1.16,\ \l
_{233} \simeq 0.64 $ in the regime of low $\tan \b = O(1)$ and $ \l_t
\simeq 0.92, \quad \l_b \simeq 0.88, \ \l ''_{323} \simeq 1.08 $ in
the regime of large $\tan \b = O(30)$.  As the couplings $ \l , \ \l '
, \ \l '' $, are successively switched on, the top quark Yukawa fixed
point coupling changes as, $ \l_t = 1.06 \to 1.06 \to 0.99 $ at small
$\tan \b $.  For large $\tan \b \simeq m_t/m_b \approx 35 $, the top
and bottom quark Yukawa fixed point couplings change as, $ \l _t =
1.00 \to 1.01 \to 0.87 ; \ \l_b = 0.92 \to 0.78 \to 0.85 $, with the
corresponding RPV fixed point couplings given by~\cite{fp2,fp1}, $\l
'_{333} = 0.71 , \ \l ''_{323} = 0.92$.  Further discussions of the
fixed point predictions can be found in Ref.~\cite{rengroup}.

The fixed point stability property is expressed by the requirement
that the matrix of beta functions derivatives in the parameter space,
$ B_{IJ} = \dh \b _I(G)/\dh G_J \vert _{G_I = G_{I \star } } $, has
all its eigenvalues at the fixed point with positive real parts.  The
stability property is a necessary condition for the validity of a
fixed point.  The presence of RPV interactions modifies the infrared
stability properties of the Yukawa coupling constants.  The study by
Ananthanaranayan and Pandita~\cite{ananpandita} relies on certain
working assumptions distinct from those used in the above quoted
studies.  For the baryon number violating interactions, the
simultaneous non-trivial fixed point for $\l ''_{323} $ and the
regular Yukawa coupling constants $\l _t $ and $ \l_b $ is found to be
stable.  On the other hand, the trivial fixed point with $\l ''_{323}
\to 0 $ and finite $\l _t , \l_b $, which corresponds to the quasi
fixed point discussed in the previous paragraph, is found to be
unstable.  The fixed point stability requirement translates under
these circumstances into the lower coupling constant bound, $\l
''_{332} > 0.98$.  We note, however, that the procedure used in this
study predicts a top quark mass, $m_t (m_t) = 70 \ \text{GeV} \times
\sin \b $, which lies well below the observed value, in contrast to
the satisfactory prediction holding in the MSSM, $m_t (m_t) = {v \over
\sqrt 2} \sin \b \l _t (m_t) \simeq 190 \ \text{GeV} \times \sin \b $.
For the lepton number violating interactions, neither the trivial nor
the non-trivial fixed point solutions for the RPV coupling constants,
$ \l _{233} , \ \l '_{333} $, are infrared stable.  A study of the
quasi-fixed point solutions in the case where several RPV coupling
constants are included simultaneously is presented by Mambrini and
Moultaka~\cite{mambrini02}.  Under certain conditions, one finds that
the fixed point RPV couplings become infrared repulsive by being
attracted to vanishing values at low energies.

\subsection{Gauge  interactions unification constraints}

The feed-back effect of the RPV interactions on the gauge and Yukawa
interactions has significant implications on the grand unification
constraints.  Using the two-loop renormalization group equations,
Allanach et al.,~\cite{review99,allner} find that the three third
generation related RPV interactions affect in a negligible way, by
less than $ 5 \% $, the value of the unified gauge coupling constant,
$\a _X ^{-1} \simeq 24.5 $, but change more significantly the value of
the unification scale, $M_X \simeq 2.3 \times 10^{16} \ \text{GeV} $,
which can get reduced by up to $ 20\% $.  For the $SU(5)$
supersymmetric grand unification scenario, involving a predominant top
quark Yukawa coupling constant at the unification scale, $ \l_t >>
(\l_b ,\ \l_\tau )$, the presence of RPV interactions induces a strong
downward shift in the top quark fixed point coupling.  The predicted
values for $\tan \b < 30 $ are~\cite{fp2,fp1}, $\l_t (m_t) =0.88, \ \l
_{233} = 0.90, \ \l '_{333} = 1.01 , \ \l ''_{323} = 1.02$.  For the
$SO(10)$ or $ E_6$ supersymmetric grand unification scenarios,
involving top and bottom quark and tau lepton Yukawa couplings of
comparable sizes at the unification scale, $ \l_t\sim \l_b \sim
\l_\tau $, the top quark fixed point coupling is pulled further
down~\cite{brahmarchi}, $\l_t (m_t) =0.65$. The large $\tan \b = O(50)
$ regime relevant to this scenario is known, however, to suffer from
severe naturalness problems~\cite{tanbeta} associated with the
electroweak symmetry breaking and generation non-universality effects.

The RPV interactions can have an important impact on the issue of
bottom-tau unification. Requiring the unified boundary condition on
the ratio of Yukawa coupling constants, $ R_{b/\tau } (M_X) \equiv
{\l_b (M_X ) \over \l_\tau (M_X ) } =1$, and the input experimental
value of the bottom quark mass, $ m_b(m_b)$, imposes a strong
correlation between the admissible values of $ m_t $ and $\tan \b$.
For the MSSM, the measured value of $ m_t$ singles out two narrow
intervals for $ \tan \b $ around $ 1 $ and $ 50 $, respectively.  Once
one includes the RPV interactions, the $m_t \ - \ \tan \b $
correlation gets significantly relaxed.  Independently of $ \tan \b $,
the Yukawa coupling constant ratio, $ R_{b/\tau } (M_X)$, can cross
unity in the allowed range of variation for the coupling
constants~\cite{review99,allner} $ \l '_{333} $ or $ \l '_{323}$.  The
implications of RPV bilinear interactions on bottom-tau unification
have been examined by Diaz et al.,~\cite{diaz99}. Switching on finite
values for the relevant third generation coupling constant and
sneutrino VEV parameters, $\mu_\tau $ and $ v_\tau $, achieves a
satisfactory top-bottom mass hierarchy at any value of $\tan \b $,
rather than selecting discrete values.  The scan over the MSSM
parameter space, subject to constraints on the $b$ quark mass, $ m_b
(m_b) \simeq 4.3 \ \text{GeV} $, and the alignment condition, $ \mu
/v_d \approx \mu_\tau / v_\tau $, aimed at bounding the neutrino mass,
constrains the sneutrino VEVs to the allowed range, $ \vert v_\tau
\vert < 50 \ \text{GeV} $.  The bounds get stronger in the
top-bottom-tau unification case, with the corresponding allowed range
given by, $ \vert v_\tau \vert < 5 \ \text{GeV} ,$ subject to the
allowed interval for the ratio of Higgs boson VEVs, $ 54 < \tan \b <
59, $ which extend somewhat above the maximal value predicted in the
MSSM case.

The familiar bottom-up approach for the MSSM, based on the
supergravity mediated supersymmetry breaking and gauge interactions
unification, allows inferring useful information on the underlying
microscopic theory from the physical constraints at the low energy
scales.  This approach has a natural generalization in the context of
broken R parity symmetry. Upon solving the renormalization group scale
evolution equations for the RPV interactions using as boundary
conditions the various existing individual coupling constant bounds,
taken one at a time, Allanach et al.,~\cite{review99,allner} find that
the corresponding bounds at the gauge unification scale are
strengthened by factors $2\ -\ 5$.  Similar studies, taking into
account the physical constraints on the superpartner particles mass
spectrum and the LSP decays, are presented in
Refs.~\cite{allanach03,godsz03}.

\subsection{Implications  on soft supersymmetry breaking interactions}

The consideration of renormalization group effects in the presence of
supersymmetry breaking leads to several novel constraints on the RPV
interactions.  Starting with finite trilinear RPV couplings at the
gauge interactions unification mass scale, one can generate at low
mass scales finite bilinear RPV couplings having the ability to
produce finite sneutrino VEVs. A useful probe of the lepton number
violating contributions is offered by the contributions to the
neutrinos Majorana masses, $ m_{\nu _i}$.  The comparison with the
experimental limits on the neutrino masses yields the coupling
constant bounds~\cite{decarlosl}, $ \l _{i33} (M_X) < O(10^{-3}) $ and
$\l '_{i33}(M_X) <O(10^{-3})$.  The radiative corrections to the
running RPV coupling constants can also give rise to indirect
contributions to the lepton and quark flavor changing radiative decay
modes. The resulting effect on the prototype decays, $ \mu ^-\to e ^-
+\g ,\ b \to s + \g $, are found~\cite{decarlosl,decarlosq} to
dominate over the distinct direct one-loop contributions discussed
previously.  The indirect renormalization effects are in fact
sub-dominant relative to the direct ones only in the case of the
neutral mesons mass splittings~\cite{decarlosq}. However, a clear
quantitative understanding of the indirect effects is still lacking,
owing to the large number of free parameters and to the cancellations
occurring between different contributions.

The RPV renormalization corrections within the supergravity grand
unification framework~\cite{decarlosq} give attractive type
contributions to the scalar particles masses which may drive certain
scalar superpartners to acquire negative (tachyonic) mass squared,
implying a potential de-stabilization of the vacuum.  The constraints
from the experimental limit on the sneutrinos masses, $ m_{ \tilde \nu
} > 37 \ \text{GeV} $, yield the coupling constant bounds, $\l '_{ijk}
(M_X)< 0.15 $, valid for the various flavor configurations. The
corresponding bounds at the electroweak mass scale read as, $\l '
_{ijk} < 0.3$.

The radiative corrections from the $\l ''$ interactions can
significantly enhance the mass splitting between the left and right
chirality squark masses.  The third generation down-squark $\tilde
d_R$ mass is reduced while the $\tilde d_L$ mass is left untouched,
thus resulting in a lowered down-stop quark mass~\cite{bastero97}.
The broken R parity symmetry explanation for the deep inelastic
scattering anomaly reported by the HERA collider Collaborations,
requires RPV coupling constants, $[ \l '_{121},\ \l '_{131}] = (0.03 -
0.04 ) / B^\ud _{ \tilde c \to d e } $, along with the squark mass
value, $m_{\tilde q} = 200 \ \text{GeV} $.  The mutual consistency of
these results within the renormalization group approach is examined by
Cheung et al.,~\cite{chedatdicus}.  While the requisite small squark
mass value is difficult to realize in the gauge mediated supersymmetry
breaking framework, it may be comfortably accommodated in the
supergravity framework.

\section{Cosmology and astrophysics}
\label{secxxx6}
The existence of a broken R parity symmetry may have far reaching
implications on the astroparticle physics in the supersymmetry
context. Our current understanding of the cosmology and the available
astrophysics experimental observations yield several useful
constraints on the RPV interactions.  The main three issues of
interest are: (1) The availability of the unstable lightest
supersymmetric particle (LSP) as a weakly interacting massive particle
(WIMP) candidate for the Universe dark matter missing mass; (2) The
dilution of the cosmic baryon or lepton asymmetries by the RPV baryon
or lepton number non conserving thermal processes; (3) The late
production of the cosmic baryon or lepton asymmetries by the RPV
interactions.  We discuss below these three issues in succession.

\subsection {\bf  Implications of LSP instability} 

The case of an absolutely stable LSP cosmic relic is not easily
distinguished from that of a long lived particle whose lifetime, $\tau
_{LSP} $, exceeds the Universe age, $\tau _{LSP}> t_0$.  The RPV
trilinear interactions can initiate the decay of the neutralinos and
sneutrinos at tree level and the radiative decay of the neutralinos at
one-loop level, in terms of the Feynman graphs G.1-G.2 and G.3-G.5 of
Figure~\ref{fig3}, respectively.  The bounds on coupling constants
implied by the non-observation of the LSP decay products in a detector
volume of $1$ meter linear dimension size are given by: $ (\sqrt 3
\vert \l ' _{ijk} \vert , \vert \l _{ijk} \vert )_{sleptons} > 10^{-8}
, \ (\sqrt 2 \vert\l ''_{ijk} \vert , \vert \l '_{ijk} \vert
)_{squarks} > 10^{-8} , \ (\sqrt 6 \vert \l ''_{ijk} \vert , \sqrt 3
\vert \l '_{ijk} \vert , \vert \l _{ijk} \vert )_{inos} > 5.\times
10^{-7} ,$ where we have quoted order of magnitude bounds using the
superpartner mass values, $ m_{\tchi } = m_{\tilde f} = O(100) $ GeV.
Since a rescaling factor on the RPV coupling constants of $
O(10^{-13})$ is involved in going from these laboratory detection
bounds to the cosmological stability bounds, one can express the
condition for the availability of an LSP dark matter candidate by the
coupling constant bounds: $ \vert \hat \l \vert = [\vert \l \vert ,
\vert \l ' \vert ,\vert \l'' \vert ] < (10^{-21} \ - \ 5. \times
10^{-20}) $.  Requiring that the cosmologically unstable LSP (with $
\vert \hat \l _{ijk} \vert > 10^{-20} $) decays before the
nucleosynthesis, so as not to disrupt the light element observed
abundances, closes the window of allowed coupling constants out to the
values, $\vert \hat \l _{ijk} \vert > 10^{-12} $.  Comparing the
maximal allowable lifetime for a neutralino LSP with the trilinear RPV
contribution to $\tau _\tchi $ leads to a lower bound on the coupling
constants of generic form~\cite{kim98}, $\vert \hat \l _{ijk} \vert >
O(10^{-12} ) .$

The terrestrial abundance of anomalous nuclear isotopes gives useful
information on the existence of stable or long lived colored
particles, such as squarks or gluinos, and on other strongly
interacting massive particles (SIMP), that might have been captured by
heavy nuclei. The most favorable cases are those of Au or Fe nuclei.
By allowing the LSP to decay, the RPV interactions have the ability to
deplete the expected abundance of the corresponding anomalous isotope
or SIMP cosmic relic.  Using the current experimental bound on the
SIMPs~\cite{javorsek02}, $\O _S < [10^ {-6}\ - \ 10^{-4}] $, along
with the information on the exposure times and nuclear capture cross
sections of squark or gluino cosmic relic particles, allows one to
infer the lower coupling constant limits, $ [\vert \l ' _{ijk} \vert
,\ \vert \l '' _{ijk} \vert ] > (10^{-21} - 10^{-20} ) \ \tilde q
^{-1/2} $, for superpartner masses inside the range, $ (2.8\ - \ 100 ) $ \
\text{GeV}.  A related constraint arises from the LEP collider
experimental data, reflecting on the absence of heavy tracks and
unexplained missing energy, which rule out stable squarks of mass,
$m_{\tilde q } < 100 $ \ \text{GeV}.  Based on the comparison with the
squarks RPV two-decay modes, Berger and Sullivan~\cite{berger03} infer
from this result the lower coupling constant bounds, $[\vert \l '
_{ijk} \vert ,\ \vert \l '' _{ijk} \vert ] > [5.\times 10^{-9} \ - \
10^{-7}] $.

The three main astrophysical observables, associated with the
photo-dissociation reactions of light nuclei, the distortion of
microwave and gamma ray background radiation energy distributions, and
the cosmic fluxes of muons and neutrinos, yield useful constraints on
the RPV interactions in the case of an unstable LSP decaying before
the present time, $\tau _{LSP}< t_0$.  The LSP decay products can
distort the primordial cosmic abundances of light nuclei through the
collisions with shower radiation and can also contribute to the flux
of upward going muons observed in underground detectors.  For a relic
particle $ X$ of mass $ m_{X}$, lifetime $ \tau _{X}$ and number
density $ n_{X}$, these two
observables~\cite{ellis91,dimoes89,lindley} establish correlations
between the mass density and lifetime of the unstable massive relic,
expressed in terms of upper bounds on the mass density variable, $
m_{X} n_{X} /n_\g $, as a function of $ \tau _{X}$.  Since the product
$ m_{X} n_{X}$ is a generically monotonic decreasing function of $
\tau _{X}$, it follows that $ n_X$ increases when $\tau_X$ decreases
and vice versa.  For the three values of the LSP number density, $
n_{\tchi }/ n^c_{\tchi } = [10^{+2}, \ 10^{0},\ 10^{-2}]$, with $
n^c_{\tchi } $ defined as the relic density for a strictly stable LSP
whose contribution to the present day Universe energy density could
have attained the critical value, $\O _ {\tchi } \simeq 1 $, the
constraints from primordial nuclei disruption by hadron showers
yield~\cite{abundance} the bounds on the neutralino LSP lifetime and
RPV coupling constants, $\tau_{\tchi } < [10^{-1} , \ 10^{0}, \ 10^3 ]
\ \ \text{s} \ \ \Longrightarrow \ \vert \hat \l \vert > 4.5 \ \times
[10^{-10.5} , \ 10^{-11}, \ 10^{-12.5} ] \ {\tilde f} ^2 \tchi ^{-5/2}
$. The constraints from the distortion of the upward going muon flux
yield~\cite{abundance} the bounds, $\tau_{\tchi } > [10 ^{19}, \ 10
^{17}, \ 10 ^{15}]\ \text{y} \ \ \Longrightarrow \ \vert \hat \l \vert
< 4.5 \ \times [ 10^{-23.5}, \ 10^{-23}, \ 10^{-22.5}] \ {\tilde f} ^2
\tchi ^{-5/2} .$

For a massive LSP neutralino cosmic relic of lifetime, $\tau_\tchi
\simeq t_0$, the comparison with the experimentally measured photon
flux and positron flux in our galaxy imposes the following bounds on
all the trilinear and bilinear RPV coupling
constants~\cite{berezinsky98}: $ \vert \hat \l _{ijk} \vert < 4.
\times 10^{-23} \ \tilde f ^{2} \tchi ^{-9/8} (1 \ \text{GeV} /m_f
)^\ud , \ \mu_i < 6. \times 10^{-23 }\ N_{1l} ^{-1} \tchi m^{-7/4} \
\text{GeV} , \ [l=3,4] $ where $ m_f$ is the emitted fermion mass and
$N_{1l}, \ [l=3,4] $ denote the higgsino field component of the lowest
lying neutralino mass eigenstate.

The RPV hadronic three-body decays of long lived massive LSP
neutralinos of lifetime, $\tau _{\tchi } > t_0$, can produce energetic
antiprotons which can be detected after diffusing through the galaxy
as a low energy (few \ \text{GeV}) component of the hadronic jets in
the cosmic particles flux.  Baltz and Gondolo~\cite{baltz98} infer
from the experimental data a correlation between the neutralino
lifetime and its cosmic energy density, $ \O_\tchi $, which translates
into allowed intervals for the RPV coupling constants, $ \tau_\tchi >
7.9 \times 10^{28} N_{jets} { \O_\tchi h^2 \over m_\tchi / \
\text{GeV} } \ \text{s} \quad \ \Longrightarrow \ 2. \times 10^{-18} <
\vert \l' _{ijk} \vert < 2.  \times 10^{-15}, \ 1.\times 10^{-18} <
\vert \l'' _{ijk} \vert < 1.  \times 10^{-15}. $ The quoted coupling
constant bounds apply to all the generation indices, with the
exception of $ \l ''_{3jk} $ in the specific case of a neutralino
lighter than the top quark.

The cosmological constraints on gravitinos in the MSSM are
known~\cite{weinberg83} to disallow the following ranges for the
gravitino mass and supersymmetry breaking scale, $ 1 \ \text{keV} <
m_{\tilde G} < (10. \ \text{ MeV} \ -\ 10^4 \ \text{GeV} ) \
\Longrightarrow \ 10^6 \ \text{GeV} < \sqrt {F_S} = (\sqrt 6 m_{\tilde
G} M_\star )^\ud <(10^8 \ \ - \ 10^{11} ) \ \text{GeV} $, where the
effective quantum gravity mass scale is related to the Planck mass by,
$ M_\star ={1\over \kappa } = {M_P\over \sqrt { 8\pi } } = 2.4 \times
10^{18} \text{GeV} $.  For a massive gravitino LSP, the problem of the
cosmic relic overabundance and late disintegration occurring between
the nucleosynthesis and recombination eras is generally resolved by
invoking an early inflationary epoch. An attractive resolution based
on an unstable gravitino might, however, be offered by R parity
symmetry violation.  The possibility rests on a comparison of the
gravitino decay rate with the Universe expansion rate.  The modified
bounds on $ m_{\tilde G}$ and $ F_S $ implied by the two-body lepton
number violating decay modes, $ {\tilde G } \to \nu + H^0_u,$ which
occur in the MSSM extension including a right-handed chirality
neutrino supermultiplet, $\nu^c $, were initially studied by
Weinberg~\cite{weinberg83} in toy models featuring explicit and
spontaneous R parity symmetry breaking.  The recent study by Takayama
and Yamaguchi~\cite{taka00} of the gravitino radiative two-body decay
mode, $ \tilde G \to \g + \nu $, initiated by the bilinear RPV
interactions, shows that the corresponding partial lifetime, obtained
by fitting the bilinear coupling constant $\mu _i$ to the atmospheric
neutrino oscillation data, is long enough to allow for an effectively
stable gravitino dark matter candidate. A different conclusion holds
if one focuses~\cite{moreau02} on the gravitino three-body
disintegration modes into quarks, leptons or a combination of quarks
and leptons initiated by the trilinear interactions.  Imposing the
current bounds on the full set of trilinear coupling constants, under
the assumption that the gravitino and scalar superpartner masses do
not exceed $ O(10) $ TeV, one finds that the gravitinos could easily
have decayed before the present epoch but fortunately not earlier than
the nucleosynthesis epoch.

The possibility that massive gravitinos might dominate the Universe
dark matter has led Hall et al.,~\cite{halltasi,dimo88} to envisage an
interesting, somewhat speculative, scenario based on a Universe that
was once baryon matter dominated.  An abundant, relatively stable
gravitino cosmic relic, could then survive until the end of
nucleosynthesis, $ t \simeq 10^3 s$, with the ability to rebuild the
light nuclear elements through its interactions with the various
particle species and decay products initiated by the lepton number
violating interactions, $\l $ and $ \l' $.  For the case of a third
generation sneutrino LSP, $\tilde \nu_\tau $, in association with a
RPV superpotential of restricted form, $ W= \l _{1 3 1} L_1 L_3 E^c _1
+ \l _{2 3 2} L_2 L_3 E^c _2 + \l '_{3mn} L_3 Q_m D^c_n$, one can
adequately suppress the gravitino baryonic decay branching fraction, $
r_B$, by restricting the ratios of coupling constants to values of
order, $ \vert \l '_{3mn} /\l _{131} \vert \simeq 0.1, \ \vert \l
'_{3mn} /\l _{232} \vert \simeq 0.1 $.

\subsection {\bf  Cosmological baryon asymmetry dilution} 

The realization of baryogenesis puts three necessary
conditions~\cite{sakharov67,cosmobarev} on the particle physics: CP
violation, $ B$ and/or $L$ number violation and an irreversible
(non-adiabatic) phase in the Universe evolution.  The reason why the
ratio of the baryon to photon number densities of the Universe has
settled today at the value, $\eta \equiv {n_B \over n_\g } \simeq (4.\
- \ 7.) \times 10^{-10 } $, has not yet found a satisfactory answer
despite numerous attempts.

We discuss first the baryon asymmetry protection against dilution
effects by the baryon and lepton number violating reactions initiated
by the RPV interactions. The non-erasure condition is approximately
expressed by the out-of-equilibrium inequality, $ \G < H \approx 20
T^2/M_P$, requiring that the baryon number violating reactions rates
stay below the Universe Hubble expansion rate.  The semiquantitative
treatment of this condition needs a careful calculation of the
transition rates associated to the various RPV $ 2 \to 1$ and $ 2 \to
2 $ body reactions.  The initial study by Bouquet and
Salati~\cite{salati}, focused on the $ 2 \to 2 $ body reactions only,
obtained the baryon number violating coupling constants bounds, $\l ''
_{ijk} < 3.1 \times 10^{-7} \ \tilde m ^\ud $.  Specializing instead
to the regime of high temperatures, for which the $ 2 \to 1 $ body
reactions are predominant, Fischler et al.,~\cite{fischler} obtained
similar bounds for the lepton number violating coupling constants,
$[\l _{ijk} , \ \l '_{ijk}] < 10^{-8} \tilde m ^\ud $.  By accounting
for the sphaleron induced $B+L$ violating anomalous reactions,
Campbell et al.,~\cite{campbell} confirmed that coupling constant
bounds of same order continue to hold true for the full set of RPV
coupling constants, $\vert \hat \l _{ijk}\vert = [\vert \l _{ijk}
\vert , \ \vert \l '_{ijk} \vert , \ \vert \l ''_{ijk} \vert ] <
10^{-8}$.  A comprehensive study of the thermal equilibrium
conditions, accounting for the particles mass effects, is presented in
the work by Dreiner and Ross~\cite{dreiross}.  The coupling constant
bounds at high temperatures, $\vert \hat \l _{ijk} \vert < 1.58 \times
10^{-7} (m _{\tilde f} /100 \ \text{GeV} ) ^\ud $, are replaced in the
regime of low temperatures, $ T< m_{\tilde f}$, by the bounds, $\vert
\hat \l _{ijk} \vert < 1.6 \times 10^{-6} (m _{\tilde f} /100 \
\text{GeV} ) ^\ud $.  The compatibility of the constraints from
non-erasure of the cosmic baryogenesis with the neutrino oscillation
data is examined in a recent work by Akeroyd et al.,~\cite{acker03}.

One must not conclude, however, that the above non-erasure bounds
apply simultaneously to all the RPV coupling constants irrespective of
the generation indices. In order to protect the cosmic baryon
asymmetry it suffices to impose the out-of-equilibrium constraints
only on the interactions violating a fixed generation lepton number,
or some linear combination of the lepton numbers, $L_i$.  The
effective conservation of the remaining lepton numbers sets additional
chemical equilibrium conditions which should prevent the asymmetry
erasure.  Explicitly, one may impose the out-of-equilibrium
constraints only on the $L_1 = L_e$ electron number violating
interactions, $ L_1 Q_j D^c_k + L_1 L_j E^c_k$, while leaving all the
remaining $L_{\mu , \tau } , \ B_{u,d,s}$ number violating
interactions unconstrained~\cite{dreiross}.

\subsection {\bf  Cosmological baryon asymmetry production} 

The broken R parity symmetry interactions in the MSSM might play an
active r\^ole in the baryogenesis process itself.  The requisite
violation of $B $ or $L $ numbers and CP would take place through the
out-of-equilibrium RPV decays into quarks and leptons of squarks,
gauginos, gravitinos, or also of the axino and saxino superpartners of
axions.  The envisaged scenarios differ in the type of relic
superparticle, which may or not be the LSP, and the mechanism
responsible for its eventual regeneration and decay.  The various
proposals focus on a late inflation while assuming a strongly first
order electroweak symmetry phase transition involving the nucleation
of vacuum bubbles.  Other specific conditions are needed in order to
produce or maintain significant abundances for the superparticle
particles close to the MeV range of temperatures appropriate to the
nucleosynthesis and neutron-proton freeze out epochs.

The proposal by Dimopoulos and Hall~\cite{halldimo,halltasi,halldimo1}
discusses a low temperature baryogenesis taking place through the RPV
decay of squarks produced during a late inflation era.  Cline and
Raby~\cite{cline91,cline291} consider instead the case of a massive
gravitino, $ m_{\tilde G} = O(10) $ TeV, with a low decay temperature
lying above the nucleosynthesis or neutron-proton freeze out
temperature, $ T_{\tilde G d} > T_{n/p} \simeq 1 $ MeV.  The
alternative mechanism invoking the decays of axino or saxino relic
particles may be more promising, in view of the lower decoupling
temperature and shorter lifetime of the axino and saxino
particles~\cite{moller92}.  A scenario for an axino cool dark matter
candidate, based on the combined violations of the Peccei-Quinn and
bilinear R parity symmetries, is discussed by Chun and
Kim~\cite{chunkim99}. The cosmic relic density of the axino, acting as
a sterile neutrino, is produced non-thermally through the
axino-neutrino resonant oscillation transition amplitude.

For an electroweak phase transition of first order kind, an efficient
production of supersymmetric particles can take place during the
collision of vacuum bubbles.  In the proposal of Masiero and
Riotto~\cite{masiero92}, the leptogenesis arises through the decay of
neutralinos produced with significant number densities in the
collisions of vacuum bubbles. Since a very large energy is released in
the bubbles collisions, one may envisage that other massive
superparticles are also produced copiously during the electroweak
phase transition.  The proposed particle production mechanism would
then hold for any superpartner particle provided it is coupled to the
Higgs bosons.  When applied to the sleptons or squarks, the same
scenario type as discussed above is found by Sarkar and
Adhikari~\cite{sarkarba} to yield an acceptably large cosmic baryon
asymmetry.  Another simple scenario has been proposed by Adhikari and
Sarkar~\cite{adhikari} for a baryogenesis occurring after the
electroweak phase transition through the decay of LSP neutralinos in
the baryon number violating three-body modes, $ \tchi^0 \to u_i+ d_j+
d_k, \ \tchi^0 \to \bar u_i + \bar d_j + \bar d_k $, initiated by the
RPV interactions.  Multamaki and Vilja~\cite{multa98} discuss a
scenario for baryogenesis in which a first order electroweak phase
transition is accompanied by a spontaneously broken R parity symmetry
realized in the context of the model of Masiero and
Valle~\cite{masierovalle}.

A successful leptogenesis can arise through finite RPV parameters, $
v_\tau , \ \mu _\tau $, associated with the weakly constrained $L_\tau
$ number violating sector of the bilinear R parity violation.  The
lepton number asymmetry originates~\cite{hamb00} from the CP-odd
interference terms between tree and one-loop contributions to the
lepton number violating decay modes, $\tilde W ^0 \to \tau _R ^\pm + H
^\mp $.  The actual realization of this scenario hinges on specific
assumptions requiring the inequality between the gaugino masses, $ M_1
>M_2 $, as needed to suppress the $\tilde B$ decay modes for
temperatures, $T< M_1$, and the presence of non-holomorphic soft
supersymmetry breaking term, $ H_u ^\dagger H_d \tilde l ^ \star _R $,
as needed to induce the requisite $\tilde \tau _R - H^-$ mixing.

\section{Discussion of indirect bounds  on trilinear couplings}
\label{secin4}


Our purpose in this section is to assess from a more global
perspective the physical relevance of the indirect bounds on the
trilinear RPV coupling constants.  For convenience, we have collected
the strongest, most robust, single and quadratic coupling constant
bounds in Table~\ref{table1} and Table~\ref{table2}, respectively.
Self-evident abbreviations are used to identify the observables used
in inferring the various coupling constant bounds.  The contents of
these tables essentially recapitulate the results already quoted in
the main discussion.  Where certain slight differences with respect to
the results quoted in the main discussion appear, the reason is due to
the fact we have selected in the tables $1 \s $ level bounds and in
the text $2 \s $ level bounds.  Our presentation is far from
exhaustive.  A much larger number of bounds are obtained in the
existing works which we have been unable to quote faithfully in the
present review.  These results can be consulted from the original
works, of course, and also from the recent
reviews~\cite{royrev,allanach03}.  We shall discuss first the
phenomenological implications of indirect bounds and present next a
general discussion of the main phenomenological constraints on the RPV
trilinear interactions.


We begin with general remarks on the uses of indirect bounds. It is
important to ask first what might be considered as the natural values
of the trilinear coupling constants.  (For convenience, we denote
these collectively as $\hat \l_{ijk}$.)  In the absence of any
symmetry, the anticipated natural values are of order unity or of same
order as the gauge interactions coupling constant, $ e =\sqrt { 4\pi
\a } $.  If one assumed instead a hierarchical structure with respect
to the quarks and leptons generations, analogous to that exhibited by
the regular R-parity conserving Yukawa interactions, an educated guess
could be, for instance, to use $\hat \l_{ijk} = O((m_im_jm_k /v
^3)^{1/3}) $, where $ m_j$ denote the relevant quarks or leptons
masses.  A global examination of Table~\ref{table1} shows that the
individual coupling constant bounds fall in an interval of values
$O(10^{-1}) \ - \ O(10^{-2} )$ roughly compatible with the above
quoted dimensional analysis estimate.  A variety of alternative
structures for the generation dependence are also suggested upon
appealing
to discrete symmetries or to models inspired by the grand or string
unification theories.


What are the general implications on theory that can one draw from the
existing bounds? The most severe constraints are clearly those arising
from single nucleon decay on the coupling constant products, $ \l
^{'\star } \l '' $ and $\l ^{\star } \l ''$, which disfavor the
simultaneous violation of $B$ and $L$ numbers.  Ready solutions to
forbid a coexistence of $B$ and $ L$ number non-conservation are
available in terms of both flavor dependent and independent discrete
symmetries.  Since the bulk part of the indirect bounds arises from
quark and lepton flavor changing processes, one might ask whether the
corresponding constraints hint at horizontal or flavor blind
symmetries.  The need for flavor symmetries is suggested by the
observation that stronger bounds are set on the coupling constants
with at least two indices belonging to the first and second light
generations.  Equivalently, one observes in general that the coupling
constants with two or three third generation indices are more weakly
constrained.  These properties might just reflect, however, the
paucity of experimental data for the heavy flavored hadrons or
leptons.  Restrictions of different nature would be imposed by the
flavor blind symmetries, which may allow for an hierarchical structure
in the RPV coupling constants to coexist with a degeneracy with
respect to the quark and lepton generations.

The indirect bounds establish a strong correlation between the RPV
coupling constants and the superpartners mass spectrum.  We have set
our reference value for the supersymmetry breaking mass parameter
uniformly at the value, $ \tilde m = 100 \ \text{GeV} $, apart from a
very few exceptions. Most experimental constraints would substantially
relax as the supersymmetry breaking mass scale reaches the TeV order.
The dependence of indirect bounds on the superpartner mass parameters
is, in general, explicitly known.  For the tree level amplitudes, the
single and or quadratic coupling constant bounds scale linearly and
quadratically with $\tilde m$, respectively.  We have exploited this
fact to explicitly include the relevant sfermion particle name in the
quoted bounds.  It is important to keep track of the superpartner
generation index in anticipation of a large mass splitting between
different sfermion generations.  With a variable sfermion generation
the bounds would vary proportionately to the generational mass
splitting.  The one-loop amplitudes have, in general, a weaker
dependence on the sfermions masses, provided one restricts to the
range, $ \tilde m = 100 \ - \ 500\ \text{GeV} $.  In the so-called
more minimal supersymmetric Standard Model~\cite{cohen}, characterized
by light third generation sfermions coexisting with nearly decoupled,
TeV scale, sfermions of the first and second generations, a large
number of the existing individual indirect bounds would become
uninteresting.  The generational structure of the chirality diagonal
and off-diagonal sfermions mass matrices, $\tilde m^2_{RR},\ \tilde
m^2_{LR}$ is also a crucial input in evaluating the RPV contributions.
Deviations from the generation universal structure can modify
substantially the predictions for the neutrinos Majorana mass and
dipole moment matrices and the $n-\bar n$ oscillation amplitude.

The suppression of RPV contributions for large sfermions masses
suffers two exceptions.  The first concerns the renormalization group
corrections originating from the resummation of large logarithms,
which are practically insensitive to the details of the mass spectrum
and depend only on the TeV order of the supersymmetry breaking mass
scale.  The relevant predictions here include the process independent
bounds discussed in Section~\ref{secin2} in connection with the
perturbative unitarity condition and the infrared quasi-fixed points
of the renormalization group scale evolution. Certain process
dependent constraints, discussed in Section~\ref{secxxx3}, also arise
from indirect flavor changing contributions initiated by the
renormalization.  The second exception concerns the physical processes
controlled by higher dimension operators, of which two prototypical
cases are the $\b \b _{0\nu }$ and the $n-\bar n$ oscillation
reactions involving dimension nine operators.  The presence of several
contributions in these amplitudes, involving the exchange of both
sfermions and gauginos, makes the inferred bounds sensitive to the
superpartners mass spectrum as a whole.

The loop dressing effects in the MSSM are very effective in deducing
bounds involving the heavier generations of quarks and leptons. Thus
strong quadratic coupling constant bounds applying to all the
generation indices, $ \vert \l ' _{ijk} \l '' _{i'j'k'} \vert <
O(10^{-9}) $, can be derived on this basis from the single nucleon
decay limit. Should a single lepton number violating coupling constant
$\l ' _{ijk} $ happens to be sizeable, one could then conclude to a
strong suppression of the complete set of coupling constants, $ \l ''
_{i'j'k'} $, and vice versa.

Proceeding now with a closer examination of the results in
Tables~\ref{table1} and \ref{table2}, it appears clearly that the most
robust bounds in order of decreasing importance are those arising from
single nucleon decay, neutrinoless double beta decay, double nucleon
decay, mixing of neutral $ K, \ B$ mesons, and lepton flavor violating
decays of leptons and hadrons.  The strongest single coupling constant
bounds arise in order of decreasing strength from the baryon number
violating $n-\bar n$ oscillation and the two nucleon decay reactions,
the universality tests of neutral and charged current reactions and
the atomic physics parity violation measurements.  The strongest
quadratic coupling constant bounds, ordered according to decreasing
strength, arise from the single nucleon reactions, $ \mu ^--e^-$
conversion, three-body leptons decays and the rare leptonic and
semileptonic decays of $ K, \ B$ mesons.


We examine next the robustness of certain single and quadratic
coupling constant bounds deduced on the basis of the single or double
coupling constant dominance hypotheses. These hypotheses are best
justified if some flavor hierarchies existed between coupling
constants for different quark and lepton generations.  Conversely, the
case of generational degeneracies may cause unexpected cancellations
which would invalidate certain deduced bounds.  The existence of some
correlation between the different RPV coupling constants is suggested
by the fact that the quadratic coupling constant product bounds are in
general more demanding than the product of the corresponding
individual coupling constant bounds. The quadratic constraints thus
appear to include a richer information on the mode of realization of R
parity symmetry than the linear constraints.

One must clearly exercise a critical eye on the model-dependent
assumptions by not treating the various bounds indiscriminately.
Several individual coupling constant bounds deduced from the
universality tests of the charge and neutral current are sensitive to
cancellation effects.  The reason is that the consideration of ratios
of related reaction rates, aimed at removing the model dependence on
hadronic structure inputs, introduces mutually canceling
contributions.  These ratios obtain corrections from different RPV
interactions which often combine together with opposite signs.  The
quadratic bounds often arise from contributions which add up
incoherently, so are less exposed to cancellation effects.
Nevertheless, a few single coupling constant bounds are immune to
invalidating cancellation effects.  These include the robust bounds
deduced from the renormalization corrections in $ G _F $ and $ m_W$,
the forward-backward asymmetry parameter $ A_{FB}$, and the atomic
physics parity violation parameter, $ C_2 (d)$.


A potentially powerful way to reduce the model dependence would be to
consider global studies of the RPV effects encompassing a large enough
body of experimental data.  As a start one could attempt fitting some
suitably chosen subset of coupling constants to an appropriately
selected subset of experimental constraints.  Such an approach is
mandatory in the case of observables depending on several coupling
constants contributing with opposite signs.  While global studies
along these lines are routinely performed in the context of higher
dimensional contact interactions~\cite{bargerz} or mirror
fermions~\cite{londonlanga}, their application to R parity violation
physics has remained so far problematic in view of the coupling
constant proliferation in that case.  Still, certain partially global
studies have been recently reported in the literature regarding fits
to the atomic physics parity violation observables~\cite{barger00} or
the $Z$-boson partial decay width observables~\cite{lebedev99}.  The
accumulated experimental information on the neutrinos oscillations
also has allowed a partial implementation of this program through
global fits to the data for the neutrinos Majorana masses based on the
RPV contributions.  With a modest assistance from theory, concerning
the generational structure of the sfermions mass matrices, these
studies yield a wealth of useful information on the RPV interactions.

\newpage



\def\NPB#1#2#3   {{\rm Nucl.~Phys.}           {\bf{B#1}}, {#3} (#2)}
\def\NPA#1#2#3   {{\rm Nucl.~Phys.}           {\bf{A#1}}, {#3} (#2)}
\def\PLB#1#2#3   {{\rm Phys.~Lett.}           {{\bf{B#1}}, {#3} (#2)}}
\def\PR#1#2#3    {{\rm Phys.~Rep.}            {\bf#1}, {#3} (#2)}
\def\PRD#1#2#3   {{\rm Phys.~Rev.}            {\bf{D#1}}, {#3} (#2)}
\def\PRC#1#2#3   {{\rm Phys.~Rev.}            {\bf{C#1}}, {#3} (#2)}
\def\PRO#1#2#3   {{\rm Phys.~Rev.}            {\bf{#1}}, {#3} (#2)}
\def\PRL#1#2#3   {{\rm Phys.~Rev.~Lett.}      {\bf{#1}}, {#3} (#2)} 
\def\ZPC#1#2#3   {{\rm Z.~Phys.}              {\bf C#1}, {#3} (#2)}  
\def\PPNP#1#2#3  {{\rm Prog.~Part.~Nucl.~Phys.} {\bf #1}, {#3} (#2)}  
\def\MPLA#1#2#3  {{\rm Mod.~Phys.~Lett.}      {\bf A#1}, {#3} (#2)}  
\def\EPJC#1#2#3  {{\rm Eur.~Phys.~J.}        {\bf C#1}, {#3} (#2)}
\def\JPG#1#2#3   {{\rm J.~Phys.}              {\bf G#1}, {#3} (#2)}
\def\AP#1#2#3    {{\rm Astropart.~Phys.}       {\bf #1}, {#3} (#2)}
\def\RMP#1#2#3   {{\rm Rev.~Mod.~Phys.}       {\bf #1}, {#3} (#2)}
\def\YF#1#2#3    {{\rm Yad.~Fiz.}             {\bf #1}, {#3} (#2)}
\def\ZETF#1#2#3  {{\rm Zh.~Eksp.~Teor.~Fiz.}  {\bf #1}, {#3} (#2)}
\def\JETP#1#2#3  {{\rm JETP~Lett.}            {\bf #1}, {#3} (#2)}
\def\SJNP#1#2#3  {{\rm   Sov.~J.~Nucl.~Phys.}   {\bf #1}, {#3} (#2)}
\def\SJNPL#1#2#3 {{\rm   Sov.~J.~Nucl.~Phys.Letters}   {\bf #1}, {#3} (#2)}
\def\ARNPS#1#2#3 {{\rm Ann.~Rev.~Nucl.~Part.~Phys.}  {\bf #1}, {#3} (#2)}
\def\JHEP#1#2#3  {{\rm JHEP}                  {\bf #1}, {#3} (#2)}
\def\NCC#1#2#3   {{\rm Nuovo~Cim.}           {\bf C#1}, {#3} (#2)}



\newpage

\begin{table}
\begin{center}
\begin{minipage}[t]{18. cm}
\caption{Single bounds for the RPV coupling constants  arranged in
order of increasing successive generation indices.  We use the
notation $ V_{ij} $ for the CKM matrix, $ R_l , \ R_{l\ l '}, \ R_D ,
\ R_{l, b} ^Z $ for various branching fractions ratios defined in the
main text, $ Q_W$ for the weak charge, $\nu q , \ \nu l$ for the
neutrino elastic scattering on quarks and leptons, $ m_\nu $ for the
neutrino Majorana mass, $ RG$ for the renormalization group, FB for
forward-backward asymmetry, APV for atomic physics parity violation,
$\b \b _{0\nu } $ for neutrinoless double beta decay, $n\bar n$ for
neutron-antineutron oscillation and $ NN$ for two nucleon nuclear
decay.  The generation indices denoted $ i, j, k$ run over the three
generations while those denoted $l, m, n$ run over the first two
generations.  The dependence on the superpartner mass follows the
notational convention $\tilde m^p = ({\tilde m \over 100 \ \text{GeV}}
)^p$. Aside from a few cases associated with one-loop effects, we use
the reference value $\tilde m = 100 \ \text{GeV} $.}
\label{table1}
\end{minipage}
\vskip 0.3cm
\begin{tabular} {|p {1cm}|p{4.5cm}|p{5.cm}|p{5.5cm}|} \hline
 & {\bf Charged Current} & {\bf Neutral Current} & {\bf Processes} \\
\hline $\vert \l_{12k} \vert$ & $ 0.04 \ \tilde e_{kR} \ [V_{ud}]$ &
$ 0.34 \ \tilde e_{kR}, \ 0.29 \ \tilde e_{k=1L}\ [\nu_\mu e ] $ & $
0.26 \ \tilde e_{kR} \ [e^-_{L,R} + A ] $ \\ 
& $ (0.14 \pm 0.05)
\tilde e_{kR}\ [G_F ]$ & $[0.10, 0.10, 0.24] \tilde \nu_{kL} \
[A_{FB}] $ & $ 0.2 \ \tilde e_{kR} \ [e^-_{L,R}  + d ] $
\\ & $ 0.05 \ \tilde e_{kR} [R_{\tau \mu }] $ & $ 0.15 \ \tilde
e_{kR} \ [Q_W (\text{Cs}) ] $ & $ 0.3 \ \tilde e_{kR} \ [e^-_{L,R}+\
^9\text{Be} ] $ \\  
&& $ 0.13 \ \tilde e_{kR} \ [\nu_\mu q ] $ & \\ \hline

$\vert \l_{13k} \vert$ & $ 0.05 \ \tilde e_{kR}\ [R_{\tau }]$ & $ \
[0.10, 0.10, 0.24] \ \tilde \nu_{kL} \ [A_{FB}] $ & \\ \hline

$\vert \l_{23k} \vert$ & $ 0.05 \ \tilde e_{kR} \ [R_{\tau }]$ & $ \
0.26\ \tilde e_{k=3L} \ [\nu_\mu e ]$ & $ 0.52\ \tilde f \ [\mu ^\g
_\mu ] $ \\ &$ 0.05 \ \tilde e_{kR}\ [R_{\tau \mu }]$ & $ 0.11 \
\tilde \nu _{iL} \ [A_{FB} ]$ & \\ \hline

$\vert \l_{233} \vert$ & & & $ 0.90 \ [RG]$ \\ \hline

$\vert \l_{i32} \vert$ & && $ 8. \ 10^{-2}\ \tilde \nu _i \ [\tau \to
3\mu ]$ \\ \hline

$\vert \l_{3j2} \vert$ & & & $ 0.52\ \tilde f \ [\mu ^\g _\mu ] $ \\
\hline

$\vert \l_{i23} \vert$ & & & $ 0.52\ \tilde f \ [\mu ^\g _\mu ] $ \\
\hline \hline

$\vert \l'_{11k} \vert$ & $ 0.01 \ \tilde d_{kR} \ [ V_{ud}]$ & $ \
0.26 \ \tilde q_{k=3L} \ [A_{FB}]$ & $ 1.2 \ 10^{-2}\ \tilde d_{kR}
\ [K\to \pi \nu \bar \nu ]$ \\ && $ 0.30 \ \tilde d_{kR} , \ 0.26\
\tilde q_{k=1L} \ [APV] $ & $ 0.11 (\tilde \nu ^\ud _{1L} ,\ \tilde
d^\ud _{kR} ) \ [ K\bar K]$ \\ && $ 0.06\ \tilde d_{kR}, \ 0.04 \
\tilde q_{k=1L } \ [Q_W (\text{Cs} )] $ & \\ \hline

$ \vert \l'_{111} \vert $ &&& $ ( 3.3 10^{-4} - 3.2 10^{-5})\ \tilde
q^2 \tchi ^\ud \ [\b \b _{0\nu }]$ \\ \hline

$ \vert \l'_{11k} \vert $ & $3. \ 10^{-2} \ \tilde d_{kR} [R_\pi ] $
&& $0.1 \ \tilde d_{kR} \ [e^-_{L,R} + \ ^{12}\text{C}  ] $ \\ &&& $ 0.29 \
\tilde d_{kR} \ [e^-_{L,R}+ d] $ \\ &&& $ 0.093 \ \tilde d_{kR} \
[e^-_{L,R} +\  ^9\text{Be}  ] $ \\ \hline

$ \vert \l'_{1j1} \vert $ &&& $0.11 \ \tilde q_{jL}\ [ e^-_{L,R} +\ 
^{12}\text{C} ] $ \\ &&& $ 0.38 \ \tilde q_{jL} \ [e^-_{L,R} +  d ] $ \\ &&& $
0.71 \ \tilde q_{jL} \ [e^-_{L,R} +\  ^9\text{Be}  ] $ \\ \hline

$ \vert \l'_{12k} \vert$ & $ 0.28\ \tilde d_{k=1,3R} [R_{D^+}] $
& $0.26 \ \tilde q_{k=3L}\ [A_{FB}]$ & $ 1.2  10^{-2}\  \tilde d_{kR}
\ [K\to \pi \nu \bar \nu ]$ \\ & $ 0.21 \ \tilde d_{k=1,3R} [R_{D^0}
]$ & $ 0.26\ \tilde q_{k=1L}\ [APV]$ & $ 0.11\ (\tilde \nu ^\ud
_{1L} , \tilde d^\ud _{kR}) \ [ K\bar K]$ \\ & $0.10 \ \tilde
d_{k=1,3R}\ [R_{D^+} ^\star ] $ && \\ \hline

$ \vert \l'_{122} \vert $ & & & $ 7. \ 10^{-2} ({m^2 _{\tilde d} /
 \tilde m_{LR} ^{d2} })^ \ud \ [m_\nu ] $ \\ \hline


$ \vert \l'_{13k} \vert$ &
& $ 0.45\ \tilde q_{k=3L} \ [A_{FB}] $ & $0.52 \ \tilde d_{kR} \
[K\to \pi \nu \bar \nu ]$ \\ && $ 0.26\ \tilde q_{k=1L}\ [APV]$ & \\
&& $ 0.63 \ \tilde m \ [R^Z_{l,b}]$ & \\ \hline

$ \vert \l '_{13n} \vert $ & && $ 0.41 \ [t\to l^+ X] $ \\ \hline

$ \vert \l'_{133} \vert $& & & $ 3.5 \ 10^{-3} ({m^2_{\tilde d} /
\tilde m_{LR} ^{d2} })^\ud \ [m_\nu ] $ \\ & & & $0.41 \ [t \to bX]
$\\ \hline

$ \vert \l'_{21k} \vert$ & $ 0.05 \ \tilde d_{kR} \ [R_\pi ]$& $ \
0.11 \ \tilde d_{kR} \ [\nu_\mu q] $ & $ 1.2 10^{-2}\ \tilde d_{kR}
\ [K\to \pi \nu \bar \nu ] $ \\ & $ 0.03 \ \tilde d_{kR} \ [R_{\tau
\pi }]$ &$ 0.22 \ \tilde d_{k=1L}\ [\nu_\mu q]$ & $ 0.11\  (\tilde \nu
^\ud _{1L} , \tilde d^\ud _{kR}) \ [ K\bar K]$ \\ \hline

\end{tabular}
\end{center}
\end{table}

\begin{table}
\begin{center}
\begin{tabular} {|p {1cm}|p{5.8cm}|p{3.5cm}|p{5.8cm}|} \hline
 & {\bf Charged Current } & {\bf Neutral Current } & {\bf Processes}
\\ \hline 

$ \vert \l'_{22k} \vert$ & $ 0.49 \ \tilde d_{k=(1,3)R}
[R_{D^+}]$ & $ 0.22 \ \tilde d_{k=2L}\ [\nu_\mu q ]$ & $ 1.2 \
10^{-2}\ \tilde d_{kR} \ [K\to \pi \nu \bar \nu ]$ \\ &$ 0.30
d_{k=(1,3)R} \ [R_{D^+} ^\star ]$ & & $ 0.11 (\tilde \nu ^\ud _{1L} ,
\tilde d^\ud _{kR} ) \ [ K\bar K]$ \\ & $ 0.13 \ \tilde d_{kR} \
[R_{D^+} ]$ & & \\ & $ 0.51 \ \tilde d_{kR} [R_{D_s} ( \tau \mu ) ]$
& & \\ \hline

$\vert \l'_{2j3} \vert$ & && $ 0.52 \ \tilde f \ [\mu _\mu ^\g ] $ \\
\hline

$\vert \l'_{23k} \vert$ & $ 0.44\  \tilde m \ [R_\mu ]$ & $ 0.22 \
\tilde d_{k=1L} \ [\nu_\mu q ] $ & $ 0.52 \ \tilde d_{kR} \ [K \to \pi
\nu \bar \nu ]$ \\ && $ 0.56 \ \tilde m \ [ R^Z_{l,b}]$ & \\ \hline

$ \vert \l'_{31k} \vert$ & $ 0.16 \ \tilde d_{k=(1,3)R} \ [R_{\tau
\pi } ]$ && $ 1.2 \ 10^{-2}\  \tilde d_{kR} \ [K\to \pi \nu \bar \nu
]$ \\ && & $0.16 \ \tilde d_{kR} \ [\tau \to \pi \nu ] $ \\ \hline

$ \vert \l'_{32k} \vert$ & $ 0.36 \ \tilde d_{k=(1,3)R}\ [R_{D_s} (
\tau \mu ) ]$ && $ 1.2  10^{-2}\  \tilde d_{kR} \ [K\to \pi \nu \bar
\nu ] $ \\ \hline

$\vert \l'_{33k} \vert$ & $ (0.12 - 0.32)  \  \tilde d_{kR} \   [B^- \to
\tau ^-\bar \nu _\tau X  ]$ & $ 0.26 \ \tilde m \ [R_\tau ^Z]$ & $
\ (0.6  - 1.3 ) \ \tilde d_{k=3R} [ B ^-\to \tau ^- \bar \nu X_q] $
\\ & & $ 0.45\ \tilde m \ [R^Z_{l,b}] $ & $ 0.52 \ \tilde d_{kR} \
[K\to \pi \nu \bar \nu ]$ \\ \hline

$\vert \l'_{333} \vert$ & $ (0.12 - 0.32) \ \tilde d_{3R} \ [B ^- \to
\tau ^- \bar \nu X] $ & & $1.3 \ [\s (t\bar t)]$ \\ &&& $1.06 \ [RG]$
\\ \hline

$ \vert \l '_{imk} \vert $ & && $ 0.11 \ (\tilde \nu_{iL} ^{1/2},
\tilde d_{kR} ^{1/2} ) \ [K - \bar K] $ \\ \hline

$ \vert \l '_{ijk} \vert $ & && $ 0.16 \ (\tilde e_ {iL} ^{1/2},
\tilde d_{kR} ^{1/2} ) \ [D-\bar D] $ \\ \hline

$ \vert \l '_{i3k} \vert $ & && $ 1.1 \ (\tilde \nu _{iL} ^{1/2},
\tilde d_{kR} ^{1/2} ) \ [B_d -\bar B_d] $ \\ \hline \hline 

$ \vert \l'' _{112} \vert $ & & & $ (5. \ 10^{-2} \ - \ 2.5 ) \ [n\bar
n ] $ \\ & & & $ 10^{-6} \ [NN\to KK] $ \\ \hline

$ \vert \l'' _{113} \vert $ & & & $ (2. \ 10^{-3} \ - \ 10^{-1} ) \
[n\bar n] $ \\ & & & $ 10^{-3} \ [NN\to KK] $ \\ \hline

$ \vert \l'' _{11k} \vert$ & & & $ 1. 10^{-6}\  \tilde g^\ud \tilde d ^\ud (\tilde m
_{RR} ^{d2} ) _{12}  \ [n\bar n ] $\\ \hline

$ \vert \l'' _{123} \vert $ & & & $ 1.25 \ [RG] $ \\ \hline

$ \vert \l'' _{212} \vert $ & & & $ 1.25 \ [RG] $ \\ \hline

$ \vert \l'' _{213} \vert $ & & & $ 1.25 \ [RG] $ \\ \hline

$ \vert \l'' _{223} \vert $ & & & $ 1.25 \ [RG] $ \\ \hline

$ \vert \l ''_{312} \vert $ & & $ 0.97 \ \tilde d_{kR} \ [R_l^Z]$ &
$ 2.1 \ 10^{-3} \ [n \bar n] $ \\ &&& $ 4.28 \ [RG]$ \\ \hline

$ \vert \l ''_{313} \vert $ & & $ 0.97 \ \tilde d_{kR} \ [R_l^Z]$ &
$ 2.6\ 10^{-3} \ [n\bar n] $ \\ & & & $ 1.12 \ [RG]$ \\ \hline

$ \vert \l ''_{323} \vert $ & & $ 0.96 \ \tilde d_{kR} \ [R_l^Z]$ &
$ 1.12 \ [RG]$ \\ \hline

$ \vert \l '' _{ijk} \vert $ & && $ [10^{-7 }\ \tilde m^{2} \ - \
10^{-9}\  \tilde m^{3}]$ \\ & && $ \times ( {m_{\tilde G } / 1 \text{eV}
} ) \ [N \to K \tilde G]$ \\ &&& $[10^{-7 }\ \tilde m^{2} \ - \
10^{-9} \tilde m^{3}] $ \\ &&& $ \times ({F_a / 10^{10} \ \text{GeV} }
) \ [N \to K \tilde a] $ \\ \hline
\end{tabular}
\end{center}
\end{table}

\begin{table}
\begin{center}
\begin{minipage}[t]{18. cm}
\caption{Quadratic coupling constant products bounds arranged in
order of increasing successive generation indices. We use the same
conventions as in Table \ref{table1}.  
The factor $ i $ is included to signal that the bound applies
to the imaginary part. }
\label{table2}
\end{minipage}
\vskip 0.3cm
\begin{tabular}{|p{2cm}|p{5.cm}|p{5.5cm}|p{3.cm}|}  \hline
 & {\bf Lepton Flavor } & {\bf Hadron Flavor } & {\bf L/B Number } \\
 \hline

$\vert \l  _{121} \l ^{\star} _{121} \vert $ & $ 5.7 \ 10^{-5} \ [\mu
  \to e \g ] $ && \\ \hline

$ \vert \l _{121} \l ^\star _{212} \vert $ & $ 0.58 \ [\mu ^\g _\nu ]
$ && \\ \hline

$\vert \l _{121} \l ^\star _{133}\vert $ & $ 1.8 \ \ 10^{-3} \ [e ^-
_i \to e ^- _j \g \g ] $ && \\ \hline

$\vert \l _{122} \l ^\star _{233}\vert $ & $ 1.8 \ \ 10^{-3} \ [e ^-
_i \to e ^- _j \g \g ] $ && \\ \hline

$\vert \l _{1j1} \l ^{\star} _{1j2} \vert \vert $ & $ O(10^{-6})\
[^{48}\text{Ti} (\mu ^-, e^-\g ) ] $ & & \\ \hline

$ \vert \l_{lm1} \l ^\star _{lm2} \vert $ & $ 2.3 \ 10^{-4 } \ \tilde
e_{L}^2 \ [e_i \to e_j \g ]$ & & \\ \hline

$\vert \l  _{131} \l ^{\star}_{131}\vert $ & $ 0.57 \ 10^{-4} \ [\mu
   \to e \g ]$ && \\ \hline

$\vert \l _{23k} \l ^{\star} _{131} \vert $ & $ 1.1\ 10^{-4} \ [\mu \to
e \g ] $ && \\ \hline

$\vert \l_{232} \l ^\star _{131} \vert $ & $ 2.2 \ 10^{-2}\  \tilde
e_{3L}^2 \ [\mu \to e\nu \bar \nu ]$ & & \\ \hline

$\vert \l  _{2jk} \l ^{\star}_{13k} \vert $ & $ O(10^{-6}) \
[^{48}\text{Ti} (\mu ^-, e^- \g ) ] $ & & \\ \hline

$\vert \l_{31n} \l ^\star _{32n} \vert $ & $ 4.6 \ 10^{-4 } (\tilde
 \nu _{L}^2, \tilde e _R^2) \ [e_i \to e_j \g ]$ & & \\ \hline

$\vert \l_{i11} \l ^\star _{i12} \vert $ & $ 7.  \ 10^{-7 } \ \tilde \nu
  _{iL}^2 \ [\mu \to 3e]$ && \\ \hline

$\vert \l_{i11} \l ^\star _{i21} \vert $ & $ 7.  \ 10^{-7 }\  \tilde \nu
  _{iL}^2 \ [\mu \to 3e]$ && \\ \hline

$ \vert \l_{i21} \l ^\star _{i12} \vert $ & $ 6.3 \ 10^{-3 }\  \tilde
\nu _{iL}^2 [Mu \overline{Mu}]$ && \\ \hline

$\vert \l_{i22} \l ^\star _{i23} \vert $ & $ 6.4 \ 10^{-3}\  \tilde \nu
  _{iL}^2 \ [\tau \to 3 \mu ]$ && \\ \hline

$\vert \l_{i22} \l ^\star _{i32} \vert $ & $ 6.4 \ 10^{-3}\  \tilde \nu
_{iL}^2 \ [\tau \to 3 \mu ]$ && \\ \hline \hline


$ \vert \l_{121} \l^{'\star }_{113} \vert $ && $ 4.5 \ 10^{-5} \ \tilde
m^2 \ [B_d ^0\to e^\pm \mu^\mp ]$ & \\ \hline

$\vert \l ^\star _{121} \l '_{123} \vert $ & $ 8.5 \ 10^{-7} \ [^{197}
  \text{Au} (\mu ^- , e ^- ) ] $ && \\ \hline

$ \vert \l_{122} \l^{'\star } _{112} \vert $ & & $ 3.8 \ 10^{-7}
\  \tilde \nu _{L}^2 \ [ K_L \to \mu \mu ]$ & \\ $ \vert \l_{122}
 \l^{'\star } _{121} \vert $ & & & \\ \hline

$ \vert \l_{122} \l^{'\star }_{113} \vert $ && $ 2.4 \ 10^{-5}\  \tilde
m^2 \ [ B_d ^0\to \mu ^+ \mu ^- ]$ & \\ \hline

$ \vert \l_{123} \l^{'\star }_{131} \vert $ && $ 6. \ 10^{-4}\  \tilde
m^2 \ [ B_d ^0 \to \mu^+ \tau^- ]$ & \\ \hline

$ \vert \l_{131} \l^{'\star }_{131} \vert $ &&$ 5. \ 10^{-4}\  \tilde
m^2 \ [ B_d ^0 \to e^+ \tau^- ]$ & \\ \hline

$\vert \l _{131} \l ^{'\star } _{133}\vert $ & $ 1.4 \ 10^{-2}\ [e ^-
_i \to e ^- _j \g \g ] $ && \\ \hline

$\vert \l_{121} \l^{'\star } _{212} \vert $ && $ 2.5 \ 10^{-8}\  \tilde
\nu _{L}^2 \ [ K_L \to ee ]$& \\ $\vert \l_{121} \l^{'\star } _{221}
\vert $ & & & \\ \hline

$\vert \l _{122} \l ^{'\star } _{211} \vert $ & $ 8.8 \ 10^{-4} \ [e
^- _i \to e ^- _j \g \g ] $ && \\ \hline

$\vert \l_{122} \l^{'\star } _{212} \vert $ && $ 2.3 \ 10^{-7}\ \tilde
\nu _{L}^2 \ [ K_L \to \mu e ]$& \\ $ \vert \l_{122} \l^{'\star }
_{221} \vert $ & & & \\ \hline

$\vert \l_{121} \l^{'\star }_{213} \vert $ && $ 4.6 \ 10^{-5}\  \tilde
m^2 \ [ B_d ^0\to e^+ e^- ]$ & \\ $\vert \l_{121} \l^{'\star }_{231}
\vert $ & & & \\ \hline

$\vert \l_{131} \l^{'\star }_{333} \vert $ && $ 7.5 \ 10^{-2}\  \tilde
e_{3L}^2 \ [ B ^-\to e ^- \bar \nu ]$ & \\ \hline

$ (\l _{1j1} ^\star \l '_{j33} ) $ & $ i
\times 6.  \ 10^{-7} \ \tilde \nu_j ^2 \ [d ^\g _e]$ && \\ \hline

$(\l _{211} \l  ^{'\star } _{233} ) $ & $ i \times 5.  \ 10^{-6} \ \tilde
\nu ^2 [d ^\g _e] $ && \\ \hline

$\vert \l _{212} \l ^{'\star } _{233} \vert $ & $ 8.5 \ 10^{-7} \ [^{197}
  \text{Au} (\mu ^- , e ^- ) ] $ && \\ \hline

$\vert \l _{312} \l ^{'\star } _{333} \vert $ & $ 8.5 \ 10^{-7} \ [^{197}
\text{Au} (\mu ^- , e ^- ) ] $ && \\ \hline

$\vert \l _{321} \l  ^{'\star } _{333} \vert $ & $ 8.5 \ 10^{-7} \ [^{197}
  \text{Au} (\mu ^- , e ^- ) ] $ && \\ \hline

\end{tabular}
\end{center}
\end{table}
\begin{table}
\begin{center}
\begin{tabular}{|p{2cm}|p{5.cm}|p{5.5cm}|p{3.cm}|}  \hline
& {\bf Lepton Flavor } & {\bf Hadron Flavor } & {\bf L/B Number } \\
\hline

$ (\l _{i21} \l ^\star _{i'12}) $ & $ i \times 5.\ 10^{-2}\ [d ^\g _\nu]$ &&
\\  \hline 

$ (\l ' _{i32} \l ^ {'\star } _{i'23}) $ & $ i\times 2.4 \ 10^{-3}\  [d
^\g _\nu]$ && \\ \hline

$\vert \l_{i31} \l^{'\star }_{i11} \vert $ & $ 6.4 \ 10^{-2} \ \tilde
\nu _{iL}^2 \ [ \tau \to e\pi ]$ & & \\ \hline

$\vert \l_{i31} \l^{'\star }_{i22} \vert $ & $ 4.5 \ 10^{-2}\  \tilde
\nu _{iL}^2 \ [ \tau \to e\eta ]$ & & \\ \hline \hline


$\vert \l _{ijk} \l ^{''\star }_{112} \vert  $ &&& $ (10^{-21}   - 10^{-16} )$  \\ 
&&& $ [ p\to K^+ e^\pm  \mu ^\mp 
\bar \nu  ]  $  \\ \hline  

$\vert \l _{ijk}  \l  ^{''\star }_{i'j'k'} \vert  $ &&&  $( 10^{-12}  - 10^{-3} )$  \\ 
&&& $  [ p\to \pi ^+ e^\pm  \mu ^\mp  \bar \nu  ]  $ \\ \hline \hline 


$\vert \l '_{112}\l^{'\star }_{121} \vert $ &&& $ 2.3 \ 10^{-6} \ [ \b
\b _{0\nu } ]$\\ \hline

$\vert \l' _{11k} \l^{'\star }_{12k} \vert $ & & $ 1.3 \ 10^{-1}  
\ \tilde d^2_{kR} \ [\L \to n e\nu ]$ & \\ && 
$ 5.3 \ 10^{-3} \ \tilde d^2_{kR} \ [\L \to n
\mu \nu ]$ & \\ & & $ 8.5 \ \ 10^{-2} \ \tilde d^2_{kR} \ [\S \to n
e\nu ]$ & \\ & & $1.6 \ 10^{-2} \ \tilde d^2_{kR} \ [\S \to n \mu \nu
]$ & \\ & & $ 1.2 \ 10^{-1}, \tilde d^2_{kR} \ [\Xi \to n e \nu ]$ &
\\ & & $ 5.0\ 10^{-2}\ \tilde d^2_{kR} \ [\Xi \to n \mu \nu ]$ & \\
\hline

$\vert \l '_{113}\l^{'\star }_{131} \vert $ & && $ 7.9 \ 10^{-8}\ [ \b
\b _{0\nu } ] $ \\ \hline

$ \vert \l '_{11k} \l^{'\star } _{21k} \vert $ & $ 3.  \ 10^{-8}\ 
\tilde d_{kR}^2 \ [\text{Ti} (\mu ^- , e ^-) ] ] $ & & \\ \hline
 
$\vert \l ' _{123} \l ^ {'\star } _{232} \vert $ & $ 0.030 \ [\mu ^\g
_\nu ] $ && \\ \hline

$ \vert \l ' _{13k} \l  ^{'\star } _{23k} \vert $ & $ 6.5 \ 10^{-2} \ [
 Z^0 \to e ^\pm  \mu ^\mp ] $ && \\ \hline

$\vert \l '_{1j1}\l^{'\star }_{1j2} \vert $ && $ 8.6 \ 10^{-5} \ [ K_L
\to ll ] $ & \\ \hline

$\vert \l '_{1j1} \l^{'\star } _{2j1} \vert $ & $ 1.6 \ 10^{-7}\  \tilde
u_{jL}^2 \ [^{48}\text{Ti} (\mu ^- , e ^-) ]$ & & \\ \hline

$\vert \l '_{1jk}\l '_{3jk} \vert $ & $ 1.2 \ 10^{-2} \ [\tau \to e \g
]$ & & \\ \hline

$\vert \l '_{2mk} \l  ^{'\star } _{1m'k}\vert $ & $ O(10^{-8}) \
[^{48}\text{Ti} (\mu ^-, e^-) ] $ & & \\ \hline

$\vert \l '_{2mk}\l^{'\star }_{1mk} \vert $ & $ 5.7 \ 10^{-4} \ [\mu
\to e \g ]$ & & \\ \hline

$ \vert \l '_{23k} \l  ^{'\star } _{33k} \vert $ & $ 1.6 \ 10^{-1} \ [
 Z^0 \to \mu ^\pm  \tau ^\mp ] $ && \\ \hline

$ \vert \l '_{2j1}\l^{'\star }_{2j2} \vert $ && $ 5.8 \ 10^{-6} \ [
K_L \to ll ]$ & \\ \hline

$\vert \l ' _{23n} \l ^{'\star }_{13n}\vert $ & $ 7.7 \ 10^{-3}\ [\mu
\to e \g ]$ & & \\ \hline

$\vert \l '_{233} \l ^{'\star } _{133}\vert $ & $ 1.0 \ 10^{-2} \ [\mu
\to e \g ]$ & & \\ \hline

$\vert \l '  _{23k} \l ^{'\star }_{1m'k} \vert $ & $ O(10^{-6}) \
[^{48}\text{Ti} (\mu ^-, e^-) ] $ & & \\ \hline

$ \vert \l '  _{33k} \l ^{'\star}_{13k} \vert $ & $ 2.0 \ 10^{-1} \ [
 Z^0 \to e ^\pm  \tau ^\mp ] $ && \\ \hline
 
$\vert \l' _{i11} \l ^{'\star } _{i13} \vert $ & & $ 2.5 \ 10^{-3} \ [
B^0 \to \pi ^+ \pi ^- ] $ & \\ \hline

$ (\l '  _{i11} \l ^{'\star }_{i33} ) $ & $ i \times 1.\ 10^{-4} \
\tilde m ^2 \ [d ^\g _n] $ && \\ \hline

$\vert \l '_{i13} \l ^{'\star } _{i12} \vert $ & & $ 5.7\ 10^{-3} \ [
B _d^\pm \to \pi^\pm K ^0] $ & \\ \hline

$\vert \l '_{i1k} \l^{'\star } _{i'2k} \vert $ && $ 4.8 \ 10^{-5}\ 
\tilde d _{kR}^2 \ [K \to \pi \nu \bar \nu ]$ &\\ \hline

$( \l '_{i21} \l^{'\star } _{i12} ) $ && $ (4.5 \ 10^{-9}, i \times 8
\ 10^{-12} )\  \tilde \nu _{iL}^2 \ [K \bar K ]$ &\\ \hline

$\vert \l '_{i23} \l ^{'\star } _{i21} \vert $ && $ 4.\ 10^{-4} \
\tilde u_{iR} ^2 \ [B^0 \to \phi \pi ] $ & \\ \hline

$\vert \l '_{i2k}\l^{'\star }_{i3k} \vert $ & & $ 0.09 \ (\tilde \nu
_{iL}^2 , \tilde d_{iR}^2) \ [B \to K \g ] $& \\ \hline

$ \vert \l '_{i31} \l^{'\star } _{i22} \vert $ && $1.  \ 10^{-4}\ 
\tilde \nu _{iL}^2 \ [K \bar K ]$ &\\ \hline

$\vert \l '_{i31} \l^{'\star } _{i23} \vert $ && $ 1. \ 10^{-4} \ [K
\bar K ]$ & \\ \hline

$\vert \l '_{i31} \l^{'\star } _{i13} \vert $ && $ 3.3 \ 10^{-8}\ 
\tilde \nu _{iL}^2 \ [B \bar B ]$ & \\ \hline

$ \vert \l '_{i31} \l^{'\star } _{i32} \vert $ && $ 7.7 \ 10^{-4} \ [K
\bar K ]$ & \\ \hline

$\vert \l '_{i31} \l^{'\star } _{i33} \vert $ && $1.3 \ 10^{-3} \ \tilde
\nu _{iL}^2 \ [B \bar B ]$ & \\ \hline

$ \vert \l' _{i32} \l ^{'\star } _{i22} \vert $ & & $ 2.3 \ 10^{-3} \
[B^0 \to MM] $ & \\ \hline

$\vert \l '_{ij2} \l^{'\star } _{i'j1} \vert $ && $ 4.8 \ 10^{-5}\ 
\tilde d _{jL}^2 \ [K \to \pi \nu \bar \nu ]$ &\\ \hline


\end{tabular}
\end{center}
\end{table}
\begin{table}
\begin{center}
\begin{tabular}{|p{2cm}|p{3.cm}|p{5.5cm}|p{5.5cm}|}  \hline
& {\bf Lepton Flavor } & {\bf Hadron Flavor } & {\bf L/B Number } \\
\hline 

$\vert \l ' _{i32} \l ^{'\star } _{i12} \vert $ && $ 4.\ 10^{-4} \
\tilde u_{iR} ^2 \ [B^0 \to \phi \pi ] $ & \\ \hline \hline
 
$\vert \l '_{ij2}\l^{'\star }_{ij3} \vert $ & & $ 3. \ 10^{-2} \
(\tilde e _{iL}^2 , \tilde d_{jL}^2) \ [B \to K \g ] $& \\ \hline

$\vert \l '_{ijk}\l^{'\star }_{i'3k} \vert $ && $ 1.1 \ 10^{-3}\ 
\tilde d _{kR}^2 \ [B ^0\to X_q \nu \bar \nu ]$ & \\ \hline

$\vert \l '_{ijk } \l  ^{'\star }  _{i'j3}\vert $ & & $ 1.1 \ 10^{-3}\
\tilde d_{jL}^2 \ [B ^0\to X_q \nu \bar \nu ]$ & \\ \hline


$\vert \l _{ijk} \l ^{'' \star } _{112} \vert $ & & & $ (10^{-22} -
10^{-16}) \ [p \to K \mu e \bar \nu ] $ \\ \hline

$\vert \l  _{ijk} \l ^{'' \star }_{i'j'k'} \vert $ & & & $ (10^{-12} -
10^{-3}) \ [p \to K \mu e \bar \nu ] $ \\ \hline \hline


$\vert \l '_{lmk} \l ^{''\star } _{11k} \vert $ & && $ 10^{-25}\  \tilde
d _{kR}^2 \ [p\to l X ] $\\ \hline

$\vert \l '_{lj1} \l ^{''\star } _{1j1} \vert $ & && $ 10^{-25}\  \tilde
d _{jL}^2 \ [p\to l X] $\\ \hline

$\vert \l '_{ijk} \l ^{''\star } _{i'j'k'} \vert $ &&& $ 10^{-9} \
[p\to l X ]$ \\ \hline

$\vert \l '_{ijk} \l ^{''\star } _{l21} \vert $ &&& $ 10^{-9} \ [p\to
l K \pi ]$ \\ \hline

$\vert \l '_{ijk} \l ^{''\star } _{l31} \vert $ &&& $ 10^{-9}\ [p\to l
K \pi ] $ \\ \hline

$\vert \l '_{ijk} \l ^{''\star } _{l32} \vert $ &&& $ 10^{-9}\ [p\to l
K \pi ] $ \\ \hline


$\vert \l '_{i1k} \l ^{''\star } _{i'2k} \vert $ && $ 4.8 \ 10^{-5}\ 
\tilde d _{kR}^2 \ [K\to \pi \nu \bar \nu ]$ &\\ \hline

$\vert \l '_{ij2} \l ^{''\star } _{i'j1} \vert $ && $ 4.8 \ 10^{-5}\ 
\tilde d _{jL}^2 \ [K\to \pi \nu \bar \nu ]$ & \\ \hline \hline


$(\l '' _{123} \l ^{ ''\star } _{113} ) $ & & $ i \times O(10^{-5} )\ 
\tilde q^2\ [ K\bar K] $ & \\ \hline

$ \vert \l ''_{232} \l ^{''\star }_{132}\vert $ & & $ 3.1 \ 10^{-3} \
\tilde s^2\ [ D\ -\ \bar D] $ & \\ \hline

$\vert \l ''_{232} \l ^{''\star } _{231} \vert $ & & $ \
3. \ 10^{-4}\  \tilde c ^2 \ [K\bar K]$ & \\ \hline

$\l '' _{313} \l ^{''\star } _{323} $ & & $ i O(10^{-8} ) \ \tilde q^2 \
[ K\bar K] $ & \\ \hline

$\vert \l ''_{332} \l ^{''\star } _{331} \vert $ && $ [ 6. \ 10^{-4}\ 
\tilde t , \ 3. \ 10^{-4} \ \tilde t ^2] \ [K\bar K]$ & \\ && $ 3.3 \
10^{-2}\ [ K\bar K] $ & \\ \hline

$ \vert \l '' _{i23} \l ^{''\star } _{i21} \vert $ && $ 6. \ 10^{-5} \
\tilde u_{iR} ^2 \ [B ^0 \to \phi \pi ^0,\ \phi \phi ]$ & \\ \hline

$\vert \l ''_{i21} \l ^{''\star } _{i31} \vert $ && $ 4. \ 10^{-3}\ 
\tilde q_i^2 \ [B \to \bar K \pi ]$ & \\ \hline
 
$\vert \l ''_{i21} \l ^{''\star } _{i32} \vert $ && $ 5. \ 10^{-3}\ 
\tilde q_i^2 \ [B \to \bar K K ]$& \\ \hline

$ ( \l ''_{213} \l _{232}^{''\star } ) $ & & $ i \times 10^{-4} \ \tilde
q ^2\ [d_n ^\g ] $ & \\ \hline

$( \l ''_{312} \l _{332}^{''\star }) $ & & $ i\times 10^{-7}\  \tilde q
^2 \ [d_n ^\g ] $ & \\ \hline

$\vert \l ''_{i2k} \l ^{''\star } _{i3k} \vert $ && $ 0.16 \ \tilde
q_{iR} ^2 \ [ B \to K \g ] $& \\ \hline \hline

\end{tabular}
\end{center}
\end{table} 



\newpage

\newpage

\end{document}